%% file: main.tex
\documentclass[11pt,b5paper,usenatbib]{thesis}
\usepackage[T1]{fontenc}
\usepackage[utf8, latin1]{inputenc}
\usepackage{graphicx}
\usepackage{times}
\usepackage{natbib}
\usepackage{fancyheadings}
\usepackage[font=small,labelfont=bf,textfont=it]{caption}
\usepackage{mathrsfs}
\usepackage{tweaklist}\renewcommand{\enumhook}{\setlength{\topsep}{6pt}\setlength{\itemsep}{2pt}}
\usepackage{amsmath, amssymb}
\usepackage{url}

\renewcommand{\theenumi}{\Roman{enumi}}
\newcommand {\moresim} {\ {\raise-.5ex\hbox{$\buildrel>\over\sim$}}\ }
\newcommand {\lessim} {\ {\raise-.5ex\hbox{$\buildrel<\over\sim$}}\ }

\begin{document}

\setlength{\baselineskip}{4.6mm}
\pagestyle{empty}
\setcounter{page}{1}

\input{subfiles/referees}

\normalsize
\input{subfiles/opening}

\input{subfiles/acknowledgements}
\tableofcontents
\pagestyle{fancy}
\lhead[\large \thepage]{\large\slshape\rightmark}
\rhead[\large\slshape \leftmark]{\large \thepage}
\cfoot[]{}
\input{subfiles/publications}
\input{subfiles/introduction}
\input{subfiles/chapters}
\input{subfiles/bibliography}

\end{document}

%% file: subfiles/referees.tex
\thispagestyle{empty}
\begin{center}
{\footnotesize TURUN YLIOPISTON JULKAISUJA} \\ 
\vspace{-1mm}{\footnotesize ANNALES UNIVERSITATIS TURKUENSIS} \\
\mbox{}\hrulefill\mbox{}\\
\vspace{0.3cm} {\footnotesize SARJA -- SER. A I OSA -- TOM. 419} \\
\vspace{-1mm}{\footnotesize ASTRONOMICA-CHEMICA-PHYSICA-MATHEMATICA} \\
\vspace{1.5cm} {\LARGE PROPERTIES OF GALAXIES AND GROUPS:} \\
\vspace{0.5cm} {\LARGE NATURE VERSUS NURTURE} \\
\vspace{1.4cm} {\LARGE by} \\
\vspace{1cm} {\LARGE Sami-Matias Niemi}
\end{center}
\vfill
\begin{center}
{\footnotesize TURUN YLIOPISTO} \\
{\footnotesize Turku 2011}
\end{center}
\newpage
\begin{flushleft}

From the Department of Physics and Astronomy\\
University of Turku\\
Turku, Finland\\

\vspace{5mm}
{\it Supervised by} \\
\vspace{4mm}

Dr.\ Pekka Hein\"am\"aki and Dr.\ Pasi Nurmi\\
Department of Physics and Astronomy\\
Tuorla Observatory\\
University of Turku\\
Turku, Finland\\

\vspace{10mm}
{\it Reviewed by} \\
\vspace{4mm}

Prof.\ Heikki Salo \\
Division of Astronomy\\
Department of Physics\\
University of Oulu\\
Oulu, Finland\\

\vspace{4mm}
{\it and} \\
\vspace{4mm}

Dr.\ Antti Tamm\\
Tartu Observatory\\
T\~oravere, Estonia\\

\vspace{10mm}
{\it Opponent} \\
\vspace{4mm}

Prof.\ Volker M\"uller\\
Leibniz-Institut f\"ur Astrophysik\\
Potsdam (AIP)\\ 
Potsdam, Germany\\

\vspace{9mm}

\vfill
\end{flushleft}
\newpage

%% file: subfiles/opening.tex
{\it
\noindent 
One could not be a successful scientist without realizing that,\\
in contrast to the popular conception supported by newspapers \\
and mothers of scientists, a goodly number of scientists are \\
not only narrow-minded and dull, but also just stupid.
\\
\rm \\Dr. James Dewey Watson\\Nobel Laureate
\newline
\newline
\newline
\newline
\newline
\newline
\newline
\newline
\newline
\newline
\newline
\newline
\newline
\newline
\newline
\newline
\newline
\newline
\newline
\newline
\newline
\newline
\newline
\newline
\newline
\newline
\newline
\newline
\newline
\newline
\newline
\newline
\newline
\newline
\newline
}

%% file: subfiles/acknowledgements.tex
\chapter*{Acknowledgments}\addcontentsline{toc}{chapter}{Acknowledgments}
\vspace{-10mm}
Countless warm and well deserved thanks are due to all the people who
contributed, and more importantly, forced me to push through the hard times when
quitting felt like the only right thing to do. These thanks are intended
especially for the benefit of those who are actually not interested in the rest
of the thesis, but are driven to read this section and this section only to find
out if their names are listed. To their convenience, I have deliberately
minimised the number of names spelled out.

First words of appreciation must go to my supervisors Pekka Hein\"am\"aki and
Pasi Nurmi who exposed me to the mysteries of cosmological simulations,
formation of large-scale structure, and cosmology. I may not have been a good
student - I was hardly ever present during my years of studies - thus, warm
thanks for sticking with me. I also want to express my gratitude to Mauri
Valtonen for his ideas in galaxy group studies. I am in debt to Henry Ferguson
and Rachel Somerville for guiding me through the last paper of the thesis. Your
guidance truly convinced me that galaxy formation and evolution is worth
spending countless long, but also fruitful, hours when others sleep.

For financial support, that made this work possible, I thank the following
contributors: the V\"ais\"al\"a foundation, Nordic Optical Telescope Science
Association, Space Telescope Science Institute, and G.J. Wulff foundation.
Special thanks go to my mother Liisa Misukka for always providing extra funds
when times weren't so great.

I would like to thank the warm staff of Nordic Optical Telescope for making my
stay in la Palma so enjoyable. I did not only enjoy eating pata negra and
drinking wine (both of which I did more than I care to admit), but I also
learned invaluable lessons about observational astronomy. For a hardcore
theoretician this might sound odd, but astronomy is actually observation driven
science and we should all make sure that we have at least seen the night sky. I
would also like to thank all the individuals who made me feel welcome to STScI
and taught me more about observations. Special thanks go to Stefano Casertano,
Danny Lennon, Harry Ferguson, and Rachel Somerville for countless career advices
and extremely useful science discussions. I would also like to thank Michael
Wolfe for reading the early draft of this manuscript and standing for my
Finglish. Last but most importantly I would like to thank Carolin Villforth for
the countless science arguments and all the wonderful moments I have been able
to share with you.

I thank the official reviewers of this thesis, Drs. Heikki Salo and Antti Tamm,
for timely reading and useful criticism. I am greatly honoured that Prof.\
Volker M\"uller has agreed to act as my opponent in the public disputation of
this dissertation. Finally, to those who wish to know the tips and tricks of
accomplishing a PhD, I confidently say that all the people mentioned here are
part of it and to them I owe my deepest gratitude.
\vspace{-2mm}
\begin{flushright}
{\it Sami-Matias Niemi}
\end{flushright}

%% file: subfiles/publications.tex
\chapter*{List of publications}
\addcontentsline{toc}{chapter}{List of publications}

\noindent
\begin{tabular}{p{0.65cm}p{12cm}}

{\large \slshape I}
& {\large \slshape Are the nearby groups of galaxies gravitationally bound
objects?} \\ & {\bfseries Niemi S.-M.}, Nurmi P.,  Hein\"am\"aki P. and
Valtonen M.\\
& {\small\sffamily MNRAS {\small\sffamily\bfseries 382}, 1864 (2007)} \\
\\
{\large \slshape II}
& {\large \slshape The origin of redshift asymmetries: how $\Lambda$CDM explains
anomalous redshift} \\ & {\bfseries Niemi S.-M.} and 
Valtonen M.\\ 
& {\small\sffamily A\&A {\small\sffamily\bfseries 494}, 857 (2009)} \\
\\
{\large \slshape III}
& {\large \slshape Formation, evolution and properties of isolated field
elliptical galaxies} \\ & {\bfseries Niemi S.-M.}, Hein\"am\"aki P.,
Nurmi P. and Saar E.\\ 
& {\small\sffamily MNRAS {\small\sffamily\bfseries 405}, 477 (2010)} \\
\\
{\large \slshape IV}
& {\large \slshape Physical properties of {\it Herschel} selected galaxies in a semi-analytical galaxy formation model} \\
& {\bfseries Niemi S.-M.}, Somerville R.S., Ferguson H.C., Huang
K.-H., Lotz J. and Koekemoer A.M.\\ 
& {\small\sffamily submitted to MNRAS} \\

\end{tabular}

%% file: subfiles/introduction.tex
\chapter{Introduction} \label{intro}

How did the Universe begin? How will it evolve? How did all the structures such
as galaxies, groups, and clusters we observe in the night sky begin to form? How
did they grow and how will they evolve? These are some of the most profound
questions the mankind have and shall seek to answer.

Observations of the cosmic microwave background (CMB) radiation from the time
when the Universe was only $\sim 380,000$\footnote{$\Lambda$Cold Dark Matter
model has been assumed, see Section \ref{s:background_cosmology}.} years old,
and significantly smaller than today, have shown that the early Universe was
extremely smooth containing only small $\sim 10^{-5}$ temperature anisotropies
\citep{Komatsu:2009p710, Hinshaw:2009p709}. Even so, we can observe a vast
amount of different type of structure in visible light and in other frequencies
\citep[e.g.][and references therein]{Blanton:2009p765} when looking at the night
sky. The smooth early Universe must therefore have gone through radical changes
when evolving from the smooth primordial gas and dark matter density field to
inhomogeneous structures such as dark matter haloes, galaxies, groups and
clusters observed, directly or indirectly, today.

In physical cosmology the global evolution of the temperature, pressure, and
density fields can be studied using Friedmann's world models. These models can
describe the evolution of an homogeneous and isotropic universe. However, as we
do not live in an empty universe, a proof for that is obvious - you are reading
this write up - we must also describe how the small anisotropies, seen in the
CMB, evolve. During the early times, in the so-called linear regime, the
evolution of the density field can be followed using linear perturbation theory
and Newtonian gravity. However, after the density perturbations grow enough they
start to collapse and their evolution turns nonlinear. Their evolution can still
be followed using analytical approximations, however, their validity is limited.
Fortunately, more accurate methods have also been developed.

Due to the inherently nonlinear nature of gravity cosmological $N$-body
simulations have become an invaluable tool when the growth of structure is being
studied and modelled closer to the present epoch. Large simulations with high
dynamical range \citep[e.g.][]{Springel:2005p595, BoylanKolchin:2009p758} have
made it possible to model the formation and growth of cosmic structure with
unprecedented accuracy. Moreover, galaxies, the basic building blocks of the
Universe, can also be modelled to good accuracy in cosmological context and
studied from their initial formation down to the present time
\citep[e.g.][]{Naab:2007p757}. For example, semi-analytical models of galaxy
formation \citep[e.g.][]{Somerville:1999p762, Somerville:2001p760,
Somerville:2008p759} allow us to populate dark matter haloes with galaxies that
are formed from baryonic matter when lacking hydrodynamical simulations
\citep[e.g.][]{Springel:2005p595, Croton:2006p249, DeLucia:2007p414}.
Reassuringly, both semi-analytical models as well as hydrodynamical simulations,
which both model for example an inflow of gas, how gas can cool and heat up
again, how stars are formed within galaxies, and how stellar populations evolve,
are in reasonable agreement with observational data at lower $(z \lessim 4)$
redshifts \citep[e.g.][]{Kitzbichler:2007p767}. Instead, at high redshifts $(z
\moresim 6)$ the small number of candidate galaxies \citep[e.g.][]{Yan:2009p782,
Labbe:2010p776, Bouwens:2011p1089} still complicates more detailed comparisons
to simulations \citep[e.g.][]{Dayal:2010p781, Razoumov:2010p777}. Even so,
simulations can be used to make predictions for different observables and aid
when interpreting observational results. The growth of cosmic structures,
cosmological $N$-body simulations, and formation of galaxies are briefly
reviewed in Chapter \ref{ch:formation}.

Despite all the simulations and successes in recent years and decades, there are
still many unanswered questions in the field of galaxy formation and evolution.
One of the longest standing issues in galaxy evolution is the significance of
the formation place and thus initial conditions to a galaxy's evolution in
respect to environment, often formulated simply as ``nature versus nurture''
like in human development and psychology. We are therefore left to ponder if the
galaxies we see today are simply the product of the primordial conditions in
which they formed, or whether experiences in the past change the path of their
evolution. Unfortunately, our understanding of galaxy evolution in different
environments is still limited, albeit the morphology-density relation
\citep[e.g.][]{Oemler:1974p646, Dressler:1980p645} has shown that the density of
the galaxy's local environment can affect its properties. For example, on
average, luminous, non-starforming elliptical and lenticular galaxies have been
found to populate denser regions than star forming spiral galaxies.
Consequently, the environment should play a role in galaxy evolution, however,
despite the efforts, the exact role of the galaxy's local environment remains
open.

A group of galaxies is the most common galaxy association in the Universe
\citep[e.g.][]{1950MeLu2.128....1H, 1956AJ.....61...97H, Turner:1976p356,
Huchra:1982p1,Ramella:1995p562, Zabludoff:1998p578}. As such, more than half of
all galaxies are found in groups and small clusters. They are therefore
important cosmological indicators of the distribution of matter in the Universe.
Moreover, groups and clusters can also provide important clues for galaxy
formation and evolution physics as the properties of galaxies can be studied as
a function of the local environment. As a result, the environmental dependency
in galaxy evolution can be understood to some extent in terms of the group
environment (e.g. \citealt{Moore:1996p716, Moore:1998p792, 2004cgpc.symp..277M,
Fujita:2004p793}, but see also \citealt{Kauffmann:2004p751, Blanton:2006p789}).
Note, however, that the debate over the exact role of group environment is far
from over. Groups of galaxies, their properties and the group environment in the
context of galaxy evolution is discussed in Chapter \ref{ch:groups}.

One fundamental question concerning groups of galaxies, and closely related to
galaxy evolution in groups, is whether the identified systems are
gravitationally bound or not. This is a valid concern because from the
observational point of view, groups and their member galaxies are not well
defined. Early studies based on simulations gave hints that not all observed
systems of galaxies are dynamically relaxed, but might be in the process of
formation \citep[e.g.][]{Diaferio:1994p278, Frederic:1995p715,
Frederic:1995p714}. Despite this, many observational studies, even today, treat
all identified groups like they were gravitationally bound structures. The work
presented in Papers I and II tackles this issue and is summarised in Chapters
\ref{s:bound?} and \ref{s:redshiftAsymmetry}, respectively. Results of this work
show that a significant fraction of systems of galaxies are gravitationally
unbound when the most often used grouping algorithm, namely Friends-of-Friends,
is being applied to simulated data. This result has several important
implications for the studies of galaxy groups and for the evolution of galaxies
in groups. For example, Paper II shows that groups with a large excess of
positive redshifts are more often gravitationally unbound than groups that do
not show any significant excess. Fortunately, this prediction can be used,
together with the methods developed for and presented in Paper I, to assess
whether observed groups are likely to be gravitationally bound or not.

Elliptical galaxies are most often found in dense environments like the cores of
groups and clusters \citep[e.g.][]{Dressler:1980p645}. Yet, observations have
shown that there is a significant population of isolated elliptical galaxies
that are found in under-dense regions with no bright nearby companions
\citep[e.g.][]{Aars:2001p564, Reda:2004p502, Smith:2004p250, Denicolo:2005p568,
Collobert:2006p580}. Whether these galaxies originally formed in under-dense
regions or if the local environment has impacted their evolution, in form of a
collapsed group, is a profound question with long reaching implications. Key
observations of galaxy evolution and the significance of the environment for
galaxy evolution are briefly reviewed in Chapter \ref{ch:isolated}.

The results of a theoretical case study of isolated field elliptical galaxies,
Paper III, are also presented and summarised in Chapter \ref{ch:isolated}. These
results show that three different yet typical formation mechanisms can be
identified, and that isolated field elliptical galaxies reside in relatively
light dark matter haloes excluding the possibility that all of them are
collapsed groups as suggested earlier. Additionally, also another case study
concerning luminous infrared galaxies, Paper IV, is discussed. The results of
this study imply a strong correlation, such that more infrared-luminous galaxies
are more likely to be merger-driven. However, the results also imply that a
significant fraction (more than half) of all high redshift infrared-luminous
galaxies detected by \textit{Herschel} Space Observatory are able to attain
their high star formation rates without enhancement by a merger. These and other
results discussed in Chapter \ref{ch:isolated} imply that both ``nature'' and
``nurture'' play a role in galaxy evolution.

%% file: subfiles/chapters.tex
\chapter{Formation of Structure}\label{ch:formation}

\begin{quote}
\footnotesize{
``\ldots the biggest blunder of my life.''
\begin{flushright}
Albert Einstein
\end{flushright}
}
\end{quote}

\section{Historical perspective}

The study of the structure formation of the Universe dates back to 1610 and
Galileo Galilei who realised that the Galaxy can be resolved into stars when
observed through a telescope. Galilei's observations can also be considered as a
starting point for early observational cosmology. However, it was Ren\'e
Descartes and Thomas Wright who were likely the first ones to speculate and
publish their cosmological views. Already around 1760s Immanuel Kant and Johann
Lambert developed the first hierarchical model of the Universe, although since
these early models it took almost two hundred years before cosmology developed
into a physical science and before the formation of structure in the Universe
could be truly appreciated and studied in a physical context. Below I briefly
mention a few key moments from the history that have lead to the structure
formation theory, as we know it today. However, many important events are not
mentioned, thus I refer the interested reader to more comprehensive reviews, see
e.g. \cite{Ratra:2008p1016}.

It was Albert Einstein and his General Theory of Relativity (GR) in 1915
\citep{1915SPAW.......844E} that gave birth to the physical cosmology as we know
it today. GR is a framework that explains one of the four fundamental forces of
the Universe, namely gravity, and enabled cosmologists to predict the behaviour
of a model universe. As a result, it became possible, for the first time in the
history of mankind, to formulate self-consistent models that describe the
Universe and large-scale structure. As a consequence, in 1917 Einstein derived
the first fully self-consistent model of the Universe
\citep{1917SPAW.......142E}. However, soon after he realised that without
modifications his field equations predicted that a static Universe was not
stable. At the time the Universe was assumed to be static, thus, Einstein
introduced a cosmological constant $\Lambda$, that enabled a static universe, to
solve the issue.

In 1924 Aleksander Friedmann and in 1927 Georges Lem\^aitre derived solutions
for expanding universes, paving the way for evolving universe models. Later,
Friedmann's models became the standard models describing the dynamics of the
Universe. In 1927 Lem\^aitre first proposed what has come to be known as the Big
Bang theory of the origin of the Universe, albeit the name was introduced by
Fred Hoyle. The framework for the Big Bang model relies on the Einstein's GR,
the Cosmological Principle, Friedmann's equations, and it is also the standard
theory for the origin of the Universe.

In 1935 Howard Percy Robertson and Arthur Geoffrey Walker derived independently
the space-time metric for all isotropic, homogeneous, uniformly expanding models
of the Universe. However, the Friedmann world models are isotropic and
homogeneous, thus, all observable structures such as galaxies and groups are
absent. Consequently, the next step towards developing more realistic models of
the Universe was to include small density perturbations and to study their
development under gravity, namely the formation of structure. Fortunately,
already in 1902, well before GR, sir James Hopwood Jeans had shown that the
stability of a perturbation depends on the competition between gravity and
pressure \citep{1902RSPTA.199....1J}: gas pressure prevents gravitational
collapse on small spatial scales and gives rise to acoustic oscillations. Jeans
showed that density perturbations can grow only if they are heavier than a
characteristic mass\footnote{Now referred to as the Jeans' mass.} scale, while
below this scale dissipative fluid effects remove energy from the acoustic
waves, which dampens them. The application of the Jeans criterion and the growth
of spherically symmetric perturbations in an expanding universe were worked out
by Lem\^aitre and Richard Tolman in the 1930s. Albeit it was not before 1946
when Evgenii Lifshitz worked out relativistic perturbation theory and started
applying it to the linear growth of cosmic structure. Finally, a general scheme
for structure formation was first outlined by Lev Davidovich Landau and Lifshitz
in the 1950s, and developed further by Phillip James Edwin Peebles during the
1970s.

However, to truly appreciate the study of the formation of structure of the Universe, observational constrains on the initial density perturbations were required. The serendipitous discovery of Cosmic Microwave Background (CMB) radiation by Arno Penzias and Robert Wilson in $1964$ \citep{1965ApJ...142..419P}, predicted already in $1948$ by Ralph Alpher, Robert Herman, and George Gamow \citep{1948PhRv...74..505G, 1948PhRv...74.1737A}, paved the way for understanding the initial conditions of large-scale structure formation. Measurements of the CMB describe the initial conditions after recombination of the very early Universe and led way to the standard model of cosmology. Furthermore, large galaxy surveys of recent years have helped to set constrains for the models of structure formation at low redshift.

\section{Observational background of cosmology}\label{s:obs_cosmology}

According to the Big Bang model, the background radiation from the sky measured today comes from the so-called last scattering surface. As the name implies the surface of last scattering is a spherical surface where the Cosmic Microwave Background photons were scattered for the last time\footnote{This however is not true for all CMB photons: some have scattered from free electrons that have become available due to reionization.} before arriving at our microwave detectors. This decoupling of photons from matter happened $t_{\star} \sim 380,000$ years after the Big Bang during the epoch of recombination when the rate of Thomson scattering became slower than the expansion of the Universe \citep[e.g.][]{Weymann:1966p1019, Peebles:1968p1018}. At that moment, photon interactions with matter became insignificant, leading to the CMB radiation. This moment also defines the ``optical'' horizon; the largest volume from which we can receive information via photons.

On the largest scales the most robust evidence for the isotropy of the Universe comes from the CMB measurements, while galaxy surveys compliment the CMB information by probing later epochs and smaller angular scales. Therefore, in the next two Sections I will briefly review what the CMB and galaxy survey observations can tell us about the formation of structure.

\subsection{Cosmic microwave background radiation}\label{s:CMB}

The epoch when the ionisation state of the intergalactic gas changed from being a fully ionised plasma to a neutral gas is known as the epoch of recombination. This is the redshift $(z \sim 1000)$ when the detailed anisotropy structure of the early Universe was imprinted onto the Cosmic Microwave Background (CMB). It can therefore provide information on the initial density perturbations and describe the Universe on the epoch well before galaxies, groups and the formation of large-scale structure.

\begin{figure}[htb] \center{
\vspace{-1mm}
\hspace{-4mm}
\includegraphics[scale = 0.34,bb=0 0 1131 644]{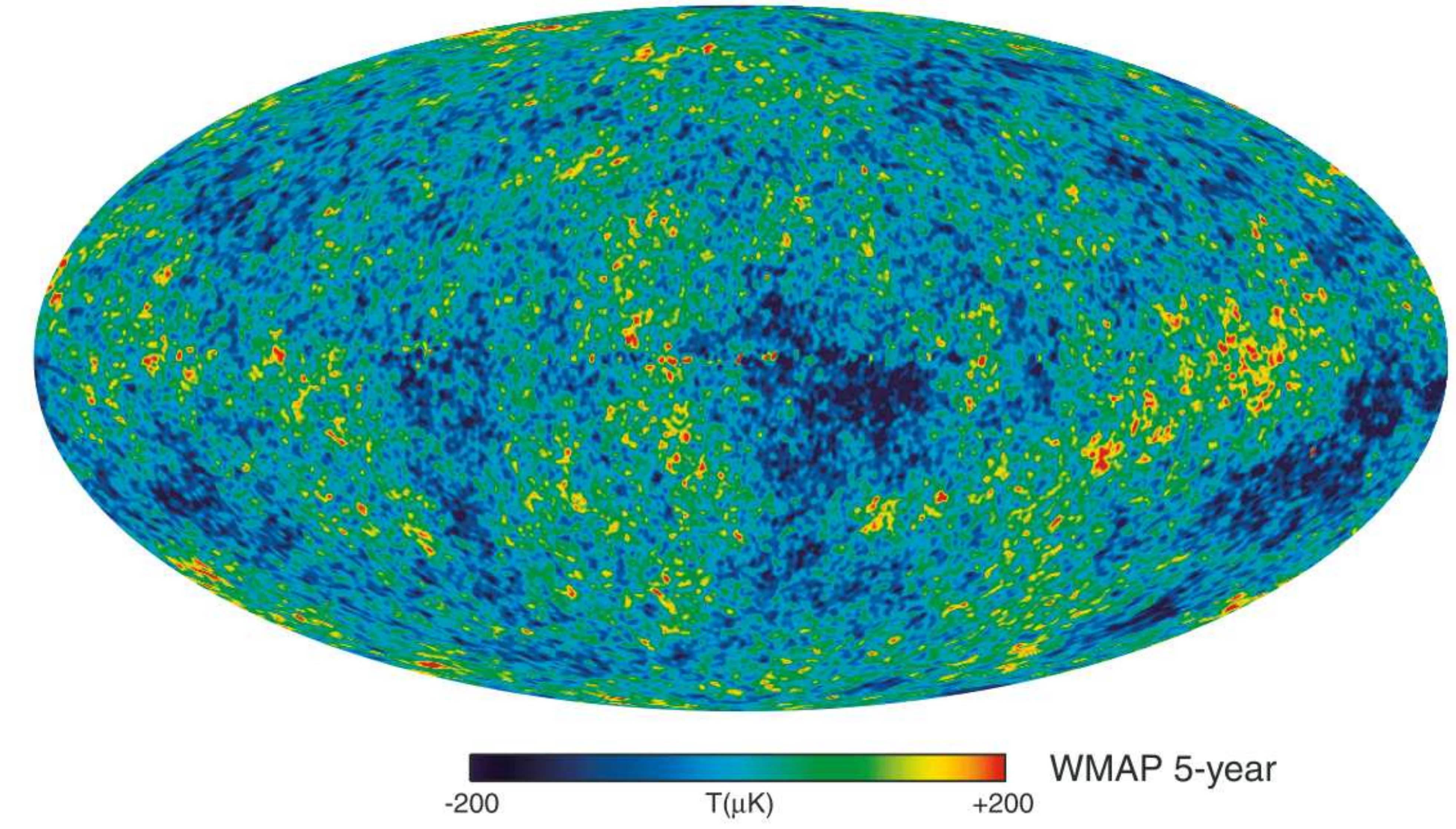}
\vspace{-3mm}
} \caption{The foreground-reduced Internal Linear Combination map based on the five year WMAP data. Image from \cite{Hinshaw:2009p709}.}
\label{fig:WMAPmap}
\end{figure}

Ever since the discovery of the CMB radiation, observations have played a key role in shaping and constraining the standard model of cosmology \citep[for a review, see e.g.][]{Hu:2002p864, Bartelmann:2010p998}. The CMB observations probe the earliest observable Universe and hence the initial conditions of the Universe, such as its homogeneity, isotropy, and flatness. Based on the CMB observations it has been established, first by Cosmic Background Explorer (COBE) and later by Wilkinson Microwave Anisotropy Probe (WMAP), that the electromagnetic spectrum of the CMB is extremely close to a thermal blackbody with a temperature $\sim 2.725$ K. Moreover, COBE and WMAP has established that the largest temperature anisotropies $\frac{\Delta T}{T}$ in the CMB are of the order of $\sim 10^{-5}$ \citep{Smoot:1992p860, Hinshaw:2009p709}, as shown in Figure \ref{fig:WMAPmap}. The absence of $\sim 10^{-3}$ K fluctuations alone show that the matter in the Universe must be dominantly something that does not interact electromagnetically \citep{Peebles:1982p1000}, i.e., ``dark''.

The angular temperature fluctuations associated with the primordial density perturbations are assumed to originate in a rather narrow range of redshifts. If this holds then the pattern of the angular temperature fluctuations in the CMB map (Fig. \ref{fig:WMAPmap}) gives us a direct snapshot of the distribution of radiation and energy at the moment of recombination. The angular scale $\Theta \sim 1\,^{\circ}$ corresponds to the Hubble radius at recombination, which can be taken as a dividing line between the small-scale perturbations that have been substantially modified by gravity and the large-scale inhomogeneities that have not changed much. The fluctuations on large angular scales $(\Theta \gg 1\,^{\circ})$ arise from inhomogeneities with wavelengths exceeding the Hubble radius at recombination. As a result, they provide pristine information about the primordial inhomogeneities. On the other hand, sub-horizon perturbations are formed by primordial sound waves.

The anisotropy of the CMB can be divided into two types: primary anisotropy, due to effects which occur at the last scattering surface and earlier; and secondary anisotropy, due to effects such as interactions of the CMB photons with hot gas or gravitational potentials (for example, the Sunyaev-Zel'dovich and integrated Sachs-Wolfe effects), between the last scattering surface and the observer. The structure of the CMB primary anisotropies is mainly determined by two effects: acoustic oscillations and, on small angular scales, photon diffusion (also known as Silk damping \citep{Silk:1968p796}). The photon diffusion damping arises from the fact that the photon-baryon-electron (PBE) fluid is not tightly coupled and the photons can diffuse through the fluid, while the acoustic oscillations result from the constructive and deconstructive interference. Overdensities in the dark matter compresses the fluid due to their gravity until the rising pressure in the coupled PBE fluid is able to counteract gravity. The cosmological importance of this is that the PBE fluid underwent acoustic oscillations, while the dark matter, being decoupled, did not.

The density fluctuations in the early Universe are assumed to be critical for structure formation, because they can provide the seeds from which the structures within the Universe can grow and eventually collapse to form the first stars and galaxies \citep[e.g.][and references therein]{1999ApJ...527L...5B, 2001PhR...349..125B, 2002Sci...295...93A, 2009Natur.459...49B}. The density perturbations of the early Universe are thought to have a very specific character when inflation is assumed: they form a Gaussian random field \citep{1986ApJ...304...15B}, which is nearly scale-invariant according to the spectral index $n$ measured by WMAP \citep{Hinshaw:2009p709, Komatsu:2009p710, Dunkley:2009p754, Jarosik:2010p770}. I will return to this in Section \ref{s:evolution_perturbations} where a more detailed discussion of the initial density perturbations is presented.

\subsubsection{Power Spectrum of the CMB}

\begin{figure}[htb]
\center{
\hspace{-1mm}
\includegraphics[scale = 0.35,bb=0 0 1044 650]{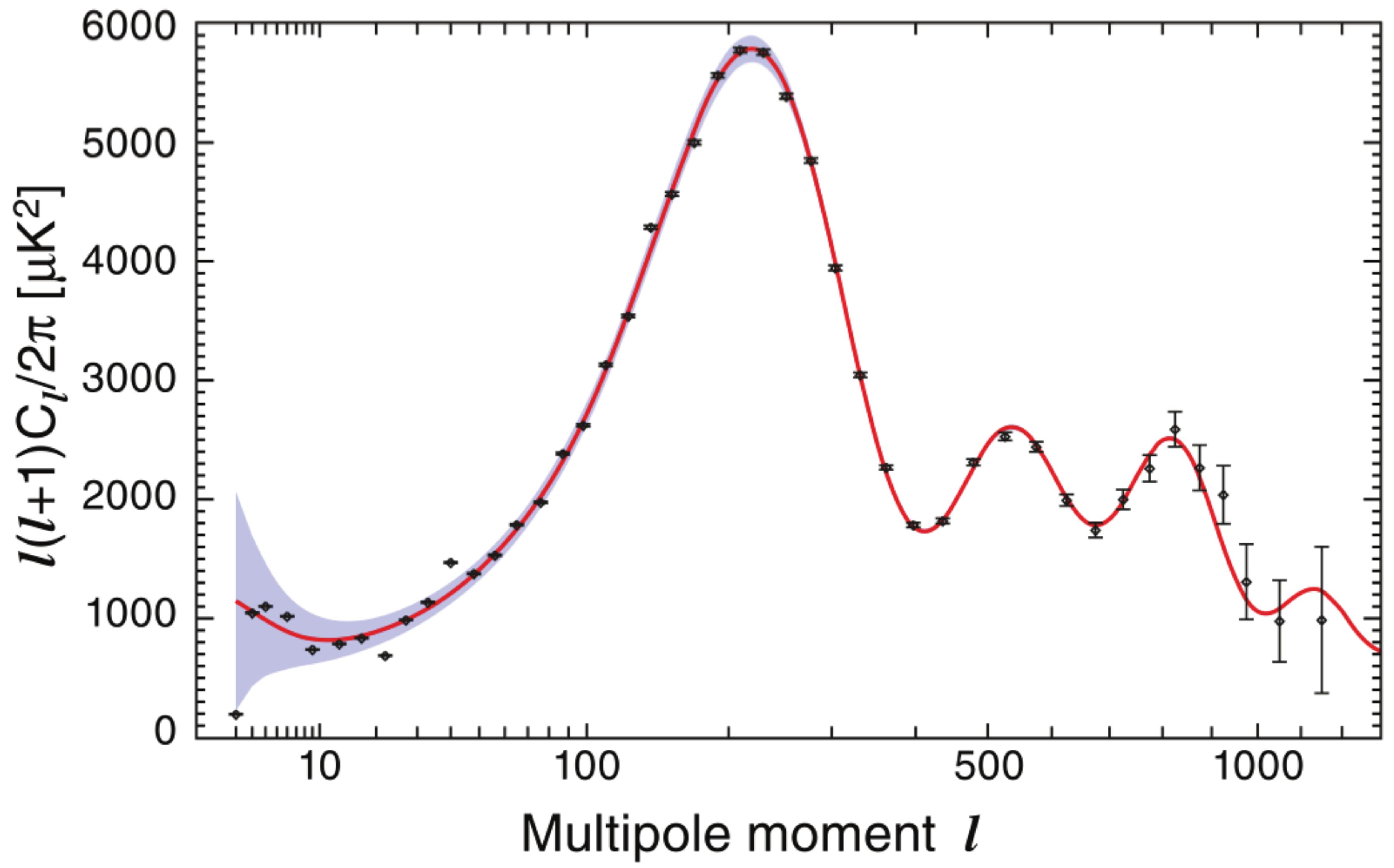}
\vspace{-1mm}
}
\caption{The temperature power spectrum for the seven-year WMAP data. The solid line shows the predicted spectrum for the best-fit flat $\Lambda$CDM model (Section \ref{s:background_cosmology}). The error bars on the data points represent measurement errors while the shaded region indicates the uncertainty in the model spectrum arising from cosmic variance. Image from \cite{Jarosik:2010p770}.}
\label{fig:WMAPmapPowerSpectrum}
\end{figure}

To maximise the information a CMB map (such as Fig. \ref{fig:WMAPmap}) can provide, the CMB information is most often presented in the form of a power spectrum $P$ in terms of the angular scale or multipole moment $l$ as shown in Figure \ref{fig:WMAPmapPowerSpectrum}. The power spectrum, which is a spherical harmonic transform\footnote{Fourier transformation is not possible on a sphere, thus, spherical harmonics which are analogous are used instead.} of the CMB map, and polarisation of the CMB radiation provide a wealth of information \citep[see e.g.][for data analysis methods]{2007RPPh...70..899T} for both constraining cosmological parameters (Table \ref{tb:lambdaCDM_values}) and for understanding the formation of the large-scale structure in the Universe. The CMB anisotropy is a powerful cosmological probe because the parameters which determine the spectrum can all be directly related to the basic cosmological parameters such as the energy densities $\Omega_{i}$, the dark energy equation of state $w$, and the Hubble parameter $H$.

The general shape of the power spectrum (Fig. \ref{fig:WMAPmapPowerSpectrum}) - a plateau at large angular scales (small $l$) and acoustic peaks at small angular scales (large $l$) - confirms that the spectrum is predominantly nearly scale-invariant and adiabatic in agreement with the basic predictions of the Big Bang and inflationary paradigm. The dominant acoustic peaks in the CMB power spectra are caused by the collapse of dark matter over-densities and the oscillation of the photon-baryon fluid into and out of these over-densities \citep{Lineweaver:2003p797}. The underlying physical notion is that the pressure of photons can erase anisotropies, whereas the gravitational attraction of baryons makes them to collapse and to form dense haloes. As a result, these effects can create acoustic oscillations, which give the CMB its characteristic peak structure (see Fig. \ref{fig:WMAPmapPowerSpectrum}). The first acoustic peak is associated with perturbations on the scale of the sound horizon at the last scattering surface $(l \sim 200 \sim 1\,^{\circ} \sim 100$ Mpc$)$, while following peaks are on the scales less than the sound horizon. Combining the information about the heights and locations of the peaks, many cosmological parameters can be determined with good accuracy independent of other observations such as galaxy surveys.

\subsection{Large galaxy surveys}\label{s:large_galaxy_surveys}

The large galaxy surveys of today such as the 2dF Galaxy Redshift Survey \citep[2dFGRS;][]{1999RSPTA.357..105C, Colless:2001p802, Percival:2001p803} and the Sloan Digital Sky Survey \citep[SDSS;][]{York:2000p799, Stoughton:2002p800, Abazajian:2003p798} have provided large, statistically significant samples of different types of galaxies. The ability to explore many dimensions of galaxy properties and scaling relations simultaneously and homogeneously has been greatly beneficial. Moreover, spectroscopic observations and multi-wavelength imaging allows galaxies to be sorted in classes and sub-populations by, e.g., morphology, environment and luminosity, while large sky coverage allows galaxies to be grouped in groups and clusters enabling studies of galaxy evolution as a function of environment \citep[e.g.][see also Chapter \ref{ch:isolated}]{Blanton:2007p790, Mateus:2007p735}. However, large galaxy surveys not only provide good statistics for galaxy properties but they can also be used to study the large-scale structure and cosmology.

\begin{figure}[htb]
\center{
\includegraphics[scale = 0.23,bb=0 0 1600 955]{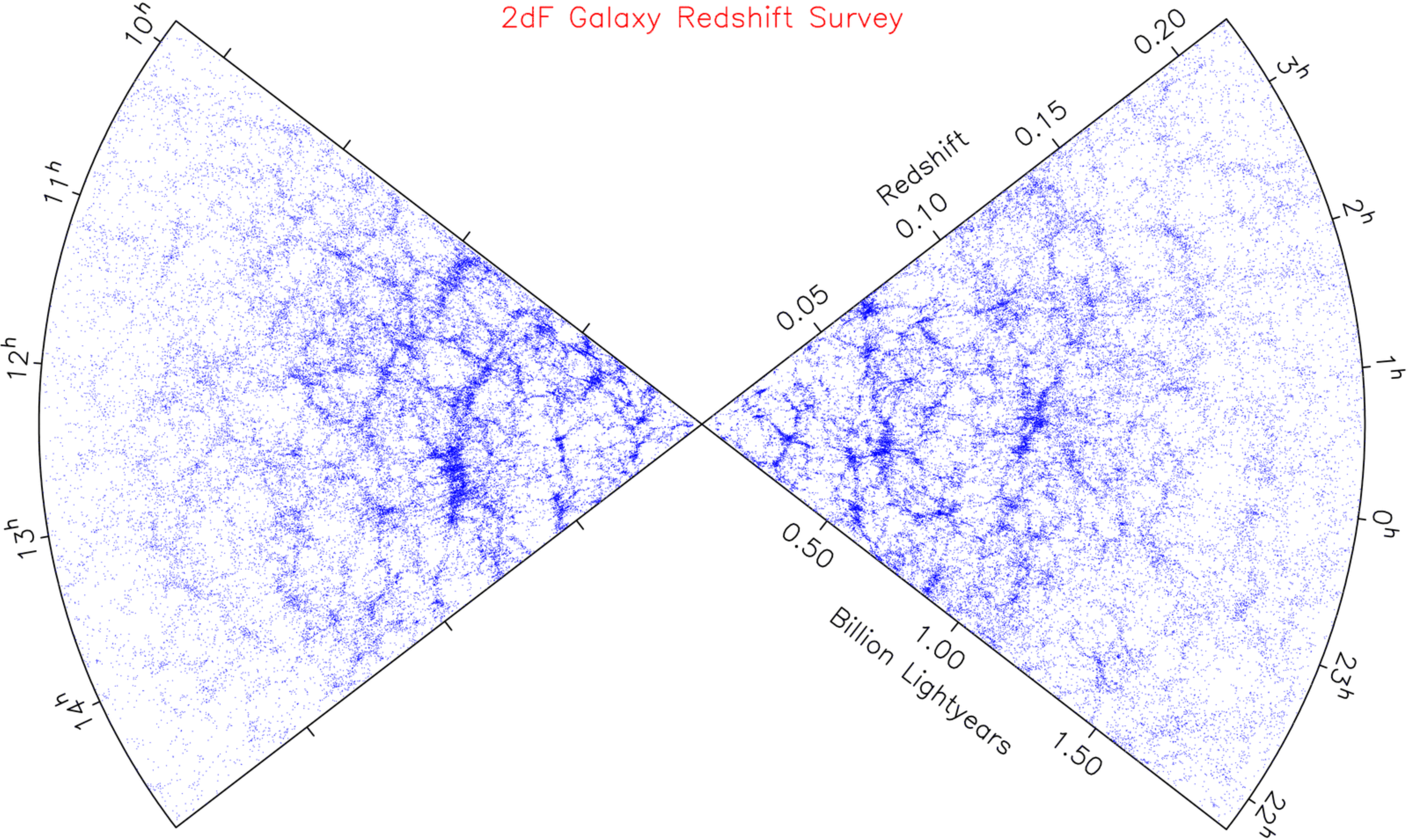}
}
\caption{The projected distribution of galaxies in the nearby Universe as a function of redshift and Right Ascension. Earth is at the centre, and each blue point represents a galaxy. Courtesy of the 2dFGRS website.}
\label{fig:2dFzcone}
\end{figure}

Despite the fact that the CMB radiation is very smooth, the visible Universe that is dominated by the light from galaxies looks highly inhomogeneous and consists of structures from the scales of isolated galaxies and voids, through groups (Chapter \ref{ch:groups}) and clusters to superclusters and to large filaments between them (see Fig. \ref{fig:2dFzcone}). One aim of large galaxy redshift surveys is therefore to map the three dimensional distribution of galaxies, in order to understand the properties of this distribution and what it implies about the contents and evolution of the Universe. Because the topology of the distribution of galaxies is closely related to the initial conditions of the Universe and to the assumption that the initial perturbations were Gaussian fluctuations with random phases on large scales this mapping can also provide information concerning the conditions in the early Universe independent from the CMB measurements. Unfortunately, the spatial distribution of galaxies, groups and clusters, depends not only on the matter distribution in the Universe, but also on how they form in the matter density field. It is therefore important to understand galaxy formation (Section \ref{galaxy_formation}) and evolution (Chapter \ref{ch:isolated}) in detail when studying the clustering of galaxies.

A representation of the large-scale distribution of galaxies on the sky in the 2dFGRS is shown in Fig. \ref{fig:2dFzcone}. From this figure alone it is obvious that galaxies form larger structures such as clusters and filaments and that the visible light is unevenly distributed in the Universe even on relatively large scales. Even though the distribution of galaxies becomes smoother and smoother when larger and larger scales are considered, non-random structure is still present in forms of superclusters and filaments between them. This non-random structure is sometimes called the cosmic web, as the long filaments of dark and baryonic matter seem to form a ``threaded" structure.

The filaments seen in large galaxy surveys are the largest known structures in the Universe and can be up to $\sim 80 - 100h^{-1}$Mpc\footnote{Here $h$ refers to the dimensionless Hubble parameter defined such that $H_{0} = 100h \, \mathrm{km}\, \mathrm{s}^{-1}\, \mathrm{Mpc}^{-1}$. Also, see equation \ref{eq:HubbleParameter} for a definition of the Hubble parameter.} long \citep{Einasto:1980p905, Batuski:1985p907, White:1987p906, Bahcall:1988p909}. Filaments are important structures for galaxy formation as it is assumed that they are the channels that carry baryons to the nodes of the filaments where clusters of galaxies are formed. Filaments can also help cool gas to avoid shock-heating \citep[e.g.][]{Dekel:2006p966, Dekel:2009p939, Keres:2009p957}, while over-densities in filaments can form gravitationally bound dark matter haloes. There is however still a debate in how gas can enter and coalesce into dark matter haloes and how it cools down to form stars and eventually galaxies \citep[see e.g.][and references therein]{Keres:2005p836, Kaufmann:2006p884}. The exact role of filaments in the formation and evolution of galaxies is therefore currently unclear.

While filaments are the largest known structures, superclusters \citep[e.g.][]{ArayaMelo:2009p908} are the largest non-percolating galaxy systems \citep{Oort:1983p984, Bahcall:1988p909, Einasto:2007p983, Einasto:2008p985}. Unlike super clusters, the scales of the largest voids are in general $\sim 30$ to $50$ times the scale of a regular relaxed cluster, i.e., up to $\sim 50h^{-1}$ Mpc, although the size measurements vary greatly \citep{Zeldovich:1982p910, Rood:1988p912, Vogeley:1994p901, Lindner:1995p900, ElAd:1997p903, Hoyle:2004p902, Ceccarelli:2006p524, vonBendaBeckmann:2008p988, Tinker:2009p904}. Because of all the structure in the cosmic web, the galaxy distribution seems to have a sponge-like topology, with both high- and low-density regions forming an interconnected network, where voids are separated from high-density regions by flattened structures called ``walls''. This raises an obvious question: how can we quantify the clustering of different types of objects and what does this tell us about the formation of the large-scale structure?

\subsubsection{Two-point correlation function and the power spectrum}\label{ss:two-point_correlation_power_spectrum}

Among the simplest methods to measure clustering properties of galaxies (or of other objects like quasars, groups, clusters, etc.) is with the spatial two-point correlation function $\xi(r)$. It describes the excess probability above Poisson of finding an object at distance $r$ from another object selected at random over that expected in a uniform, random distribution \citep[see][for a complete discussion]{1980lssu.book.....P}. We can now write the probability to find galaxies in infinitesimal small volumes $\mathrm{d}V_{1}$ and $\mathrm{d}V_{2}$ as follows
\begin{equation}
P = (1 + \xi(r)) \bar{n}^{2} \mathrm{d}V_{1} \mathrm{d}V_{2} \quad ,
\end{equation}
where $\bar{n}$ is the mean galaxy density. In practise, it is however often more convenient to derive the two-point correlation function $\xi(r)$ using, for example, the \cite{Landy:1993p1081} estimator.

Large datasets provided by galaxy surveys have been used to study, for example, the spatial correlation functions \citep[e.g.][]{Connolly:2002p806, Scranton:2002p809, Zehavi:2004p805, Masjedi:2006p920}, clustering of matter, galaxies \citep[e.g.][]{Coil:2007p1080} and groups \citep[e.g.][]{Coil:2006p1079} in the Universe. Consequently, allowing to set constrains for the structure formation and cosmological parameters \citep{2004LRR.....7....8L, Tegmark:2004p885}. In galaxy surveys, redshifts of galaxies are usually used as distances, thus the correlation function is said to work in redshift-space. However, because of peculiar velocities\footnote{The peculiar velocity is the velocity that remains after subtracting off the contribution due to the Hubble expansion.}, an isotropic distribution in real-space will appear anisotropic in redshift-space and vice versa. The redshift-space correlation function therefore differs from the real-space correlation function. This effect is known as the redshift-space distortion  \citep[for observational studies, see e.g.][]{Hamilton:1997p994, Tegmark:2004p812}. It is important to note that redshift-space distortions due to peculiar velocities along the line of sight will introduce systematic effects to the estimate of $\xi(r)$. For example, at small separations, random motions within a virialized overdensity cause an elongation along the line of sight (dubbed as ``fingers of God"). On the other hand, on large scales, coherent infall of galaxies into forming structures causes an apparent contraction of structure along the line of sight (dubbed as the ``Kaiser effect").

On scales smaller than $\sim 10h^{-1}$ Mpc the real-space correlation function is well approximated by a power law
\begin{equation}
\xi(r) = \left (\frac{r}{r_{0}} \right)^{- \gamma} \quad ,
\end{equation}
where the slope $\gamma \sim 1.8$ and $r_{0} \sim 5h^{-1}$ Mpc is the correlation length. This shows that galaxies are strongly clustered on scales $\lessim 5h^{-1}$ Mpc, and the amplitude of clustering becomes weak on scales much larger than $> 10h^{-1}$ Mpc. Note, however, that the exact values of $\gamma$ and $r_{0}$ are found to depend on the properties of the galaxies. Particularly, brighter and redder galaxies are more strongly clustered than fainter and bluer ones \citep{Norberg:2001p1006, Zehavi:2005p1007, Coil:2006p1078, Wang:2008p1005}. Additionally, early-type galaxies have been found to be much more clustered on small scales, leading to a morphology-density relation \citep[][see also Chapter \ref{ch:isolated}]{Peacock:2002p995}.

The galaxy correlation function is a measure of the degree of clustering in either the spatial $(\xi(r))$ or the angular distribution $(w(\theta))$ of galaxies. The spatial two-point correlation (or autocorrelation) function $\xi (r)$ and the power spectrum $P(k)$ forms a Fourier-transform pair, i.e.
\begin{eqnarray}
P(\mathbf{k}) & = & \frac{1}{V} \int \xi(\mathbf{r}) e^{-i \mathbf{k} \cdot
\mathbf{r}} \mathrm{d}^{3}r \label{eq:PowerSpectrum} \\
\xi(\mathbf{r}) & = & \frac{V}{8 \pi^{3}} \int P(\mathbf{k}) e^{i \mathbf{k}
\cdot \mathbf{r}} \mathrm{d}^{3}k \label{eq:two_point_correlation} \quad .
\end{eqnarray}
Here $V$ is the volume within which $\xi(r)$ is defined. Assuming isotropy and that the two-point correlation function is spherically symmetric $(\mathrm{d}^{3}k = 4 \pi k^{2} \mathrm{d}k)$ leads to
\begin{eqnarray}
P(k) & = & 4 \pi \int_{0}^{\infty} \xi(r) \frac{\sin kr}{kr} r^{2}
\mathrm{d}r \label{eq:PowerSpectrum_isot} \\
\xi(r) & = & \frac{1}{2 \pi^{2}} \int_{0}^{\infty} P(k) \frac{\sin kr}{kr} k^{2}
\mathrm{d}k \label{eq:two_point_correlation_isot} \quad .
\end{eqnarray}
Note that the function $\sin kr \ (kr)^{-1}$ allows only wave-numbers $k \leq r^{-1}$ to contribute to the amplitude of the fluctuations on the scale $r$.

\begin{figure}[htb]
\center{
\includegraphics[scale = 0.29]{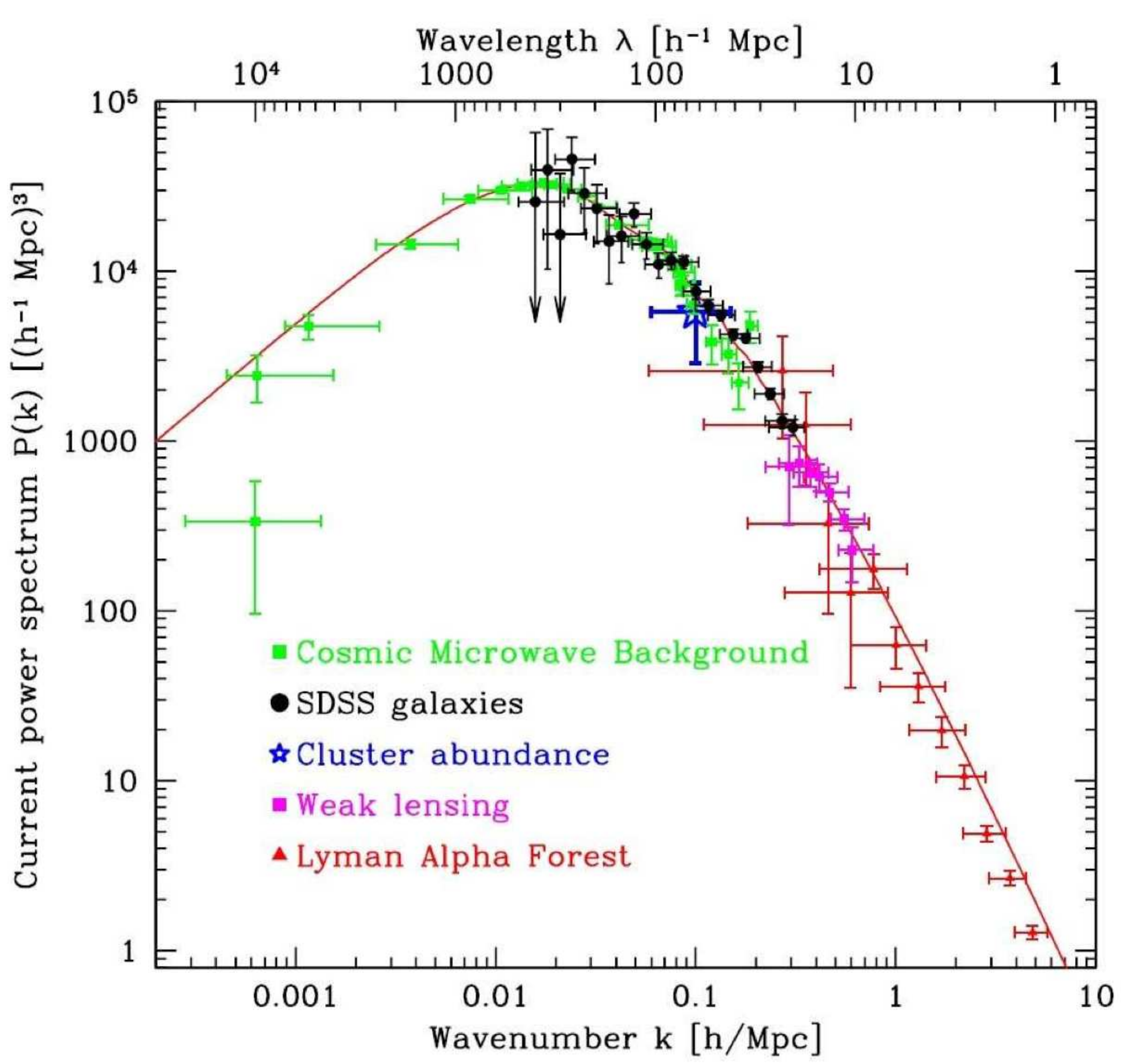}
}
\caption{The matter power spectrum $P(k)$ as a function of wave-number $k$ at the present time. Note the turnover at $k_{eq} \sim 0.01h^{-1}$ Mpc. The plot combines data from different scales: CMB, large galaxy surveys, weak lensing and Ly$\alpha$ forest, in order of decreasing co-moving wavelength. In addition, there is a single data point for galaxy clusters. Figure from \cite{Tegmark:2004p812}.}
\label{fig:matter_power_spectrum}
\end{figure}

Figure \ref{fig:matter_power_spectrum} shows a matter power spectrum at the present time. According to the figure on large scales (small wave-numbers $k < 0.01h^{-1}$ Mpc) the current matter power spectrum still has its primordial shape. This shape corresponds to a power law dependence on scale; $P(k) \propto k^{n}$, where $n$ is the so-called spectral index, assumed to be close to unity (for theoretical background, see Section \ref{s:initial_fluctuations} and for the reference value, see Table \ref{tb:lambdaCDM_values}). The horizon scale at the epoch when the matter and radiation densities are equal (the matter-radiation equality, $z_{eq}$, see Table \ref{tb:lambdaCDM_values} for a value) is imprinted upon the power spectrum as the scale at which the spectrum turns over. Hence, the peak position in the spectrum corresponds to the Jeans length (Eq. \ref{eq:jeans_length}) at matter-radiation equality (I will return to this in Section \ref{s:evolution_perturbations}). The position of this turnover corresponds to a physical scale determined by the matter $(\Omega_{m}h^{2})$ and radiation densities $(\Omega_{r}h^{2})$. Moreover, the shape of the observed power spectrum $P(k)$ depends on the amount and the nature of the matter in the Universe, providing constrains for cosmology. For example, if all of the dark matter were hot then the matter power spectrum would fall off sharply to zero to the right of the peak.

Large redshift surveys can be used not only to study the power spectrum of
galaxy clustering but also the presence of the acoustic oscillations of baryons
\citep[e.g.][]{Tegmark:2004p812, Cole:2005p848, Eisenstein:2005p808,
Percival:2007p810}. The physics of these oscillations are analogous to those of
the CMB acoustic oscillations. The amplitude of baryonic acoustic oscillations
(BAOs) is however suppressed in comparison to CMB because not all the matter in
the Universe is composed of baryons. In practise, many studies of BAOs have
taken advantage of the clustering of luminous red galaxies \citep[LRGs;
e.g.][]{Padmanabhan:2007p1036, Sanchez:2009p1035}. The clustering of LRGs also
allows the values of cosmological parameters to be derived independently from
the CMB measurements. Furthermore, the observed power spectrum allows to
constrain on both the amplitude and the scale dependence of the galaxy bias
\citep[e.g.][]{Padmanabhan:2007p1036}. This ultimately links the galaxy power
spectrum to the matter power spectrum. Despite this connection, large galaxy
surveys can provide constrains for structure formation models independent of the
CMB measurements.


A viable structure formation model has to therefore simultaneously explain both the smoothness of the CMB and the clear evidence from the galaxy surveys that the assumptions of isotropy and homogeneity do not hold on smaller scales, but galaxies and other inhomogeneities such as groups do form. Moreover, large galaxy surveys together with the CMB observations can be used to set strict constrains for cosmological parameters (see Section \ref{s:background_cosmology} and Table \ref{tb:lambdaCDM_values}). Importantly, these observations can provide information about structures of different sizes and times from the recombination to the present epoch. A successful structure formation model must therefore be able to explain all the current observations. However, before we start looking into structure formation models in more detail, some tools to study the dynamical evolution of an expanding universe that is isotropic and homogeneous must be introduced. In the following Sections I shall give a minimalistic overview of the dynamics of an expanding universe on top of which the structure formation theory can be build upon. For more detailed treatment, I refer the interested readers to great textbooks of e.g. \cite{1980lssu.book.....P, 2003moco.book.....D, TheGreatBook, Mo_big_book}.

\section{Dynamics of an expanding universe}\label{s:dynamics_of_universe}

Being able to model the dynamical evolution of an expanding universe is a basic requirement for any structure formation model. Because space-time can be curved and is not static, we must rely on General Relativity when deriving the equations that govern the evolution of the background universe.

\subsection{Einstein equation}\label{s:Einstein_Equation}

Einstein's General Relativity (GR) enabled self-consistent models of the Universe to be constructed as it relates matter and energy to the geometrical properties of the Universe. In GR, the gravity field is described by Einstein's field equation:
\begin{equation} \label{eq:Einstein}
G^{\alpha}_{\beta} = 8 \pi G T^{\alpha}_{\beta} \quad ,
\end{equation}
where $G^{\alpha}_{\beta}$ is the Einstein tensor, $G$ is Newton's gravitational constant and $T^{\alpha}_{\beta}$ is the energy-momentum tensor. Note that matter is incorporated in Einstein's equation through the energy-momentum tensor. On large scales, matter can be approximated as a perfect fluid characterised by an energy density $\epsilon$, pressure $P$ and four velocity $u^{\alpha}$. Now the energy-momentum tensor may be written as
\begin{equation}\label{eq:energy_momentum}
T^{\alpha}_{\beta} = (\epsilon +P)u^{\alpha}u_{\beta} - P
\delta^{\alpha}_{\beta} \quad ,
\end{equation}
where the equation of state $P = P(\epsilon)$ depends on the properties of matter. Often in cosmologically interesting cases $P = \mathrm{constant} \times \epsilon$ (or more general as in Eq. \ref{eq:equation_of_state}). Equation \ref{eq:energy_momentum} shows that in GR the strength of the gravitational field depends not only on the energy density $\epsilon$, but also on the pressure $P$.

It is also possible to write the Einstein equation in a form that explicitly shows the cosmological constant $\Lambda$, now
\begin{equation}\label{eq:withLambda}
G^{\alpha}_{\beta} = R^{\alpha}_{\beta} - \frac{1}{2}\delta^{\alpha}_{\beta}R -
\Lambda \delta^{\alpha}_{\beta} \quad ,
\end{equation}
where $R^{\alpha}_{\beta}$ is the Ricci tensor, $R$ is the scalar curvature, $\Lambda$ is the cosmological constant, and $\delta$ is the unit tensor\footnote{Note that later in this Chapter $\delta$ will refer to the density perturbation field.} defined by the metric such that $g^{\alpha \gamma} g_{\gamma \beta} = \delta^{\alpha}_{\beta}$. As the cosmological constant can be interpreted as the contribution of vacuum energy to the Einstein equation it can also be included in the energy-momentum tensor. 

According to the Cosmological Principle \citep{1933ZA......6....1M} the Universe is homogeneous and isotropic, at least on large enough scales, thus space-time can be described by the Robertson-Walker metric:
\begin{equation}\label{eq:RWmetric}
\mathrm{d}s^{2} = c^{2}\mathrm{d}t - a^{2}(t) \left( \frac{\mathrm{d}r^{2}}{1-Kr^{2}}
+ r^{2}(\mathrm{d}\theta^{2}
+ \sin^{2}\theta \mathrm{d}\phi^{2}) \right) \quad ,
\end{equation}
where the spatial positions are described by spherical coordinates $(r, \phi, \theta)$, $c$ is the speed of light and $a(t)$ is the scale factor. The scale factor and the changing scale of the Universe causes the cosmological redshift $z$, which can be defined as
\begin{equation}
z = \frac{a(t_{0})}{a(t_{em})} - 1 \quad .
\end{equation}
Here $a(t_{0})$ is the present value of the scale factor and $a(t_{em})$ is the value of the scale factor at the time when the light was emitted. As a result, the redshift $z$ can be used to parameterise the history of the Universe; a given $z$ corresponds to a time when our Universe was $1 + z$ times smaller than now. The importance of the metric \ref{eq:RWmetric} is that it allows to define the invariant interval $\mathrm{d}s^{2}$ between events at any epoch or location in an expanding universe, and thus determines the metric, Riemann, and Ricci curvature tensors.

The Einstein equation describes the geometry of space which is curved by matter and energy. It is also the basic equation of GR that the dynamical variables characterising the gravitational field must follow. The cosmological evolution of relativistic matter can therefore be derived from the Einstein equation (\ref{eq:Einstein}) when the metric of the space-time and the energy-momentum tensor are fixed.

\subsection{Friedmann equations}\label{s:Friedmann_equations}

Alexander Friedmann was the first to derive a pair of equations that can describe the expansion rate of a homogeneous and isotropic universe. The first Friedmann equation can be derived directly from Einstein's field equation (\ref{eq:Einstein}), which can be simplified using the Robertson-Walker metric (Eq. \ref{eq:RWmetric}). Now, Friedmann's first equation can be written as
\begin{equation}\label{eq:Friedmann1}
H^{2} + \frac{K c^{2}}{a^{2}} = \frac{8\pi G}{3}\epsilon + \frac{\Lambda
c^{2}}{3} \quad ,
\end{equation}
where $G$ is Newton's gravitational constant, $K$ is the spatial curvature ($\pm 1$ or $0$), $\epsilon$ is the sum over all energy densities (e.g. baryons, photons, neutrinos, dark matter and dark energy), and $H$ is the Hubble parameter that describes the rate of the expansion:
\begin{equation}\label{eq:HubbleParameter}
H(t) = \frac{\dot{a}(t)}{a(t)} \quad .
\end{equation}
Here $\dot{a}$ denotes the time derivative of the scale factor. Note that the dynamical content of the metric is encoded in the function $H(a,t)$. Friedmann's second equation can be derived from the trace of Einstein's field equation and can be written as:
\begin{equation}\label{eq:Friedmann2}
\frac{\ddot{a}}{a} = - \frac{4\pi G}{3}\left(\epsilon + \frac{3P}{c^{2}}\right)
+ \frac{\Lambda c^{2}}{3} \quad .
\end{equation}
Now the $\ddot{a}$ is the second time derivative of the scale factor and $P$ is the pressure.

The importance of the two Friedmann equations (\ref{eq:Friedmann1} and \ref{eq:Friedmann2}) is that they determine the two unknown functions; the scale factor $a(t)$ and the energy density $\epsilon (t)$. Consequently, the Friedmann equations can describe the dynamics of a homogeneous, isotropic and expanding universe, because the scale factor completely describes the time evolution of such a universe. The Friedmann equations therefore determine the expansion rate of the Universe, based on the density of material within it and the curvature of space. Note that, a Robertson-Walker metric (\ref{eq:RWmetric}) whose scale factor $a$ satisfies Friedmann's equations is called the Friedmann-Lemai\^tre-Robertson-Walker metric and the cosmological standard model upholds that the Universe at large is described by such a metric. 

Construction of a cosmological model requires solving the Friedmann equations, resulting to the expansion rate as a function of time, and hence the size of the Universe. However, solving the Friedmann equations alone is not enough for a cosmological model, it is also essential to know how the energy density $\epsilon$ changes as a function of time.

\subsection{Evolution of energy density: the fluid equation}\label{s:fluid_equation}

The evolution of energy density in an expanding universe can be described with the so-called fluid equation. The fluid equation, which holds only for adiabatic processes, can be obtained from the Friedmann's equations when equation \ref{eq:Friedmann2} is rewritten using equation \ref{eq:Friedmann1} resulting in:
\begin{equation}\label{eq:fluid_equation}
\dot{\epsilon} + 3 \frac{\dot{a}}{a} \left ( \epsilon + \frac{P}{c^{2}} \right) 
= 0 \quad ,
\end{equation}
where $\epsilon$ denotes the energy density and $P$ is the pressure, as defined earlier. The fluid equation expresses the conservation of mass-energy, thus it is also known as the energy conservation or the continuity equation. The second term in the fluid equation corresponds to the loss in energy because the pressure of the material has done work as the volume of the Universe increased. However, because the energy is conserved the energy lost from the fluid via the work done goes into gravitational potential energy. To generalise, for an adiabatically expanding volume the entropy per unit co-moving volume is conserved, and the expansion of the Universe causes an increase or decrease of its internal energy depending on whether the pressure $P$ is smaller or larger than zero.

For a given equation of state, $P(\epsilon)$, the fluid equation gives the density and pressure as a function of the scale factor $a$. The equation of state, which describes the energy content of the Universe, is often parametrized with $w_{i}$ in the following way:
\begin{equation}\label{eq:equation_of_state}
P_{i}(\epsilon) = w_{i} \epsilon_{i} c^{2} \quad .
\end{equation}
Here the subscript $i$ denotes the species of the material. Note, however, that this so-called perfect fluid hypothesis holds only for material whose pressure is directly related to its density. Finally, if $w$ is time-independent, then substituting Eq. \ref{eq:equation_of_state} into \ref{eq:fluid_equation} gives the time evolution of the mean energy density of the Universe as follows:
\begin{equation}\label{eq:density_in_Universe}
\epsilon_{i} \propto a^{-3(1 + w_{i})} \quad .
\end{equation}

The fluid equation together with the equation of state allow the derivation of the mean energy density, temperature and pressure of the Universe at any redshift from their values at the present time. At early times the Universe is assumed to be radiation dominated thus we can approximate that the energy content of the Universe is dominated by an ultra-relativistic radiation fluid for which $w_{rad} = \frac{1}{3}$. As a result, the mean energy density evolves proportional to $a^{-4}$. After the nucleosynthesis, but before the recombination, at  $z_{eq} \sim 3196$\footnote{The exact time depends on the matter density of the Universe and the Hubble constant. The redshift given is for the reference values, see Table \ref{tb:lambdaCDM_values}.}, the densities of non-relativistic matter and relativistic radiation are equal. However, after this point the Universe turns into a matter dominated. Now, a non-relativistic gas can be approximated with a fluid of zero pressure\footnote{sometimes referred to as dust} and we take $w_{mat} = 0$. As a consequence, the mean energy density evolves $\propto a^{-3}$. Finally, at recent epochs the energy density seems to have become dominated by vacuum energy. In order to keep a constant energy density as the Universe expands, the pressure must be negative, therefore, for vacuum energy we take $w_{vac} = -1$. Note that $w = -1$ can also be taken for the cosmic inflation.

Table \ref{tb:thermodynamics_of_universe} summarises the evolution of energy density, pressure, and temperature as a function of the scale parameter $a$. Note, however, that these scaling relations only hold if the equation of state remains constant with respect to time, while in reality such a simplification may not hold on all times. It should also be kept in mind that although the contribution of baryons and photons to the present day energy budget is small, they make an important contribution to shaping the matter power spectrum. Moreover, in realistic cases the Universe is not made out of a single material component as presented above. Fortunately, each material, baryons, photons, dark matter, neutrinos, dark energy, etc., obey their own fluid equation containing the appropriate expression for its pressure if the fluids are non-interacting. One can therefore take a linear combination of the terms and substitute that into the Friedmann equations in case of more realistic models.

\begin{table}[htb]
\vspace{+2mm}
\caption{Thermodynamics of a homogenous and isotropic universe.}
\label{tb:thermodynamics_of_universe}
\vspace{-5mm}
\begin{center}
\begin{tabular}{lrccc}
\hline 
Dominant component & $w$ & Energy density $\epsilon$ & Pressure $P$ &
Temperature $T$ \\
\hline\hline
Radiation 	& $\frac{1}{3}$ & $a^{-4}$ & $a^{-4}$ & $a^{-1}$ \\
Matter 		& $0$			& $a^{-3}$ & $a^{-5}$ & $a^{-2}$ \\
Vacuum energy & $-1$		& $1$	& $1$ & $-$ \\
\hline\hline
\end{tabular}
\end{center}
\end{table}

\section{Ingredients of structure formation}\label{s:structure_formation}


The ultimate goal of a structure formation theory is to describe how the phase transition progressed from almost perfectly homogeneous initial fields to all the structure we observe. To get closer to achieving this goal - to model the formation of structure in an evolving background universe - we must next concentrate on small density perturbations. A realistic structure formation model must be able to describe the evolution of the density field in the Universe with time when the field contains small fluctuations. The usual approach is to model the fluctuations as a perturbation to a smooth background which we assume is homogeneous and isotropic.


The key idea of any structure formation model is that if there are small perturbations, i.e., fluctuations in the energy density of the early Universe, then gravitational instability can amplify them leading to virialized structures such as the galaxies, groups, and clusters we observe today. To model the formation of structure in a realistic, self-consistent, and physical way several ingredients are required \citep[][]{2002coec.book.....C}:
\begin{enumerate}
  \item a background cosmology (Section \ref{s:dynamics_of_universe}),
  \item an initial fluctuation spectrum (Section \ref{s:initial_fluctuations}),
  \item a choice of fluctuation mode and a statistical distribution of fluctuations,
  \item a Transfer function (Section \ref{s:transfer_function}),
  \item a recipe for the nonlinear evolution (Sections \ref{s:non_linear_evolution_analytical_methods} and \ref{s:cosmological_simulations}), and
  \item a prescription to relate mass fluctuations to observable light (Section \ref{galaxy_formation}).
\end{enumerate}
As can be seen from the comprehensive list above, detailed modelling of structure formation involves several different ingredients. All of which interact in a complicated manner. To complicate the matter even further, most of the above items involve one or more assumptions \citep[see e.g.][]{2002coec.book.....C}. It should, however, be kept in mind that most of the assumptions are physically motivated, albeit this does not exclude the possibility that they are inaccurate or even incorrect.

\begin{figure}[htb]
\center{
\includegraphics[scale = 0.5]{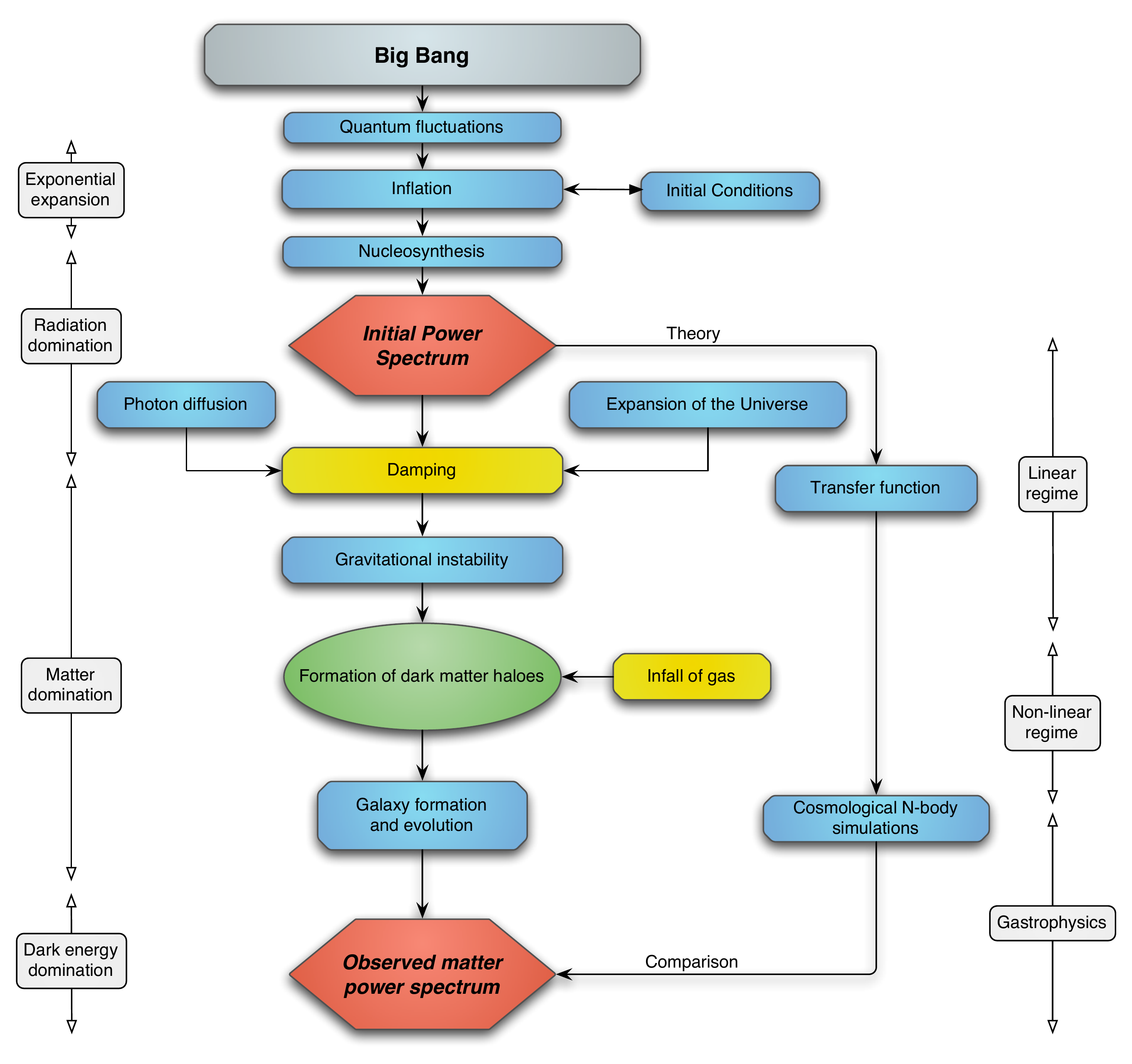}
}
\vspace{-5mm}
\caption{A logic flow chart for the formation of structure. The exponential expansion refers to the expansion of the Universe during inflation. The radiation, matter and dark energy domination refers to the dominant energy content during a given time period, while the linear and nonlinear regimes refer to the evolution of the density perturbation field. The gastrophysics indicate the regime where hydrodynamics and also forces other than gravity play a significant role. Red hexagons show the initial and the observed power spectrum. Galaxy formation is described in more details in Figure \ref{fig:galaxy_formation_flow_chart}.}
\label{fig:structure_formation_flow_chart}
\end{figure}

Figure \ref{fig:structure_formation_flow_chart} shows a logical flow chart describing the formation of structure. The chart shows how structures start to form from small initial density field fluctuations after the Big Bang and proceed through various steps to galaxy formation and to the matter power spectrum observed, for example, in large galaxy surveys (Section \ref{s:large_galaxy_surveys}). At present, the physical mechanism that can best describe the initial density field is inflation \citep{Guth:1981p37, Mukhanov:1981p1009, 1982PhLB..108..389L, Narlikar:1991p858}. The initial conditions of the Universe are thought to arise from the scale-invariant quantum-mechanical zero-point fluctuations of the scalar field that drove the inflation in the very early Universe \citep{Guth:1982p859, 1982PhLB..115..295H, Baumann:2007p979}. Inflation predicts, for example, that the initial fluctuations are adiabatic (i.e. perturbations are in thermal equilibrium) and behave as a Gaussian random field with a nearly scale invariant spectrum. In the following Sections I will therefore concentrate on adiabatic fluctuations with a Gaussian random phase and leave out isocurvature perturbations and non-Gaussian random fields (this is the structure formation ingredient III).

The first ingredient of the structure formation describes the global evolution of the background universe (Section \ref{s:dynamics_of_universe}), and is usually done using Friedmann equations (Section \ref{s:Friedmann_equations}). The rest of the ingredients are related to the formation and evolution of density perturbations under gravity in an expanding background universe. These will be described in the following sections in more detail: the spectrum of the initial fluctuations and the Transfer function are discussed in Section \ref{s:evolution_perturbations}, while the nonlinear evolution using both analytical methods and cosmological $N-$body simulations is explored in Sections \ref{s:non_linear_evolution_analytical_methods} and \ref{s:cosmological_simulations}, respectively. Finally, to form realistic galaxies that can be compared to the galaxies observed in large galaxy surveys at least hydrodynamics, baryonic physics, and star formation \citep[e.g.][]{McKee:2007p763} must be considered. One realisation for modelling gas, star formation, and feedback processes is the semi-analytical models of galaxy formation, which are discussed in Section \ref{s:SAMs}. However, before exploring the theory of structure formation, lets introduce the most successful structure formation model so far in more detail.

\subsection{The $\Lambda$ Cold Dark Matter model}\label{s:background_cosmology}

The $\Lambda$ Cold Dark Matter ($\Lambda$CDM) -model is nowadays accepted by the majority of astronomers as a standard model of Big Bang cosmology and cosmological structure formation.  The success of $\Lambda$CDM is widely recognised and is due to its simplicity, yet at the same time, it has the capability to simultaneously explain several profound observations of the Universe. $\Lambda$CDM can explain the structure and existence of the cosmic microwave background, the large-scale structure of galaxy groups and clusters, weak and strong gravitational lensing, and the accelerated expansion of the Universe inferred from type Ia supernovae observations \citep[e.g.][]{Narlikar:2001p870}. The statistical analysis of observations (Section \ref{s:obs_cosmology}) strongly support the flat $\Lambda$CDM cosmological model with the total energy density equal to the critical density. Table \ref{tb:lambdaCDM_values} summarises the energy budget and the values of the basic $\Lambda$CDM parameters based on WMAP results \citep{Hinshaw:2009p709, Komatsu:2009p710, Dunkley:2009p754, Jarosik:2010p770}.

The $\Lambda$ term of the standard cosmology stands for the Einstein's cosmological constant, often assumed to be the vacuum energy of space, and dubbed as dark energy due to its unknown origin \citep{Narlikar:2001p870, Frieman:2008p764}. The energy budget of the model is dominated by this unknown dark energy; the $\Lambda$-term constitutes almost $75$ per cent of the total energy composition. The remaining $\sim 25$ per cent of the energy budget is matter, however, about $85$ per cent of the matter is assumed to be in form of non-baryonic dark matter. Because the dark matter particles are assumed to be non-relativistic, the dark matter is said to be ``cold''. The term cold dark matter was introduced by Peebles and Richard Bond in $1982$ to cover the wide range of particles that were (and have been) suggested for the origin of this unknown gravitating material. Note that the coldness of the dark matter particles is actually required by the large-scale structure: hot dark matter, i.e., relativistic particles, does not predict enough structure on small scales.

The shape of the $\Lambda$CDM power spectrum is such that structures form from smaller to larger structures, i.e., ``bottom-up'', with galaxies forming first followed by the formation of groups and clusters. More general, in all CDM-models, independent of the $\Lambda$-term, the initial density fluctuations have larger amplitudes on smaller scales, thus the CDM-models are hierarchical; larger structures form by clustering of smaller objects via gravitational instability \citep[e.g.][]{Davis:1985p845, Frenk:1988p874, White:1991p851, Bullock:2001p893}. In CDM-models the collapse of matter happens when a local perturbation starts to turn around while the Universe is expanding. The process of collapsing continues until the internal velocity of system's components are large enough to hold the system against more collapse. As a result, a dark matter halo is formed. Note that unlike baryonic matter, the behaviour of dark matter does not depend on the scale of the system since dark matter only interacts gravitationally. Thus, in the $\Lambda$CDM-model dark matter haloes with different sizes and masses are scaled versions of each other.

\begin{table}[htb]
\vspace{+7mm}
\caption{Values of the basic $\Lambda$CDM parameters based on the WMAP
7-year results \citep{Jarosik:2010p770}.}
\label{tb:lambdaCDM_values}
\vspace{-5mm}
\begin{center}
\begin{tabular}{lcl}
\hline 
Quantity & Symbol & Value $\pm$ Error\\
\hline\hline
Total Density & $\Omega_{t}$ & $1.080^{+0.093}_{-0.071}$\\
Dark Energy Density & $\Omega_{\Lambda}$ & $ 0.734 \pm 0.029$ \\
Matter Energy Density& $\Omega_{m}$ & $0.258 \pm 0.030$ \\
Dark Matter Density & $\Omega_{dm}h^{2}$ & $0.1109 \pm 0.0056$ \\
Baryonic Matter Density & $\Omega_{b}h^{2}$ & $0.02258 \pm 0.00057$ \\
Hubble Parameter & $h$ & $0.710^{+0.013}_{-0.014}$ \\
Power Spectrum Normalisation & $\sigma_{8}$ &  $0.801 \pm 0.030$ \\
Scalar Spectral Index & $n_{s}$ & $ 0.963 \pm 0.014$ \\
Redshift of Matter-radiation Equality & $z_{eq}$ & $3196^{+134}_{-133}$\\
Redshift of Decoupling & $z_{\star}$ & $1090.79^{+0.94}_{-0.92}$\\
Age of Decoupling & $t_{\star}$ & $379164^{+5187}_{-5243}$ yr\\
Sound Horizon at Decoupling & $r_{s}(z_{\star})$ & $146.6^{+1.5}_{-1.6}$ Mpc \\
Redshift of Reionization & $z_{re}$ & $10.5 \pm 1.2$ \\
Reionization Optical Depth & $\tau$ & $0.088 \pm 0.015$ \\
Age of the Universe & $t_{0}$ & $13.75 \pm 0.13$ Gyr\\
\hline\hline
\end{tabular}
\end{center}
\end{table}

Even though the $\Lambda$CDM-model is widely accepted, there are still many unknowns. For example, several particle candidates exist for dark matter \citep[see e.g.][]{2004astro.ph.12170B, 2004IJMPA..19.3093M, Bertone:2004pz}, however, none has been observed thus far. One of the leading candidates for the dark matter particle is the lightest stable supersymmetric particle called neutralino, which is weakly interacting and massive, but several other candidates, for example Axions, exist. Hence, the type of the dark matter particles is yet to be confirmed. The nature of dark energy is even more mysterious, though the leading candidate is the vacuum energy of space \citep{Peebles:1988p1017, Frieman:2008p764}. As both dark matter and energy reveal themselves only via gravity all attempts to detect them directly have been unsuccessful thus far. A significant amount of work remains therefore to be done before the formation of structure and cosmology can be considered as fully understood.

\section{Evolution of initial perturbations: structure formation}\label{s:evolution_perturbations}

The present consensus in cosmology, as seen in the previous Sections, is that the observed structures developed from small initial perturbations of the physical fields (density, velocity, gravitational potential, etc.) resulting from the instability of the Friedmann models for small perturbations. Hence, to model structure formation we must model how the density perturbation field evolves. A full treatment would require GR and would proceed by perturbing the background metric and the energy-momentum tensor \citep[see e.g.][]{WeinbergCosmology}. However, a fully relativistic treatment is beyond the scope of this introduction to the formation of structure, thus I will describe a Newtonian method, which gives an excellent approximation. Even so, it should be kept in mind that the following analysis holds only for perturbations on scales much smaller than the Hubble radius (i.e. on sub-horizon scale), because the Newtonian description assumes instantaneous gravity (i.e. the speed of gravity has been assumed to be infinite).

\subsection{Growth of small perturbations in an expanding universe}\label{s:linear_pertrubation_theory}

The hierarchy of cosmic structures is assumed to have grown from primordial density seed fluctuations, which can be described with the density contrast $\delta(\mathbf{x}, t)$. The density contrast, as a function of the co-moving coordinates $\mathbf{x}$, motivates the study of the density perturbation field. It can be defined as
\begin{equation}\label{eq:density_contrast}
\delta(\mathbf{x}, t) = \frac{\rho(\mathbf{x}, t) -
\bar{\rho}(t)}{\bar{\rho}(t)} \quad ,
\end{equation}
where $\bar{\rho}(t)$ is the mean density. A critical feature of the $\delta$ field is that it inhabits a universe that is isotropic and homogeneous in its large-scale properties.

In order to describe the structure formation in an expanding universe we must follow the evolution of the initial perturbation field as a function of time, while gravitation magnifies the perturbations in both the baryonic and dark matter distribution. After recombination the amplitude of the density fluctuations is $\sim 10^{-3}$. Thus, the onset of structure formation happens well within the linear regime where $\delta \ll 1$. Consequently, a linear perturbation theory can be used as long as the density field does not turn nonlinear $(\delta \gtrsim 1)$ \citep[for a comprehensive presentation, see e.g.][]{2002coec.book.....C}.

On large scales matter can be described with a perfect fluid approximation. Thus, at any given time matter can be characterised by the energy density distribution $\epsilon (\mathbf{x}, t)$, the entropy per unit mass $S (\mathbf{x}, t)$, and the vector field of three-velocities $\mathbf{v} (\mathbf{x}, t)$. These quantities satisfy the hydrodynamical equations that allow the study of the behaviour of small perturbations in a homogeneous, isotropic background. The equations of motion for a non-relativistic fluid are the continuity, Euler, and Poisson equation. The continuity equation states that the change in the mass inside an element of the fluid equals to the mass convected into the element, i.e., it defines the conservation of mass. On the other hand, the Euler equation states that the acceleration of a small fluid element is due to the difference in pressure acting on opposite sides of the element, while the Poisson's equation describes the relation between the potential fluctuations and the density perturbations causing them.

In their basic form \citep[see e.g.][]{2002coec.book.....C, TheGreatBook} the continuity, Euler and Poisson equation hold for a smooth background. However, in this Section we derive the evolution of the density perturbations and hence we must consider small perturbations to the background. All the quantities involved must be then written as a sum of the smooth background and the perturbed quantity, e.g., in case of pressure: $P = \bar{P} + \delta P$. Here, $\bar{P}$ corresponds to the smooth background pressure field, while $\delta P$ is a small perturbation to this smooth component. The perturbed quantities can then be substituted to the basic continuity, Euler and Poisson equation. After ignoring terms higher than the first order in the perturbation and subtracting the zero order equations one obtains the linear perturbation forms of these equations. Thus, it is possible to describe the evolution of the density, velocity, potential and pressure fields in an expanding universe with Newtonian gravity by using the perturbed continuity, Euler, and Poisson equations, together with the conservation of entropy \citep[for a full derivation, see e.g.][]{1980lssu.book.....P, TheGreatBook, Mo_big_book}.

The linear perturbation theory, briefly described above, holds that, during the matter-dominated era, the density field $\delta$ of sub-horizon perturbations can be described with the growth equation as follows
\begin{equation}\label{eq:gravit_instability}
\frac{\partial^{2} \delta(\mathbf{x}, t)}{\partial t^{2}} +
2H(t)\frac{\partial \delta(\mathbf{x}, t)}{\partial t} = 
4 \pi G \bar{\rho} \delta(\mathbf{x}, t) + \frac{c_{s}^{2}}{a^{2}(t)}
\nabla^{2} \delta (\mathbf{x}, t)  \quad.
\end{equation}
Here $H(t)$ is the Hubble parameter (defined in Eq. \ref{eq:HubbleParameter}), $c_{s}$ is the speed of sound, $\nabla$ denotes the differentials with respect to co-moving coordinates, and $\frac{\partial}{\partial t}$ is a partial time derivative. The above second-order growth equation has been written in a general\footnote{Setting $c_{s} = 0$ gives the often shown form of the growth equation.} but single-fluid form as a function of cosmic time $t$ and co-moving coordinate $\mathbf{x}$. This gives a linear approximation for the growth of density perturbations in an expanding universe. Note that the second term on the left-hand side is the so-called Hubble drag term, which tends to suppress perturbation growth due to the expansion of the Universe. On the other hand, the first term on the right-hand side is the gravitational term, which causes perturbations to grow via gravitational instability, while the last term on the right-hand side is a pressure term and is due to the spatial variations in density.

The solution to the growth equation (\ref{eq:gravit_instability}) apply to the evolution of a single Fourier mode of the density field. However, in the linear regime, the equations governing the evolution of the perturbations are all linear in perturbation quantities. It is then useful to expand the perturbation fields in chosen mode functions. If the curvature $K$ of the Universe can be neglected, the mode function can be chosen to be plane waves. Now the perturbation fields can be represented by their Fourier transforms. We therefore seek a wave solution for $\delta$ of the form
\begin{equation}
\delta(\mathbf{x}, t) = \displaystyle \sum_{\mathbf{k}} \delta_{\mathbf{k}} (t)
e^{i \mathbf{k} \cdot \mathbf{x}} \quad .
\end{equation}
We can now write a wave equation for $\delta$ after taking a Fourier transform of equation \ref{eq:gravit_instability}. Because each $k$ mode is assumed to evolve independently, we can write
\begin{equation}\label{eq:gravit_instability_four}
\ddot{\delta}_{\mathbf{k}} + 
2H\dot{\delta}_{\mathbf{k}} +
\left (\frac{k^{2} c_{s}^{2}}{a^{2}(t)} - 
4 \pi G \bar{\rho} \right ) \delta_{\mathbf{k}} = 
0 \quad .
\end{equation}
The equation \ref{eq:gravit_instability_four}, sometimes called the Jeans equation, describes the evolution of each of the individual modes $\delta_{\mathbf{k}}(t)$, corresponding to $\delta(\mathbf{x}, t)$. Note that because $\mathbf{x}$ (in Eq. \ref{eq:gravit_instability}) is in co-moving units, the wave-vectors $\mathbf{k}$ are also and that the derivatives (denoted by dots) are time derivatives, because $\delta_{\mathbf{k}}$ does not explicitly depend on a spatial position.

If the dark matter is cold and collisionless we can neglect the pressure term in Eq. \ref{eq:gravit_instability_four}. This allows us to write a general solution in a form of two linearly independent power laws, i.e., 
\begin{equation}\label{eq:growing_decaying_solution}
\delta_{\mathbf{k}}(t) =
A_{1}D_{+}(\mathbf{k}, t) + 
A_{2}D_{-}(\mathbf{k}, t) \quad ,
\end{equation}
where $A_{1}$ and $A_{2}$ are constants to be determined by initial conditions. The growth (or Jeans) equation therefore has two solutions: a growing $(+)$ and decaying $(-)$ mode. The latter is hardly interesting for structure formation, thus, hereafter we concentrate on the growing mode. The growing mode is described by the growth factor $D_{+}$ defined such that the density contrast at the time $t$ is related to the density contrast today $\delta(t_{0})$ by 
\begin{equation}
\delta(t) = \delta(t_{0}) \frac{D_{+}(t)}{D_{+}(t_{0})} \quad .
\end{equation}

It is useful to note that the solution to equation
\ref{eq:gravit_instability_four} can either grow or decrease depending on the
sign of the
\begin{equation}
k_{J} = \left (\frac{k^{2}c_{s}^{2}}{a^{2}(t)} - 4 \pi G \bar{\rho} \right )
\quad .
\end{equation}
The density perturbations can grow only if the second term in the above equation dominates, while the transition takes place at the wave-number for which the two terms are equal, at
\begin{equation}\label{eq:jeans_wavenumb}
k_{J} = \frac{a(t)}{c_{s}}\sqrt{4 \pi G \bar{\rho}}
\quad .
\end{equation}
We can now write \ref{eq:jeans_wavenumb} in terms of the physical wavelength using the simple relation $k = 2 \pi a \lambda^{-1}$, and by doing so obtain the Jeans length:
\begin{equation}\label{eq:jeans_length}
\lambda_{J} = c_{s} \sqrt{\frac{\pi}{G \bar{\rho}}} \quad ,
\end{equation}
which defines a scale length on which structures can grow. In general, on scales smaller than the Jeans length, i.e., $\lambda < \lambda_{J}$ $($or $ k > k_{J})$, the solution to the Jeans equation corresponds to a sinusoidal sound wave; so pressure can counter gravity. Due to the damping caused by the Hubble drag term, there is no growth of structure for sub-Jeans scales, but the solution is oscillatory. Instead, on scales longer than the Jeans length, but smaller than the horizon, the pressure can no longer support the gravity and the solution can grow. As the Jeans length is time dependent in an expanding universe, for example, before the recombination $\lambda_{J} \sim 2500$ Mpc, while after the Jeans length is only $\sim 10$ kpc, a given mode $\lambda$ may switch between periods of growth and stasis governed by the evolution of $\lambda_{J}$. But what does all this mean for the growth of perturbations?

The growth rate of a density perturbation depends on epoch or more precisely on what component dominates the global expansion, whether a perturbation $k$-mode is super- or sub-horizon, and the Jeans length. As already noted in the case of the evolution of the energy density (Section \ref{s:fluid_equation}), the early Universe was assumed to be radiation dominated until the time of matter-radiation equality $z_{eq}$. During the epoch of radiation domination the Universe can be taken as flat and for $k \to 0$ the growth equation can be solved by $\delta \propto t^{n}$ \citep[for a detailed derivation, see e.g.][]{2002coec.book.....C}. Thus, for all perturbations, the growing mode (here we ignore the decaying mode) outside the horizon (on super-horizon scales) grows as $\delta \propto t \propto a^{2}$. Instead, on the sub-horizon\footnote{Note that when the universe was radiation dominated $c_{s} = c (\sqrt{3})^{-1}$ and thus the Jeans length is always close to the size of the horizon.} scales, the cold and collisionless dark matter, which has no pressure $(c_{s} = 0)$ of its own and is not coupled to photons, grows at most logarithmically $\delta_{\mathrm{dm}} \propto ln (a)$. After the matter-radiation equality, matter begins to dominate the dynamics. On the super-horizon scale all perturbations (dark matter, baryons, and photons) grow as $\delta \propto a$. Dark matter, being pressureless, grows with the same rate $(\delta_{\mathrm{dm}} \propto a \propto t^{\frac{2}{3}} \propto (1 + z)^{-1})$ \citep[e.g.][]{2002coec.book.....C} also on sub-horizon scales. However, baryons are still coupled to the radiation until the time of decoupling $z_{\star}$. As a result, the sub-horizon perturbations in the baryons cannot grow but instead oscillate. Finally, at the time of decoupling $z_{\star}$, the rate of collisional ionization does not dominate any longer and the baryons can decouple from the photons. At this point the baryonic perturbations can start to grow as $\propto a \propto t^{\frac{2}{3}}$ on scales $\lambda > \lambda_{J}$. On the smaller scales they instead continue to oscillate. Finally, at the latest times when the $\Lambda$-term is assumed to be dominant, the growing mode solution is $\propto \mathrm{constant}$.

Note, however, that in general the presentation in this Section applies only to adiabatic perturbations in a non-relativistic fluid with a single component. Fortunately though, the Newtonian perturbation theory is valid even with the presence of relativistic energy components, such as radiation and dark energy, as long as they can be considered smooth and their perturbations can be ignored. In this case they contribute only to the background solution.

\subsection{Statistical description of the initial fluctuations}\label{s:initial_fluctuations}

In the previous Section an equation (\ref{eq:gravit_instability}) describing the evolution of the density perturbation field (Eq. \ref{eq:density_contrast}) was presented. However, as was noted, it is often convenient to consider the density perturbation field by the superposition of many modes. The natural tool for achieving this is via Fourier analysis in case the co-moving geometry is flat or can be approximated as such (Eq. \ref{eq:gravit_instability_four}). In such case the power spectrum can be defined as
\begin{equation}
P(k) = \langle |\delta_{k}|^{2} \rangle \quad ,
\end{equation}
where the angle brackets indicate an average. As a result, the power spectrum describes how much the density field varies on different scales. Note that in an isotropic universe, the density perturbation spectrum cannot contain any preferred direction. Thus, we must have an isotropic power spectrum and we can write it simply as a function of wave-number $k$ rather than a vector. Even with such simplification, the power spectrum provides a complete statistical characterisation of a particular kind of stochastic process: a Gaussian random field \citep{1986ApJ...304...15B}.

Thus, the power spectrum can characterise the statistical properties of the cosmological perturbations. This is highly useful in order to be able to relate theory to observations. For example, results of large galaxy surveys (e.g., correlation functions, Section \ref{s:large_galaxy_surveys}) suggest that the spectrum of the initial fluctuations must have been very broad with no preferred scales. Because power spectrum $P(k)$ is related to a two-point correlation function by a Fourier transform (Eqs. \ref{eq:PowerSpectrum} and \ref{eq:two_point_correlation}), it is then natural to assume that the power spectrum of the initial fluctuations generated in the early phases of the Big Bang is of a power-law form. Thus,
\begin{equation}
	 P(k) = A k^{n} \quad ,
\end{equation}
where $A$ is the amplitude, $k$ is a wave-number in physical units $h^{-1}$Mpc and $n$ is a free parameter. Hence, the power spectrum also describes the normalisation of the spectrum of density perturbations on large physical scales.

Because there is yet no theory for the origin of the cosmological perturbations, the amplitude $A$ of the power spectrum has to be fixed by observations. The amplitude $A$ is therefore set using either the so-called COBE normalisation\footnote{In the COBE normalisation the amplitude of the large scale temperature anisotropies in the CMB are used to constrain the amplitude.} \citep[e.g.][]{Bunn:1997p1008} or by the variance of the density fluctuations within spheres of $8h^{-1}$ Mpc radius, $\sigma_{8}$ (for the reference model value, see Table \ref{tb:lambdaCDM_values}). The $\sigma_{8}$ parameter can be measured, for example, using the cosmic-shear autocorrelation function, the abundance and evolution of the galaxy-cluster population, the statistics of Lyman-$\alpha$ forest lines \citep{Seljak:2006p1025}, or by counting the number of hot X-ray emitting clusters in the local universe \citep[see e.g.][and references therein]{Bartelmann:2010p998}.

According to equation \ref{eq:two_point_correlation_isot} the power-law form of the power spectrum
\[
P(k) \propto k^{n}
\]
corresponds to a two-point correlation function of form
\begin{equation}
	\xi (r) \propto r^{-n -3} \quad .
\end{equation}
The mass $M$ within a fluctuation is $\propto \rho r^{3}$, thus, the spectrum of a mass fluctuation is
\begin{equation}
	\xi (M) \propto M^{\frac{-n-3}{3}} \quad .
\end{equation}
Finally, the root-mean-square (rms) density fluctuation at mass scale $M$ can be written as
\begin{equation}
\delta_{\mathrm{rms}} \propto M^{\frac{-n-3}{6}} \quad .
\end{equation}
As can be seen from the equations above, the spectral index $n$ has a significant role in the structure formation. If $n \neq 1$, the power spectrum is called tilted: a tilted spectrum is called ``red'' if $n < 1$ and ``blue'' if $n > 1$ \citep[for a review, see e.g.][]{1984NuPhB.244..541A, 1985PhRvD..32.1316L}. A red spectrum shows that there is more structure at large scales, while a blue spectrum describes that there is more structure at small scales. The special case is found when the spectral index $n$ equals unity.

\subsubsection{The Harrison-Zel'dovich power spectrum}

Simple inflationary theories predict that right after inflation the matter power
spectrum would have a simple power-law form. Consequently, the primordial, or 
Harrison-Zel'dovich \citep{1970PhRvD...1.2726H, 1972MNRAS.160P...1Z}, power
spectrum can be written as $P(k) = Ak^{n}$ with the spectral index $n$ equals
unity. The simple power-law form, $P(k) = Ak$, now results the spectrum of
density perturbations to have a following form
\begin{equation}
\delta_{\mathrm{rms}} \propto M^{- \frac{2}{3}} \quad ,
\end{equation}
while the two-point correlation function takes the form:
\begin{equation}
\xi (r) \propto r^{-4} \quad .
\end{equation}
The importance of the Harrison-Zel'dovich spectrum is the property that it is scale-invariant: the density contrast $\Delta (M)$ had the same amplitude $(\sim 10^{-4})$ on all scales when the perturbations came through their particle horizons during the radiation dominated era. Interestingly, the scale-invariant spectrum corresponds to a metric that is a fractal, leading to a fractal nature of the Universe \citep[e.g.][]{Jones:1988p944, Balian:1989p938}.

The current large-scale observations (see Section \ref{s:large_galaxy_surveys} and the citations therein) are reasonably well fit by an $n = 1$ scale-invariant primordial spectrum of perturbations. Theoretically the spectral index $n$ would be precisely unity if inflation lasts forever. As this is obviously not the case, the spectral index must however deviate slightly from the unity. It can be shown \citep[for a detailed discussion, see][]{2000cils.book.....L} that $n$ must be slightly smaller than unity, in agreement with the $\Lambda$CDM -model value ($n_{s}$ in Table \ref{tb:lambdaCDM_values}). It can also be shown that the spectral index of the temperature fluctuations (Fig. \ref{fig:WMAPmap}) as a function of angular scale depends only upon the spectral index $n$ of the initial power spectrum. Thus, for the Harrison-Zel'dovich power spectrum, the amplitude is independent of the angular scale.

\subsection{The Transfer function}\label{s:transfer_function}

If we wish to model how the form of the power spectrum evolves as a function of time, the statistical description of the initial fluctuations described by the initial power spectrum $P_{0}(k)$ must be evolved. During the evolution the form can be modified by several physical phenomena. For example, radiation and relativistic particles can cause kinematic suppression of growth of the initial perturbations. Moreover, the imperfect coupling of photons and baryons may also cause dissipation of perturbations. On the other hand, gravity will amplify the perturbations and eventually leads to collapsed and bound structures. Thus, real power spectra result from modification of any primordial power by a variety of processes: growth under self-gravity, the effects of pressure, and dissipative processes. In general, however, modes of short wavelengths have their amplitude reduced relative to those of long wavelengths.

A possible way to quantify how the shape of the initial power spectrum is modified by different physical processes as a function of time is to use a simple function of a wave-number, namely the Transfer function $T(k)$. For statistically homogeneous initial Gaussian fluctuations, the shape of the original power spectrum is changed by physical processes and the processed power spectrum $P(k)$ is related to its primordial form $P_{0}(k)$ via the Transfer function as follows
\begin{equation}
P(k, t) = \langle |\delta(k, t)|^{2} \rangle = P_{0}(k)T^{2}(k) D_{+}^{2}(t)
\quad .
\end{equation}
Here $D_{+}(t)$ is the solution of the linearised density perturbations equation (\ref{eq:gravit_instability_four}), i.e., the growth factor. Hence, once the Transfer function is known, one can calculate the post-recombination power spectrum from the initial conditions.


The form of the Transfer function is a function of the amount and type of the dark matter particles. As Section \ref{s:background_cosmology} described, the currently favoured dark matter particles are non-relativistic. Damping processes can also effect $T(k)$ during the linear evolution. As noted already, the cold dark matter does not suffer from strong dissipation, but on scales less than the horizon size at matter-radiation equality there is a kinematic suppression of growth on small scales. Additional complication to the form of the Transfer function arises from having a mixture of matter (both collisionless dark matter and baryonic plasma) and relativistic particles (collisional photons and collisionless neutrinos). One more complication for the shape of the Transfer function arises from the fact that sub-horizon perturbations grow differently during the radiation and matter dominated eras (Section \ref{s:linear_pertrubation_theory}). Due to the complicated form of the Transfer function it therefore must, in general, be calculated using an approximation formula, e.g., by \cite{Bond:1984p989, 1986ApJ...304...15B} or more precisely numerically using publicly available programs such as CMBfast\footnote{\url{http://www.cmbfast.org}} \citep{Seljak:1996p990}.



\subsection{Evolution of the initial power spectrum to the present time}

Section \ref{s:obs_cosmology} showed that the current matter power spectrum is far from its initial form, even though at scales $\lambda > 1000h^{-1}$ Mpc a Harrison-Zel'dovich power law is a good approximation. Thus, if one assumes that the initial power spectrum has a Harrison-Zel'dovich form after the inflation, it must have evolved significantly. As described above, the Transfer function can describe how the shape of the initial power spectrum evolves through the epochs of horizon crossing and radiation-matter equality. For the largest scales $(\gg 10h^{-1}$ Mpc$)$ the perturbations are still small even today, and one can use the Transfer function. However, for smaller scales such as galaxies, groups and clusters, the inhomogeneities have become so large at later times $(z < 100)$ that the physics of structure growth has become nonlinear. 

The caveats in mind, we can still provide an approximation for the form of the present-day power spectrum. If the mass fluctuations inside the co-moving horizon radius at matter-radiation equality are independent of time, and assuming that the dark matter is cold, the present day power spectrum can be approximated with a functional form
\[
P(k) \propto \left\{
\begin{array}{l l}
k^{n}			& \quad \mbox{when $k \ll k_{eq}$} \\
k^{n-4}\ln k 	& \quad \mbox{when $k \gg k_{eq}$} \\
\end{array}
\right. \quad ,
\]
where the co-moving wave-number $k_{eq} \sim 0.01h^{-1}$ Mpc. By and large this form is in agreement with Figure \ref{fig:matter_power_spectrum}, however, the real power spectrum has a smooth maximum, which is caused by the different rates of growth before and after the matter-radiation equality. Note that the co-moving wave-number describes the location of the peak and is set by the co-moving horizon radius at matter-radiation equality. The steep decline for structures smaller than the horizon radius reflects the suppression of structure growth during radiation domination.

The evolution of the matter power spectrum from its initial to the current form (see Fig \ref{fig:matter_power_spectrum}) can be summarised as follows. Before the matter-radiation equality, the matter distribution followed mostly that of the radiation. Because radiation has significant pressure perturbations were forced to oscillate on sub-horizon scales. Instead, the largest perturbations were too large for radiation pressure to be able to hold back the collapse, thus, perturbations on scales larger than the horizon scale were able to grow (with the rate given in Section \ref{s:evolution_perturbations}). This caused the matter power spectrum to increase in height and started to induce a bump at small scales (larger $k$). As the times went on, the horizon scale increased and larger scales were able to oscillate, causing the bump to shift to larger scales (smaller $k$). Instead, after matter-radiation equality, $z_{eq}$, the dark matter was able to grow with a rate given in Section \ref{s:evolution_perturbations}. Because the dark matter is assumed to be cold and collisionless it does not have significant pressure, and hence there were practically no more acoustic oscillations. During the matter dominated epoch the lack of pressure in dark matter allows it to continue to collapse, causing the whole of the matter power spectrum to increase. As a result, the turn-over point of the matter power spectrum is frozen into the power spectrum, which corresponds to the co-moving horizon radius at matter-radiation equality. Finally, closer to the current time, the small scale perturbations (high $k$) have turned nonlinear. Consequently, their growth is fast causing the matter power spectrum to rise on smaller scales. Quantitatively, such an evolution has been noted to lead to the current form of the matter power spectrum.

Unfortunately, the power spectrum can only describe the statistical properties of the density contrast, not the evolution of the individual density fluctuations. As a result, the evolution of individual perturbations as well as the nonlinear evolution on small scales must be studied using some other techniques. In such a case one must resort to, for example, analytical techniques or cosmological $N$-body simulations.

\section{Nonlinear evolution: I. Analytical methods}\label{s:non_linear_evolution_analytical_methods}

As the density contrast $\delta$ (Eq. \ref{eq:density_contrast}) approaches unity, the evolution of the density fluctuations becomes nonlinear. In the course of nonlinear evolution, overdensities contract, causing matter to flow from larger to smaller scales causing the power spectrum to deform. Even though the linear perturbation theory (Section \ref{s:linear_pertrubation_theory}) fails for $\delta \gtrsim 1$, the onset of nonlinear evolution can still be described analytically.

\subsection{Spherical top-hat model}\label{s:top_hat}

The simplest analytical model for the nonlinear evolution of a discrete perturbation is called the spherical top-hat model \citep[for a textbook review, see e.g.][]{1993sfu..book.....P}. In this approximation the perturbation evolves according to Birkhoff's theorem; in a spherically symmetric situation, matter external to the sphere will not influence its evolution. An evolving density perturbation will therefore eventually stop expanding, turns around, and collapses. Because in this model the perturbation has no internal pressure, it collapses to infinite density. Interpreting this literally leads to a conclusion that all spherical perturbations would result to black holes. In reality, this however has obviously not happened, and thus, one should be aware of its limitations.

In the spherical top-hat model a perturbation reaches a maximum size $r_{max}$ at the time of a turnaround $t_{turn}$. According to the model, the perturbation will then collapse at $2t_{turn}$. In a realistic case during the collapse the gravitational potential energy must be converted into kinetic energy of the particles involved in the collapse because a collisionless system cannot dissipate energy. This can be achieved, for example, via the process of violent relaxation \citep{LyndenBell:1967p1030}. The collapsed object will therefore eventually relax to a structure supported by random motions and satisfy the virial theorem, in which the internal kinetic energy of the system is equal to half of its gravitational potential energy (see Section \ref{ss:virial_theorem} and equation \ref{eq:virial_theorem}).

If one assumes that the relaxed object virialises at $t_{vir} = 2t_{turn}$, then a mean overdensity within a virial radius $r_{vir} = r_{max}/2$ can be derived using the virial theorem. The mean overdensity $\Delta_{vir}$ within the virial radius at the time of virialisation can now be written as
\begin{equation}\label{eq:mean_overdensity}
\Delta_{vir} = \frac{\rho(t_{turn})}{\bar{\rho}(t_{turn})}
\frac{\bar{\rho}(t_{turn})}{\bar{\rho}(t_{vir})}
\left ( \frac{r_{max}}{r_{vir}} \right )^{3}
- 1 \quad ,
\end{equation}
where $\bar{\rho}(t)$ is the background density at time $t$. In the case of a non-zero cosmological constant and for a flat universe, an approximation for the mean overdensity can be written
\begin{equation}\label{eq:mean_overdensity_approximation}
\Delta_{vir} \approx \frac{18 \pi^{2} + 82x -39x^{2}}{\Omega_{m}(t_{vir})} \quad ,
\end{equation}
where $x = \Omega_{m}(t_{vir}) - 1$ \citep{Bryan:1998p1029}. This simple approximation can be used to derive an average density of a virialized object formed through gravitational collapse in an expanding universe. In the simplest approximation $(\Omega_{m} = 1)$, this leads to $\Delta_{vir} \sim 18\pi^{2} \sim 178$. Note also that the virial theorem and the spherical collapse model can also be used to estimate the redshift at which the object became virialized.

One of the shortcomings of the spherical top-hat model is however the assumption that the perturbations were exactly spherically symmetric. Hence, a more general approximation should be considered.

\subsection{Zel'dovich approximation}\label{s:Zeldovich_approximation}

Given the fact that fluctuations of early times were small, it is reasonable to assume that at later epochs only the growing mode has a significant amplitude. Now, if one assumes that the density field grows self-similarly with time, the onset of nonlinear evolution can be described by the so-called Zel'dovich approximation \citep{ZelDovich:1970p999}. The Zel'dovich approximation is a form of the linear perturbation theory and is applicable to a pressureless fluid. The basic assumptions of the Zel'dovich approximation are as follows: $1)$ the scales of interest are much smaller than the size of horizon; $2)$ the universe is dominated by the matter component; and $3)$ the curvature of the universe is zero. The Zel'dovich approximation does therefore not assume spherical symmetry, like the top-hat model.

The Zel'dovich approximation is a Lagrangian description for the growth of perturbations and it specifies the growth of structure by giving the displacement and the peculiar velocity of each mass element in terms of the initial position \citep{ZelDovich:1970p999, Shandarin:1989p1001}. Furthermore, the Zel'dovich approximation is a kinematic approximation by nature: particle trajectories are straight lines. The first nonlinear structures to form in this approximation will be two-dimensional sheets, also called the Zel'dovich pancakes. However, the approximation is not valid after the formation of the pancakes when shell crossing will start to occur. Thus, other techniques to follow the evolution of density perturbations further into the nonlinear regime are required.

\subsection{The Press-Schechter formalism}

The Press-Schechter \citep[PS;][]{Press:1974p694} theory, which was derived heuristically using the linear growth theory and the spherical top-hat model (Section \ref{s:top_hat}) provides an analytical description for the evolution of gravitational structure in a hierarchical universe. In the PS formalism, an early universe is assumed to be well-described by an isotropic random Gaussian field of small density perturbations. Moreover, the phases of fluctuations are assumed to be random so that the field is entirely defined by its power spectrum \citep{Bower:1991p695}. The basic idea of the PS theory is to imagine smoothing the cosmological density field at any epoch $z$ on a given scale $R$ so that the mass scale of virialized objects of interest satisfies $M = \frac{4 \pi}{3}\bar{\rho}(z)R^{3}$. However, as noted earlier, the growth of the density perturbations can only be followed with simple analytical techniques until they become nonlinear. The PS formalism circumvents this difficulty by assuming that the region collapses rapidly and independently of its surroundings once it has turned nonlinear. As a result, the collapsed region can be described as a single large body to the rest of the universe. This simplification allows the linear equations to be applied, however, one must still take into account the nonlinear single body objects when modelling the formation of large-scale structure.

The PS formalism allows the modelling of the growth of cosmic structure in a highly simplified universe \citep{Cole:1991p857}. Perhaps more importantly, it also allows to estimate the mass function of the collapsed objects. The PS formalism leads to a halo mass function, which has the form of a power law multiplied by an exponential. While the PS formalism gives a reasonable approximation to the numerical data, it has been shown to underestimate the number of massive systems, while over-predicting the number of ``typical'' mass objects \citep[e.g.][]{Governato:1999p1028, Sheth:1999p1027, Jenkins:2001p1011}. Thus, a more realistic and detailed modelling of the nonlinear growth of structure in an expanding universe is required for more vigorous comparisons. This can be achieved, for example, by using numerical simulations.

\section{Nonlinear evolution: II. Cosmological simulations}\label{s:cosmological_simulations}

\subsection{Background}

Cosmological $N$-body simulations provide a robust method to study large-scale structure and the formation and growth of cosmic structure of a universe on the nonlinear regime. This is possible because the equations of motions are integrated numerically \citep[e.g.][and the references therein]{Springel:2001p591}. The basic idea of cosmological simulations was founded in the 1960s and it owes its existence to early few body simulations. The first cosmological simulations were run with a modest number of particles and only the formation of a few galaxies were followed. Instead, today large cosmological simulations use billions of particles, however, many simulations still use pure dark matter and no baryons. To overcome the issue that in dark matter only simulations galaxies must be placed by hand using, e.g., semi-analytical methods (Section \ref{s:SAMs}) hydrodynamics can also be modelled. Hydrodynamical simulations \citep{Katz:1991p1084, Navarro:1991p1083, Katz:1992p1082, Cen:1992p896, Rosswog:2009p772} that contain gas particles are gaining popularity with increasing computing power, but their volumes are still modest compared to dark matter only simulations \citep{Frenk:1999p881, Teyssier:2002p873, SommerLarsen:2003p889, Springel:2003p867, Dave:2010p945, Razoumov:2010p777, Tantalo:2010p813}. Moreover, even in hydrodynamical simulations some key (sub-grid) physics, such as star formation, is not modelled directly from the first principles but using similar prescriptions as in semi-analytical models.

Observations of the cosmic microwave background, discussed in Section \ref{s:CMB}, show that the perturbations of the gravitational potential are caused by non-relativistic material \citep[e.g.][]{Hu:2002p864}. Therefore, cosmological $N$-body simulations can usually operate on Newtonian limit without the framework of General Relativity. Even so, the expansion of the Universe must be taken into account. This is often done using a co-moving coordinate system, which moves as a function of time as the Universe ages and expands. Consequently, cosmological $N$-body simulation codes essentially follow the evolution of the density field $\delta$ in an expanding background by following the motions of particles caused via gravity by integrating the equations of motions numerically.

The basic assumptions of a modern cosmological $N$-body simulation code can be summarised as follows:
\begin{itemize}
  \item mass content of the Universe is build up mainly from dark matter;
  \item gravity is the only notable force on large scales;
  \item each dark matter particle in the simulation volume represents several particles of the real Universe and they are collisionless;
  \item a simulation starts from initial conditions with all modes well within the linear regime;
  \item periodic boundary conditions are adopted, resulting in that no particle can be lost during the simulation.
\end{itemize}
These notions provide the basic assumptions that most simulations obey. It is also noteworthy that the assumption that the dark matter particles are collisionless also means that the evolution of the Universe is driven by the mean gravitational potential rather than two-body interactions. In what follows, I briefly describe a general idea of generating initial conditions and how to follow the motions of particles.

\subsection{Setting up the initial conditions}\label{s:cosmo_initial_coditions}

For most galaxy formation and large-scale structure problems, setting up the initial conditions of a cosmological $N$-body simulation can be split into three parts:
\begin{enumerate}
	\item generating a power spectrum;
	\item generating a Gaussian random density field using the power spectrum;
	\item imposing density perturbation field on the particle distribution.  
\end{enumerate}
The first step, generation of a power spectrum, defines the dark matter model. The power spectrum can be generated, for example, by taking a primordial power spectrum (Section \ref{s:initial_fluctuations}) and then multiplying it with the Transfer function (Section \ref{s:transfer_function}) of a chosen cosmology. The second step sets up a ``smooth'' distribution of particles \citep[for a technical description, see e.g.][]{Martel:2005p1002} by generating a single realisation of the density field in $k$-space. The third step is to impose density perturbations with the desired characteristics, i.e., the assignment of displacements and velocities to particles. A suitable particle distribution, i.e., a linear fluctuation distribution can be generated using, for example, the Zel'dovich approximation \citep[Section \ref{s:Zeldovich_approximation}, but see also][]{Efstathiou:1985p1003}. Note that when following this technique the matter density and velocity fluctuations are initialised at the starting redshift chosen usually such that all modes in the simulation volume are still within the linear regime.

After the initial conditions have been set and the starting redshift has been chosen, the simulation can be evolved towards the current epoch by using a cosmological $N$-body code. In general, the code allows the time evolution of the simulated particles to be followed by integrating the equations of motions.

\subsection{Equations of motions}

Several different techniques to follow the gravitational evolution of the density field in cosmological $N$-body simulations exist. In this Section the basics are briefly introduced, while more detailed descriptions of different techniques can be found from the literature \citep[see e.g.][and references therein]{Efstathiou:1985p1003, Barnes:1986p1014, Couchman:1991p1012, Cen:1992p896, Xu:1995p1015, Kravtsov:1997p1013, Teyssier:2002p873, 2003gnbs.book.....A, Springel:2003p867, Bagla:2004p1004}.

In usual cases, dark matter (and stars if applicable) can be modelled as a self-gravitating collisionless fluid in cosmological simulations. Since the number of dark matter particles is large, two-body scattering events are assumed to be seldom. As a result, it is convenient to describe the system in terms of the single particle distribution function $f = f(\mathbf{x}, \mathbf{\dot{x}}, t)$ in phase space. Now, if we make a reasonable assumption that there are no collisions between particles, the evolution of the distribution function $f$ of the fluid follows, in the co-moving coordinates $\mathbf{x}$, the collisionless Boltzmann equation:
\begin{equation}
\frac{\partial f}{\partial t} + \dot{\mathbf{x}} \frac{\partial f}{\partial \mathbf{x}}
- \frac{\partial \Phi}{\partial \mathbf{r}} \frac{\partial f}{\partial \mathbf{\dot{x}}}
= 0 \quad ,
\end{equation}
where the self-consistent potential $\Phi$ is the solution of Poisson's equation
\begin{equation}
\nabla^{2} \Phi (\mathbf{r}, t) = 4 \pi G \int f(\mathbf{r}, \mathbf{\dot{x}},
t)\mathrm{d}\mathbf{\dot{x}} \quad .
\end{equation}
Here $f(\mathbf{r},\mathbf{\dot{x}},t)$ is the mass density of the single-particle phase space. Unfortunately though, the coupled equation pair consisting of the collisionless Boltzmann and Poisson equation is difficult to solve directly. Thus, simulations often follow the so-called $N$-body approach, where the smooth phase fluid $f$ is represented by $N$ particles which are integrated along the characteristic curves of the collisionless Boltzmann equation. Consequently, the problem is conveniently reduced to a task of following Newton's equations of motion for a large number of particles under their own self-gravity \citep{Springel:2001p591}.

The dynamics of $N$ particles can be described by the Hamiltonian:
\begin{equation}
\mathscr{H} = \displaystyle \sum_{i} \frac{\mathbf{p}_{i}^{2}}{2m_{i}a(t)^{2}} +
 \frac{1}{2} \displaystyle \sum_{i, j} \frac{m_{i}m_{j}\phi(\mathbf{x}_{i} - \mathbf{x}_{j})}{a(t)}
 \quad ,
\end{equation}
where $\mathbf{x}_{i}$ and $\mathbf{x}_{j}$ are the co-moving coordinate vectors, the corresponding canonical momenta are given by $\mathbf{p}_{i} = a^{2}m_{i}\mathbf{\dot{x}}_{i}$, and the $\phi(\mathbf{x})$ is the interaction potential. Note that the time dependency of the Hamiltonian $\mathscr{H}$ is caused by the time dependency in the scale parameter $a = a(t)$. Before the equations of motions of simulated particles can be derived the interaction potential has to be solved. When periodic boundary conditions are assumed the interaction potential can be solved from equation:
\begin{equation}
\nabla^{2}\phi(\mathbf{x}) = 4 \pi G \left[- \frac{1}{L^{3}} + \displaystyle
\sum_{n} \hat{\delta}(\mathbf{x} - L\mathbf{n})\right] \quad ,
\end{equation}
where $L$ is the side length of the simulation volume, $\mathbf{n} = (n_{1}, n_{2}, n_{3})$, and $\hat{\delta}$ is the particle density distribution function. Finally, after the interaction potential has been solved the Hamilton's equations of motions
\begin{eqnarray}
\frac{\partial \mathscr{H}}{\partial \mathbf{x}} &=& - \mathbf{\dot{p}} \\
\frac{\partial \mathscr{H}}{\partial \mathbf{p}} &=& \mathbf{\dot{x}}
\end{eqnarray}
can be derived. To follow the time evolution of the simulated particles the derived equations of motions must be integrated, after making a small variation $\delta t$ to time, via e.g. the ``leapfrog integration scheme'' \citep[e.g.][]{2008SSRv..134..229D}. Note that in most cases, the particle motion integrals are time-integrals and require integrating the scale parameter $a(t)$.

\subsection{Identifying dark matter haloes}

\begin{figure}[htb]
\center{
\includegraphics[scale = 0.41]{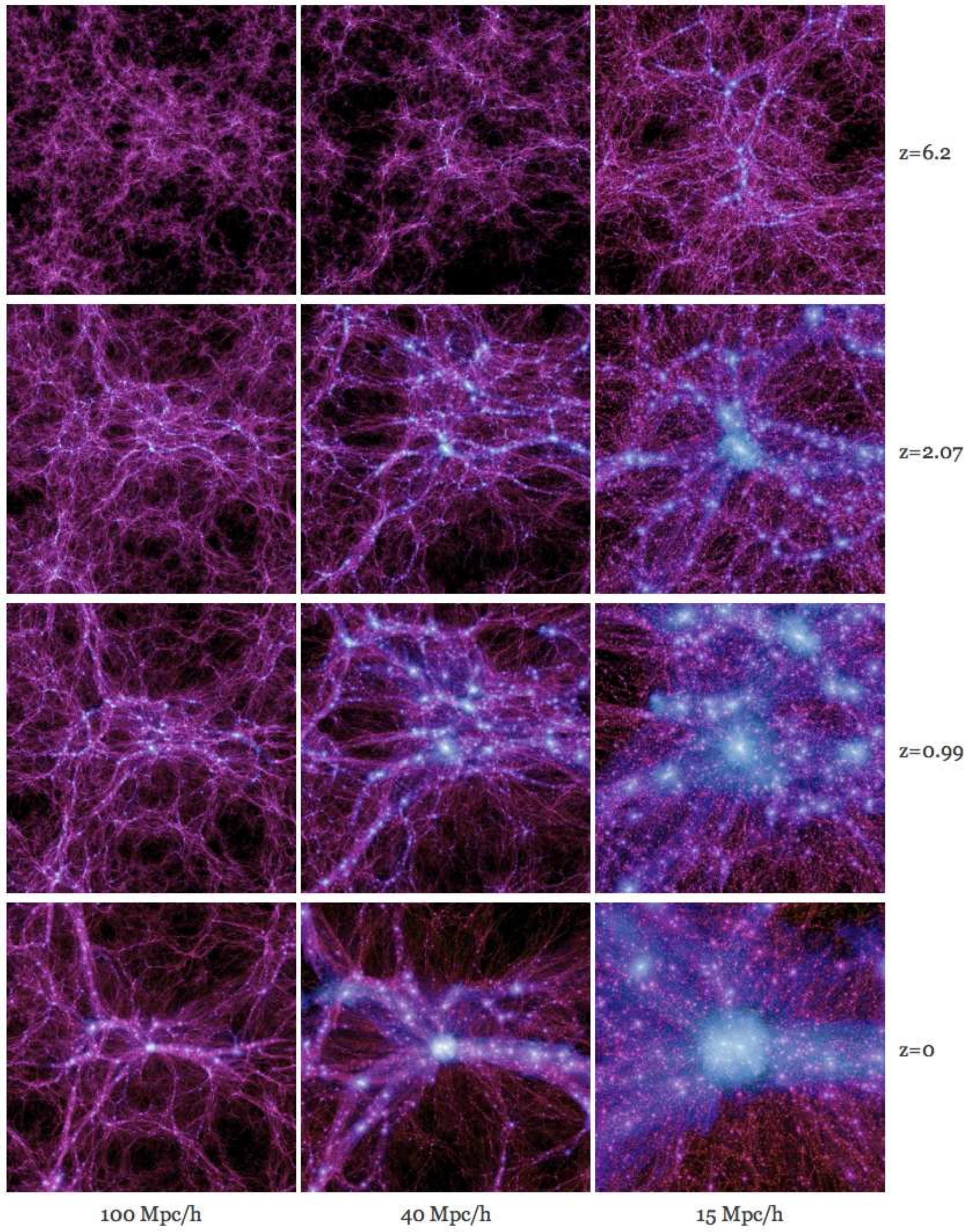}
}
\caption{Example of structure formation as a function of time in the Millennium-II simulation. Courtesy of Michael Boylan-Kolchin and the Millennium-II.}
\label{fig:millenniumTime}
\end{figure}

Far in the nonlinear regime, towards the end of a simulation run, bound structures start to form (for an example, see Fig. \ref{fig:millenniumTime}). After their formation they grow in mass either by accretion or by merging with other bound structures. These bound structures can be identified from simulations by using the so-called halo finders that search for collections of dark matter particles that are gravitationally bound. The bound structures of particles are called dark matter haloes due to their relatively spherical nature.

The most popular algorithm to identify virialized haloes is likely the so-called Friends-of-Friends halo finder \citep{Davis:1985p845}. This simple algorithm links all particles with distances less than a linking length to a single halo. In general, the linking length is set to correspond to the mean virialisation overdensity $\Delta_{vir}$ (Eq. \ref{eq:mean_overdensity_approximation}) derived using the spherical top-hat model (Section \ref{s:top_hat}). The linear linking length of the Friends-of-Friends halo finder is a free parameter, often taken to be a fraction, e.g. $0.2$, of the mean particle separation. For other halo finding algorithms, see for example \cite{Eisenstein:1998p1033, Neyrinck:2005p1034, Kim:2006p424, Knollmann:2009p1032} and references therein.

\begin{figure}[htb]
\center{
\includegraphics[scale = 0.29]{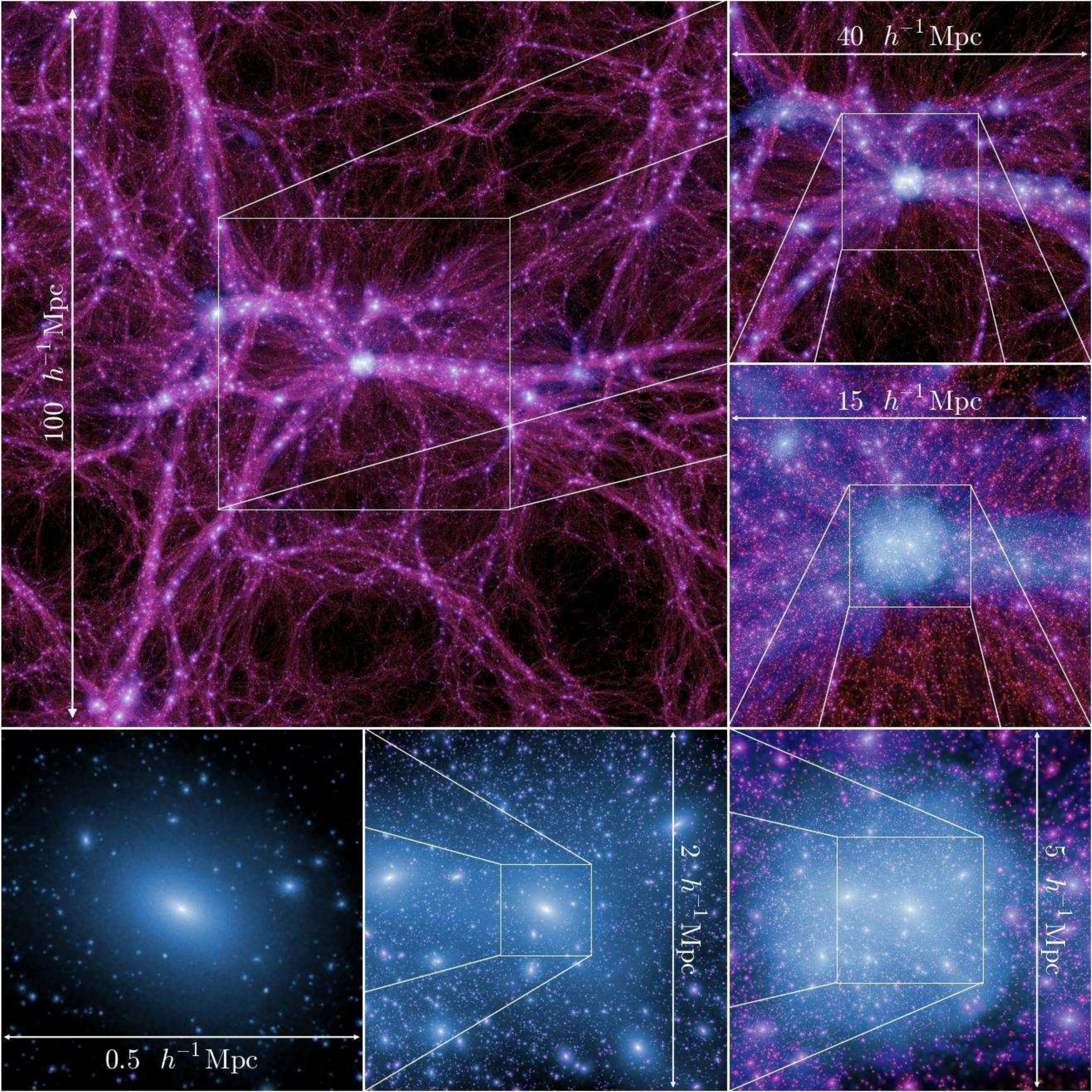}
}
\caption{A zoom sequence from $100$ to $0.5h^{-1}$ Mpc into the most massive halo in the Millennium-II simulation volume at redshift zero. Courtesy of Michael Boylan-Kolchin and the Millennium-II.}
\label{fig:millenniumZoom}
\end{figure}

Figure \ref{fig:millenniumZoom} shows an extraction from the Millennium-II simulation \citep{BoylanKolchin:2009p758}. The Figure shows a zoom sequence from $100$ to $0.5h^{-1}$ Mpc into the most massive halo in the simulation at redshift zero visualising the dynamical range of the simulation. Figure \ref{fig:millenniumTime} shows the structure formation as a function of time. This set of $12$ images shows the growth of the most massive halo over the cosmic time. The left column is $100h^{-1} \times 100h^{-1}$ Mpc, the centre column is $40h^{-1} \times 40h^{-1}$ Mpc, and the right is $15h^{-1} \times 15h^{-1}$ Mpc in co-moving units. From top to bottom, the regions plotted are at redshifts $6, 2, 1,$ and $0$.

\subsection{Resolution effects}

The simplified scheme of collisionless particles used in cosmological $N$-body simulations leads to finite mass- and force-resolution. In an ideal case the number of simulation particles should be as large as possible to enable a detailed study of formation and growth of cosmic structure in all scales ranging from the smallest dwarf galaxies to the largest clusters and filaments. However, in reality this choice is in general limited by the available computing resources. Moreover, the number of particles must also be balanced with the choice of a simulation volume size to compete against the cosmic variance.

The mass resolution $R_{m}$ of a simulation can be derived when the simulation volume and the number of dark matter particles have been chosen. The mass resolution, in units of solar mass $M_{\odot}$, describes the mass of a single dark matter particle and can be derived from
\begin{equation}\label{eq:mass_resolution}
R_{m} = C \frac{L^{3}}{N} \quad .
\end{equation}
Here $L$ is the side length of the simulation volume in Mpc, $N$ is the number of particles in the volume $L^{3}$, and $C \sim 7.496 \times 10^{10}$ is a constant. Note that the mass resolution sets a hard limit: no object below the mass resolution can form in a simulation. In reality, however, the smallest structures to form and which are identifiable must be made out of several tens of particles. This renders the effective mass resolution at least an order of magnitude worse than indicated by equation \ref{eq:mass_resolution}. Furthermore, a finite mass resolution also limits the ability to study the internal structures of dark matter haloes with masses close to the resolution limit.

On the other hand, a finite force resolution arises from the fact that the gravitational force between two particles diverges as their distance approaches to zero. In reality, however, the gravitational force between two extended objects is finite. The force resolution, which is more subtle effect than mass resolution, is more complicated to quantify because it depends on the simulation code used. In a mesh-based simulation code, the force is automatically softened on the scale of a chosen mesh. Instead, in particle-particle algorithms, the force softening is often applied artificially by modifying Newton's gravity law by writing it as:
\begin{equation}
F = G \frac{M^{2}}{r^{2} + \epsilon^{2}} \quad ,
\end{equation}
where $M$ is the mass of a single particle, $r$ is the distance between the two particles, and $\epsilon$ is the gravitational softening length. Non-zero softening length now guarantees that the force does not diverge even when $r \rightarrow 0$. However, by doing so the $\epsilon$ simultaneously sets a limit for the highest density contrast that can be resolved.

\section{Galaxy formation and evolution}\label{galaxy_formation}

\subsection{Background}


So far I have briefly shown how the formation and growth of large scale dark matter structure can be modelled from the CMB down to the current epoch. However, observations such as large galaxy surveys (Section \ref{s:large_galaxy_surveys}) cannot yet directly probe dark matter, thus one should also try to model the luminous component or baryonic matter, i.e. the galaxy formation and evolution. A detailed discussion of the theory of galaxy formation, is however beyond the scope of this introduction. Instead, in what follows I try briefly to summarise the main concepts of galaxy formation, illustrated in Figure \ref{fig:galaxy_formation_flow_chart}, to provide background for the following chapters. For more detailed presentations, I refer the interested reader to the great text books of \cite{TheGreatBook} and \cite{Mo_big_book}.

The previous sections showed how to model dark matter haloes. If we now assume that galaxies form and reside in dark matter haloes, it becomes obvious that the properties of the galaxy population are related to the cosmological density field and to the dark matter halo population. One can therefore try to link the properties of dark matter haloes to the properties of observed galaxies by using statistical arguments. As a result, it can be shown that the properties of the galaxy population depend on the properties of the dark matter halo and subhalo populations \citep[see e.g.][for detailed discussion]{Mo_big_book}. This allows, for example, the galaxy luminosity function to be compared to the dark matter halo mass function. The correspondence of light and mass is important, for example, when the mass power spectrum is being derived from observations (Section \ref{s:large_galaxy_surveys}), because a correlation between the observed light and the underlying mass must be assumed.

\subsection{Linking halo mass to galaxy luminosity}

To overcome the difficulty of linking dark matter haloes to luminous galaxies, \cite{Vale:2004p26} \citep[see also][]{Oguri:2006p863} proposed that a galaxy's luminosity can be related to its host dark matter halo's virial mass $M_{vir}$ via a simple relation as follows:
\begin{equation} 
	L(M_{vir}) = 5.7 \times 10^{9} h^{-2} L_{\odot} \frac{M_{11}^{p}}{\left[ q + 
	M_{11}^{s(p-r)} \right]^{1/s}} \quad , 
\end{equation} 
where $q$, $p$, $r$, and $s$ are free parameters and $M_{11}$ is scaled such that
\begin{equation} 
	M_{11} = \frac{M_{vir}}{10^{11}h^{-1}M_{\odot}} \quad . 
\end{equation} 
However, \cite{Cooray:2005p697} showed that the relation between the mass of a dark matter halo and the luminosity of the galaxy it hosts is not straightforward due to the complicated baryonic physics involved.

The baryonic content of a dark matter halo becomes dynamically important in the nonlinear regime when dark matter haloes are forming. Consequently, to model realistic galaxies, such as the Galaxy we live in, baryons must be modelled, albeit they do not contribute to the structure formation as much as dark matter. Hydrodynamical effects such as heating and cooling processes of gas, shocks, star formation, and feedback processes must all be taken into account when the formation of baryonic structure is being considered. One solution is to use hydrodynamical simulations to model gas directly, but the lack of fundamental theories for physical processes involved in the formation and evolution of galaxies, such as star formation, render them less than optimal. As a result, the hydro-simulations also require simple prescriptions for the so-called sub-grid physics, which cannot be modelled directly. What can be modelled then?

\subsection{Modelling of galaxy formation and evolution}\label{s:galaxyFormation}

Galaxy formation \citep[for a comprehensive view, see e.g.][]{TheGreatBook, Mo_big_book} is expected to proceed via a two-stage process originally outlined by \cite{White:1978p755}, but see also \cite{Hoyle:1953p1023, Binney:1977p1021, Rees:1977p1024, Silk:1977p1022} for early development. In this paradigm, the gravitational instability acting on collisionless dark matter results in the formation of self-gravitating dark matter haloes (as already noted earlier). Because baryons, initially well mixed with the dark matter, are assumed to ``feel'' the dark matter via gravity, they also participate in this collapse after the dark matter haloes have started to form. However, unlike the dark matter, the gas is not collisionless, but can dissipate. As a result, in a very simplified picture, the gas can be assumed to be heated by shocks to the virial temperature of the dark matter halo during this infall. After which, the hot gas can cool radiatively, on a time scale set by atomic physics.

During the collapse and cooling, the gas is assumed to condense to the cores of collapsed dark matter haloes. However, it is assumed that this process is not the same in all haloes. In smaller structures such as galaxy host haloes, the dominant physical process is cooling, which allows baryons to be more centrally concentrated than dark matter. In contrast, in larger structures, baryons experience a deeper gravitational potential and can therefore gain potential energy as they fall to the centre of a halo. This process heats baryonic matter and increases its temperature via shocks \citep[e.g.][]{Birnboim:2003p922}. As a result, baryonic matter experiences pressure forces which do not let them to be as concentrated as its host dark matter. In large-scale structures the process of cooling will therefore be highly important in the dense cores of dark matter where the gas can cool. Finally, the cold gas can fragment into stars, and a galaxy is born.

\begin{figure}[htb]
\center{
\includegraphics[scale = 0.365]{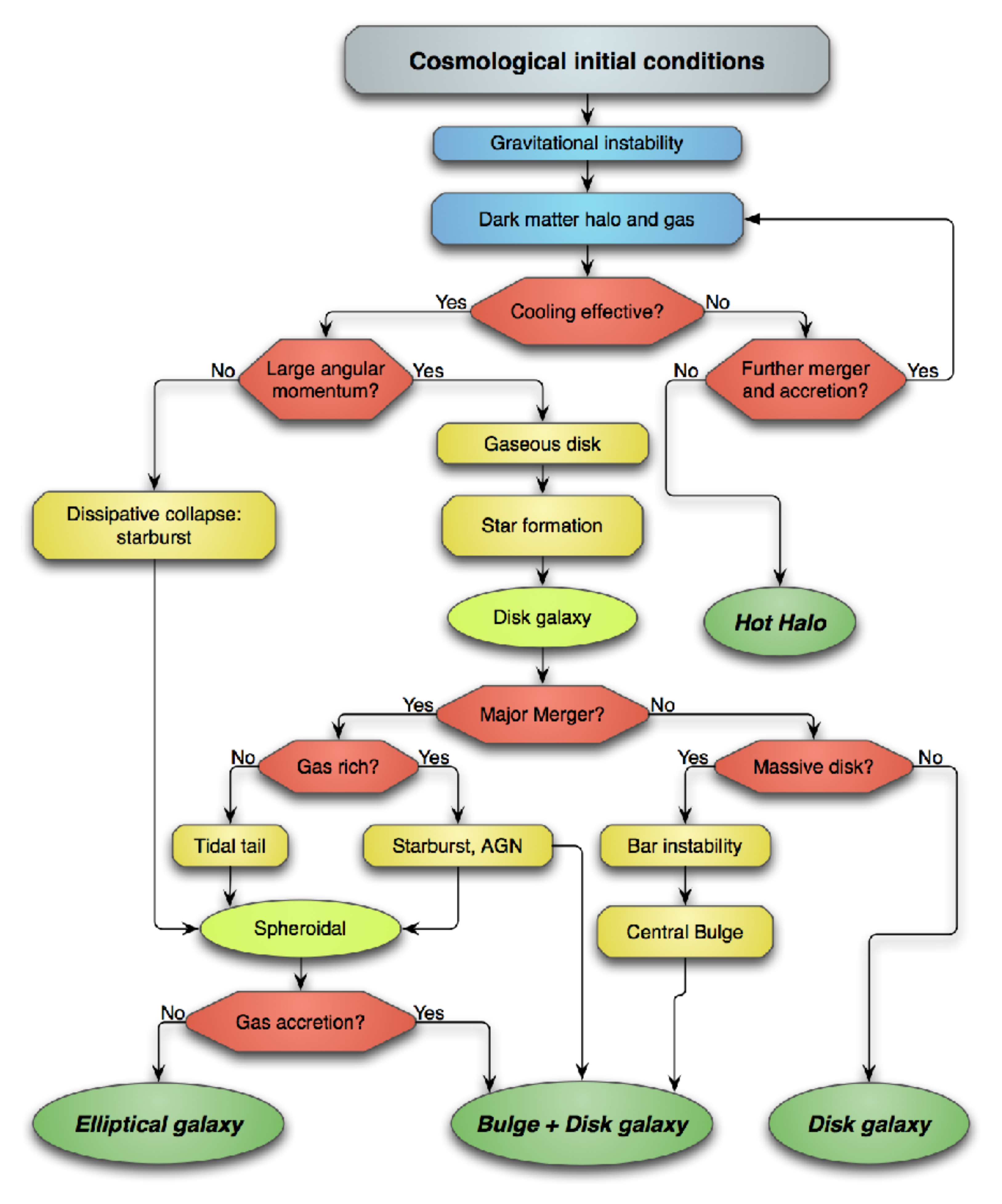}
}
\vspace{-3mm}
\caption{A logic flow chart for galaxy formation. The dominantly gravitational processes have been coloured with blue, while mainly hydrodynamical processes are coloured with yellow. The red hexagons indicate a binary choice, while dark green ellipses note the end products of the simplified processes described in the flow chart. The figure is an adaptation of Fig. 1.1 from \cite{Mo_big_book}.}
\label{fig:galaxy_formation_flow_chart}
\end{figure}

To simplify, in hierarchical models, such as the $\Lambda$CDM (Section \ref{s:background_cosmology}), the galaxy formation involves at minimum the following three stages:
\begin{enumerate}\addtolength{\itemsep}{-0.02\baselineskip}
  \item the hierarchical formation of dark matter haloes,
  \item the accretion of gas into the haloes, and
  \item the cooling and fragmentation of the hot gas into stars.
\end{enumerate}
Figure \ref{fig:galaxy_formation_flow_chart} shows a logic flow chart for galaxy formation\footnote{The flow chart is by no means a complete description of all gas physics that may play a role in galaxy formation, but rather tries to capture the main aspects.}. The paths leading to the formation of various galaxies (green ellipses) are drawn from the initial conditions set by the cosmological framework, discussed in earlier Sections and outlined in Figure \ref{fig:structure_formation_flow_chart}. Note that the flow chart does not include any feedback effects, which have been found to be significant and will be discussed later.

\subsection{Semi-analytical models of galaxy formation}\label{s:SAMs}

Due to the complicated physics related to galaxy formation and evolution (as seen in Fig. \ref{fig:galaxy_formation_flow_chart}), simple rules that can be easily varied to study the importance of different physical processes are highly useful. Semi-analytical models (SAMs) of galaxy formation try to fill this void by encoding simplistic rules for the formation and evolution of galaxies within a cosmological framework. A SAM is a collection of physical recipes that describe an inflow of gas, how gas can cool and heat up again, how stars are formed within galaxies, how stellar populations evolve and how black holes grow using simplified physics \citep[][]{Cole:1991p857, White:1991p851, Lacey:1991p842, Kang:2005p724, Baugh:2006p734, DeLucia:2007p414}. SAMs can also easily include different feedback effects: stellar winds, active galactic nuclei (AGN) or supernovae (SNe) feedbacks \citep[e.g.][]{Croton:2006p249, Somerville:2008p759, 2010arXiv1004.3289R}, for example. Hence, SAMs try to describe all the gas physics that goes into galaxy formation and evolution, but is not modelled in the dark matter only simulation.

Due to their nature SAMs can be used to explore ideas of galaxy formation and evolution and to understand which physical processes are the most important in the life of a galaxy by changing the recipes describing the physics. SAMs can also be applied to the so-called sub-grid physics that operates below the resolution of a simulation. As it is not yet possible to simulate all star formation processes \citep{McKee:2007p763, 2011arXiv1101.5172K} in a cosmological context, sub-grid physics must be modelled with simplified physics even when hydrodynamics is involved.

\begin{figure}[htb]
\center{
\includegraphics[scale = 0.55]{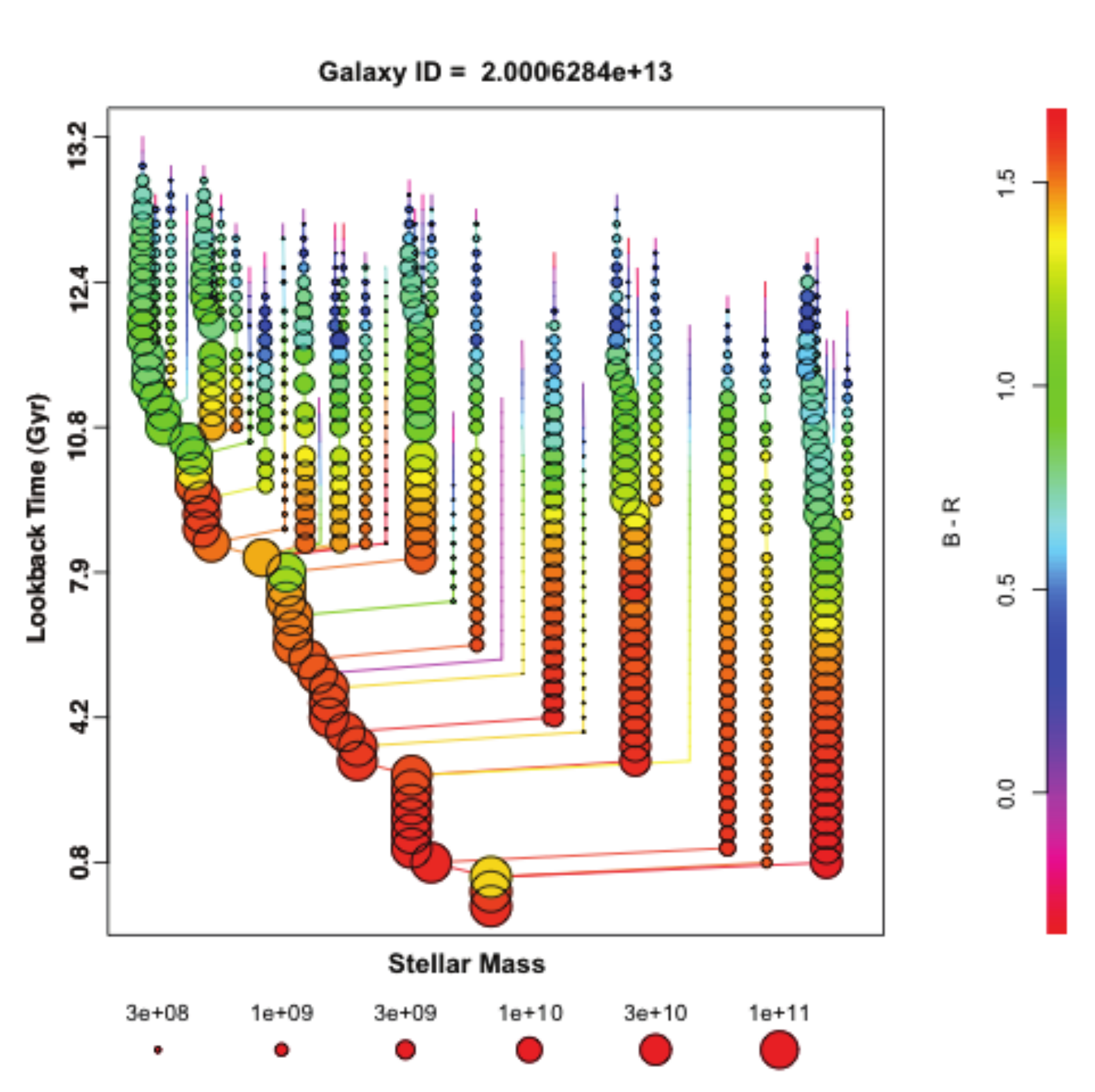}
}
\caption{Example of a merger tree as a function of the lookback time. Symbols are colour-coded as a function of the $B - R$ colour and their area scales with the stellar mass. Only progenitors more massive than $10^{8}h^{-1}$M$_{\odot}$ are shown.}
\label{fig:MergerTree}
\end{figure}

The backbone of a SAM is the evolution of dark matter haloes. Often, this evolution is parameterised with dark matter halo merger trees (Fig. \ref{fig:MergerTree}) that allow the hierarchical nature of gravitational instabilities to be explicitly taken into account \citep{Baugh:1998p847}. Dark matter merger trees describe how the dark matter haloes form via mergers of smaller haloes. They provide the backdrop for the introduction of the baryonic component which reacts gravitationally to the growing network of dark matter potential wells. Even though modern studies derive merger trees directly from simulations, this is by means not necessary as they can be derived also by using Monte Carlo techniques. In such case the extended Press-Schechter (EPS) theory \citep{Bond:1991p899,Lacey:1993p898} is often used. The Monte Carlo techniques provide a fast method of generating merger trees, however, they have been found to be less than reliable in some cases \citep[e.g.][]{Cole:2008p599}.

After a cosmological model has been chosen and the merger trees have been generated the baryonic processes must be taken into account. A SAM typically consists of the following steps: 1) follow the three baryonic components: hot and cold gas, and stars and adopt a recipe for disk formation; 2) specify a recipe for the conversion rate between the three components, including star formation and feedback effects; 3) keep track of the metallicity of each component; 4) convert the star formation history and metallicity of the stellar populations into luminosities; and 5) adopt a recipe for galaxy-galaxy mergers. In what follows, I briefly describe these steps in turn.

\subsubsection{Heating and cooling of gas}


Modern galaxy formation theories assume that the gas density profile follows that of dark matter. When gas falls into the potential well of a dark matter halo it is assumed to be shock-heated \citep{White:1978p755} to the virial temperature of the halo, given by
\begin{equation}
T_{vir} = \frac{\mu m_{H}GM_{halo}}{2k r_{vir}} \quad .
\end{equation}
Here $\mu$ is the mean molecular mass of the gas, $m_{H}$ is the mass of a hydrogen atom, $k$ is the Boltzmann's constant, $M_{halo}$ is the mass of the halo and $r_{vir}$ is the virial radius, within which the mean density is 200 times the critical density.

Before stars can form from the cold $(T < 100$ K$)$ molecular clouds the shock-heated hot gas must cool radiatively \citep[e.g.][]{Helly:2003p965}. The cooling time can be defined, for example, as the ratio of the thermal energy density and the cooling rate per unit volume. In this case, the cooling time can be written as
\begin{equation}
t_{cool}(r) = \frac{\frac{3 \bar{\mu}m_{p} k T_{vir}}{2\mu
m_{H}}}{\rho_{gas}(r)\Lambda(T_{vir}, Z_{gas})} \quad ,
\end{equation}
where $\bar{\mu}m_{p}$ is the mean particle mass, $k$ refers to the Boltzmann constant, $\rho_{gas}(r)$ is the hot gas density, and $\Lambda(T_{vir}, Z_{gas})$ is the cooling function. Note that the cooling rate is a function of metallicity of the gas $Z_{gas}$ and the virial temperature of the halo $T_{vir}$, and thus, the cooling is in general more effective in higher density regions. Additionally, in highly simplified scenarios, the more metal-enriched gas tends to cool faster.

Different cooling mechanisms (inverse Compton scatter, molecular and atomic cooling, and bremsstrahlung) can however be dominant at different times and temperatures complicating the gas cooling modelling significantly. For example, in massive haloes, where $T_{vir} \gtrsim 10^{7}$ K, gas is fully collisionally ionised and cools mainly through bremsstrahlung emission from free electrons. In the temperature range $10^{4}$ K $< T_{vir} < 10^{6}$ K excitation and de-excitation mechanisms dominate, while in haloes with $T_{vir} < 10^{4}$ K gas is mainly neutral and the cooling processes are therefore suppressed. However, in a simplified scenario the gas is assumed to be able to cool if cooling time is shorter than some characteristic timescale, which is model dependent.

As the cold condensed gas accumulates in the central regions of the dark matter haloes it can be identified as the ISM of the protogalaxy. What is assumed to follow after the cold gas has settled down is a disk formation.

\subsubsection{Disk formation}

The disk formation may be modelled after, for example, the amount of cold gas and the properties of the host halo. The underlying physical notion is that when structures grow and collapse in the early universe, they exert tidal torques on each other. This provides each collapsing dark matter halo with some angular momentum. As a result, SAMs typically assume that the cold gas will form a disk with the same specific angular momentum as the dark matter halo, while the size and rotation velocity are determined by the spin parameter:
\begin{equation}
\Gamma = \frac{J |E|^{\frac{1}{2}}}{GM^{\frac{5}{2}}} \quad .
\end{equation}
Here $J$ is the angular momentum of the halo, $E$ refers to the total energy of the halo, and $M$ is the mass of the dark matter halo. When adopting this simple prescription, what follows is that the mass and spin of the disk is tightly coupled to the mass and spin of the dark matter halo.

The dimensionless angular momentum $\Gamma$ is a measure of the degree of rotational support of the galaxy. Note, however, that the typical values $(\sim 0.05)$ of $\Gamma$ of collapsed dark matter haloes have been found to be significantly smaller than that of the largely flattened centrifugally supported disk galaxies we observe today $($with $\Gamma \sim 0.4 - 0.5)$. Hence, a considerable amount of dissipation must have occurred to produce the observed disks.

\subsubsection{Star formation}

In a typical SAM star formation \citep[for a general review, see][]{1998ARA&A..36..189K} is assumed to take place in the disks of galaxies, while the actual onset of star formation is assumed to occur once the surface density of cold gas exceeds a critical density \citep{1998ApJ...498..541K, Kennicutt:1989p976} in a molecular cloud \citep{2011arXiv1101.5172K}. Ideally, the star formation law should be derived from the first principles as a function of the physical conditions, such as density, temperature, metallicity, radiation and magnetic fields of the ISM, however, the detailed physics involved in the fragmentation of the cold gas, collapse and onset of a protostar, and the physical conditions of the ISM are not yet well understood. I will, however, return to the importance of the ISM in Section \ref{s:HerschelGalaxies}.

Due to the complicated physics involved, SAMs often derive the star formation rate (SFR) of a galaxy using a simple relation
\begin{equation}
\dot{\rho}_{\star} = \epsilon_{\textrm{SF}} \frac{\rho_{\textrm{cold}}}{\tau_{\star}} \quad ,
\end{equation}
where $\rho_{\textrm{cold}}$ is the density of the cold gas, $\tau_{\star}$ is the characteristic timescale, and $\epsilon_{\textrm{SF}}$ is a measure for the efficiency of star formation. Note, however, that several different forms of the above equation have been developed. These recipes for the star formation efficiency range from simple models that assume that $\tau_{\star}$ is a constant to models that are proportional to the dynamic time of the galaxy and take into account, for example, the circular velocity and/or the radius of the disc. In any event, the above star formation law is a variant of the empirical \cite{1959ApJ...129..243S} law, for which it has been assumed that the SFR is controlled by the self-gravity of the gas.

Closely tied to the star formation in galaxies is the number of stars of a given mass that forms, that is the the initial mass function \citep[IMF; e.g.][]{Kroupa:2001p974, Chabrier:2003p876}. SAMs typically assume that the IMF of stellar populations is universal when modelling star formation. Note, however, that theoretical arguments \citep[e.g.][]{Dave:2008p949} and indirect observational evidence suggest that the stellar IMF may evolve with, e.g., time \citep[][and references therein]{vanDokkum:2008p913} or environment, casting a shadow over the assumption of universality.

\subsubsection{Feedback effects}

The early SAMs, using recipes similar to the above ones, were however, not able to reproduce the observed form of the galaxy luminosity function (LF; Eq. \ref{eq:Schechter_LF}). They often over-predicted the number of both faint and bright galaxies, especially in the infrared \citep[e.g.][]{Benson:2003p964, Croton:2006p249, Benson:2010p962}. To alleviate the discrepancy i.e. to limit the number of faint and bright galaxies a feedback mechanism was introduced. To regulate the star formation in both light and massive dark matter haloes the feedback was divided to two separate mechanisms that operate on different mass regimes.

Modern SAMs model active galactic nuclei (AGN) feedback, which can suppress the cooling flow in high mass systems \citep[e.g.][]{Silk:1998p996, Croton:2008p600, Lagos:2008p528}. The AGN feedback \citep{2007MNRAS.380..877S} provides additional energy that can suppress the cooling of hot gas generating a sharp cut-off to the high-luminosity end of the LF. In many models the strength of the AGN feedback depends directly on the mass accretion of the black hole $\dot{M}_{\mathrm{BH}}$, thus, the modified cooling rate can be written, for example, as
\begin{equation}
\dot{M}_{\mathrm{cool}}^{\mathrm{mod}} = \dot{M}_{\mathrm{cool}} - \eta \frac{\dot{M}_{\mathrm{BH}}c^{2}}{\frac{1}{2}V^{2}_{\mathrm{vir}}} \quad .
\end{equation}
Here, $\eta$ refers to the black hole accretion efficiency. The additional energy from the AGN can prevent gas from cooling and is more important in later times when galaxies have more massive black holes. The AGN feedback can therefore help to regulate the SFR at later epochs and to prevent the overproduction of very massive galaxies. Another effect of AGN feedback is seen on the ages of high stellar mass systems, which are significantly older for AGN feedback models \citep{Khalatyan:2008p429}.

While the AGN feedback affects mainly massive galaxies, the supernovae (SN) feedback is important for lighter galaxies and for the metal enrichment of the inter-stellar and possibly even the inter-galactic medium \citep{DallaVecchia:2008p396}. The SN feedback helps to self-regulate the process of star formation throughout the galaxies' history. In the absence of the SN feedback star-formation rates (SFRs) are extremely high in early times and fairly low in more recent epochs. However, with the SN feedback, SFRs are initially lower so more gas is available for later periods of star formation (see also Section \ref{s:evolution}).

The SN feedback can function because in the models it is assumed that a SN can blow gas out of a star forming disc. Moreover, it is also assumed that the rate of mass ejection is proportional to the total mass of stars formed. As a result, the re-heating of gas due to the feedback can then be modelled as follows
\begin{equation}
\dot{M}_{\mathrm{reh}} = \epsilon_{0}\frac{\eta_{\mathrm{SN}} E_{\mathrm{SN}}}{V_{c}^{2}}\dot{M}_{\star} \quad
,
\end{equation}
where $E_{\mathrm{SN}}$ is the energy injected by the SN, $\eta_{\mathrm{SN}}$ is the number of SN per solar mass, and $V_{c}$ is the circular velocity of the dark matter halo. In dwarf galaxies with low circular velocities the energy from SNe can efficiently heat the inter-stellar media (ISM) which can then escape the halo through a cool wind. Note that hot metals expelled in SNe explosions are much less bound to the galaxy than the cold ISM and can therefore escape to the intra-galactic media (IGM), causing the enrichment of the IGM. Hence, the energy from SNe not only suppresses the star formation in light galaxies, but is also partially responsible for the chemical evolution in galaxies (see also Section \ref{s:chemical_evolution_of_galaxies}).

\subsubsection{Galaxy mergers}\label{s:mergers}

The above definitions for the amount of cool gas, disk formation, star formation, metal enrichment, and feedback effects have assumed a static dark matter halo. Dark matter haloes are not, however, static, but can interact and merge with other haloes \citep[e.g.][]{Kauffmann:1993p946, Fakhouri:2008p730} as described by their merger trees. The mergers are assumed to be important to the extent that, for example, \cite{Li:2007p1098} showed that each dark matter halo, virtually independent of its mass, experiences about $3 \pm 2$ major\footnote{Note that \cite{Li:2007p1098} defined a major merger as a merger with a progenitor mass ratio greater than 1:3.} mergers since its main progenitor has acquired one per cent of the final halo mass. Consequently, a realistic SAM must also take mergers into account.

When dark matter haloes merge, galaxies share the same potential well, but do not immediately merge \citep[e.g.][]{Stewart:2009p614}. Instead, the two galaxies are expected to orbit in a common halo. However, with enough time tidal interactions and dynamical friction remove orbital energy and cause the orbit of the subhalo (and its galaxy) to decay, transporting it towards the centre. The orbit decay time has been shown to depend on the mass ratio of the merging haloes, eccentricity of the orbits, and the mass loss due to tidal stripping \citep{Colpi:1999p1100, BoylanKolchin:2008p1099}. The details of dark matter halo (and galaxy) mergers are therefore less than simple. For example, if the angular momentum is high and the orbital energy is not low enough, the merger cannot happen in a Hubble time.

Despite the complications, SAMs often apply simple calculations for the dynamical friction time. If however $N-$body merger trees are used instead of the EPS formalism, the spatial information of the simulated haloes can be used. In any event, after the haloes merge SAMs usually assume that the hot halo gas is shock-heated to a new virial temperature, while the cold gas is attached to the centre. The hot gas that has not escaped from the new halo is assumed to form a reservoir in the halo, while galaxies residing inside the halo will merge within a dynamical friction time. It is obvious that the physics of such description of mergers is highly simplified, but even so, the simplified merger scenario can lead to galaxies that can statistically match observations reasonably well (see, for example, Paper IV). 

Several different types of mergers have been identified: major \citep[e.g.][]{Springel:2005p916} and minor \cite[e.g.][]{Bournaud:2007p925} that are related to the mass ratio of the merging pair, and the so-called wet \citep[e.g.][]{Lin:2008p924} and dry mergers \citep[e.g.][]{Bell:2006p919, Khochfar:2009p725}, which are related to the gas richness of the merging pair. The remnants of mergers between two galaxies are therefore assumed to depend primarily on four properties: 1) the progenitor mass ratio; 2) the gas mass fraction of the progenitors; 3) the orbital properties; and to some extent 4) the morphologies of the progenitors. It is then usual to assume that the different types of mergers require different physical prescriptions (see the lower part of Fig. \ref{fig:galaxy_formation_flow_chart}).

Unfortunately, the relevant mass and gas ratios are often rather arbitrarily defined in SAMs, rendering the current treatment of the different types of galaxy mergers less physical. In case of a minor merger, the simplified schemes of mergers often transfer stars from the lighter galaxy to the bulge of the more massive one, generating a spherical component to the merger remnant. Instead, in a case of major merger, the models may also take into account the fact that major mergers can induce rapid star formation, a starburst, changing the stellar population and the integrated colour of the newly formed galaxy rapidly. The possible starburst is however often tied up to the gas fraction of the merging pair and to the final surface density of cold gas. Thus, the treatment of starbursts is often model dependent. Mergers may also induce AGN activity, which has been taken into account in some models \citep[e.g.][]{Somerville:2008p759}, while some have implemented a variable IMF \citep[e.g.][]{Lacey:2010p1041}. Independent of the adopted model, it is however clear that due to the mergers of dark matter haloes and galaxies, a detailed analysis of the formation history and the possible environmental effects must be taken into account when studying galaxy formation and evolution. I will return to this in Chapter \ref{ch:isolated} when discussing the evolution of galaxies.

\chapter{Groups of Galaxies}\label{ch:groups}

\begin{quote}
\footnotesize{
``Someone told me that each equation I included in the book would halve the
sales\ldots''
\begin{flushright}
Stephen Hawking
\end{flushright}
}
\end{quote}

\section{The overall picture}

Chapter \ref{ch:formation} briefly described the formation of structure and how observable structures such as galaxies form. It was also mentioned that large galaxy surveys have shown that galaxies (Fig. \ref{fig:galaxies}) can be considered to be the basic building blocks of the visible Universe. However, as Figure \ref{fig:2dFzcone} implies, galaxies are found to be located in larger structures such as groups (Fig. \ref{fig:group}) and clusters more often than in isolation. To be more precise, more than half of all galaxies are found to be part of larger structures. This has been found to be true to the extent that most galaxies with luminosities less than the characteristic luminosity $L^{\star}$ (for a definition, see Eq. \ref{eq:Schechter_LF}) have been found to be located in an environment comparable to the Local Group. Thus, to understand galaxy formation and evolution - and to understand the Universe - one must understand groups and clusters of galaxies as the evolution of most galaxies takes place in these systems. Additionally, groups of galaxies are important cosmological indicators of the distribution of matter in the Universe, making them important also for cosmology. But what defines a galaxy group?

A galaxy group is a concentration of galaxies, assumed to be embedded in an extended dark matter halo\footnote{This definition applies also to galaxy clusters.}. Ideally, the galaxies forming a group are physically bound together due to their mutual gravitational attraction and the presence of the dark matter halo. However, from the observational point of view, group members are not easy to define because dark matter haloes cannot be observed directly. Thus, not all observed groupings of galaxies are real physical and gravitationally bound systems as they can be a result of chance superpositions of galaxies at different distances (see, e.g., Fig. \ref{fig:group}) or galaxies within filaments that are viewed edge-on. Such systems are gravitationally unbound (sometimes phrased as spurious or pseudo-groups) rather than real gravitationally bound groups of galaxies \citep[e.g.][]{Hernquist:1995p717, Ramella:1997p718}.

Groups of galaxies typically contain fewer than $\sim 50$ members and are often dominated by spiral galaxies. When larger groups are considered, the main constituent of galaxies usually shifts from spirals to lenticulars, but no clear cut-off in number of members exists between groups and clusters. The number of member galaxies is in general a problematic property \citep[see e.g.][]{Paz:2006p489}, especially in magnitude limited observations, where deeper observations can reveal new members who were previously too faint to be detected. This is true even for the Local Group (LG). Thus, better quantities to discriminate between groups and clusters are mass and size, though, these do not provide clear cut-off values either. Typical groups are $\sim 1-2$ Mpc in a diameter and their total mass $M_{DM} \sim 10^{12.5 - 14.0}h^{-1}$M$_{\odot}$ \citep{Huchra:1982p1}, while typical clusters are about an order of magnitude more massive and a few times larger \citep[e.g.][]{Einasto:2003p871, Einasto:2005p872, Koester:2007p588}. Typical groups are therefore smaller and less massive than clusters of galaxies, but larger and more massive than binary galaxy systems.

Most of the stellar mass in the present Universe is in groups similar to the LG with masses a few times $10^{12}$M$_{\odot}$ and only $\sim 2$ per cent is in clusters with total mass $> 5 \times 10^{14}$M$_{\odot}$ \citep{Eke:2006p605}. Groups have been found to be present already at redshifts $z > 1$ \citep[e.g.][]{Francis:1996p1038, Moller:1998p1039, Andreon:2009p1037, Bielby:2010p1077} and their environment density is intermediate between that of isolated galaxies and that of the cores of rich clusters. The study of groups may therefore provide clues to the processes that create the observed dependency of galaxy morphology on environment \citep{Postman:1984p719, AllingtonSmith:1993p723, Whitmore:1993p720, Zabludoff:1996p722, Hashimoto:1998p721}. I return to this question in the next Chapter.

The birth of the study of galaxy groups (and clusters) can be dated back to $1933$ and to Fritz Zwicky who was the first to apply the virial theorem to the Coma cluster \citep{1933AcHPh...6..110Z}. In his early work Zwicky derived a dynamical mass for the cluster that seemed to be significantly larger than if all of the mass came from visible galaxies. This was interpreted as a first sign of the yet-to-be-seen invisible matter holding the cluster together. Note that the early interpretation of Zwicky's dynamical mass statement holds only when the object is at least gravitationally bound, if not in virial equilibrium, but this was taken for granted at the time.

Since $1933$ several authors have studied groups of galaxies and concluded that the majority of galaxies in the Universe lie in groups \citep{1950MeLu2.128....1H, 1956AJ.....61...97H, Turner:1976p356, Huchra:1982p1, Geller:1983p576, Nolthenius:1987p357, Ramella:1989p4, Karachentsev:2005p447}. Large astronomical redshift surveys such as the 2dFGRS (2dF Galaxy Redshift Survey) and SDSS (Sloan Digital Sky Survey) have also produced large group catalogues \citep{Eke:2004p509, Eke:2004p507, Balogh:2004p748, Merchan:2005p495, Tago:2006p490, Eke:2006p605, Berlind:2006p484, Yang:2007p506, Knobel:2009p1075} with up to ten thousand groups \citep[e.g.][]{Eke:2004p509}. Even so, the physical processes operating in groups of galaxies are still poorly understood, albeit groups are known to be important for several reasons. For example, as Chapter \ref{ch:formation} described, dark matter haloes and even galaxies merge together (Section \ref{s:mergers}). This is obviously more probable in denser environments where relative velocities are lower and dynamical friction is higher as in groups. Moreover, mergers are only effective in systems with a velocity dispersion smaller than or comparable to the internal velocities of galaxies \citep[for detailed discussion, see e.g.][]{Mo_big_book}. Hence galaxy mergers are assumed to take place effectively in groups of galaxies.

Before groups can be used to study the evolution of galaxies or even before any group properties can robustly be measured, groups of galaxies must be correctly identified and discriminated from other forms of structures like binary galaxy pairs, large clusters of galaxies and most importantly from chance alignments. Interestingly, this has been found to be not as simple as one might na\"{\i}vely first assume (Papers I and II and references therein).

\section{Group identification}

There are generally three basic pieces of information available in observations for the study of galaxy distribution: position, luminosity and the redshift of each galaxy. Although the apparent luminosity is important as a measure of the object's visibility, it is usually a poor criterion for group membership. Moreover, the apparent luminosity can also be misleading when applied in group studies as will be shown in Section \ref{s:redshiftAsymmetry}. Observations are therefore usually left with only information describing the position of the galaxies in redshift-space (see, however, the caveats mention in Section \ref{s:large_galaxy_surveys}).

In recent years a number of different grouping algorithms have been developed and applied \citep[e.g.][]{Turner:1976p356, Materne:1978p353, Huchra:1982p1, 2004MNRAS.349..425B, Goto:2002p585, Kim:2002p498, Bahcall:2003p227, Gerke:2005p691, Koester:2007p588, Yang:2007p506} to identify groups using the redshift-space information. Despite the vast number of grouping algorithms, the Friends-of-Friends\footnote{Note that although the name is unfortunately the same as in case of the most popular halo finder, these two algorithms are not identical.} (FoF; \citealt{Huchra:1982p1}) percolation algorithm remains the most frequently applied one. The FoF algorithm, or slightly modified versions of it \citep[see e.g.][]{Knobel:2009p1075}, are the most widely used grouping algorithms, even for modern day galaxy surveys like SDSS and COSMOS. Hence, it is important to understand its functionality and limitations.

\subsection{The Friends-of-Friends algorithm}\label{s:FoF}

The Friends-of-Friends (FoF) group finding algorithm \citep{Huchra:1982p1} takes advantage of two often available quantities in observational galaxy catalogues: the projected separation in the sky and the velocity difference in the redshift space. The grouping method itself begins with the selection of a galaxy, which has not been previously assigned to any of the existing groups. After choosing a galaxy the next step is to search for companions with the projected separation $D_{12}$ smaller or equal to the separation $D_{L}$:
\begin{equation}
	D_{12} = 2 \sin \left( \frac{\theta}{2} \right) \frac{V}{H_{0}} \leq
	D_{L}(V_{1}, V_{2}, m_{1}, m_{2}) \quad ,
\end{equation}
where the mean cosmological expansion velocity
\begin{equation} 
	V = \frac{V_{1} + V_{2}}{2} \quad , 
\end{equation} 
and the velocity difference $V_{12}$ is smaller or equal to the velocity $V_{L}$: 
\begin{equation} 
	V_{12} = |V_{1} - V_{2}| \leq V_{L}(V_{1}, V_{2}, m_{1}, m_{2}) \quad .
\end{equation} 
Here, $V_{1}$ and $V_{2}$ refer to the velocities (redshifts) of the galaxy and its companion, $m_{1}$ and $m_{2}$ are their magnitudes, and $\theta$ is their angular separation in the sky. If no companions are found, the galaxy is entered on a list of isolated galaxies, while all companions found are added to the list of group members. The surroundings of each companion are then searched by using the same method. This process is repeated until no further members are found and all potential group members have been searched.

There is a variety of prescriptions for $D_{L}$ and $V_{L}$ in the literature (for references, see Paper I). However, the original method assumes that the luminosity function (LF) is independent of distance and position and that at larger distances only the fainter galaxies are missing. For each pair we therefore take
\begin{equation}\label{Dl} 
	D_{L} = D_{0} \left ( \frac{\int_{-\infty}^{\mathscr{M}_{12}} \Phi(\mathscr{M})\mathrm{d}\mathscr{M}}{\int_{-\infty}^{\mathscr{M}_{lim}} \Phi(\mathscr{M})\mathrm{d}\mathscr{M}} \right ) ^{-\frac{1}{3}} \quad ,
\end{equation}
where the integration limits can be calculated from equations:
\begin{equation}\label{i1}
	\mathscr{M}_{lim} = m_{lim} - 25 - 5 \log (D_{F}) 
\end{equation}
and
\begin{equation}\label{i2}
	\mathscr{M}_{12} = m_{lim} - 25 - 5 \log(V) \quad .
\end{equation}
In Eq. \ref{Dl} $\Phi(\mathscr{M})$ is the differential galaxy luminosity function for the sample, and $D_{0}$ is the projected separation in Mpc chosen at some fiducial distance $D_{F}$. The limiting velocity difference can be scaled in the same way as the distance $D_{L}$, i.e.,
\begin{equation}\label{Vl}
	V_{L} = V_{0} \left ( \frac{\int_{- \infty}^{\mathscr{M}_{12}} \Phi(\mathscr{M})\mathrm{d}\mathscr{M}}{\int_{-\infty}^{\mathscr{M}_{lim}} \Phi(\mathscr{M})\mathrm{d}\mathscr{M}} \right )^{-\frac{1}{3}} \quad ,
\end{equation}
where the fiducial velocity $V_{0}$ is often taken to be $\sim 200 - 400$ km s$^{-1}$ and the integration limits are given by Eqs. \ref{i1} and \ref{i2}.

For simplicity, it is often convenient to assume that the differential galaxy luminosity function can be described in form of a \citet{Schechter:1976p592} LF:
\begin{equation}\label{eq:Schechter_LF}
	\Phi (\mathscr{M}) = \frac{2}{5}\Phi^{\star} \ln 10 \left( 10^{\frac{2}{5}(\mathscr{M}^{\star}-\mathscr{M})} \right)^{\alpha + 1} e^{-10^{\frac{2}{5}(\mathscr{M}^{\star}-\mathscr{M})}} \quad , 
\end{equation}
where $\mathscr{M}$ is the absolute magnitude of the object and $\alpha$, $\mathscr{M}^{\star}$, and $\Phi^{\star}$ parametrize the Schechter luminosity function. In the Schechter formalism the $\mathscr{M}^{\star}$ and $L^{\star}$ refer to the characteristic absolute magnitude and luminosity, respectively, and mark a point in which the luminosity function exhibits a sudden change in the slope. For example, the characteristic luminosity in $B$-band $L^{\star}_{B} \sim 2 \times 10^{10}$L$_{\odot}$ is comparable to the brightness of the Galaxy \citep{Driver:2007p587}, but the exact value depends on the environment and on the dark matter halo mass \citep{Cooray:2005p697}. Note also that, the galaxy LF has been found to evolve significantly as a function of redshift, especially at early cosmic epochs \citep[see e.g.][]{Bouwens:2011p1089}. Nevertheless, the luminosity function is a powerful tool and provides information on the relative frequency of galaxies with a given luminosity.

The theory of gravitational instability (see Section \ref{s:evolution_perturbations}) predicts that the number of virialized systems that have formed at any given time depends on their mass, with more massive systems being less abundant than less massive ones. Therefore, by coupling together the information on the luminosity and the number of galaxies, the LF can provide information on the formation and evolution of both the structural and visible components of galaxies. However, for identifying galaxy groups, the exact shape of the luminosity function is less important than the adopted values of $D_{0}$ and $V_{0}$, as was noticed during the study of Paper I. Hence, I will not concentrate more on the LF in the context of groups, I will however return to it in Chapter \ref{ch:isolated}.

\section{Different species of groups: a brief overview}

Grouping algorithms, such as the FoF algorithm, produce catalogues of galaxy systems with a vast amount of different properties. This has resulted in a classification of systems with similar properties. For example, systems of galaxies can be classified as loose \cite[e.g.][]{Ramella:1995p562, Ramella:1997p718, Tucker:2000p453, Einasto:2003p590}, poor \cite[e.g.][]{Zabludoff:1998p578, Mahdavi:1999p581}, compact \cite[e.g.][]{1973Afz.....9..495S, Hickson:1982p571, Hickson:1989p573, Diaferio:1994p278, Barton:1996p269, Tovmassian:2006p286} or fossil groups \cite[e.g.][]{1994Natur.369..462P, Jones:2003p594, Santos:2007p217, vonBendaBeckmann:2008p598} depending on common properties. However, the vast number of different classes can also be interpreted as a sign that the relationship among systems of galaxies is not yet well understood.

The key idea behind the above classification rises from the diversity of evolutionary stages within which a galaxy group can be found. For example, compact groups are assumed to be observed briefly before they are about to merge, while fossil groups, which are often dominated by a large central elliptical galaxy, are possibly the end product of such a merging. Therefore, a general unification of different classes might be possible in the future when the evolution of galaxy groups is better understood. However, due to the historical reasons I briefly summarise each class and how they have been defined in the following Sections. When appropriate, I will also describe some selected properties of each class, their evolutionary stage and the possible significance for galaxy evolution in general.

\subsection{Loose groups}\label{s:looseg}

Loose groups of galaxies with a space density of $\sim 10^{-5}$ Mpc$^{-3}$ \citep{Nolthenius:1987p357} represent the most common class of groups and are often simply referred to as groups. They comprise $\sim 50$ members, including large number of faint dwarf galaxies, and the whole group typically extends to a diameter of up to $\sim 1.5$ Mpc. Thus, loose groups are, as the name implies, an intermediate in scale between compact groups (Section \ref{s:hickson}) and rich clusters. Consequently, it has been argued that their dynamics is important for the study of the distribution of dark matter \citep{Oemler:1988p699}.

\cite{Tucker:2000p453} finds a median line-of-sight velocity dispersion of 164 km s$^{-1}$ and median virial mass $\sim 1.9 \times 10^{13}h^{-1}$M$_{\odot}$ for loose groups. However, \cite{Einasto:2003p590} argue that loose groups in the neighbourhood of a rich cluster are typically $2.5$ times more massive and $1.6$ times more luminous than loose groups on average. Furthermore, \cite{Einasto:2003p590} find that these groups have velocity dispersions of about $1.3$ times larger than the loose groups on average. The immediate neighbourhood of a loose group can therefore have a significant impact on the group properties and dynamics. Hence, a nearby large cluster can enhance the evolution of the neighbouring loose group making it difficult to draw common values for the properties of loose groups that would apply to all of them.

Due to their relatively small velocity dispersions and intermediate sizes, loose groups can be important when galaxies that may merge in the future are being identified and studied. For example, \cite{Mamon:1986p701} was the first to suggest that compact groups might be transient unbound cores of loose groups. Indeed, further studies have confirmed that loose groups are associated with compact groups, and that loose groups are often the birth places of compact groups \citep{Vennik:1993p702, Ramella:1994p703, Diaferio:1994p278}. Thus, loose groups can host some sort of an association with a compact group providing a link between their evolutionary stages.

\subsection{Poor groups}

Most galaxies in the local Universe, including the Galaxy, belong to a poor group of galaxies \citep{Zabludoff:1998p578}. The historical definition of a poor group dictates that poor groups typically contain fewer than five bright $(L \sim L^{\star})$ galaxies \citep{Zabludoff:1998p578} and that the total number of group members is less than in a typical loose group. Deeper redshift surveys have however identified new and faint galaxies around poor groups bringing them closer to the definition of loose groups. For example, surveys such as the SDSS have helped to identify new faint dwarf galaxies that belong to the Local Group (LG) \citep[e.g.][]{Willman:2005p1072, Irwin:2007p707, Walsh:2007p706, Belokurov:2008p705, Belokurov:2010p756}, raising the total number of LG members to $\sim 40$. Obviously the number of brighter $L^{\star}$-galaxies in the LG has not changed and satisfy the criterion of less than five, but as the total number of members is reaching $50$ it is questionable whether one can talk about a poor group any longer.

As mentioned earlier, the total number of companions may however not be all that important for groups. Instead, the fraction of early-type galaxies in the poor groups has been found to vary significantly, ranging from that characteristic of the field ($\sim 25$ per cent) to that of rich clusters ($\sim 55$ per cent) \citep{Zabludoff:1998p578} in disagreement with many loose groups. The relatively high early-type fraction in poor groups is indeed surprising because most poor groups have low galaxy number density, thus, the effects of disruptive mechanisms such as galaxy harassment \citep{Moore:1996p716} are assumed to be weaker than in rich or compact groups. In contrast, however, the kinematics of poor groups makes them preferred sites for galaxy-galaxy mergers, which may alter the morphologies and star formation histories of some group members \citep{Zabludoff:1998p578}, providing a possible explanation for the relatively large number of elliptical galaxies.

The dynamical status of poor groups has also been questioned in several studies \citep[e.g.][and references therein]{Zabludoff:1998p578, Mahdavi:1999p581}. The higher galaxy densities than in the field and lower velocity dispersions than in cluster cores, make them favourable sites for galaxy-galaxy mergers \citep{Barnes:1985p708}. Consequently, one would assume that if poor groups are old structures several mergers should have taken place. Thus, if some or even all poor groups are gravitationally bound, why do we observe them at all? One possible explanation is that bound poor groups are collapsing for the first time, and in such, they will eventually face the same destiny as loose groups, that is being eventually associated with compact groups.

\subsection{Compact groups}\label{s:hickson}

A compact group (CG) of galaxies can be loosely defined as a group of galaxies with a small number of members in which the typical intergalactic separation is of the order of the scale of the galaxies. Historically, CGs have been studied by several authors \cite[e.g.][]{1973Afz.....9..495S, Rose:1977p692}, however, probably the most well-studied CG catalogue is the Hickson Compact Groups (HCGs) \citep[see e.g.][]{Hickson:1982p571, Hickson:1988p358, Hickson:1989p573}.

\begin{figure}[htb]
\center{
\includegraphics[scale = 0.30]{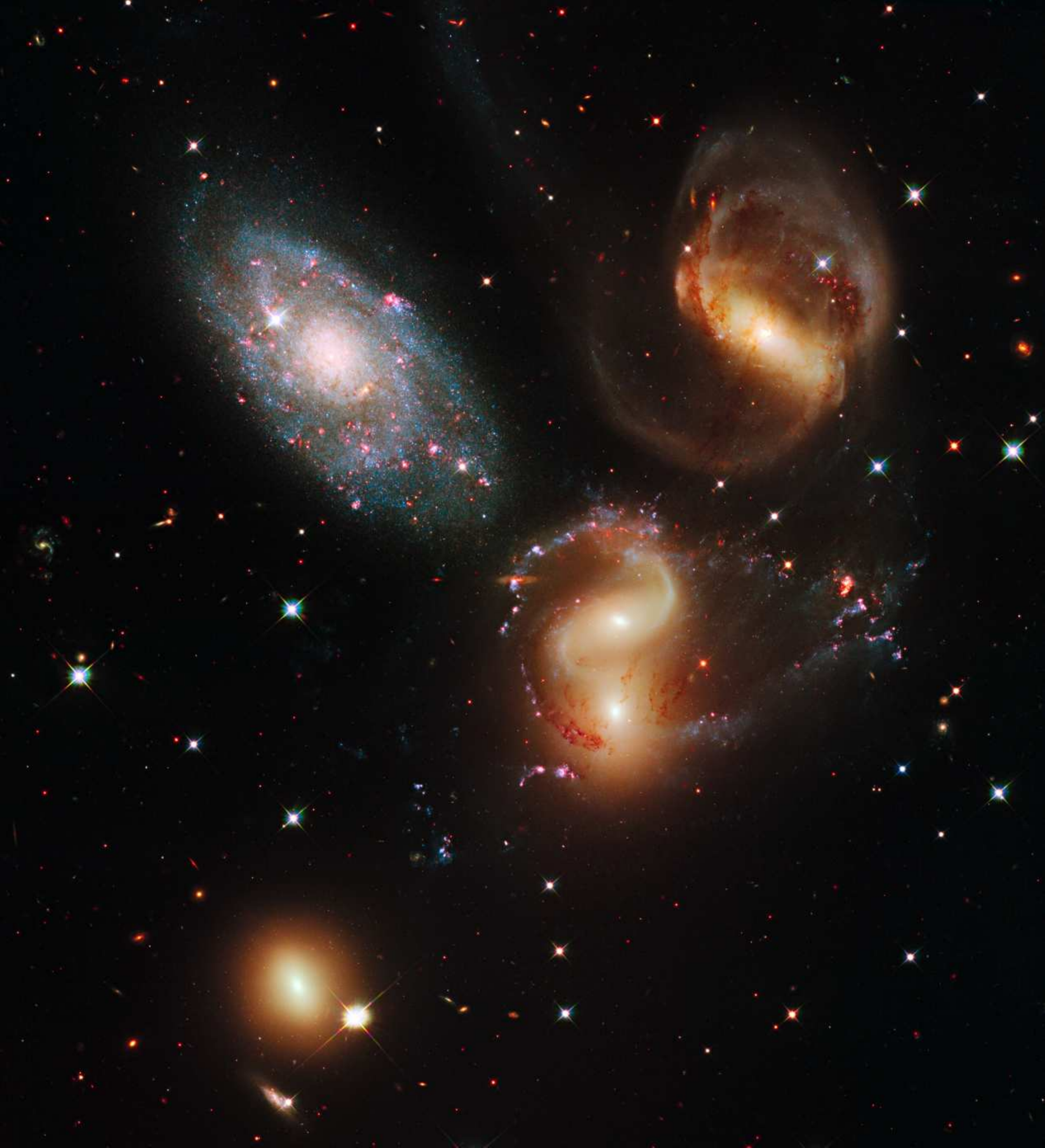}
}
\caption{The Hickson Compact Group 92 or Stephan's Quintet, as the name implies, is a group of five galaxies. Note, however, that NGC $7320$ (at upper left) is actually a foreground galaxy not a real group member. Courtesy of NASA, ESA and the Hubble SM4 ERO Team.}
\label{fig:group}
\end{figure}

HCGs are compact configurations of relatively isolated systems of typically four or five galaxies in close proximity to one another \citep[see Fig. \ref{fig:group}, and][and references therein]{Hickson:1989p573}. They have also been found to show peculiarities in terms of morphology or kinematics, starbursts or even AGN activity \citep[for a complete review, see][]{Hickson:1997p693}. Furthermore, HCGs have been found to contain large quantities of diffuse gas and to be dynamically dominated by dark matter \citep[for predictions of X-ray properties, see ][]{Diaferio:1995p268}. They have also been found to trace the large-scale structure, but to prefer low-density environments. As already mentioned in Section \ref{s:looseg}, HCGs may form as subsystems within looser galaxy associations and evolve by gravitational processes. Thus, while compact groups are associated with loose groups and filaments, these tend to be low-density and sparsely populated systems. Even so, \cite{Walker:2010p1074} speculate that due to their mid-infrared colours the compact group environment fosters accelerated evolution of galaxies.

The fraction of late-type galaxies has been found to be significantly lower in compact groups than in the field. However, a given CG is also more likely to contain galaxies of a similar type than would be expected for a random distribution \citep[][and references therein]{Hickson:1997p693}. \cite{Zepf:1993p704} found that elliptical galaxies in compact groups tend to have lower internal velocity dispersions than do ellipticals of similar properties in other environments. Moreover, elliptical galaxies of CGs have been found to lie off the fundamental plane defined by ellipticals in other environments. This suggests that the velocity dispersion is of greater physical relevance to the formation and evolution of galaxies in CGs, than is the apparent physical density.

The strong galaxy interactions in CGs, see Fig. \ref{fig:group}, are expected to induce mergers of group members (for early simulations, see \cite{Mamon:1987p1095}). The dynamical timescales of CGs have however been argued and range from relatively short $(\sim 0.1H_{0}^{-1})$ \citep[e.g.][]{Diaferio:1994p278} to longer ones $\sim H^{-1}_{0}$ \citep[e.g.][]{Governato:1996p287, Athanassoula:1997p1096}. If however the merger timescales are somewhere between the two extremes, it is likely that CGs are rather short lived. To explain this, it has been suggested that CGs may be continuously replenished through dynamical evolution of loose groups \citep{Diaferio:1994p278, Ramella:1994p703}. This provides a reasoning for the existence of CGs, but what will they become in the course of evolution? One potential scenario was described by \cite{Borne:2000p1090} who suggest that the evolutionary progression from CGs can lead first to pairs followed by ultra-luminous infrared galaxies (ULIRGs) and finally to elliptical galaxies. Thus, it is possible that CGs are the progenitors for ULIRGs (see also Section \ref{s:HerschelGalaxies}). Note, however, that it has also been suggested that the eventual demise of the CG due to mergers could lead to the formation of a fossil group \citep[e.g.][]{Vikhlinin:1999p696, Mulchaey:1999p563}.

\subsection{Fossil groups}

The definition of a fossil group is often based on the following criteria of \cite{Jones:2003p594}. A fossil system is defined as a spatially extended X-ray source with an X-ray luminosity from diffuse, hot gas of $L_{\mathrm{X, bol}} \geq 10^{42}h^{-2}_{50}$ erg s$^{-1}$, while the optical counterpart is a system of galaxies with $\Delta m_{12} \geq 2.0$, where $\Delta m_{12}$ is the magnitude gap between the brightest and the second brightest galaxy in the $R$-band within half the projected virial radius of the group centre. The reasoning for the optical criterium is that it is supposed to guarantee that the system is dominated by an E or cD type galaxy and that other members of the system can only cause small perturbations to the total potential well of the system. Consequently, rendering fossil groups as systems of galaxies that are dominated mainly by a single massive galaxy.

The first fossil group was discovered by \cite{1994Natur.369..462P} using R\"ontgensatellit (ROSAT) X-ray data. Since the discovery of fossil groups \cite{Khosroshahi:2007p97} compiled a list of seven fossil groups based on Chandra X-ray Observatory data and \cite{Santos:2007p217} used SDSS data to identify 34 candidates. Fossil groups have also been studied theoretically using cosmological $N$-body simulations \citep[e.g.][]{DOnghia:2005p128, DOnghia:2007p388, Sales:2007p380, DiazGimenez:2008p617}. Interestingly, based on the results of different studies fossil groups have been interpreted in different ways.

\cite{Jones:2003p594} describe fossil groups as old, undisturbed systems which have avoided infall into galaxy clusters, but where galaxy merging of most of the $L^{\star}$ galaxies has occurred. \cite{Khosroshahi:2007p97} suggest that fossil groups have formed early, while \cite{Vikhlinin:1999p696} and \cite{Mulchaey:1999p563} suggest that fossils can be the result of galaxy merging within a compact group. In contrast, it has also been suggested that fossil groups are the remnants of what was initially a poor group of galaxies that has been transformed to this old stage of galaxy evolution in low density environments with compact groups acting as likely way station in this evolution \citep{Eigenthaler:2009p698}.  In cosmological $N$-body simulations fossil groups represent undisturbed, early forming systems in which large and massive galaxies have merged to form a single dominant elliptical galaxy \citep{Dariush:2007p265, vonBendaBeckmann:2008p598}. Given the different interpretations, it is clear that more work is required before fossil groups can be considered as fully understood.

\section{Dynamical properties of groups}

The earlier discussion about the properties of different group classes implied
that the dynamical state and properties can be helpful when studying the effects
of the environment involved in galaxy evolution. The first basic property of a
galaxy group to consider is therefore the velocity dispersion $\sigma_{v}$. The
velocity dispersions of groups usually range up to a few hundred km s$^{-1}$,
while clusters of galaxies can show dispersions up to about a thousand km
s$^{-1}$. Due to the difficulties in measuring the true three dimensional
motions of galaxies, the velocity dispersion is however often measured in radial
direction. Moreover, the velocity dispersion of a group is only a meaningful
quantity if the group is a gravitationally bound system, otherwise some of the
group members are participating in the pure Hubble flow and their radial
velocities are biased \citep[e.g.][]{Baryshev:2001p969, Maccio:2005p971,
Teerikorpi:2008p589, Chernin:2009p968}. The assumption of boundness is however
not straightforward, and there is no guarantee that, e.g., the FoF algorithm has
identified only groups that are bound structures. This assumption and its
implications will be discussed in more detail as described in the next Section.
Despite the potential complications, studies (Paper I, and references therein)
have shown that observed velocity dispersions are in general in agreement with
those of cosmological $N$-body simulations when a $\Lambda$CDM cosmology is
adopted.

The velocity dispersion of a group is obviously connected to its internal dynamics and to the depth of the potential well of the host dark matter halo. Thus, for a gravitationally bound system, the velocity dispersion can be used to derive the total dynamical mass of the group. For example, one simple yet often applied method links the group's velocity dispersion and size to its mass in a following way
\begin{equation}
M_{\mathrm{obs}} \propto \sigma_{v}^{2}R_{H} \quad .
\end{equation}
Here $R_{H}$ is the mean harmonic radius, i.e., the size of the group. When this simple relation is applied to an observed group the assumption of boundness is often taken for granted or argued based on the small value of the virial crossing time
\begin{equation}\label{eq:crossingTime}
t_{c} = \frac{3R_{H}}{5^{3/2}\sigma_{v}} \quad ,
\end{equation} 
which is written in units of the Hubble time $H_{0}^{-1}$.

The virial crossing time is usually assumed to describe the group's dynamical status. If a crossing time is short compared to the Hubble time, the group must be bound, otherwise it would have dispersed long ago. However, as one may choose various definitions for the velocity dispersion and for the group size, which in combination define the crossing time, these choices introduce significant and systematic biases in the final value of the crossing time. Moreover, the inclusion of non-members and the existence of galaxy pairs, both of which increase the mean projected velocity, can systematically bias the velocity dispersion. The crossing time is therefore not a robust indicator for defining a group's dynamical status and to discriminate between gravitationally bound and unbound groups \citep[e.g.][Paper I]{Diaferio:1994p278}.

\section{Are observed groups gravitationally bound?}\label{s:bound?}

\subsection{Background}

Due to the uncertainties, for example, in the virial crossing times, it is unclear, which observed groupings of galaxies, if any, are gravitationally bound structures. However, as Chapter \ref{ch:formation} described, galaxies reside in large dark matter haloes. Thus, if group members are required to belong to the same dark matter halo, most groups of such type can readily be taken as gravitationally bound structures. However, as observations cannot directly observe dark matter haloes, and therefore identify galaxies that belong to the same halo, the problem of identifying real group members remains. Moreover, it is possible that some substructure of a given halo has higher velocity than the required escape velocity, complicating the matter even further.

Fortunately, cosmological $N$-body simulations provide a tool to study whether grouping algorithms, such as the FoF, can identify groups of galaxies that are gravitationally bound systems. In simulations it is simple to mimic observations by placing the observer inside the simulation volume, either to an arbitrary location or a specially selected environment that mimics the observed surroundings of the Local Group. After choosing the origin it is then relatively straightforward to use simulation data to generate mock group catalogues that mimic observations (for a detailed description, see Papers I and II). Finally, such catalogues can be used to make theoretical predictions about the properties of groups. 

\subsection{Virial theorem}\label{ss:virial_theorem}

As cosmological $N$-body simulations provide detailed dynamical information of each dark matter halo, and even each particle that forms the halo, the virial theorem can be used to measure the dynamical status of a galaxy group by relating its kinetic energy to that of the potential. In general, the total kinetic energy of a galaxy group may be written as
\begin{equation}\label{eq:kineticE}
	T = \frac{1}{2M} \sum_{i < j} M_{i}M_{j}(\mathbf{\dot{x}}_{i} -
	\mathbf{\dot{x}}_{j} )^{2} \quad ,
\end{equation}
while the total potential energy is
\begin{equation}\label{eq:potentialE}
	U = G \sum_{i < j} \frac{M_{i}M_{j}}{R_{i,j}} \quad .
\end{equation}
Here $M_{i}$ and $M_{j}$ are the masses of the two galaxies, $\mathbf{\dot{x}}_{i}$ and $\mathbf{\dot{x}}_{j}$ are their velocities, and $R_{i,j}$ is the distance between them, while $M$ corresponds to the total mass of the system. It is now trivial to find the dynamical status of the system; if the group fulfils the simple criterion
\begin{equation}\label{eq:virial_ratio}
	T - U < 0 \quad ,
\end{equation}
it can be considered gravitationally bound. Note, however, that this criteria is looser than the requirement of equilibrium or virialisation, which are also used in literature, and which require that
\begin{equation}\label{eq:virial_theorem}
| T | = \frac{|U|}{2} \quad .
\end{equation}

\subsection{The fraction of gravitationally bound groups}

Based on the results of simulations of different cosmologies (Paper I) it is clear that a significant fraction of groupings returned by the FoF are gravitationally unbound according to Eq. \ref{eq:virial_ratio}. Importantly, this result is mostly independent of the value of the cosmological constant or the apparent magnitude limit of the mimicked observations as Papers I and II show. In Paper II stricter parameter values for $D_{0}$ and $V_{0}$ (Section \ref{s:FoF}) were adopted, however, this did not change the fact that almost half of the groupings returned by the FoF algorithm were found to be unbound. In Paper I two different parameter sets for the luminosity function were used, but that did not have any significant effect either: the large number of gravitationally unbound systems remained.

The large fraction of unbound groupings returned by the FoF algorithm should not be disregarded lightly without further consideration. The fact that a significant fraction of identified groups are in reality gravitationally unbound implies that the majority of current group catalogues have overestimated group masses, when, e.g., the virial theorem has been erroneously applied to unbound systems. Furthermore, Paper I finds that the virial crossing time, in a form of Eq. \ref{eq:crossingTime}, is not a good measure to discriminate unbound pseudo-groups from real bound systems. Paper I \citep[see also][]{Diaferio:1994p278} shows that there is no correlation between Eq. \ref{eq:crossingTime} and the virial ratio, which is the ratio between the group's potential (Eq. \ref{eq:potentialE}) and kinetic energy (Eq. \ref{eq:kineticE}). This calls into question the crossing time as an estimator of gravitationally bound systems, albeit it has been widely accepted amongst majority of observational astronomers.

\subsection{Identifying bound structures}

If the virial crossing time does not provide a good and unbiased estimate for the dynamical status of a group, can we find another quantity that can be used instead? Unfortunately, there is no simple relation that would readily tell if a group is gravitationally bound or even virialized. The problem lies in the difficulty of observing dark matter haloes and estimating their masses in observations. Currently there is no any easy and accurate way to measure the depth of the potential well of a group, complicating the matter even further. Hence, the question remains: is there a method to estimate if a group of galaxies is gravitationally bound?

The prime focus of Paper I was to answer this question. Its Figures 12 and 13 show how the probability of a galaxy group being unbound depends on two observable parameters: the velocity dispersion and the pairwise separation. However, the complication here lies in the fact that both of these quantities had to be normalised with the total mass of the group, before clear trends could be seen. Thus, one must obtain a reasonable estimate for the total mass of the group before the method described in Paper I can be applied. At the moment, the most robust mass estimates are likely provided by X-ray observations of the hot intergalactic medium that is assumed to describe the depth of the potential well. If information describing the group's mass is available, the method developed in Paper I can be used to provide information about a group's dynamical status that should prove to be more robust than currently favoured practise of applying the virial crossing time.

\section{Discordant redshifts}\label{s:redshiftAsymmetry}

\subsection{Background}

An excess of higher redshift galaxies with respect to the group centre was discovered by \citet{1970Natur.225.1033A, Arp:1982p577} and it was studied in detail by \citet{1971Natur.234..534J}. Since then many authors have found a statistically significant excess of high redshift companions relative to the group centre. \citet{Bottinelli:1973p352} extended the study of \citet{1970Natur.225.1033A} to nearby groups of galaxies, and \citet{Sulentic:1984p341} found a statistically significant excess of positive redshifts while studying spiral-dominated (i.e. the central galaxy is a spiral galaxy) groups in contrast to the E/S0-dominated (i.e. the central galaxy is an elliptical or lenticular galaxy) groups that showed a minor blueshift excess. Also \citet{Girardi:1992p339} found discordant redshifts while studying nearby small groups identified by \citet{1988ngc..book.....T} in the Nearby Galaxy Catalogue. However, the conventional theory holds that the distribution of redshift differentials for galaxies moving under the gravitational potential of a group should be evenly distributed. Even systematic radial motions within a group would be expected to produce redshift differentials that are evenly distributed. To solve this discrepancy between observations and the conventional theory even new physics was suggested \citep[see e.g.][]{1970Natur.225.1033A}.

Since the first observations of discordant redshifts in the 1970s, multiple theories have been suggested to explain the observed redshift excess. \citet{Sulentic:1984p341} listed some possible origins for the observed redshift excess, while \citet{Byrd:1985p342} and \citet{Valtonen:1986p340} argued that this positive excess is mainly due to the unbound expanding members and the fact that the dominant members of these groups are sometimes misidentified. Opposite to this, \citet{Girardi:1992p339} argued that the positive excess may be explained if groups are still collapsing and contain dust in the intragroup medium. \citet{Hickson:1988p358} ran Monte Carlo simulations and concluded that the effects caused by the random projection can explain discordant redshifts, and \citet{Iovino:1997p338} similarly found that projection effects alone can account for the high incidence of discordant redshifts. However, studies by \citet{Hickson:1988p358} and \citet{Iovino:1997p338} dealt only with Hickson Compact Groups that have only a few members in close proximity (as noted in Section \ref{s:hickson}). \citet{Zaritsky:1992p516} studied asymmetric distribution of satellite galaxy velocities with Monte Carlo simulations and concluded that observational biases partially explain the observed redshift asymmetry, but cannot account for the whole magnitude of it. Despite, and partially because of, the vast number of explanations none of the explanations were found to be truly satisfactory and some even contradict one another. As a result, the problem of discordant redshifts remained open.

\subsection{The origin of redshift asymmetries}

\citet{Byrd:1985p342} and \citet{Valtonen:1986p340} were the first ones to propose that redshift asymmetries should arise in nearby groups of galaxies, if a large fraction of the group population is unbound to the group. They argued that the redshift asymmetry explains the need for "missing matter", the dark matter that was at the time supposed to exist at the level of the closing density of the Universe in groups of galaxies. These authors argued that if the group as a whole is not virialized, there is no need for excessive amounts of binding matter and it might also lead to the observed asymmetries. Despite their efforts, a confirmation was never obtained. Fortunately, cosmological $N$-body simulations (Section \ref{s:cosmological_simulations}) provide an invaluable tool to study projection effects as the origin of the observer can be chosen freely. Simulations also provide information about the dark matter halo dynamics and substructure, while the group dynamics could be studied with the tools developed and introduced in Paper I and briefly summarised in the previous Section. Thus, cosmological simulations provided an excellent tool to test the explanation of \citet{Byrd:1985p342} and \citet{Valtonen:1986p340}. This was the main topic of Paper II.

Paper II finds that gravitationally bound groups of simulated galaxies do not show any statistically significant redshift excess. This result is in agreement with the conventional theory, where it is expected that the distribution of redshift differentials are evenly distributed. A simulated group catalogue that includes only gravitationally bound groups of galaxies show an equal number of galaxies relative to the brightest member within statistical fluctuations. It is therefore important for galaxy group studies to be able to accurately identify gravitationally bound structures and exclude change alignments and spurious groups from the group catalogue (as noted already in Section \ref{s:bound?} and in Paper I).

A detailed study of simulated groups shows that when the dominant members of groups are identified by using their absolute $B$-band magnitudes a small blueshift excess arises. This is mainly due to the magnitude limited observations that can miss the faint background galaxies in groups. Moreover, $B$-band is also problematic because of another reason; it tends to make star-forming spirals, which have a larger number of hot O and B-type stars and which do not have much dust, brighter than regular red elliptical galaxies. Thus, the brightest galaxy in $B$-band may not accurately mark the centre of the potential well of a group and can create an artificial imbalance between the front and the back part of the group. This, together with the current inability to accurately measure the relative distances inside groups of galaxies, except for a few of the nearest ones \citep[e.g.][]{Karachentsev:1997p584, Jerjen:2001p583, Rekola:2005p586, Rekola:361p330R, Teerikorpi:2008p589} complicates the identification of the group centre if no X-ray detection of possible hot IGM is available.

The misidentification of the group centre can lead to a redshift excess, since it is more likely that the apparently brightest galaxy is in the front part of the group than in the back part of it. Paper II shows that when the group centre is not correctly identified, it can cause the majority of the observed redshift excess. If simultaneously the group is also gravitationally unbound, the level of the redshift excess becomes as high as in observations. The explanation of \citet{Byrd:1985p342} and \citet{Valtonen:1986p340} for the origin of the redshift excess was therefore verified in Paper II. This further rendered the need to introduce any ``anomalous" redshift mechanism futile in order to explain the redshift excess first noted by \citet{1970Natur.225.1033A}. The result also underlines the importance of robust group centre and member identification. Interestingly though, this fact, together with the method of Paper I, can also be used as an advantage; that is, to estimate if a given group is likely to be gravitationally bound system.

\section{Summaries of Papers I and II}

\subsection{Summary of Paper I}

The ability to identify gravitationally bound groups of galaxies from observational data is of great importance (as discussed in this Chapter). Paper I shows that a large fraction of groupings are gravitationally unbound when the most often used grouping algorithm (Friends-of-Friends) is applied to simulated data. This result was found to be mostly independent of the cosmology and values of the free parameters adopted as all tested cases led to large fractions of gravitationally unbound systems. We can therefore conclude that the results of Paper I imply that all group catalogues based on the FoF are likely to hold a significant fraction of spurious groups.

To mitigate the difficulty of identifying gravitationally bound groups a method was developed in Paper I that can be used to estimate the probability that an observed group is a gravitationally bound system. This method however requires an estimate for the total group mass, which may not be straightforward to obtain from the available observational data. Even so, this method has been successfully applied, for example, in \cite{Mendel:2008p1040}.

\subsection{Summary of Paper II}

Paper II extended the study and the results of Paper I. It explains a long-standing phenomenon of discordant redshifts that was found from observational galaxy group catalogues. The explanation of redshift asymmetries was found to be two fold: 1) gravitationally unbound groups and 2) the misidentification of the group centre. Together these two difficulties can comprise a redshift asymmetry that is as large as found from observational group catalogues. Thus, observational bias together with the $\Lambda$CDM cosmology can naturally provide a satisfactory explanation for the discordant redshifts without the need for a new physics or more complicated explanations, which had earlier been suggested in literature.

\chapter{Galaxy Evolution}\label{ch:isolated}

\begin{quote}
\footnotesize{
``I hate tennis, hate it with all my heart, and still keep playing, keep hitting all morning, and all afternoon, because I have no choice.''
\begin{flushright}
Andre Agassi
\end{flushright}
}
\end{quote}

\section{Preamble}

Chapter \ref{ch:formation} briefly described the theory of structure formation, while Section \ref{galaxy_formation} concentrated on the theory of galaxy formation and evolution. It was noted that galaxy formation and evolution are the results of a complex sequence of events that occurred during the structure formation. However, how does the theory relate to observations and what can the properties of galaxies tell us about their evolution? Thanks to the finite speed of light, galaxies at earlier cosmic epochs can be studied simply by observing distant galaxies. Hence, observations can help to set constraints for evolution of galaxies. In this Chapter I therefore briefly summarise the observational constraints on galaxy evolution and present two theoretical case studies of the evolution of galaxies.

The galaxies we observe in the Universe at the current epoch (Fig. \ref{fig:galaxies}) exhibit an enormous variety of properties such as morphologies, colours, luminosities, masses and dynamics \citep[e.g.][and references therein]{Blanton:2009p765}. For example, most regular elliptical galaxies have very low atomic gas content, albeit they often have significant hot ionised gas \cite[]{Mathews:2003p787}. Instead, spiral galaxies frequently show star formation, which is fuelled by cool molecular hydrogen. Thus, the galaxy population is vast in their properties, but why do galaxies show such a variety? Is a given property an indication of an evolutionary phase in the life of a galaxy, or have galaxies with different properties formed and evolved in completely different ways?

A goal of galaxy evolution studies is to reconstruct back in time the physical mechanisms that led to the present-day galaxies and to explain them. One of the main questions in the study of galaxy evolution is whether initial conditions of the formation place (and therefore time) will govern the galaxy evolution or if the surrounding environment will shape galaxies when they mature. This question is often dubbed shortly as 'nature versus nurture' dichotomy and has been a long standing puzzle in modern astrophysics. I will return to this important question in Section \ref{s:nature_nurture}, but before delving into it, I shall briefly describe the basic properties of galaxies and how they have evolved as a function of cosmic time.

\section{Properties of galaxies}
\subsection{Morphologies and the colour distribution}

The observed galaxy population, both locally and out to redshift $\sim 1$, is found to be effectively described as a combination of two distinct galaxy types: red, early-type (elliptical or lenticular) galaxies lacking much star formation and blue, late-type\footnote{The nomenclature of early- and late-type is historical and does not reflect the current understanding of galaxy evolution.} (spiral or irregular) galaxies with active star formation \citep[e.g.][]{Strateva:2001p746, Balogh:2004p749, Baldry:2004p745, Bell:2004p747, Kauffmann:2004p751, Croton:2005p404}. In general, bluer galaxies are dominated by emission from young hot stars, while red galaxies contain old stars and/or more attenuating dust. Note, however, that the integrated colour of a galaxy does not necessarily correlate directly with age or the rate in which a given galaxy is forming stars, but can also be due to other effects such as dust or chemical evolution. I will return to this important matter in the following Sections.

\begin{figure}[htb]
\center{
\includegraphics[width=\textwidth]{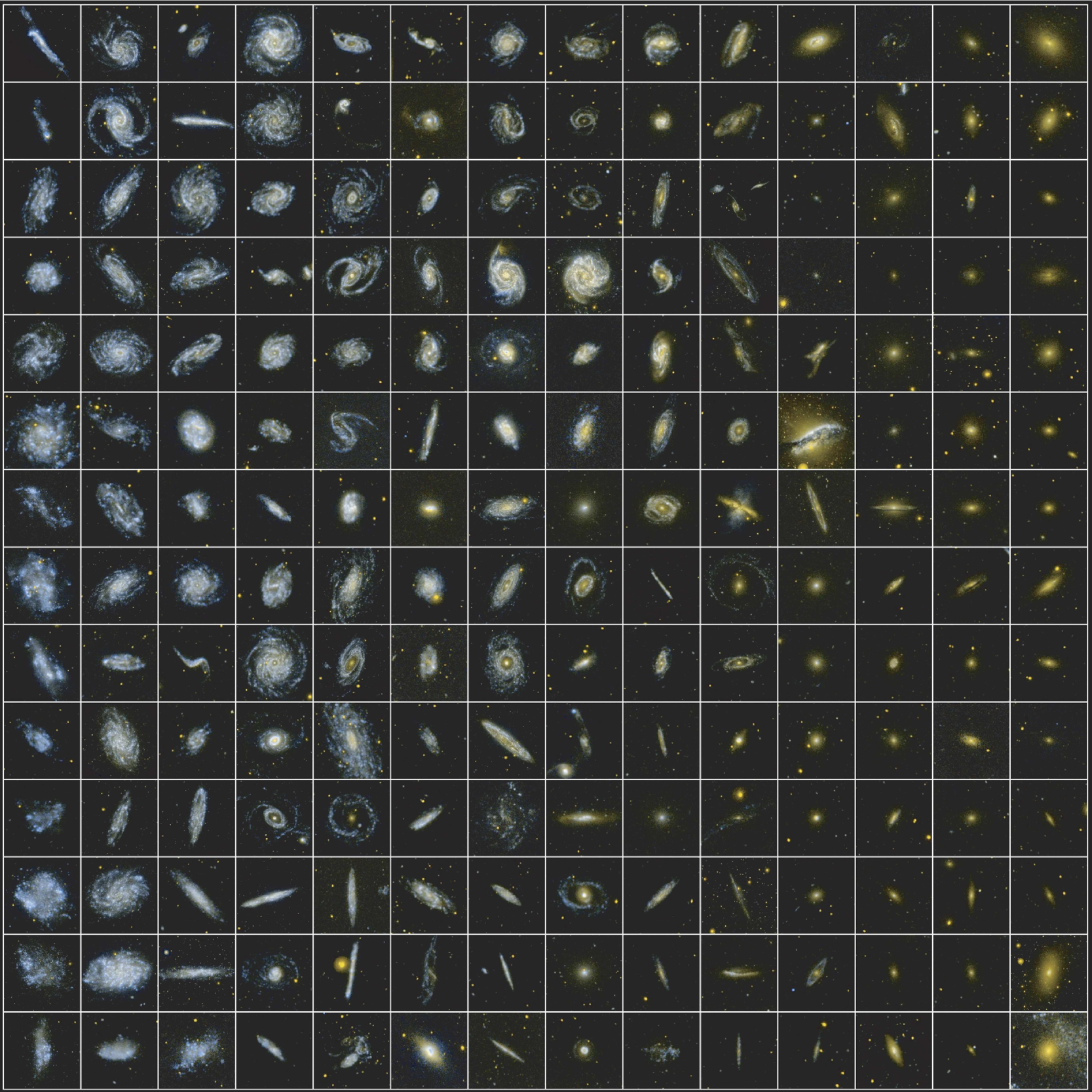}
}
\caption{Galaxies are arranged in bins of increasing ultraviolet colour (the difference between far and near ultraviolet flux). Those with relatively strong far ultraviolet emission appear blue, toward the left, and those with relatively strong near ultraviolet emission appear red, toward the right. Each colour bin is sorted vertically by far ultraviolet luminosity, with the most luminous objects at the top. Courtesy of NASA/JPL-Caltech.}
\label{fig:galaxies}
\end{figure}

The two sequences of galaxies can also be identified in the luminosity function (LF; for a definition, see Eq. \ref{eq:Schechter_LF}): the faint end of the luminosity function is dominated by the blue cloud galaxies, while the bright end is dominated by the red, passively evolving galaxies \citep[e.g.][]{Bell:2004p747}. It has also been noted that while red galaxies constitute roughly one-fifth of the population, they produce about two-fifths of the total cosmic galaxy luminosity density \citep[]{Hogg:2002p788}. More importantly, observations imply that the total stellar mass density of galaxies on the red sequence has roughly doubled over the last $6-8$ Gyr while that of blue galaxies has remained almost constant \citep[e.g.][]{Bell:2004p747, 2007ApJ...665..265F}. If we assume that new stars form mostly in blue galaxies, this suggests that galaxies are being transformed from the blue cloud to the red sequence. This transformation is assumed to take $\sim 2 $ Gyr after the star formation of the blue cloud galaxy has been truncated \citep{Mo_big_book}. Thus, significant evolution between the two sequences is assumed to take place as a function of cosmic time. As a result, it should be kept in mind that even though the galaxy distribution of many properties is bimodal, all galaxies, irrespectively of morphology or colour, show great variations in the amounts and spatial distribution of gas, dust, stars, and metals as well as their luminosities, surface brightnesses, and masses. This is illustrated in Fig. \ref{fig:galaxies}, where a small sample of local galaxies is being presented.

\subsection{The morphology-density relation}

The galaxy population is frequently phrased today in terms of the morphology-density or colour-density relation, while the morphological diversity of galaxies is often dubbed as the galaxy zoo. As first noted by \cite{Hubble:1931p741} and later quantified by several other authors \cite[e.g.][]{Oemler:1974p646, Davis:1976p743, Dressler:1980p645}, the morphology-density relation holds that star-forming, disc-dominated spiral galaxies tend to reside in regions of lower galaxy density relative to those of 'red and dead' elliptical galaxies \citep{Croton:2008p600}. This implies that the environment can affect at least galaxy morphologies and star formation history. Moreover, in disk galaxies, which are rotationally supported, the galaxy sizes are a measure of their specific angular momenta, while in case of elliptical galaxies, which are supported by random motions, the sizes are a measure of the amount of dissipation during their formation. Hence, the morphologies and dynamical differences of galaxies have been interpreted as evidence for different evolutionary histories \citep{1998ARA&A..36..189K}.

In general, it is assumed that the environment local to a galaxy is most fundamental in determining its morphology. The fact that the morphology of a galaxy is closely related to the density of galaxies in its vicinity, also implies that the local environment is of key importance in the formation and evolution of a galaxy. Because groups (see Chapter \ref{ch:groups}) form on time scales of gigayears, the morphology-density relation implies that at least some elliptical and lenticular galaxies are likely to be the result of mergers of spirals. Mergers are therefore probably responsible for a lot of the enhanced early-type fraction in groups and clusters. However, secular evolution \citep[][]{Sellwood:1994p935, Norman:1996p950, Kormendy:2004p958} and disk instabilities can also change the morphology even in absence of mergers \citep[e.g.][but see also Figure \ref{fig:galaxy_formation_flow_chart}]{Raha:1991p943, Dekel:2009p948}. Disk instabilities form when the disk of the galaxy grows large enough and becomes unstable. Thus, morphology alone does not describe the formation history of a galaxy well, though, it can give hints from the possible past merging activity.

The origin of the morphology-density relation can be a combination of several processes (i.e., ram-pressure and tidal stripping, strangulation, galaxy harassment, etc., see for example \cite{Tasca:2009p737}). Thus, whether the influence of the local environment is felt at the time of formation or when the galaxy evolves, or during both phases, is not yet clear. I will, however, return to the question of environment in Sections \ref{s:nature_nurture} and \ref{s:IfEIntro}.

\subsection{Star formation}\label{s:evolution}

One of the main questions in the study of galaxy evolution is related to the star formation rate and history of galaxies and to the physics that triggers star formation \citep[e.g.][]{Kennicutt:1989p976, Madau:1996p891, Madau:1998p752, 1998ARA&A..36..189K, Bouwens:2010p865}. Section \ref{galaxy_formation} showed that as the gas in a dark matter halo cools its self-gravity will eventually dominate over that of the dark matter. It was noted that in general star formation of a galaxy can be assumed to be limited by the availability of hydrogen gas that can cool, fragment and form new stars. Interestingly though, the efficiency of the conversion of molecular gas into stars has been argued to be nearly independent of the galaxy type, its large-scale environment, or the particular local conditions within the galaxy \citep[e.g.][]{Rownd:1999p977, Leroy:2008p975}. So, what controls the star formation?

Based on observations \citep[e.g.][and references therein]{1998ARA&A..36..189K}, it has been argued that star formation takes place in two modes: quiescent star formation in gas disks and circumnuclear starburst. In the disks of spiral galaxies star formation has been observed to proceed at a relatively low pace but in a continuous fashion. Since gas-rich spirals are relatively common in the nearby Universe, disks must be replenished with infall of gas, or otherwise the present time must mark the end of the epoch of star forming disk galaxies. It is also likely that the galaxies that are forming stars at present epoch have obtained cool molecular hydrogen relatively recently, because the typical star formation timescales are $\sim 3 - 4$ Gyr. Indeed, observations \citep[e.g.][]{1980gmcg.work...41S, 1984ApJ...276..182S, 1998ApJ...506L...7F, 2001Natur.409...58P, 2008ApJ...685L...5F} suggest that many spiral galaxies, such as the Galaxy, have a hydrogen gas reservoir around them. This reservoir is likely to consist of three parts: 1) gas that is just falling onto the halo; 2) gas that has already fallen in and been shock heated, but which has not yet cooled radiatively; and 3) gas that has been reheated and expelled from the galaxy due to feedback processes. As a consequence, galaxies with such reservoir can form new stars from the gas that infalls from the reservoir to the disk \citep{Kormendy:2004p958}. Additional to infall, also gas rich (the so-called wet) mergers can bring new gas to a galaxy and induce star formation, even at late cosmic times. Moreover, mergers of galaxies of a roughly equal size (shortly major mergers), may also trigger an exceptionally high rate of star formation or starburst \citep[e.g.][]{Mihos:1996p1093}. Unfortunately though, the exact role of mergers, the composition of molecular gas reservoir, the form and potential evolution of the IMF (briefly mentioned in Section \ref{s:SAMs}), and consequently the detailed physics of star formation are not yet well understood. If the exact physical conditions for star formation are not well known, can the observations tell something about the global star formation rate evolution instead?

\begin{figure}[htb]
\center{
\includegraphics[scale = 0.25]{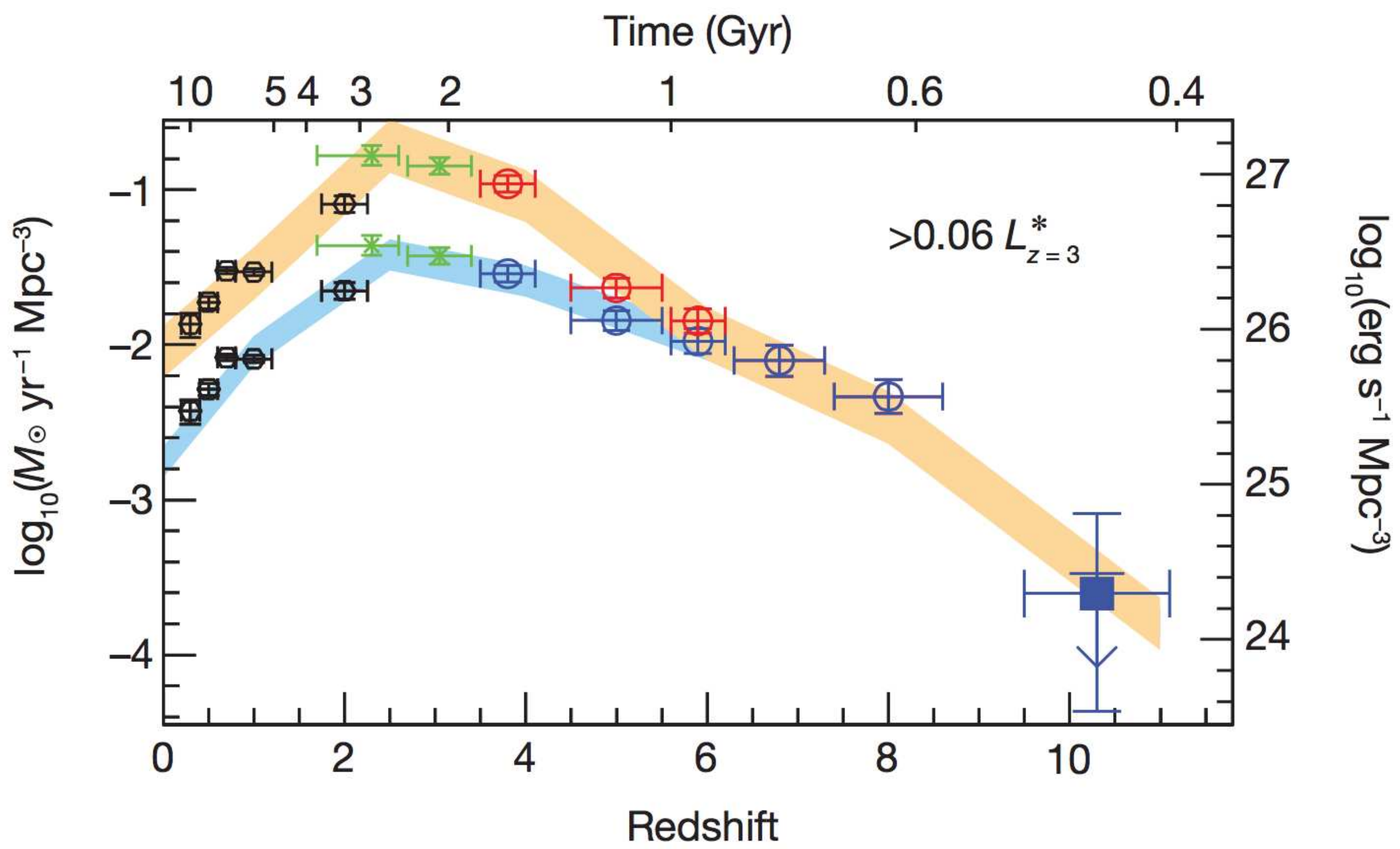}
}
\caption{The star formation rate density (left axis) and luminosity density
(right axis) as a function of cosmic time. The lower set of points (and the blue
region) shows the SFR density determination inferred directly from the UV light,
and the upper set of points (the orange region) shows what one would infer using
dust corrections inferred from the UV-continuum slope measurements. The
conversion from ultraviolet luminosity to star formation rate assumes a Salpeter
initial mass function. Image from \cite{Bouwens:2011p1089}.}
\label{fig:StarFormationHistory}
\end{figure}

The cosmic star formation rate density (SFRD) of the Universe has been observed to evolve significantly with redshift and to peak at $z \sim 2$ \citep[e.g.][]{Madau:1998p752, Giavalisco:2004p753, Bouwens:2010p865, Bouwens:2011p1089}. This is therefore the epoch when the majority of galaxies were growing most vigorously and forming stars fastest. This is likely true for today's elliptical galaxies that show little star formation at current epochs and are therefore often considered to be 'red and dead'. However, this is also true globally as the current SFRD is about a factor of ten smaller than at $z \sim 2$ \citep{Bouwens:2011p1089} as Fig. \ref{fig:StarFormationHistory} shows. Unfortunately, the obscuring effects of dust cast serious uncertainty over the interpretation of data. As a result, the star formation rate density as a function of cosmic time is somewhat uncertain as illustrated by the blue and orange regions in the Figure. Despite the complication due to dust, the decline between $z \sim 0$ and $z \sim 2$ in the global trend of SFRD is obvious. On the other hand however, the rate of decline at high redshifts $(z > 6)$ is more uncertain. This is mainly due to the fact that currently there are only a handful of galaxy candidates at such high redshifts \citep[see e.g.][]{Yan:2009p782, Labbe:2010p776, Bouwens:2011p1089}. Moreover, usually these candidates have not been confirmed spectroscopically, but rely on the dropout technique, which may suffer from contamination. Thus, the SFRD of the early Universe will likely remain an active area of research for years to come. 

Despite the uncertainties in SFRD, it has been argued that the observations indicate that disk galaxies at higher redshifts have higher specific star formation rates (SSFRs). This can be interpreted such that the SFRD is not primarily driven by evolution in the frequency of starburst, but rather reflects a decline in the typical SSFRs of star-forming galaxies \cite[for a detailed discussion, see e.g.][]{Mo_big_book}. I will return to this matter in Section \ref{s:HerschelGalaxies}.

\subsection{Chemical evolution and the age of a galaxy}\label{s:chemical_evolution_of_galaxies}

As already pointed out, blue galaxies are dominantly made out of hot, young, blue stars. As these stars are short lived, one could na\"ively assume that blue galaxies are on average younger than red galaxies. Unfortunately though, another element closely related to star formation, which is not related to the age of the stellar populations, can affect the colour of a galaxy, namely the chemical evolution of stellar populations \citep[e.g.][]{1980FCPh....5..287T, Fall:1993p980, Pei:1995p981, 1997nceg.book.....P, Pipino:2004p947, Tantalo:2010p813, Tumlinson:2010p868}, that is the production of heavy elements in stars. As a result, the chemical properties of galaxies reflect the amount of gas that has been processed and returned by stars. As such the metallicities of galaxies, and stars forming the galaxies, can serve as a fossil record of their evolution.

The chemical evolution of the ISM and IGM \citep[e.g.][]{Matteucci:1986p956, Kauffmann:1998p960, Recchi:2001p959, Oppenheimer:2006p952} is important for several reasons. For example, it dictates that stars that form later in cosmic times have, in a statistical sense, higher abundances of metals\footnote{In Astronomy every element more massive than helium is called a metal.}. As a result, galaxies that are forming stars at the current epoch, late in cosmic time scales, have on average stellar populations that are more metal rich (Population I stars) than stellar populations of non-starforming galaxies (Population II stars). Note, however, that if a galaxy can obtain pristine gas, that is gas that has not yet been polluted by metals, and such gas is not mixed with the existing ISM, the galaxy can form metal poor stars also at late cosmic epochs. In general though, this may not be so simple, because the IGM is also assumed to be metal-enriched.

Usually Population I stars have been found to populate the plane of the disk in spiral galaxies, while Population II stars are located in the spheroidal component of galaxies \citep[for a review of stellar populations in the Galaxy, see][]{1982ARA&A..20...91M}. What does this tell about the evolution of galaxies? Stellar evolutionary models show that typical Pop II stars are old and of low metallicity. As Pop II stars are usually older than $\sim 10$ Gyr, they must have formed early in the formation history of a galaxy. Consequently, elliptical galaxies that are mainly made out of Pop II stars, must have formed most of their stars early. Note, however, that this does not necessarily imply that elliptical galaxies have also assembled at high redshifts. I will return to the formation and assembly times of elliptical galaxies in Section \ref{s:evolutionIfEs}. For now, we can conclude that it is likely that the stellar population that forms an elliptical galaxy has formed a long time ago. How about disk galaxies then?

A fraction of Pop I O and B stars are as young as $\sim 10^{6}$ yr, implying that the disks of spiral galaxies assemble by relatively continuous infall of gas. The existence of Pop I stars could also lead us to conclude that spiral galaxies are predominantly blue and contain a young stellar population. While the latter is at least partially true, the former statement is more complicated due to the chemical evolution. One should note that a confusion in the colours and ages of galaxies arise from the fact that metal rich stars, i.e., Population I stars, have lower temperatures than metal poor stars of the same age and mass, and thus, they end up looking redder. This results in the so-called age-metallicity degeneracy that complicates the age estimates of all galaxies.

The chemical evolution of the gas and stars in galaxies is important also for other reasons. The cooling efficiency of gas has been found to depend strongly on its metallicity, as noted in Section \ref{s:SAMs}. As the metal-enriched gas can cool faster, the chemical evolution of galaxies may offset the star formation rates in these galaxies. Additionally, the extinction by dust (Section \ref{s:dustExtinction}), which is believed to be produced in the envelopes of AGB stars and injected into the ISM through stellar winds, is assumed to depend on, for example, the chemical composition of dust grains \citep{Mo_big_book}. Consequently, the amount of dust in the ISM is assumed to scale with its metallicity. I will return to the importance of ISM and dust when discussing the case study of luminous infrared galaxies in Section \ref{s:HerschelGalaxies}.

\section{Environmental dependence: nature versus nurture}\label{s:nature_nurture}

From the theoretical perspective the environmental dependency can be divided into two parts: those related to the host dark matter halo and those of larger scales (filaments, sheet, etc.). As discussed in Section \ref{s:galaxyFormation}, the properties of a galaxy are assumed to depend strongly on the properties of the host halo, however, the effect of larger scales are less clear. One complication is that the dynamical times on scales significantly larger than dark matter haloes are large. As a result, the gravitational processes have had only little time to make an impact. It is, however, possible that non-gravitational effects may play a role, so we should not exclude the option lightly. Independent of the scale, the complication of studying environmental dependencies arises from the fact that a correlation between galaxy properties and an environmental property does not necessarily imply causality.

Chapter \ref{ch:formation} described how dark matter haloes form from the initial quantum density fluctuations of the early Universe. It is therefore natural to assume that the formation place in space affects the halo collapse. As a consequence, the initial conditions and the formation place of the dark matter halo should also have an effect on the galaxy that forms in the centre of the halo. The part of galaxy evolution, which is imprinted to the initial conditions of dark matter halo formation is often dubbed as 'nature'. Observations have shown that on average more massive dark matter haloes host more massive galaxies \citep[e.g.][]{2002MNRAS.335..311G, 2003ApJ...598..260P, 2004ApJ...606...67H, 2007ApJ...654..153C, 2009MNRAS.394..929C}. One simple explanation for such a correlation can be provided by the argument that more massive haloes form from the larger volumes, which contain more baryons. It has also been noted that, on average, central galaxies in more massive haloes are also redder and more centrally concentrated \citep{Mo_big_book}. These findings provide support for the idea that the immediate environment, i.e. the dark matter halo, can affect the evolution of a galaxy residing in it. Thus, 'nature' should be considered an important part of galaxy formation and evolution.

In Section \ref{s:galaxyFormation} it was noted that dark matter haloes and galaxies are not static nor do they remain unaltered throughout the cosmic evolution of Universe. Observations \citep[e.g.][]{vanDokkum:2005p854, Bell:2006p927, Bell:2006p919, Lin:2008p924, Tacconi:2008p914, Bundy:2009p915, Kormendy:2009p659}, theory, and simulations \citep[e.g.][]{Hausman:1978p921, Lacey:1993p898, Hernquist:1995p917, Springel:2005p916, BoylanKolchin:2006p926, Naab:2006p923, Bournaud:2007p925, Fakhouri:2008p730, Khochfar:2009p725, Stewart:2009p614} have shown that galaxies and dark matter haloes can merge in dense environments. The galaxy properties can therefore be affected by environment through physical mechanisms acting on galaxies. As a result, the surrounding environment of a galaxy and a dark matter halo it resides in should also have an effect on their evolution. The effects directly linked to the physical mechanisms caused by the environment (briefly mentioned in Chapter \ref{ch:groups}) are often described in a collective manner with a word 'nurture'. It therefore seems that also 'nurture' plays a role at least in the formation and evolution of some galaxies.

It is obvious that the properties of galaxies are correlated with their environment, as was noted when discussing the morphology-density relation, but what physical processes are likely to drive these correlations? Interestingly, much of the environmental dependency can be understood in terms of the group environment, discussed briefly in Chapter \ref{ch:groups}. For example, observations seem to favour an important role for preprocessing of galaxies in groups, possibly by mergers \citep[e.g.][]{2004cgpc.symp..277M}, by gas-dynamic interactions with warm or hot gas \citep[e.g.][]{Fujita:2004p793}, or by tidal harassment or ram pressure stripping \citep[e.g.][]{Moore:1996p716, Moore:1998p792}. However, some evidence \citep[e.g.][]{Kauffmann:2004p751, Blanton:2006p789} have accumulated lately that seem to support the opposite conclusion that the large-scale density field appears to be less important than what mass halo hosts the galaxy and what its position is within the halo \citep[e.g.][and references therein]{Blanton:2007p790}. Thus, it has been argued that environmental effects may be relatively local even at low density \citep[][]{Blanton:2009p765}. Hence, the debate over the roles of nature and nurture and their physical mechanisms continues. The environmental effects will be discussed next in a context of a case study of isolated elliptical galaxies. 

\section{A case study: isolated field elliptical galaxies}\label{s:IfEIntro}

\subsection{Background}

The merger hypothesis \citep{Toomre:1972p619} suggests that the product of the merger of two spiral galaxies will be an elliptical galaxy. If this holds, then the probability to find an elliptical galaxy is larger in environments with high densities and low velocity dispersions, i.e., groups of galaxies. This is in agreement with the morphology-density relation that has shown that majority of elliptical galaxies are located in dense regions. However, observations \citep[e.g.][and references therein]{Smith:2004p250, Reda:2004p502, HernandezToledo:2008p648, Norberg:2008p652} have shown that elliptical galaxies can also be found from the field. These galaxies are often dubbed as isolated field elliptical galaxies (IfEs) and are considered an unusual class among the galaxy zoo, because of their "misplacement". Hence, a study that identifies and describes their properties in detail is important if we are to understand the big picture of the formation and evolution of elliptical galaxies and the effects of the environment and formation place.

\subsection{Observational evidence}\label{s:IfE_observations}

In the past two decades several studies based on observations have identified and studied the properties of isolated field elliptical galaxies \citep[e.g.][and references therein]{Smith:2004p250, Reda:2004p502, HernandezToledo:2008p648, Norberg:2008p652}. In many of these studies different possible formation scenarios have been proposed \citep[see e.g.][]{Mulchaey:1999p563, Reda:2004p502, Reda:2007p559}, ranging from a clumpy collapse at an early epoch to multiple merging events. Also equal-mass mergers of two massive galaxies or collapsed groups have been suggested.

Observational studies have also shown that several IfEs reveal a number of features such as tidal tails, dust, shells, discy and boxy isophotes and rapidly rotating discs \citep[e.g.][]{Reduzzi:1996p622, Reda:2004p502, Reda:2005p561, Hau:2006p557, HernandezToledo:2008p648} indicating recent merger and/or accretion events. Some observational evidence suggests that some isolated elliptical galaxies may have suffered late dry mergers \citep{HernandezToledo:2008p648}, while others could have formed via a major merger of two massive galaxies \citep{Reda:2004p502, Reda:2005p561}. A collapsed poor group of a few galaxies has also been suggested as a possible formation scenario \citep{Mulchaey:1999p563}, but other studies \citep[e.g.][]{Marcum:2004p572} have concluded that isolated systems are underluminous by at least a magnitude compared with objects identified as merged group remnants. In general, studies based on observations \citep[e.g.][]{Reda:2007p559} have concluded very broadly that mergers at different redshifts of progenitors of different mass ratios and gas fractions are needed to reproduce the observed properties of IfEs.

Despite the evidence, some IfEs do not show any signs of recent merging activity \citep[e.g.][]{Aars:2001p564, Denicolo:2005p568} complicating the picture of IfE formation even further. It is possible that the merging events have happened in distant past, and all signs of these events have been wiped out. For some IfEs this is even likely as merger remnants usually appear morphologically indistinguishable from a ``typical'' elliptical $\leq 1$ Gyr after the galaxies merged \citep{Combes:1995p663, Mihos:1995p596}. In reality, however, the merger observability time-scales depend on the method used to identify the merger as well as the gas fraction, pericentric distance and relative orientation of the merging galaxies \citep{Lotz:2008p1076}, rendering the time-scale in which a merger can still be observed uncertain. As a consequence, the lack of observable evidence does not exclude the possibility of mergers. On the other hand, it is also possible that some or even all of these galaxies have initially formed in underdense regions and developed quietly without any major mergers and disturbances. It is therefore unclear from the observational point of view how IfEs form and evolve and whether nature or nurture, or both, are important for the evolution of IfEs. These questions were the main driver of Paper III, which took advantage of a semi-analytical galaxy formation model to study the physical properties, evolution, and formation of IfEs. Before discussing the evolution of IfEs in detail, lets briefly look into their basic properties and what hints these properties may give about the evolution of IfEs.

\subsection{The basic properties of IfEs}

It was noted in Paper III that the number density of isolated field elliptical galaxies is as low as $\sim 8.0 \times 10^{-6}h^{3}$ Mpc$^{-3}$. That is a few orders of magnitude lower than that of local luminous spiral galaxies, however, as Paper III shows the number density of IfEs is tightly coupled to the identification criteria adopted. For example, up to an order of magnitude more IfEs can be identified if the criteria are relaxed. This however will affect the immediate environment from which the IfEs are identified, resulting in a sample of less isolated galaxies that can have relatively massive companions. Hence, one can readily conclude that IfEs are relatively rare objects: more isolated ones being the rarest.

Observational studies \citep[e.g.][]{Marcum:2004p572} have found that IfEs show global blue colours compared to other elliptical galaxies. Blue colours can imply that IfEs have formed on average later than group and cluster elliptical galaxies, however, a recent merging event or a gas reservoir could have supplied new gas and induced star formation rendering the integrated colours of the IfEs bluer. Moreover, due to the colour-metallicity degeneracy it is unclear whether the blue global colours should be interpreted as a measure of age or metallicity or both. Hence, from the observational point of view it is unclear if blue colours mean that IfEs are younger than cluster ellipticals, if their metallicities are different, or both.

The median mass weighted age of simulated IfEs was found to be $\sim 8.8$ Gyr (Paper III). The number density of young (mass weighted age $< 5$ Gyr) IfEs is extremely low, giving a lower limit for the age of the IfE population. These age estimates are in agreement with some observations \citep[e.g.][]{Collobert:2006p580}, while others \citep[e.g.][]{Reda:2005p561,Proctor:2005p558} have quoted age estimates around $4$ Gyr. Note, however, that the observational age estimates use a different definition (luminosity weighted age) than the values quoted for simulated IfEs, rendering a direct comparison less conclusive. In any event, \cite{Collobert:2006p580} derived a broad range of stellar ages for their IfEs; ranging from $\sim 2$ to $15$ Gyr in a modest agreement with the simulated IfEs of Paper III. The big scatter in age estimates suggest that the formation of IfEs is not concentrated at a fixed epoch, which further implies that there may also be a large scatter in the basic properties of IfEs.

Paper III shows that on average IfEs have a few $(\sim 5$ - $20)$ companion galaxies. This result alone indicates that IfEs should reside in relatively low density environments, unlike groups (see Chapter \ref{ch:groups}) and clusters, which usually contain $\sim 50$ and $\sim 200$ to $2000$ galaxies, respectively. Moreover, when the number of IfE companions is $\lesssim 20$ as in the case of the most IfEs, most of the surrounding galaxies were found within $\sim 0.5h^{-1}$ Mpc from the IfE. If IfEs were to reside in cluster-type dark matter haloes and environment, companion galaxies should also be found further away from the IfE. Even though for a given IfE the number of companions is small, their properties may still provide some information for the evolution of IfEs and their dark matter haloes. For example, observations \citep[e.g.][]{Reda:2004p502} have found that only the very faint dwarf galaxies ($\mathscr{M}_{R} \geq -15.5$) appear to be associated with isolated ellipticals. On the contrary, simulated IfEs show companion galaxies with a broad range of magnitudes. Deeper observations and larger samples are therefore required to identify more dwarf companions that should surround IfEs as predicted by simulations based on the $\Lambda$CDM cosmology.

The simulated isolated field elliptical galaxies are mainly found to reside in dark matter haloes that are lighter than $7 \times 10^{12}h^{-1}M_{\odot}$, while even the most massive dark matter halo hosting an IfE is lighter than $2.2 \times 10^{13}h^{-1}M_{\odot}$. These masses are comparable to a dark matter halo of a small group thus rendering it impossible that IfEs were collapsed rich groups. \cite{Memola:2009p642} calculated the total masses of two of their isolated ellipticals NGC 7052 and NGC 7785 from X-ray observations and quote values $\sim 5 \times 10^{11}M_{\odot}$ and $\sim 1.9 \times 10^{12}M_{\odot}$, respectively. These mass estimates agree well with the masses of dark matter haloes of simulated IfEs, giving support for the idea that IfEs reside inside relatively light dark matter haloes. In general, galaxies in massive dark matter haloes tend to form the bulk of their stars already at a very early cosmic epoch, rendering it likely that some IfEs contain younger stellar populations than those in cluster ellipticals.

\subsection{Formation and evolution of IfEs}\label{s:evolutionIfEs}

The large scatter in the ages of IfEs suggest that they have formed at different times and under different circumstances. Is this interpretation correct? To quantify the evolution of IfEs one must first define a general set of times related to the formation and evolution of dark matter haloes and galaxies that reside in the haloes. Briefly, the different times that are of interest can be defined as follows \citep[followed by][]{DeLucia:2007p414}:
\begin{itemize}
  \item Assembly time ($z_{a}$) is the redshift when $50$ per cent of the final stellar mass is already present in a single galaxy of the merger tree;
  \item Identity time ($z_{i}$) is the redshift when the last major (the two progenitors both contain at least $20$ per cent of the stellar mass of the descendant galaxy) merger occurred;
  \item Formation time ($z_{f}$) is the redshift when $50$ per cent of the mass of the stars in the final galaxy at $z = 0$ have already formed;
  \item Last merging time ($z_{l}$) is the redshift when the last merger occurred.
\end{itemize}
Using these definitions, it can be shown (for details, see Paper III) that
isolated field elliptical galaxies have assembled at lower redshifts than other
elliptical galaxies. On average, the stars of an IfE assemble to the main halo
later than the stars of group and cluster ellipticals. This is in agreement with
the conventional theory, which states that higher density areas collapse earlier
than less dense areas. Note, however, that in the hierarchical merger scenario,
the star formation history (related to the formation time) and assembly history
of an elliptical galaxy can be very different. For example, in some cases the
stars that will eventually form a simulated IfE galaxy are present already at
high redshifts (high $z_{f}$), in agreement with some observational findings
\cite[e.g.][]{Reda:2005p561}, even though the assembly of the final galaxy will
happen late (small $z_{a}$). The results of Paper III also show that IfEs
undergo their possible major merging events at significantly lower redshifts
than group and cluster ellipticals. This gives further support for the
conclusion that IfEs assemble late.

The mass accretion of IfEs differs from the mass accretion of other elliptical
galaxies according to the findings of Paper III. In general, IfEs seem to form
(or accrete) stars more efficiently than group and cluster ellipticals. However,
oppose to this, IfEs seem to accrete dark matter slightly slower than the
comparison galaxies. Nevertheless, simulations confirm that IfEs form the bulk
of their stars at $z > 2$, as suggested in \cite{Reda:2005p561}, and show that
at a redshift of one IfEs have formed over half of their stars (stellar mass)
and have gathered as much as $80$ per cent of their final dark matter.
Similarly, the group and cluster ellipticals have, on average, accreted roughly
the same fraction of dark matter as IfEs at $z \sim 1$, however, IfEs continue
to accrete dark matter till $z = 0$, while the group and cluster ellipticals
have gathered $\sim 99$ per cent of their final dark matter already at $z \sim
0.5$. All these results point towards a different formation mechanism for
isolated and cluster elliptical galaxies and also suggest that late merging or
accretion events are likely to be a significant part of the evolution of an IfE.

\subsubsection{Formation classes}

Early theoretical studies predicted that isolated ellipticals are formed in relatively recent mergers of spiral galaxy pairs, while large isolated ellipticals may be the result of the merging of a small group of galaxies \citep[e.g.][]{Jones:2000p624, DOnghia:2005p128}. However, these results are in disagreement with some observations (Section \ref{s:IfE_observations}). Moreover, the result that the majority of IfEs reside in light dark matter haloes disagrees with the hypothesis of merging of a group. Hence, the obvious question is if some IfEs are descendants of merged groups while others have developed more quietly, and what are the possible formation mechanisms.

In Paper III three different yet typical formation mechanisms were identified. In the first formation mechanism, named ``solitude'', the IfE develops quietly without any major or minor mergers. Solitude IfEs start to form later than other IfEs and their dark matter haloes are usually lighter than haloes of IfEs that form via different mechanisms. The second formation scenario, called ``coupling'', comprise of IfEs that have undergone at least one ``equal'' sized merger during their evolution. The third and the last formation mechanism identified in Paper III is called ``cannibalism''. IfEs of this formation class undergo several minor and also possibly major mergers during their evolution. The evolution of the cannibal IfEs is significantly impacted by the merging events and their dark matter haloes are usually the most massive ones among all IfEs. Consequently, they can be collapsed poor groups as suggested by, e.g., \cite{Mulchaey:1999p563}. Interestingly, all three formation scenarios are in agreement with selected observations. This shows that the formation of an IfE is not a simple process that could be quantified using one or two simple quantities. Most importantly, Paper III also shows that an IfE can be formed without major mergers. This clearly implies that disk instabilities should be significant in the evolution of some isolated field elliptical galaxies.

Paper III also discusses possible observational techniques that could distinguish between each formation class. Future observational studies can therefore try to identify which formation and evolution process might have been the driving force behind the evolution of an observed IfE. Identification of a formation class can further aid in the quest of understanding the galaxy evolution in different environments and help to attack the fundamental dichotomy of nature vs nurture.

\section{A case study: luminous infrared galaxies}\label{s:HerschelGalaxies}

\subsection{Background}

It was noted in Section \ref{galaxy_formation} that galaxies are not made solely of stars. Instead, a picture was painted in which the stars that form a galaxy are more likely embedded in an interstellar medium (ISM) consisting not only of hot and cold gas, but also dust. Consequently, one of the most important discoveries from extragalactic observations at mid- and far-infrared has been the identification of luminous and ultra-luminous infrared bright galaxies (LIRGs; $L_{\mathrm{IR}} > 10^{11}L_{\odot}$ and ULIRGs; $L_{\mathrm{IR}} > 10^{12}L_{\odot}$, respectively) and they have been studied extensively in the literature \citep[e.g.][]{Rieke:1972p1101,Harwit:1987p1111, Sanders:1988p1131, Sanders:1991p1151, Condon:1991p1127,Sanders:1996p1085, Auriere:1996p1135, Duc:1997p1147, Genzel:1998p1155,Lutz:1998p1152, Rigopoulou:1999p1114, RowanRobinson:2000p1143, Genzel:2001p1106,Colina:2001p1123, Colbert:2006p1150, Dasyra:2006p1119, HernanCaballero:2009p1137, Magdis:2011p1146}. These objects emit more energy in the infrared ($\sim 5 - 500 \ \mu$m) than at all other wavelengths combined. Even though the (ultra-) luminous infrared galaxies are reasonably rare objects in the Universe, reasonable assumptions about the lifetime of the infrared phase suggest that a substantial fraction of all galaxies pass through a stage of intense infrared emission \citep[][and references therein]{Sanders:1996p1085}. Consequently, the majority of the most luminous galaxies in the Universe emit the bulk of their energy in the far-infrared, rendering IR extremely interesting wavelength regime to study, especially in the context of galaxy formation and evolution.

Understanding the IR bright galaxies is especially important because light from bright, young blue stars is often attenuated by dust. The rest-frame ultra-violet (UV) light of a galaxy may therefore provide a biased view on, for example, star formation rate in the galaxy. For example, the global star formation rate required to explain the far-infrared and sub-millimetre background appears to be higher than that inferred from the data in the UV-optical \citep[e.g.][]{Mo_big_book}. The dust attenuated light from blue stars is however not lost, but assumed to be re-radiated thermally at IR wavelengths. As a result, a large fraction of radiation of the cosmic star formation is radiated not at UV but at IR rest-frame wavelengths. This has been found to be true to the extent that the majority of a luminous star-forming galaxy's energy is emitted in the IR \citep[for a review of IR bright galaxies, see][]{Sanders:1996p1085}. It is therefore important to understand the properties and evolution of IR bright galaxies when trying to understand the cosmic star formation history, and galaxy formation and evolution. However, before we can concentrate on these questions, the modelling of dust attenuation and emission should be briefly discussed.

\subsection{Modelling of dust attenuation and emission}\label{s:dustExtinction}

The ISM is assumed to be complex structure \citep[see, for example, the textbooks of][]{1970inme.book.....K, 1989agna.book.....O}, complicating an accurate modelling of it and its interactions with the interstellar radiation. It is usually assumed, however, that most of the dust in the ISM is produced by AGB stars and injected into the ISM through stellar winds. It is therefore likely that the extinction depends on the physical properties of dust grains, which may vary even within a galaxy. Obviously, it is possible to try to quantify the dust extinction by using a statistical description and simplified physics. For example, in semi-analytical models of galaxy formation (Section \ref{s:SAMs}) the dust extinction by ISM is often assumed to follow simply from the diffuse dust in the disc of a galaxy and from a second component, which is associated with the dense `birth clouds' surrounding young star forming regions \citep{Charlot:2000p1068}. Such simplification now allows the computation of the total fraction of the energy emitted by stars that is absorbed by dust, over all wavelengths. If one then assumes that all of this absorbed energy is re-radiated in the IR (hereby neglecting scattering), one can thereby compute the total IR luminosity $L_{\mathrm{IR}}$ of each galaxy. Finally, it is then possible to make use of dust emission templates to determine the spectral energy distribution (SED) of the dust emission, based on the hypothesis that the shape of the dust SED is well-correlated with $L_{\mathrm{IR}}$. The underlying physical notion is that the distribution of dust temperatures is set by the intensity of the local radiation field; thus more luminous or actively star forming galaxies should have a larger proportion of warm dust, as observations \citep[e.g.][]{Sanders:1996p1085} seem to imply.

There are two basic kinds of approaches for constructing these sorts of templates. The first is to use a dust model along with either numerical or analytic solutions to the standard radiative transfer equations to create a library of templates, calibrated by comparison with local prototypes. This approach was pioneered by \cite{Desert:1990p1069} who posited three main sources of dust emission: polycyclic aromatic hydrocarbons (PAHs), very small grains and big grains. The latter are composed of graphite and silicates, with small and big grains probably dominated by graphite and silicate respectively. The thermal properties of each species are determined by the size distribution and thermal state. Big grains are assumed to be in near thermal equilibrium, and their emission can be modelled as a modified black-body spectrum. However, small grains and PAHs are probably in a state that is intermediate between thermal equilibrium and single photon heating. They are therefore subject to temperature fluctuations and their emission spectra are much broader than a modified black-body spectrum. In the \cite{Desert:1990p1069} type approach, the detailed size distributions are modelled using free parameters, which are calibrated by requiring the model to fit a set of observational constraints, such as the extinction or attenuation curves, observed IR colours and the IR spectra of local galaxies. The second approach is to make direct use of observed SEDs \citep[e.g.][]{Chary:2001p1086, Dale:2001p1087, Rieke:2009p1088} for a set of prototype galaxies and to attempt to interpolate between them, allowing to determine the SED of the dust emission. Finally, after a proper SED of the dust emission has been derived, it can be used together with the total IR luminosity of a given galaxy to compute the flux of the galaxy at any given IR band.

\subsection{The physical properties of luminous infrared galaxies}

Although the brightness of a galaxy at a given band is relatively simple to derive from observations, the physical properties are often more important quantities for galaxy formation and evolution. However, when deriving physical properties from observational data, several assumptions are usually required. A priori predictions are therefore often useful when interpreting observational results and drawing conclusions. Predictions for physical properties such as sizes of late-type galaxies, stellar masses, star formation rates, and merger activity of IR bright galaxies are therefore briefly discussed.

The simulated galaxies of Paper IV imply that on average more massive disk galaxies have larger disks. This trend, however, seems to be driven mainly by the galaxies with less massive stellar populations $(\log_{10}(M_{\star}/M_{\odot}) < 10.5)$. Consequently, the trend mostly disappears when limiting to only IR bright galaxies (with $250 \ \mu$m flux $S_{250} > 5$ mJy). The late-type high-redshift galaxies contain stellar disks which on average are $\sim 2.2$ kpc in size. This is significantly larger than the mean disk size ($\sim 0.9$ kpc) of all late-type galaxies in the same redshift range $(2 \leq z < 4)$. This small disk size for all high-redshift late-type galaxies is however driven by the galaxies with the lightest stellar disks (as their number density is highest), which, on average, contain the smallest disks. However, even if the stellar masses are matched the average size of all disk galaxies is almost a factor of two smaller $(\sim 1.3$ kpc$)$ than the mean disk size of the IR bright galaxies. This implies that at high-redshift IR bright galaxies contain on average larger stellar disks than their IR faint counterparts. Interestingly thought, the largest disks are not always associated with galaxies with the highest stellar masses. Instead, they seem to be distributed rather equally for all stellar masses.  Thus, the formation of stellar material, while important in general, is not the only quantity important for sizes of stellar disks of IR bright galaxies. Even so, Paper IV shows that at high redshift $(2 \leq z < 4)$ the IR observations are likely to probe the galaxies with the highest stellar masses. This will obviously bias the observational results if not taken into account, as only the tip of the iceberg is being probed.

Paper IV also shows that at lower redshifts $(z < 0.5)$ the IR bright galaxies
can be found to reside in dark matter haloes as light as
$\log_{10}(M_{\mathrm{DM}}/M_{\odot}) \sim 11.0$. Instead, at higher redshifts
$(2 \leq z < 4)$ Paper IV predicts that all galaxies with $S_{250} > 5$ mJy
reside in  relatively massive dark matter haloes. Even so, it should be noted
that the masses of dark matter haloes cover a broad range from
$\log_{10}(M_{\mathrm{DM}}/M_{\odot}) \sim 11.5$ to $13.5$ at higher redshifts.
Despite the broad range the bulk of IR bright galaxies can be found to reside in
dark matter haloes with masses $\log_{10}(M_{\mathrm{DM}}/M_{\odot}) \sim 12.5$.
Interestingly, the simulated galaxies show evidence for a weak correlation
between the dark matter halo mass and the 250 micron IR flux. Statistically
speaking galaxies residing in more massive dark matter haloes emit a higher
median IR flux, however, this trend was noted to be weak. Moreover, for the
high-redshift galaxies $(2 \leq z < 4)$ with $S_{250} > 5$ mJy no statistically
significant correlation can be confirmed. Thus, something else than what is
directly linked to the mass of a dark matter halo must drive the IR flux.

If dark matter halo masses and stellar disk sizes show only weak correlations
with the IR fluxes at best, how about star formation? It is hardly surprising
that on average the 250 micron IR flux correlates well with a star formation
rate (SFR): the higher the median $S_{250}$ flux the higher the star formation
rate (e.g. Paper IV and references therein). When moving towards earlier cosmic
times it becomes clear that higher and higher star formation rates are required
for galaxies to be detected in currently available IR observations. At $z > 2$
the galaxies directly detectable, for example, with \textit{Herschel} have
median star formation rates $> 30 M_{\odot} \mathrm{yr}^{-1}$, while the average
SFR is $\sim 150 M_{\odot} \mathrm{yr}^{-1}$. This is significantly higher than
for all galaxies in the same redshift range, for which Paper IV predicts a mean
SFR of merely $\sim 2 M_{\odot} \mathrm{yr}^{-1}$. For high-redshift IR bright
galaxies the highest star formation rates predicted are as high as a few
thousand $M_{\odot} \mathrm{yr}^{-1}$, raising an interesting question: what can
cause and fuel such high star formation rates?

\subsection{Merger activity}

Paper IV shows that at high redshift $(2 \leq z < 4)$ the majority of IR bright
($S_{250} > 5$ mJy) galaxies has experienced a merger during their formation
process. To be precise, out of all high redshift IR bright galaxies $\sim 85$
per cent have merged with another galaxy at some point in their formation
history. This alone, however, does not yet imply that a merger activity would be
behind the high star formation rates. When only concentrating on
major\footnote{A major merger is here defined as a galaxy-galaxy merger in which
the mass ratio of the merging pair is $> 1:4$.} mergers that are assumed to
trigger starburst (Section \ref{s:galaxyFormation}), it was noted that about
half of the IR bright galaxies had experienced a major merger. These results
show that a large fraction of all IR bright galaxies have experienced a major
merger and as a consequence experienced a starburst. Could this be the reason
for high SFRs?

As starbursts are usually relatively short lived phenomenon, more important
quantity than the fraction of mergers to look for, is the time since the last
merger. Paper IV shows that about $84$ $(53)$ per cent of high-redshift galaxies
with $S_{160} > 10$ mJy have experienced a (major) merger during their lifetime.
Note, however, that the fraction drops to about $34$ percent if we concentrate
on major mergers that have taken place less than $250$ Myr ago, which are the
mergers that are likely to be causally linked to high star formation rates. For
the $250$ micron band, the results are very similar; about $86$ $(57)$ per cent
of high-redshift galaxies with $S_{250} > 20$ mJy have experienced a (major)
merger. If we again concentrate only on recent mergers in which the major
merging event took place less than $250$ Myr ago, the merger fraction drops to
$\sim 42$ per cent. Thus the model used in Paper IV makes an interesting
prediction:  a significant fraction (half or more) of IR-luminous galaxies at
high redshift ($z>2$) have not experienced a recent merger. This implies that
the high gas accretion rates and efficient feeding via cold flows predicted by
cosmological simulations at high redshift can fuel a significant fraction of the
galaxies detected by \textit{Herschel}. Interestingly, this appears consistent
with preliminary observational results of \cite{Sturm:2010p1055} who
concluded that their two galaxies add support to recent results which indicate
that with an increased gas reservoir star forming galaxies at high redshifts can
achieve ultra-high luminosities without being major mergers. However, as the
samples are still small, more observational evidence is required for more robust
conclusions.

\section{Summaries of Papers III and IV}
\subsection{Summary of Paper III}

In Paper III isolated field elliptical galaxies (IfEs) were studied in detail by using cosmological $N$-body simulations. Elliptical galaxies are usually found to be red and located in dense environments such as cores of groups and clusters of galaxies, unlike IfEs. Hence, the obvious question Paper III sought answer to was how do elliptical galaxies form in underdense regions?

Paper III shows that the formation of isolated field elliptical galaxies is not linked to fossil groups, albeit this has been suggested before, as they reside in significantly lighter dark matter haloes. This also renders the argument, also suggested in literature, that the majority of IfEs are progenitors of collapsed groups, impossible. Instead, simulations show that three different yet typical formation scenarios can lead to an IfE all in agreement with current observations. It was also noted in Paper III that IfEs have high baryon to dark matter fractions, making IfEs therefore good candidates for studies of dark matter poor haloes and their host galaxies.

The most significant result of Paper III is a prediction of a previously
unobserved population of faint and blue isolated field elliptical galaxies. This
population comprises $\sim 25$ per cent of all IfEs, thus, a significant number
of IfEs identified from simulations seem to be missing in observations
\citep[although, see][for possible candidates]{2009AJ....138..579K}. Deeper
observations of the surroundings of faint and blue elliptical galaxies are
required if we are to identify this population. Moreover, these galaxies show
significant star formation and can therefore be important for the study of
evolution of elliptical galaxies as a whole. Thus, an effort to identify them in
the future should be made using, for example, large redshift surveys or
dedicated observations.

\subsection{Summary of Paper IV}

With the \textit{Herschel} space observatory, we can finally probe the infrared
light of faint galaxies all the way from 70 to 500 microns. Unfortunately, at
such long wavelengths contamination and crowding becomes often a limiting factor
rather than the depth of the observations. This is especially true at higher
redshifts. Interpretation of IR observations can therefore be complicated and
less than robust. To help to overcome such complications a theoretical study of
high-redshift $(2 \leq z < 4)$ infrared luminous galaxies was undertaken in
Paper IV. The simulated galaxies of Paper IV were generated using a
semi-analytical galaxy formation model and they were used to study the evolution
of galaxies in the \textit{Herschel} IR bands and to make detailed predictions
for the physical properties and evolution of luminous IR galaxies in hope that
the study could help to shed some light on galaxy evolution. Specifically, the
goal of Paper IV was to present quantitative predictions for the relationship
between the observed SPIRE 250 micron flux and physical quantities such as halo
mass, stellar mass, cold gas mass, star formation rate, and total infrared
luminosity, at different redshifts. The goal was also to quantify the
correlation between SPIRE 250 micron flux and the probability that a galaxy has
experienced a recent major or minor merger.

Results of Paper IV imply, for example, that SPIRE detectable galaxies are on
average significantly larger than IR faint galaxies with comparable stellar
masses. The results also show that in case of luminous IR galaxies the largest
stellar disks are not always associated with the galaxies of the highest stellar
masses. If so, how has all the stellar mass formed then? Not surprisingly Paper
IV shows that at high redshift $(2 \leq z < 4)$ the average star formation rate
of luminous IR galaxies is relatively high ($\sim 150 M_{\odot}
\mathrm{yr}^{-1}$). For high-redshift galaxies the highest SFRs found from the
simulation are as high as a few thousand $M_{\odot} \mathrm{yr}^{-1}$, leading
to the question what fuels such a rapid star formation.

Paper IV tries to answer this question, and shows that about $80$ per cent of
the simulated high-redshift IR bright galaxies have experienced a recent merger.
Almost $40$ per cent of these mergers have been major events that are assumed to
trigger starburst, emphasising the importance of major mergers for IR
brightness. Furthermore, Paper IV finds a fairly strong trend between the 250
micron flux and the probability that a galaxy has had a recent merger,
indicating that brighter galaxies are more likely to be merger driven. However,
also the interesting result was found that in the model, a significant fraction
(half or more) of IR-luminous galaxies at high redshift ($z>2$) have not
experienced a recent major merger. This implies that high gas accretion rates
and efficient feeding via cold flows may fuel a significant fraction of the
galaxies detected by \textit{Herschel}.

%% file: subfiles/bibliography.tex

\addcontentsline{toc}{chapter}{Bibliography}

\setlength{\baselineskip}{15pt}

\bibliographystyle{mn2e} 
\bibliography{subfiles/references}

%% file: main.bbl
\begin{thebibliography}{424}
\expandafter\ifx\csname natexlab\endcsname\relax\def\natexlab#1{#1}\fi

\bibitem[{Aars {et~al.}(2001)Aars, Marcum, \& Fanelli}]{Aars:2001p564}
Aars C.~E., Marcum P.~M., Fanelli M.~N., 2001, AJ, 122, 2923

\bibitem[{{Aarseth}(2003)}]{2003gnbs.book.....A}
{Aarseth} S.~J., 2003, {Gravitational N-Body Simulations, by Sverre J.~Aarseth,
  pp.~430. ISBN 0521432723.~Cambridge, UK: Cambridge University Press, November
  2003.}

\bibitem[{Abazajian {et~al.}(2003)Abazajian, Adelman-McCarthy, Ag{\"u}eros,
  Allam, Anderson, Annis, Bahcall, Baldry, Bastian, Berlind, Bernardi, Blanton,
  Blythe, Bochanski, Boroski, Brewington, Briggs, Brinkmann, Brunner,
  Budav{\'a}ri, Carey, Carr, Castander, Chiu, Collinge, Connolly, Covey,
  Csabai, Dalcanton, Dodelson, Doi, Dong, Eisenstein, Evans, Fan, Feldman,
  Finkbeiner, Friedman, Frieman, Fukugita, Gal, Gillespie, Glazebrook,
  Gonzalez, Gray, Grebel, Grodnicki, Gunn, Gurbani, Hall, Hao, Harbeck, Harris,
  Harris, Harvanek, Hawley, Heckman, Helmboldt, Hendry, Hennessy, Hindsley,
  Hogg, Holmgren, Holtzman, Homer, Hui, ichi Ichikawa, Ichikawa, Inkmann,
  Ivezi{\'c}, Jester, Johnston, Jordan, Jordan, Jorgensen, Juri{\'c},
  Kauffmann, Kent, Kleinman, Knapp, Kniazev, Kron, Krzesi{\'n}ski, Kunszt,
  Kuropatkin, Lamb, Lampeitl, Laubscher, Lee, Leger, Li, Lidz, Lin, Loh, Long,
  Loveday, Lupton, Malik, Margon, McGehee, McKay, Meiksin, Miknaitis, Moorthy,
  Munn, Murphy, Nakajima, Narayanan, Nash, Neilsen, Newberg, Newman, Nichol,
  Nicinski, Nieto-Santisteban, Nitta, Odenkirchen, Okamura, Ostriker, Owen,
  Padmanabhan, Peoples, Pier, Pindor, Pope, Quinn, Rafikov, Raymond, Richards,
  Richmond, Rix, Rockosi, Schaye, Schlegel, Schneider, Schroeder, Scranton,
  Sekiguchi, Seljak, Sergey, Sesar, Sheldon, Shimasaku, Siegmund, Silvestri,
  Sinisgalli, Sirko, Smith, Smol{\v c}i{\'c}, Snedden, Stebbins, Steinhardt,
  Stinson, Stoughton, Strateva, Strauss, SubbaRao, Szalay, Szapudi, Szkody,
  Tasca, Tegmark, Thakar, Tremonti, Tucker, Uomoto, Berk, Vandenberg, Vogeley,
  Voges, Vogt, Walkowicz, Weinberg, West, White, Wilhite, Willman, Xu, Yanny,
  Yarger, Yasuda, Yip, Yocum, York, Zakamska, Zehavi, Zheng, Zibetti, \&
  Zucker}]{Abazajian:2003p798}
Abazajian K., Adelman-McCarthy J.~K., Ag{\"u}eros M.~A., Allam S.~S., Anderson
  S.~F., Annis J., Bahcall N.~A., Baldry I.~K., Bastian S., Berlind A.,
  Bernardi M., Blanton M.~R., Blythe N., Bochanski J.~J., Boroski W.~N.,
  Brewington H., Briggs J.~W., Brinkmann J., Brunner R.~J., Budav{\'a}ri T.,
  Carey L.~N., Carr M.~A., Castander F.~J., Chiu K., Collinge M.~J., Connolly
  A.~J., Covey K.~R., Csabai I., Dalcanton J.~J., Dodelson S., Doi M., Dong F.,
  Eisenstein D.~J., Evans M.~L., Fan X., Feldman P.~D., Finkbeiner D.~P.,
  Friedman S.~D., Frieman J.~A., Fukugita M., Gal R.~R., Gillespie B.,
  Glazebrook K., Gonzalez C.~F., Gray J., Grebel E.~K., Grodnicki L., Gunn
  J.~E., Gurbani V.~K., Hall P.~B., Hao L., Harbeck D., Harris F.~H., Harris
  H.~C., Harvanek M., Hawley S.~L., Heckman T.~M., Helmboldt J.~F., Hendry
  J.~S., Hennessy G.~S., Hindsley R.~B., Hogg D.~W., Holmgren D.~J., Holtzman
  J.~A., Homer L., Hui L., ichi Ichikawa S., Ichikawa T., Inkmann J.~P.,
  Ivezi{\'c} {\v Z}., Jester S., Johnston D.~E., Jordan B., Jordan W.~P.,
  Jorgensen A.~M., Juri{\'c} M., Kauffmann G., Kent S.~M., Kleinman S.~J.,
  Knapp G.~R., Kniazev A.~Y., Kron R.~G., Krzesi{\'n}ski J., Kunszt P.~Z.,
  Kuropatkin N., Lamb D.~Q., Lampeitl H., Laubscher B.~E., Lee B.~C., Leger
  R.~F., Li N., Lidz A., Lin H., Loh Y.-S., Long D.~C., Loveday J., Lupton
  R.~H., Malik T., Margon B., McGehee P.~M., McKay T.~A., Meiksin A., Miknaitis
  G.~A., Moorthy B.~K., Munn J.~A., Murphy T., Nakajima R., Narayanan V.~K.,
  Nash T., Neilsen E.~H., Newberg H.~J., Newman P.~R., Nichol R.~C., Nicinski
  T., Nieto-Santisteban M., Nitta A., Odenkirchen M., Okamura S., Ostriker
  J.~P., Owen R., Padmanabhan N., Peoples J., Pier J.~R., Pindor B., Pope
  A.~C., Quinn T.~R., Rafikov R.~R., Raymond S.~N., Richards G.~T., Richmond
  M.~W., Rix H.-W., Rockosi C.~M., Schaye J., Schlegel D.~J., Schneider D.~P.,
  Schroeder J., Scranton R., Sekiguchi M., Seljak U., Sergey G., Sesar B.,
  Sheldon E., Shimasaku K., Siegmund W.~A., Silvestri N.~M., Sinisgalli A.~J.,
  Sirko E., Smith J.~A., Smol{\v c}i{\'c} V., Snedden S.~A., Stebbins A.,
  Steinhardt C., Stinson G., Stoughton C., Strateva I.~V., Strauss M.~A.,
  SubbaRao M., Szalay A.~S., Szapudi I., Szkody P., Tasca L., Tegmark M.,
  Thakar A.~R., Tremonti C., Tucker D.~L., Uomoto A., Berk D. E.~V., Vandenberg
  J., Vogeley M.~S., Voges W., Vogt N.~P., Walkowicz L.~M., Weinberg D.~H.,
  West A.~A., White S. D.~M., Wilhite B.~C., Willman B., Xu Y., Yanny B.,
  Yarger J., Yasuda N., Yip C.-W., Yocum D.~R., York D.~G., Zakamska N.~L.,
  Zehavi I., Zheng W., Zibetti S., Zucker D.~B., 2003, AJ, 126, 2081

\bibitem[{{Abbott} \& {Wise}(1984)}]{1984NuPhB.244..541A}
{Abbott} L.~F., {Wise} M.~B., 1984, Nuclear Physics B, 244, 541

\bibitem[{{Abel} {et~al.}(2002){Abel}, {Bryan}, \&
  {Norman}}]{2002Sci...295...93A}
{Abel} T., {Bryan} G.~L., {Norman} M.~L., 2002, Science, 295, 93

\bibitem[{Allington-Smith {et~al.}(1993)Allington-Smith, Ellis, Zirbel, \&
  Oemler}]{AllingtonSmith:1993p723}
Allington-Smith J.~R., Ellis R., Zirbel E.~L., Oemler A., 1993, ApJ, 404, 521

\bibitem[{{Alpher} \& {Herman}(1948)}]{1948PhRv...74.1737A}
{Alpher} R.~A., {Herman} R.~C., 1948, Physical Review, 74, 1737

\bibitem[{Andreon {et~al.}(2009)Andreon, Maughan, Trinchieri, \&
  Kurk}]{Andreon:2009p1037}
Andreon S., Maughan B., Trinchieri G., Kurk J., 2009, A{\&}A, 507, 147

\bibitem[{Araya-Melo {et~al.}(2009)Araya-Melo, Reisenegger, Meza, van~de
  Weygaert, D{\"u}nner, \& Quintana}]{ArayaMelo:2009p908}
Araya-Melo P.~A., Reisenegger A., Meza A., van~de Weygaert R., D{\"u}nner R.,
  Quintana H., 2009, MNRAS, 399, 97

\bibitem[{{Arp}(1970)}]{1970Natur.225.1033A}
{Arp} H., 1970, Nature, 225, 1033

\bibitem[{Arp(1982)}]{Arp:1982p577}
Arp H., 1982, ApJ, 256, 54

\bibitem[{Athanassoula {et~al.}(1997)Athanassoula, Makino, \&
  Bosma}]{Athanassoula:1997p1096}
Athanassoula E., Makino J., Bosma A., 1997, MNRAS, 286, 825

\bibitem[{Auriere {et~al.}(1996)Auriere, Hecquet, Coupinot, Arthaud, \&
  Mirabel}]{Auriere:1996p1135}
Auriere M., Hecquet J., Coupinot G., Arthaud R., Mirabel I.~F., 1996, A{\&}A,
  312, 387

\bibitem[{{Bagla} \& {Padmanabhan}(1997)}]{Bagla:2004p1004}
{Bagla} J.~S., {Padmanabhan} T., 1997, Pramana, 49, 161

\bibitem[{Bahcall(1988)}]{Bahcall:1988p909}
Bahcall N.~A., 1988, Annual Review of Astronomy and Astrophysics, 26, 631

\bibitem[{Bahcall {et~al.}(2003)Bahcall, McKay, Annis, Kim, Dong, Hansen, Goto,
  Gunn, Miller, Nichol, Postman, Schneider, Schroeder, Voges, Brinkmann, \&
  Fukugita}]{Bahcall:2003p227}
Bahcall N.~A., McKay T.~A., Annis J., Kim R. S.~J., Dong F., Hansen S., Goto
  T., Gunn J.~E., Miller C.~J., Nichol R.~C., Postman M., Schneider D.~P.,
  Schroeder J., Voges W., Brinkmann J., Fukugita M., 2003, ApJS, 148, 243

\bibitem[{Baldry {et~al.}(2004)Baldry, Glazebrook, Brinkmann, Ivezi{\'c},
  Lupton, Nichol, \& Szalay}]{Baldry:2004p745}
Baldry I.~K., Glazebrook K., Brinkmann J., Ivezi{\'c} {\v Z}., Lupton R.~H.,
  Nichol R.~C., Szalay A.~S., 2004, ApJ, 600, 681

\bibitem[{Balian \& Schaeffer(1989)}]{Balian:1989p938}
Balian R., Schaeffer R., 1989, A\&A, 226, 373

\bibitem[{Balogh {et~al.}(2004{\natexlab{a}})Balogh, Eke, Miller, Lewis, Bower,
  Couch, Nichol, Bland-Hawthorn, Baldry, Baugh, Bridges, Cannon, Cole, Colless,
  Collins, Cross, Dalton, Propris, Driver, Efstathiou, Ellis, Frenk,
  Glazebrook, Gomez, Gray, Hawkins, Jackson, Lahav, Lumsden, Maddox, Madgwick,
  Norberg, Peacock, Percival, Peterson, Sutherland, \&
  Taylor}]{Balogh:2004p748}
Balogh M., Eke V., Miller C., Lewis I., Bower R., Couch W., Nichol R.,
  Bland-Hawthorn J., Baldry I.~K., Baugh C., Bridges T., Cannon R., Cole S.,
  Colless M., Collins C., Cross N., Dalton G., Propris R.~D., Driver S.~P.,
  Efstathiou G., Ellis R.~S., Frenk C.~S., Glazebrook K., Gomez P., Gray A.,
  Hawkins E., Jackson C., Lahav O., Lumsden S., Maddox S., Madgwick D., Norberg
  P., Peacock J.~A., Percival W., Peterson B.~A., Sutherland W., Taylor K.,
  2004{\natexlab{a}}, MNRAS, 348, 1355

\bibitem[{Balogh {et~al.}(2004{\natexlab{b}})Balogh, Baldry, Nichol, Miller,
  Bower, \& Glazebrook}]{Balogh:2004p749}
Balogh M.~L., Baldry I.~K., Nichol R., Miller C., Bower R., Glazebrook K.,
  2004{\natexlab{b}}, ApJ, 615, L101

\bibitem[{{Baltz}(2004)}]{2004astro.ph.12170B}
{Baltz} E.~A., 2004, arXiv:astro-ph/0412170

\bibitem[{{Bardeen} {et~al.}(1986){Bardeen}, {Bond}, {Kaiser}, \&
  {Szalay}}]{1986ApJ...304...15B}
{Bardeen} J.~M., {Bond} J.~R., {Kaiser} N., {Szalay} A.~S., 1986, ApJ, 304, 15

\bibitem[{{Barkana} \& {Loeb}(2001)}]{2001PhR...349..125B}
{Barkana} R., {Loeb} A., 2001, Physics Reports, 349, 125

\bibitem[{Barnes(1985)}]{Barnes:1985p708}
Barnes J., 1985, MNRAS, 215, 517

\bibitem[{Barnes \& Hut(1986)}]{Barnes:1986p1014}
Barnes J., Hut P., 1986, Nature, 324, 446

\bibitem[{Bartelmann(2010)}]{Bartelmann:2010p998}
Bartelmann M., 2010, Reviews of Modern Physics, 82, 331

\bibitem[{Barton {et~al.}(1996)Barton, Geller, Ramella, Marzke, \&
  da~Costa}]{Barton:1996p269}
Barton E., Geller M., Ramella M., Marzke R.~O., da~Costa L.~N., 1996, AJ, 112,
  871

\bibitem[{Baryshev {et~al.}(2001)Baryshev, Chernin, \&
  Teerikorpi}]{Baryshev:2001p969}
Baryshev Y.~V., Chernin A.~D., Teerikorpi P., 2001, A{\&}A, 378, 729

\bibitem[{Batuski \& Burns(1985)}]{Batuski:1985p907}
Batuski D.~J., Burns J.~O., 1985, ApJ, 299, 5

\bibitem[{Baugh(2006)}]{Baugh:2006p734}
Baugh C.~M., 2006, Reports on Progress in Physics, 69, 3101

\bibitem[{Baugh {et~al.}(1998)Baugh, Cole, Frenk, \& Lacey}]{Baugh:1998p847}
Baugh C.~M., Cole S., Frenk C.~S., Lacey C.~G., 1998, ApJ, 498, 504

\bibitem[{{Baumann}(2007)}]{Baumann:2007p979}
{Baumann} D., 2007, ArXiv:astro-ph/0710.3187

\bibitem[{Bell {et~al.}(2006{\natexlab{a}})Bell, Naab, McIntosh, Somerville,
  Caldwell, Barden, Wolf, Rix, Beckwith, Borch, H{\"a}ussler, Heymans, Jahnke,
  Jogee, Koposov, Meisenheimer, Peng, Sanchez, \& Wisotzki}]{Bell:2006p919}
Bell E.~F., Naab T., McIntosh D.~H., Somerville R.~S., Caldwell J. A.~R.,
  Barden M., Wolf C., Rix H.-W., Beckwith S.~V., Borch A., H{\"a}ussler B.,
  Heymans C., Jahnke K., Jogee S., Koposov S., Meisenheimer K., Peng C.~Y.,
  Sanchez S.~F., Wisotzki L., 2006{\natexlab{a}}, ApJ, 640, 241

\bibitem[{Bell {et~al.}(2006{\natexlab{b}})Bell, Phleps, Somerville, Wolf,
  Borch, \& Meisenheimer}]{Bell:2006p927}
Bell E.~F., Phleps S., Somerville R.~S., Wolf C., Borch A., Meisenheimer K.,
  2006{\natexlab{b}}, ApJ, 652, 270

\bibitem[{Bell {et~al.}(2004)Bell, Wolf, Meisenheimer, Rix, Borch, Dye,
  Kleinheinrich, Wisotzki, \& McIntosh}]{Bell:2004p747}
Bell E.~F., Wolf C., Meisenheimer K., Rix H.-W., Borch A., Dye S.,
  Kleinheinrich M., Wisotzki L., McIntosh D.~H., 2004, ApJ, 608, 752

\bibitem[{Belokurov {et~al.}(2008)Belokurov, Walker, Evans, Faria, Gilmore,
  Irwin, Koposov, Mateo, Olszewski, \& Zucker}]{Belokurov:2008p705}
Belokurov V., Walker M.~G., Evans N.~W., Faria D.~C., Gilmore G., Irwin M.~J.,
  Koposov S., Mateo M., Olszewski E., Zucker D.~B., 2008, ApJ, 686, L83

\bibitem[{{Belokurov} {et~al.}(2010){Belokurov}, {Walker}, {Evans}, {Gilmore},
  {Irwin}, {Just}, {Koposov}, {Mateo}, {Olszewski}, {Watkins}, \&
  {Wyrzykowski}}]{Belokurov:2010p756}
{Belokurov} V., {Walker} M.~G., {Evans} N.~W., {Gilmore} G., {Irwin} M.~J.,
  {Just} D., {Koposov} S., {Mateo} M., {Olszewski} E., {Watkins} L.,
  {Wyrzykowski} L., 2010, ApJL, 712, L103

\bibitem[{Benson {et~al.}(2003)Benson, Bower, Frenk, Lacey, Baugh, \&
  Cole}]{Benson:2003p964}
Benson A.~J., Bower R.~G., Frenk C.~S., Lacey C.~G., Baugh C.~M., Cole S.,
  2003, ApJ, 599, 38

\bibitem[{Benson \& Devereux(2010)}]{Benson:2010p962}
Benson A.~J., Devereux N., 2010, MNRAS, 402, 2321

\bibitem[{Berlind {et~al.}(2006)Berlind, Frieman, Weinberg, Blanton, Warren,
  Abazajian, Scranton, Hogg, Scoccimarro, Bahcall, Brinkmann, Gott, Kleinman,
  Krzesinski, Lee, Miller, Nitta, Schneider, Tucker, \&
  Zehavi}]{Berlind:2006p484}
Berlind A.~A., Frieman J., Weinberg D.~H., Blanton M.~R., Warren M.~S.,
  Abazajian K., Scranton R., Hogg D.~W., Scoccimarro R., Bahcall N.~A.,
  Brinkmann J., Gott J.~R., Kleinman S.~J., Krzesinski J., Lee B.~C., Miller
  C.~J., Nitta A., Schneider D.~P., Tucker D.~L., Zehavi I., 2006, ApJS, 167, 1

\bibitem[{Bertone {et~al.}(2005)Bertone, Hooper, \& Silk}]{Bertone:2004pz}
Bertone G., Hooper D., Silk J., 2005, Phys. Rept., 405, 279

\bibitem[{Bielby {et~al.}(2010)Bielby, Finoguenov, Tanaka, McCracken, Daddi,
  Hudelot, Ilbert, Kneib, F{\`e}vre, Mellier, Nandra, Petitjean, Srianand,
  Stalin, \& Willott}]{Bielby:2010p1077}
Bielby R.~M., Finoguenov A., Tanaka M., McCracken H.~J., Daddi E., Hudelot P.,
  Ilbert O., Kneib J.~P., F{\`e}vre O.~L., Mellier Y., Nandra K., Petitjean P.,
  Srianand R., Stalin C.~S., Willott C.~J., 2010, A{\&}A, 523, 66

\bibitem[{Binney(1977)}]{Binney:1977p1021}
Binney J., 1977, ApJ, 215, 483

\bibitem[{Birnboim \& Dekel(2003)}]{Birnboim:2003p922}
Birnboim Y., Dekel A., 2003, MNRAS, 345, 349

\bibitem[{Blanton(2006)}]{Blanton:2006p789}
Blanton M.~R., 2006, ApJ, 648, 268

\bibitem[{Blanton \& Berlind(2007)}]{Blanton:2007p790}
Blanton M.~R., Berlind A.~A., 2007, ApJ, 664, 791

\bibitem[{Blanton \& Moustakas(2009)}]{Blanton:2009p765}
Blanton M.~R., Moustakas J., 2009, Annual Review of Astronomy and Astrophysics,
  47, 159

\bibitem[{Bond {et~al.}(1991)Bond, Cole, Efstathiou, \& Kaiser}]{Bond:1991p899}
Bond J.~R., Cole S., Efstathiou G., Kaiser N., 1991, ApJ, 379, 440

\bibitem[{Bond \& Efstathiou(1984)}]{Bond:1984p989}
Bond J.~R., Efstathiou G., 1984, ApJ, 285, L45

\bibitem[{Borne {et~al.}(2000)Borne, Bushouse, Lucas, \&
  Colina}]{Borne:2000p1090}
Borne K.~D., Bushouse H., Lucas R.~A., Colina L., 2000, ApJ, 529, L77

\bibitem[{Bottinelli \& Gouguenheim(1973)}]{Bottinelli:1973p352}
Bottinelli L., Gouguenheim L., 1973, A{\&}A, 26, 85

\bibitem[{{Botzler} {et~al.}(2004){Botzler}, {Snigula}, {Bender}, \&
  {Hopp}}]{2004MNRAS.349..425B}
{Botzler} C.~S., {Snigula} J., {Bender} R., {Hopp} U., 2004, MNRAS, 349, 425

\bibitem[{Bournaud {et~al.}(2007)Bournaud, Jog, \& Combes}]{Bournaud:2007p925}
Bournaud F., Jog C.~J., Combes F., 2007, A{\&}A, 476, 1179

\bibitem[{Bouwens {et~al.}(2011)Bouwens, Illingworth, Labbe, Oesch, Trenti,
  Carollo, Dokkum, Franx, Stiavelli, Gonz{\'a}lez, Magee, \&
  Bradley}]{Bouwens:2011p1089}
Bouwens R.~J., Illingworth G.~D., Labbe I., Oesch P.~A., Trenti M., Carollo
  C.~M., Dokkum P. G.~V., Franx M., Stiavelli M., Gonz{\'a}lez V., Magee D.,
  Bradley L., 2011, Nature, 469, 504

\bibitem[{Bouwens {et~al.}(2010)Bouwens, Illingworth, Oesch, Stiavelli, Dokkum,
  Trenti, Magee, Labb{\'e}, Franx, Carollo, \& Gonzalez}]{Bouwens:2010p865}
Bouwens R.~J., Illingworth G.~D., Oesch P.~A., Stiavelli M., Dokkum P.~V.,
  Trenti M., Magee D., Labb{\'e} I., Franx M., Carollo C.~M., Gonzalez V.,
  2010, ApJL, 709, L133

\bibitem[{Bower(1991)}]{Bower:1991p695}
Bower R.~G., 1991, MNRAS, 248, 332

\bibitem[{Boylan-Kolchin {et~al.}(2006)Boylan-Kolchin, Ma, \&
  Quataert}]{BoylanKolchin:2006p926}
Boylan-Kolchin M., Ma C.-P., Quataert E., 2006, MNRAS, 369, 1081

\bibitem[{Boylan-Kolchin {et~al.}(2008)Boylan-Kolchin, Ma, \&
  Quataert}]{BoylanKolchin:2008p1099}
---, 2008, MNRAS, 383, 93

\bibitem[{Boylan-Kolchin {et~al.}(2009)Boylan-Kolchin, Springel, White,
  Jenkins, \& Lemson}]{BoylanKolchin:2009p758}
Boylan-Kolchin M., Springel V., White S. D.~M., Jenkins A., Lemson G., 2009,
  MNRAS, 398, 1150

\bibitem[{{Bromm} {et~al.}(1999){Bromm}, {Coppi}, \&
  {Larson}}]{1999ApJ...527L...5B}
{Bromm} V., {Coppi} P.~S., {Larson} R.~B., 1999, ApJL, 527, L5

\bibitem[{{Bromm} {et~al.}(2009){Bromm}, {Yoshida}, {Hernquist}, \&
  {McKee}}]{2009Natur.459...49B}
{Bromm} V., {Yoshida} N., {Hernquist} L., {McKee} C.~F., 2009, Nature, 459, 49

\bibitem[{Bryan \& Norman(1998)}]{Bryan:1998p1029}
Bryan G.~L., Norman M.~L., 1998, ApJ, 495, 80

\bibitem[{Bullock {et~al.}(2001)Bullock, Kravtsov, \&
  Weinberg}]{Bullock:2001p893}
Bullock J.~S., Kravtsov A.~V., Weinberg D.~H., 2001, ApJ, 548, 33

\bibitem[{Bundy {et~al.}(2009)Bundy, Fukugita, Ellis, Targett, Belli, \&
  Kodama}]{Bundy:2009p915}
Bundy K., Fukugita M., Ellis R.~S., Targett T.~A., Belli S., Kodama T., 2009,
  ApJ, 697, 1369

\bibitem[{Bunn \& White(1997)}]{Bunn:1997p1008}
Bunn E.~F., White M., 1997, ApJ, 480, 6

\bibitem[{Byrd \& Valtonen(1985)}]{Byrd:1985p342}
Byrd G.~G., Valtonen M., 1985, ApJ, 289, 535

\bibitem[{{Cacciato} {et~al.}(2009){Cacciato}, {van den Bosch}, {More}, {Li},
  {Mo}, \& {Yang}}]{2009MNRAS.394..929C}
{Cacciato} M., {van den Bosch} F.~C., {More} S., {Li} R., {Mo} H.~J., {Yang}
  X., 2009, MNRAS, 394, 929

\bibitem[{Ceccarelli {et~al.}(2006)Ceccarelli, Padilla, Valotto, \&
  Lambas}]{Ceccarelli:2006p524}
Ceccarelli L., Padilla N.~D., Valotto C., Lambas D.~G., 2006, MNRAS, 373, 1440

\bibitem[{Cen(1992)}]{Cen:1992p896}
Cen R., 1992, ApJS, 78, 341

\bibitem[{Chabrier(2003)}]{Chabrier:2003p876}
Chabrier G., 2003, PASP, 115, 763

\bibitem[{Charlot \& Fall(2000)}]{Charlot:2000p1068}
Charlot S., Fall S.~M., 2000, ApJ, 539, 718

\bibitem[{Chary \& Elbaz(2001)}]{Chary:2001p1086}
Chary R., Elbaz D., 2001, ApJ, 556, 562

\bibitem[{Chernin {et~al.}(2009)Chernin, Teerikorpi, Valtonen, Dolgachev,
  Domozhilova, \& Byrd}]{Chernin:2009p968}
Chernin A.~D., Teerikorpi P., Valtonen M.~J., Dolgachev V.~P., Domozhilova
  L.~M., Byrd G.~G., 2009, A{\&}A, 507, 1271

\bibitem[{Coil {et~al.}(2006{\natexlab{a}})Coil, Gerke, Newman, Ma, Yan,
  Cooper, Davis, Faber, Guhathakurta, \& Koo}]{Coil:2006p1079}
Coil A.~L., Gerke B.~F., Newman J.~A., Ma C.-P., Yan R., Cooper M.~C., Davis
  M., Faber S.~M., Guhathakurta P., Koo D.~C., 2006{\natexlab{a}}, ApJ, 638,
  668

\bibitem[{Coil {et~al.}(2007)Coil, Hennawi, Newman, Cooper, \&
  Davis}]{Coil:2007p1080}
Coil A.~L., Hennawi J.~F., Newman J.~A., Cooper M.~C., Davis M., 2007, ApJ,
  654, 115

\bibitem[{Coil {et~al.}(2006{\natexlab{b}})Coil, Newman, Cooper, Davis, Faber,
  Koo, \& Willmer}]{Coil:2006p1078}
Coil A.~L., Newman J.~A., Cooper M.~C., Davis M., Faber S.~M., Koo D.~C.,
  Willmer C. N.~A., 2006{\natexlab{b}}, ApJ, 644, 671

\bibitem[{Colbert {et~al.}(2006)Colbert, Teplitz, Francis, Palunas, Williger,
  \& Woodgate}]{Colbert:2006p1150}
Colbert J.~W., Teplitz H., Francis P., Palunas P., Williger G.~M., Woodgate B.,
  2006, ApJ, 637, L89

\bibitem[{Cole(1991)}]{Cole:1991p857}
Cole S., 1991, ApJ, 367, 45

\bibitem[{Cole {et~al.}(2008)Cole, Helly, Frenk, \& Parkinson}]{Cole:2008p599}
Cole S., Helly J., Frenk C.~S., Parkinson H., 2008, MNRAS, 383, 546

\bibitem[{Cole {et~al.}(2005)Cole, Percival, Peacock, Norberg, Baugh, Frenk,
  Baldry, Bland-Hawthorn, Bridges, Cannon, Colless, Collins, Couch, Cross,
  Dalton, Eke, Propris, Driver, Efstathiou, Ellis, Glazebrook, Jackson,
  Jenkins, Lahav, Lewis, Lumsden, Maddox, Madgwick, Peterson, Sutherland, \&
  Taylor}]{Cole:2005p848}
Cole S., Percival W.~J., Peacock J.~A., Norberg P., Baugh C.~M., Frenk C.~S.,
  Baldry I., Bland-Hawthorn J., Bridges T., Cannon R., Colless M., Collins C.,
  Couch W., Cross N. J.~G., Dalton G., Eke V.~R., Propris R.~D., Driver S.~P.,
  Efstathiou G., Ellis R.~S., Glazebrook K., Jackson C., Jenkins A., Lahav O.,
  Lewis I., Lumsden S., Maddox S., Madgwick D., Peterson B.~A., Sutherland W.,
  Taylor K., 2005, MNRAS, 362, 505

\bibitem[{{Coles} \& {Lucchin}(2002)}]{2002coec.book.....C}
{Coles} P., {Lucchin} F., 2002, Cosmology: The Origin and Evolution of Cosmic
  Structure, Second Edition, by Peter Coles, Francesco Lucchin, pp.~512.~ISBN
  0-471-48909-3.~Wiley-VCH , July 2002.

\bibitem[{Colina {et~al.}(2001)Colina, Borne, Bushouse, Lucas, Rowan-Robinson,
  Lawrence, Clements, Baker, \& Oliver}]{Colina:2001p1123}
Colina L., Borne K., Bushouse H., Lucas R.~A., Rowan-Robinson M., Lawrence A.,
  Clements D., Baker A., Oliver S., 2001, ApJ, 563, 546

\bibitem[{{Colless}(1999)}]{1999RSPTA.357..105C}
{Colless} M., 1999, Royal Society of London Philosophical Transactions Series
  A, 357, 105

\bibitem[{Colless {et~al.}(2001)Colless, Dalton, Maddox, Sutherland, Norberg,
  Cole, Bland-Hawthorn, Bridges, Cannon, Collins, Couch, Cross, Deeley,
  Propris, Driver, Efstathiou, Ellis, Frenk, Glazebrook, Jackson, Lahav, Lewis,
  Lumsden, Madgwick, Peacock, Peterson, Price, Seaborne, \&
  Taylor}]{Colless:2001p802}
Colless M., Dalton G., Maddox S., Sutherland W., Norberg P., Cole S.,
  Bland-Hawthorn J., Bridges T., Cannon R., Collins C., Couch W., Cross N.,
  Deeley K., Propris R.~D., Driver S.~P., Efstathiou G., Ellis R.~S., Frenk
  C.~S., Glazebrook K., Jackson C., Lahav O., Lewis I., Lumsden S., Madgwick
  D., Peacock J.~A., Peterson B.~A., Price I., Seaborne M., Taylor K., 2001,
  MNRAS, 328, 1039

\bibitem[{Collobert {et~al.}(2006)Collobert, Sarzi, Davies, Kuntschner, \&
  Colless}]{Collobert:2006p580}
Collobert M., Sarzi M., Davies R.~L., Kuntschner H., Colless M., 2006, MNRAS,
  370, 1213

\bibitem[{Colpi {et~al.}(1999)Colpi, Mayer, \& Governato}]{Colpi:1999p1100}
Colpi M., Mayer L., Governato F., 1999, ApJ, 525, 720

\bibitem[{Combes {et~al.}(1995)Combes, Rampazzo, Bonfanti, Prugniel, \&
  Sulentic}]{Combes:1995p663}
Combes F., Rampazzo R., Bonfanti P.~P., Prugniel P., Sulentic J.~W., 1995,
  A\&A, 297, 37

\bibitem[{Condon {et~al.}(1991)Condon, Huang, Yin, \& Thuan}]{Condon:1991p1127}
Condon J.~J., Huang Z.-P., Yin Q.~F., Thuan T.~X., 1991, ApJ, 378, 65

\bibitem[{Connolly {et~al.}(2002)Connolly, Scranton, Johnston, Dodelson,
  Eisenstein, Frieman, Gunn, Hui, Jain, Kent, Loveday, Nichol, O'Connell,
  Postman, Scoccimarro, Sheth, Stebbins, Strauss, Szalay, Szapudi, Tegmark,
  Vogeley, Zehavi, Annis, Bahcall, Brinkmann, Csabai, Doi, Fukugita, Hennessy,
  Hindsley, Ichikawa, Ivezi{\'c}, Kim, Knapp, Kunszt, Lamb, Lee, Lupton, McKay,
  Munn, Peoples, Pier, Rockosi, Schlegel, Stoughton, Tucker, Yanny, \&
  York}]{Connolly:2002p806}
Connolly A.~J., Scranton R., Johnston D., Dodelson S., Eisenstein D.~J.,
  Frieman J.~A., Gunn J.~E., Hui L., Jain B., Kent S., Loveday J., Nichol
  R.~C., O'Connell L., Postman M., Scoccimarro R., Sheth R.~K., Stebbins A.,
  Strauss M.~A., Szalay A.~S., Szapudi I., Tegmark M., Vogeley M.~S., Zehavi
  I., Annis J., Bahcall N., Brinkmann J., Csabai I., Doi M., Fukugita M.,
  Hennessy G.~S., Hindsley R., Ichikawa T., Ivezi{\'c} {\v Z}., Kim R. S.~J.,
  Knapp G.~R., Kunszt P., Lamb D.~Q., Lee B.~C., Lupton R.~H., McKay T.~A.,
  Munn J., Peoples J., Pier J., Rockosi C., Schlegel D., Stoughton C., Tucker
  D.~L., Yanny B., York D.~G., 2002, ApJ, 579, 42

\bibitem[{{Conroy} {et~al.}(2007){Conroy}, {Prada}, {Newman}, {Croton}, {Coil},
  {Conselice}, {Cooper}, {Davis}, {Faber}, {Gerke}, {Guhathakurta}, {Klypin},
  {Koo}, \& {Yan}}]{2007ApJ...654..153C}
{Conroy} C., {Prada} F., {Newman} J.~A., {Croton} D., {Coil} A.~L., {Conselice}
  C.~J., {Cooper} M.~C., {Davis} M., {Faber} S.~M., {Gerke} B.~F.,
  {Guhathakurta} P., {Klypin} A., {Koo} D.~C., {Yan} R., 2007, ApJ, 654, 153

\bibitem[{Cooray \& Milosavljevi{\'c}(2005)}]{Cooray:2005p697}
Cooray A., Milosavljevi{\'c} M., 2005, ApJ, 627, L89

\bibitem[{Couchman(1991)}]{Couchman:1991p1012}
Couchman H. M.~P., 1991, ApJ, 368, L23

\bibitem[{Croton \& Farrar(2008)}]{Croton:2008p600}
Croton D.~J., Farrar G.~R., 2008, MNRAS, 386, 2285

\bibitem[{Croton {et~al.}(2005)Croton, Farrar, Norberg, Colless, Peacock,
  Baldry, Baugh, Bland-Hawthorn, Bridges, Cannon, Cole, Collins, Couch, Dalton,
  Propris, Driver, Efstathiou, Ellis, Frenk, Glazebrook, Jackson, Lahav, Lewis,
  Lumsden, Maddox, Madgwick, Peterson, Sutherland, \& Taylor}]{Croton:2005p404}
Croton D.~J., Farrar G.~R., Norberg P., Colless M., Peacock J.~A., Baldry
  I.~K., Baugh C.~M., Bland-Hawthorn J., Bridges T., Cannon R., Cole S.,
  Collins C., Couch W., Dalton G., Propris R.~D., Driver S.~P., Efstathiou G.,
  Ellis R.~S., Frenk C.~S., Glazebrook K., Jackson C.~A., Lahav O., Lewis I.,
  Lumsden S., Maddox S.~J., Madgwick D., Peterson B.~A., Sutherland W., Taylor
  K., 2005, MNRAS, 356, 1155

\bibitem[{Croton {et~al.}(2006)Croton, Springel, White, Lucia, Frenk, Gao,
  Jenkins, Kauffmann, Navarro, \& Yoshida}]{Croton:2006p249}
Croton D.~J., Springel V., White S. D.~M., Lucia G.~D., Frenk C.~S., Gao L.,
  Jenkins A., Kauffmann G., Navarro J.~F., Yoshida N., 2006, MNRAS, 365, 11

\bibitem[{Dale {et~al.}(2001)Dale, Helou, Contursi, Silbermann, \&
  Kolhatkar}]{Dale:2001p1087}
Dale D.~A., Helou G., Contursi A., Silbermann N.~A., Kolhatkar S., 2001, ApJ,
  549, 215

\bibitem[{Dariush {et~al.}(2007)Dariush, Khosroshahi, Ponman, Pearce,
  Raychaudhury, \& Hartley}]{Dariush:2007p265}
Dariush A., Khosroshahi H.~G., Ponman T.~J., Pearce F., Raychaudhury S.,
  Hartley W., 2007, MNRAS, 382, 433

\bibitem[{Dasyra {et~al.}(2006)Dasyra, Tacconi, Davies, Naab, Genzel, Lutz,
  Sturm, Baker, Veilleux, Sanders, \& Burkert}]{Dasyra:2006p1119}
Dasyra K.~M., Tacconi L.~J., Davies R.~I., Naab T., Genzel R., Lutz D., Sturm
  E., Baker A.~J., Veilleux S., Sanders D.~B., Burkert A., 2006, ApJ, 651, 835

\bibitem[{Dav{\'e}(2008)}]{Dave:2008p949}
Dav{\'e} R., 2008, MNRAS, 385, 147

\bibitem[{Dav{\'e} {et~al.}(2010)Dav{\'e}, Finlator, Oppenheimer, Fardal, Katz,
  Kere{\v s}, \& Weinberg}]{Dave:2010p945}
Dav{\'e} R., Finlator K., Oppenheimer B.~D., Fardal M., Katz N., Kere{\v s} D.,
  Weinberg D.~H., 2010, MNRAS, 404, 1355

\bibitem[{Davis {et~al.}(1985)Davis, Efstathiou, Frenk, \&
  White}]{Davis:1985p845}
Davis M., Efstathiou G., Frenk C.~S., White S. D.~M., 1985, ApJ, 292, 371

\bibitem[{Davis \& Geller(1976)}]{Davis:1976p743}
Davis M., Geller M.~J., 1976, ApJ, 208, 13

\bibitem[{{Dayal} {et~al.}(2011){Dayal}, {Maselli}, \&
  {Ferrara}}]{Dayal:2010p781}
{Dayal} P., {Maselli} A., {Ferrara} A., 2011, MNRAS, 410, 830

\bibitem[{Dekel \& Birnboim(2006)}]{Dekel:2006p966}
Dekel A., Birnboim Y., 2006, MNRAS, 368, 2

\bibitem[{Dekel {et~al.}(2009{\natexlab{a}})Dekel, Birnboim, Engel, Freundlich,
  Goerdt, Mumcuoglu, Neistein, Pichon, Teyssier, \& Zinger}]{Dekel:2009p939}
Dekel A., Birnboim Y., Engel G., Freundlich J., Goerdt T., Mumcuoglu M.,
  Neistein E., Pichon C., Teyssier R., Zinger E., 2009{\natexlab{a}}, Nature,
  457, 451

\bibitem[{Dekel {et~al.}(2009{\natexlab{b}})Dekel, Sari, \&
  Ceverino}]{Dekel:2009p948}
Dekel A., Sari R., Ceverino D., 2009{\natexlab{b}}, ApJ, 703, 785

\bibitem[{del P~Lagos {et~al.}(2008)del P~Lagos, Cora, \&
  Padilla}]{Lagos:2008p528}
del P~Lagos C., Cora S.~A., Padilla N.~D., 2008, MNRAS, 388, 587

\bibitem[{Denicol{\'o} {et~al.}(2005)Denicol{\'o}, Terlevich, Terlevich,
  Forbes, Terlevich, \& Carrasco}]{Denicolo:2005p568}
Denicol{\'o} G., Terlevich R., Terlevich E., Forbes D.~A., Terlevich A.,
  Carrasco L., 2005, MNRAS, 356, 1440

\bibitem[{Desert {et~al.}(1990)Desert, Boulanger, \& Puget}]{Desert:1990p1069}
Desert F.-X., Boulanger F., Puget J.~L., 1990, A\&A, 237, 215

\bibitem[{Diaferio {et~al.}(1994)Diaferio, Geller, \&
  Ramella}]{Diaferio:1994p278}
Diaferio A., Geller M.~J., Ramella M., 1994, AJ, 107, 868

\bibitem[{{Diaferio} {et~al.}(1995){Diaferio}, {Geller}, \&
  {Ramella}}]{Diaferio:1995p268}
{Diaferio} A., {Geller} M.~J., {Ramella} M., 1995, AJ, 109, 2293

\bibitem[{D{\'\i}az-Gim{\'e}nez {et~al.}(2008)D{\'\i}az-Gim{\'e}nez, Muriel, \&
  Oliveira}]{DiazGimenez:2008p617}
D{\'\i}az-Gim{\'e}nez E., Muriel H., Oliveira C. M.~D., 2008, A{\&}A, 490, 965

\bibitem[{{Dodelson}(2003)}]{2003moco.book.....D}
{Dodelson} S., 2003, Modern cosmology by Scott Dodelson.~Amsterdam
  (Netherlands): Academic Press.~ISBN 0-12-219141-2, 2003, XIII + 440 p.

\bibitem[{{Dolag} {et~al.}(2008){Dolag}, {Borgani}, {Schindler}, {Diaferio}, \&
  {Bykov}}]{2008SSRv..134..229D}
{Dolag} K., {Borgani} S., {Schindler} S., {Diaferio} A., {Bykov} A.~M., 2008,
  Space Science Reviews, 134, 229

\bibitem[{{D'Onghia} {et~al.}(2007){D'Onghia}, {Maccio'}, {Lake}, {Stadel}, \&
  {Moore}}]{DOnghia:2007p388}
{D'Onghia} E., {Maccio'} A.~V., {Lake} G., {Stadel} J., {Moore} B., 2007,
  ArXiv:astro-ph/0704.2604

\bibitem[{D'Onghia {et~al.}(2005)D'Onghia, Sommer-Larsen, Romeo, Burkert,
  Pedersen, Portinari, \& Rasmussen}]{DOnghia:2005p128}
D'Onghia E., Sommer-Larsen J., Romeo A.~D., Burkert A., Pedersen K., Portinari
  L., Rasmussen J., 2005, ApJ, 630, L109

\bibitem[{Dressler(1980)}]{Dressler:1980p645}
Dressler A., 1980, ApJ, 236, 351

\bibitem[{Driver {et~al.}(2007)Driver, Allen, Liske, \&
  Graham}]{Driver:2007p587}
Driver S.~P., Allen P.~D., Liske J., Graham A.~W., 2007, ApJ, 657, L85

\bibitem[{Duc {et~al.}(1997)Duc, Mirabel, \& Maza}]{Duc:1997p1147}
Duc P.-A., Mirabel I.~F., Maza J., 1997, A{\&}A Supplement Series, 124, 533

\bibitem[{Dunkley {et~al.}(2009)Dunkley, Komatsu, Nolta, Spergel, Larson,
  Hinshaw, Page, Bennett, Gold, Jarosik, Weiland, Halpern, Hill, Kogut, Limon,
  Meyer, Tucker, Wollack, \& Wright}]{Dunkley:2009p754}
Dunkley J., Komatsu E., Nolta M.~R., Spergel D.~N., Larson D., Hinshaw G., Page
  L., Bennett C.~L., Gold B., Jarosik N., Weiland J.~L., Halpern M., Hill
  R.~S., Kogut A., Limon M., Meyer S.~S., Tucker G.~S., Wollack E., Wright
  E.~L., 2009, ApJS, 180, 306

\bibitem[{Efstathiou {et~al.}(1985)Efstathiou, Davis, White, \&
  Frenk}]{Efstathiou:1985p1003}
Efstathiou G., Davis M., White S. D.~M., Frenk C.~S., 1985, ApJS, 57, 241

\bibitem[{Eigenthaler \& Zeilinger(2009)}]{Eigenthaler:2009p698}
Eigenthaler P., Zeilinger W.~W., 2009, Astronomische Nachrichten, 330, 978

\bibitem[{Einasto {et~al.}(2007)Einasto, Einasto, Tago, Saar, H{\"u}tsi,
  J{\~o}eveer, Liivam{\"a}gi, Suhhonenko, Jaaniste, Hein{\"a}m{\"a}ki,
  M{\"u}ller, Knebe, \& Tucker}]{Einasto:2007p983}
Einasto J., Einasto M., Tago E., Saar E., H{\"u}tsi G., J{\~o}eveer M.,
  Liivam{\"a}gi L.~J., Suhhonenko I., Jaaniste J., Hein{\"a}m{\"a}ki P.,
  M{\"u}ller V., Knebe A., Tucker D., 2007, A{\&}A, 462, 811

\bibitem[{Einasto {et~al.}(2003{\natexlab{a}})Einasto, H{\"u}tsi, Einasto,
  Saar, Tucker, M{\"u}ller, Hein{\"a}m{\"a}ki, \& Allam}]{Einasto:2003p871}
Einasto J., H{\"u}tsi G., Einasto M., Saar E., Tucker D.~L., M{\"u}ller V.,
  Hein{\"a}m{\"a}ki P., Allam S.~S., 2003{\natexlab{a}}, A{\&}A, 405, 425

\bibitem[{Einasto {et~al.}(1980)Einasto, Joeveer, \& Saar}]{Einasto:1980p905}
Einasto J., Joeveer M., Saar E., 1980, MNRAS, 193, 353

\bibitem[{Einasto {et~al.}(2005)Einasto, Tago, Einasto, Saar, Suhhonenko,
  Hein{\"a}m{\"a}ki, H{\"u}tsi, \& Tucker}]{Einasto:2005p872}
Einasto J., Tago E., Einasto M., Saar E., Suhhonenko I., Hein{\"a}m{\"a}ki P.,
  H{\"u}tsi G., Tucker D.~L., 2005, A{\&}A, 439, 45

\bibitem[{Einasto {et~al.}(2003{\natexlab{b}})Einasto, Einasto, M{\"u}ller,
  Hein{\"a}m{\"a}ki, \& Tucker}]{Einasto:2003p590}
Einasto M., Einasto J., M{\"u}ller V., Hein{\"a}m{\"a}ki P., Tucker D.~L.,
  2003{\natexlab{b}}, A{\&}A, 401, 851

\bibitem[{Einasto {et~al.}(2008)Einasto, Saar, Mart{\'\i}nez, Einasto,
  Liivam{\"a}gi, Tago, Starck, M{\"u}ller, Hein{\"a}m{\"a}ki, Nurmi, Paredes,
  Gramann, \& H{\"u}tsi}]{Einasto:2008p985}
Einasto M., Saar E., Mart{\'\i}nez V.~J., Einasto J., Liivam{\"a}gi L.~J., Tago
  E., Starck J.-L., M{\"u}ller V., Hein{\"a}m{\"a}ki P., Nurmi P., Paredes S.,
  Gramann M., H{\"u}tsi G., 2008, ApJ, 685, 83

\bibitem[{{Einstein}(1915)}]{1915SPAW.......844E}
{Einstein} A., 1915, Sitzungsberichte der K{\"o}niglich Preu{\ss}ischen
  Akademie der Wissenschaften (Berlin), Seite 844-847., 844

\bibitem[{{Einstein}(1917)}]{1917SPAW.......142E}
---, 1917, Sitzungsberichte der K{\"o}niglich Preu{\ss}ischen Akademie der
  Wissenschaften (Berlin), Seite 142-152., 142

\bibitem[{Eisenstein \& Hut(1998)}]{Eisenstein:1998p1033}
Eisenstein D.~J., Hut P., 1998, ApJ, 498, 137

\bibitem[{Eisenstein {et~al.}(2005)Eisenstein, Zehavi, Hogg, Scoccimarro,
  Blanton, Nichol, Scranton, Seo, Tegmark, Zheng, Anderson, Annis, Bahcall,
  Brinkmann, Burles, Castander, Connolly, Csabai, Doi, Fukugita, Frieman,
  Glazebrook, Gunn, Hendry, Hennessy, Ivezi{\'c}, Kent, Knapp, Lin, Loh,
  Lupton, Margon, McKay, Meiksin, Munn, Pope, Richmond, Schlegel, Schneider,
  Shimasaku, Stoughton, Strauss, SubbaRao, Szalay, Szapudi, Tucker, Yanny, \&
  York}]{Eisenstein:2005p808}
Eisenstein D.~J., Zehavi I., Hogg D.~W., Scoccimarro R., Blanton M.~R., Nichol
  R.~C., Scranton R., Seo H.-J., Tegmark M., Zheng Z., Anderson S.~F., Annis
  J., Bahcall N., Brinkmann J., Burles S., Castander F.~J., Connolly A., Csabai
  I., Doi M., Fukugita M., Frieman J.~A., Glazebrook K., Gunn J.~E., Hendry
  J.~S., Hennessy G., Ivezi{\'c} Z., Kent S., Knapp G.~R., Lin H., Loh Y.-S.,
  Lupton R.~H., Margon B., McKay T.~A., Meiksin A., Munn J.~A., Pope A.,
  Richmond M.~W., Schlegel D., Schneider D.~P., Shimasaku K., Stoughton C.,
  Strauss M.~A., SubbaRao M., Szalay A.~S., Szapudi I., Tucker D.~L., Yanny B.,
  York D.~G., 2005, ApJ, 633, 560

\bibitem[{Eke {et~al.}(2006)Eke, Baugh, Cole, Frenk, \& Navarro}]{Eke:2006p605}
Eke V.~R., Baugh C.~M., Cole S., Frenk C.~S., Navarro J.~F., 2006, MNRAS, 370,
  1147

\bibitem[{Eke {et~al.}(2004{\natexlab{a}})Eke, Baugh, Cole, Frenk, Norberg,
  Peacock, Baldry, Bland-Hawthorn, Bridges, Cannon, Colless, Collins, Couch,
  Dalton, Propris, Driver, Efstathiou, Ellis, Glazebrook, Jackson, Lahav,
  Lewis, Lumsden, Maddox, Madgwick, Peterson, Sutherland, \&
  Taylor}]{Eke:2004p509}
Eke V.~R., Baugh C.~M., Cole S., Frenk C.~S., Norberg P., Peacock J.~A., Baldry
  I.~K., Bland-Hawthorn J., Bridges T., Cannon R., Colless M., Collins C.,
  Couch W., Dalton G., Propris R.~D., Driver S.~P., Efstathiou G., Ellis R.~S.,
  Glazebrook K., Jackson C.~A., Lahav O., Lewis I., Lumsden S., Maddox S.~J.,
  Madgwick D., Peterson B.~A., Sutherland W., Taylor K., 2004{\natexlab{a}},
  MNRAS, 348, 866

\bibitem[{Eke {et~al.}(2004{\natexlab{b}})Eke, Frenk, Baugh, Cole, Norberg,
  Peacock, Baldry, Bland-Hawthorn, Bridges, Cannon, Colless, Collins, Couch,
  Dalton, Propris, Driver, Efstathiou, Ellis, Glazebrook, Jackson, Lahav,
  Lewis, Lumsden, Maddox, Madgwick, Peterson, Sutherland, \&
  Taylor}]{Eke:2004p507}
Eke V.~R., Frenk C.~S., Baugh C.~M., Cole S., Norberg P., Peacock J.~A., Baldry
  I.~K., Bland-Hawthorn J., Bridges T., Cannon R., Colless M., Collins C.,
  Couch W., Dalton G., Propris R.~D., Driver S.~P., Efstathiou G., Ellis R.~S.,
  Glazebrook K., Jackson C.~A., Lahav O., Lewis I., Lumsden S., Maddox S.~J.,
  Madgwick D., Peterson B.~A., Sutherland W., Taylor K., 2004{\natexlab{b}},
  MNRAS, 355, 769

\bibitem[{El-Ad \& Piran(1997)}]{ElAd:1997p903}
El-Ad H., Piran T., 1997, ApJ, 491, 421

\bibitem[{{Faber} {et~al.}(2007){Faber}, {Willmer}, {Wolf}, {Koo}, {Weiner},
  {Newman}, {Im}, {Coil}, {Conroy}, {Cooper}, {Davis}, {Finkbeiner}, {Gerke},
  {Gebhardt}, {Groth}, {Guhathakurta}, {Harker}, {Kaiser}, {Kassin},
  {Kleinheinrich}, {Konidaris}, {Kron}, {Lin}, {Luppino}, {Madgwick},
  {Meisenheimer}, {Noeske}, {Phillips}, {Sarajedini}, {Schiavon}, {Simard},
  {Szalay}, {Vogt}, \& {Yan}}]{2007ApJ...665..265F}
{Faber} S.~M., {Willmer} C.~N.~A., {Wolf} C., {Koo} D.~C., {Weiner} B.~J.,
  {Newman} J.~A., {Im} M., {Coil} A.~L., {Conroy} C., {Cooper} M.~C., {Davis}
  M., {Finkbeiner} D.~P., {Gerke} B.~F., {Gebhardt} K., {Groth} E.~J.,
  {Guhathakurta} P., {Harker} J., {Kaiser} N., {Kassin} S., {Kleinheinrich} M.,
  {Konidaris} N.~P., {Kron} R.~G., {Lin} L., {Luppino} G., {Madgwick} D.~S.,
  {Meisenheimer} K., {Noeske} K.~G., {Phillips} A.~C., {Sarajedini} V.~L.,
  {Schiavon} R.~P., {Simard} L., {Szalay} A.~S., {Vogt} N.~P., {Yan} R., 2007,
  ApJ, 665, 265

\bibitem[{Fakhouri \& Ma(2008)}]{Fakhouri:2008p730}
Fakhouri O., Ma C.-P., 2008, MNRAS, 386, 577

\bibitem[{Fall \& Pei(1993)}]{Fall:1993p980}
Fall S.~M., Pei Y.~C., 1993, ApJ, 402, 479

\bibitem[{Francis {et~al.}(1996)Francis, Woodgate, Warren, Moller, Mazzolini,
  Bunker, Lowenthal, Williams, Minezaki, Kobayashi, \&
  Yoshii}]{Francis:1996p1038}
Francis P.~J., Woodgate B.~E., Warren S.~J., Moller P., Mazzolini M., Bunker
  A.~J., Lowenthal J.~D., Williams T.~B., Minezaki T., Kobayashi Y., Yoshii Y.,
  1996, ApJ, 457, 490

\bibitem[{{Frayer} {et~al.}(1998){Frayer}, {Ivison}, {Scoville}, {Yun},
  {Evans}, {Smail}, {Blain}, \& {Kneib}}]{1998ApJ...506L...7F}
{Frayer} D.~T., {Ivison} R.~J., {Scoville} N.~Z., {Yun} M., {Evans} A.~S.,
  {Smail} I., {Blain} A.~W., {Kneib} J., 1998, ApJL, 506, L7

\bibitem[{Frederic(1995{\natexlab{a}})}]{Frederic:1995p715}
Frederic J.~J., 1995{\natexlab{a}}, ApJS, 97, 275

\bibitem[{Frederic(1995{\natexlab{b}})}]{Frederic:1995p714}
---, 1995{\natexlab{b}}, ApJS, 97, 259

\bibitem[{Frenk {et~al.}(1999)Frenk, White, Bode, Bond, Bryan, Cen, Couchman,
  Evrard, Gnedin, Jenkins, Khokhlov, Klypin, Navarro, Norman, Ostriker, Owen,
  Pearce, Pen, Steinmetz, Thomas, Villumsen, Wadsley, Warren, Xu, \&
  Yepes}]{Frenk:1999p881}
Frenk C.~S., White S. D.~M., Bode P., Bond J.~R., Bryan G.~L., Cen R., Couchman
  H. M.~P., Evrard A.~E., Gnedin N., Jenkins A., Khokhlov A.~M., Klypin A.,
  Navarro J.~F., Norman M.~L., Ostriker J.~P., Owen J.~M., Pearce F.~R., Pen
  U.-L., Steinmetz M., Thomas P.~A., Villumsen J.~V., Wadsley J.~W., Warren
  M.~S., Xu G., Yepes G., 1999, ApJ, 525, 554

\bibitem[{Frenk {et~al.}(1988)Frenk, White, Davis, \&
  Efstathiou}]{Frenk:1988p874}
Frenk C.~S., White S. D.~M., Davis M., Efstathiou G., 1988, ApJ, 327, 507

\bibitem[{Frieman {et~al.}(2008)Frieman, Turner, \& Huterer}]{Frieman:2008p764}
Frieman J.~A., Turner M.~S., Huterer D., 2008, Annual Review of Astronomy and
  Astrophysics, 46, 385

\bibitem[{{Frye} {et~al.}(2008){Frye}, {Bowen}, {Hurley}, {Tripp}, {Fan},
  {Holden}, {Guhathakurta}, {Coe}, {Broadhurst}, {Egami}, \&
  {Meylan}}]{2008ApJ...685L...5F}
{Frye} B.~L., {Bowen} D.~V., {Hurley} M., {Tripp} T.~M., {Fan} X., {Holden} B.,
  {Guhathakurta} P., {Coe} D., {Broadhurst} T., {Egami} E., {Meylan} G., 2008,
  ApJL, 685, L5

\bibitem[{Fujita(2004)}]{Fujita:2004p793}
Fujita Y., 2004, PASJ, 56, 29

\bibitem[{{Gamow}(1948)}]{1948PhRv...74..505G}
{Gamow} G., 1948, Physical Review, 74, 505

\bibitem[{Geller \& Huchra(1983)}]{Geller:1983p576}
Geller M.~J., Huchra J.~P., 1983, ApJS, 52, 61

\bibitem[{Genzel {et~al.}(1998)Genzel, Lutz, Sturm, Egami, Kunze, Moorwood,
  Rigopoulou, Spoon, Sternberg, Tacconi-Garman, Tacconi, \&
  Thatte}]{Genzel:1998p1155}
Genzel R., Lutz D., Sturm E., Egami E., Kunze D., Moorwood A. F.~M., Rigopoulou
  D., Spoon H. W.~W., Sternberg A., Tacconi-Garman L.~E., Tacconi L., Thatte
  N., 1998, ApJ, 498, 579

\bibitem[{Genzel {et~al.}(2001)Genzel, Tacconi, Rigopoulou, Lutz, \&
  Tecza}]{Genzel:2001p1106}
Genzel R., Tacconi L.~J., Rigopoulou D., Lutz D., Tecza M., 2001, ApJ, 563, 527

\bibitem[{Gerke {et~al.}(2005)Gerke, Newman, Davis, Marinoni, Yan, Coil,
  Conroy, Cooper, Faber, Finkbeiner, Guhathakurta, Kaiser, Koo, Phillips,
  Weiner, \& Willmer}]{Gerke:2005p691}
Gerke B.~F., Newman J.~A., Davis M., Marinoni C., Yan R., Coil A.~L., Conroy
  C., Cooper M.~C., Faber S.~M., Finkbeiner D.~P., Guhathakurta P., Kaiser N.,
  Koo D.~C., Phillips A.~C., Weiner B.~J., Willmer C. N.~A., 2005, ApJ, 625, 6

\bibitem[{Giavalisco {et~al.}(2004)Giavalisco, Dickinson, Ferguson,
  Ravindranath, Kretchmer, Moustakas, Madau, Fall, Gardner, Livio, Papovich,
  Renzini, Spinrad, Stern, \& Riess}]{Giavalisco:2004p753}
Giavalisco M., Dickinson M., Ferguson H.~C., Ravindranath S., Kretchmer C.,
  Moustakas L.~A., Madau P., Fall S.~M., Gardner J.~P., Livio M., Papovich C.,
  Renzini A., Spinrad H., Stern D., Riess A., 2004, ApJ, 600, L103

\bibitem[{Girardi {et~al.}(1992)Girardi, Mezzetti, Giuricin, \&
  Mardirossian}]{Girardi:1992p339}
Girardi M., Mezzetti M., Giuricin G., Mardirossian F., 1992, ApJ, 394, 442

\bibitem[{Goto {et~al.}(2002)Goto, Sekiguchi, Nichol, Bahcall, Kim, Annis,
  Ivezic, Brinkmann, Hennessy, Szokoly, \& Tucker}]{Goto:2002p585}
Goto T., Sekiguchi M., Nichol R.~C., Bahcall N.~A., Kim R. S.~J., Annis J.,
  Ivezic Z., Brinkmann J., Hennessy G.~S., Szokoly G.~P., Tucker D.~L., 2002,
  AJ, 123, 1807

\bibitem[{Governato {et~al.}(1999)Governato, Babul, Quinn, Tozzi, Baugh, Katz,
  \& Lake}]{Governato:1999p1028}
Governato F., Babul A., Quinn T., Tozzi P., Baugh C.~M., Katz N., Lake G.,
  1999, MNRAS, 307, 949

\bibitem[{Governato {et~al.}(1996)Governato, Tozzi, \&
  Cavaliere}]{Governato:1996p287}
Governato F., Tozzi P., Cavaliere A., 1996, ApJ, 458, 18

\bibitem[{{Guth}(1981)}]{Guth:1981p37}
{Guth} A.~H., 1981, Physical Review D, 23, 347

\bibitem[{Guth \& Pi(1982)}]{Guth:1982p859}
Guth A.~H., Pi S.-Y., 1982, Physical Review Letters, 49, 1110

\bibitem[{{Guzik} \& {Seljak}(2002)}]{2002MNRAS.335..311G}
{Guzik} J., {Seljak} U., 2002, MNRAS, 335, 311

\bibitem[{{Hamilton}(1998)}]{Hamilton:1997p994}
{Hamilton} A.~J.~S., 1998, in Astrophysics and Space Science Library, Vol. 231,
  The Evolving Universe, {D.~Hamilton}, ed., pp. 185--+

\bibitem[{{Harrison}(1970)}]{1970PhRvD...1.2726H}
{Harrison} E.~R., 1970, Physical Review D, 1, 2726

\bibitem[{Harwit {et~al.}(1987)Harwit, Houck, Soifer, \&
  Palumbo}]{Harwit:1987p1111}
Harwit M., Houck J.~R., Soifer B.~T., Palumbo G. G.~C., 1987, ApJ, 315, 28

\bibitem[{Hashimoto {et~al.}(1998)Hashimoto, Oemler, Lin, \&
  Tucker}]{Hashimoto:1998p721}
Hashimoto Y., Oemler A., Lin H., Tucker D.~L., 1998, AJ, 499, 589

\bibitem[{Hau \& Forbes(2006)}]{Hau:2006p557}
Hau G. K.~T., Forbes D.~A., 2006, MNRAS, 371, 633

\bibitem[{Hausman \& Ostriker(1978)}]{Hausman:1978p921}
Hausman M.~A., Ostriker J.~P., 1978, ApJ, 224, 320

\bibitem[{{Hawking}(1982)}]{1982PhLB..115..295H}
{Hawking} S.~W., 1982, Physics Letters B, 115, 295

\bibitem[{Helly {et~al.}(2003)Helly, Cole, Frenk, Baugh, Benson, Lacey, \&
  Pearce}]{Helly:2003p965}
Helly J.~C., Cole S., Frenk C.~S., Baugh C.~M., Benson A., Lacey C., Pearce
  F.~R., 2003, MNRAS, 338, 913

\bibitem[{Hern{\'a}n-Caballero {et~al.}(2009)Hern{\'a}n-Caballero,
  P{\'e}rez-Fournon, Hatziminaoglou, Afonso-Luis, Rowan-Robinson, Rigopoulou,
  Farrah, Lonsdale, Babbedge, Clements, Serjeant, Pozzi, Vaccari,
  Montenegro-Montes, Valtchanov, Gonz{\'a}lez-Solares, Oliver, Shupe,
  Gruppioni, Vila-Vilar{\'o}, Lari, \& Franca}]{HernanCaballero:2009p1137}
Hern{\'a}n-Caballero A., P{\'e}rez-Fournon I., Hatziminaoglou E., Afonso-Luis
  A., Rowan-Robinson M., Rigopoulou D., Farrah D., Lonsdale C.~J., Babbedge T.,
  Clements D., Serjeant S., Pozzi F., Vaccari M., Montenegro-Montes F.~M.,
  Valtchanov I., Gonz{\'a}lez-Solares E., Oliver S., Shupe D., Gruppioni C.,
  Vila-Vilar{\'o} B., Lari C., Franca F.~L., 2009, MNRAS, 395, 1695

\bibitem[{Hern{\'a}ndez-Toledo {et~al.}(2008)Hern{\'a}ndez-Toledo,
  V{\'a}zquez-Mata, Mart{\'\i}nez-V{\'a}zquez, Reese, M{\'e}ndez-Hern{\'a}ndez,
  Ortega-Esbr{\'\i}, \& N{\'u}{\~n}ez}]{HernandezToledo:2008p648}
Hern{\'a}ndez-Toledo H.~M., V{\'a}zquez-Mata J.~A., Mart{\'\i}nez-V{\'a}zquez
  L.~A., Reese V.~A., M{\'e}ndez-Hern{\'a}ndez H., Ortega-Esbr{\'\i} S.,
  N{\'u}{\~n}ez J. P.~M., 2008, AJ, 136, 2115

\bibitem[{Hernquist {et~al.}(1995)Hernquist, Katz, \&
  Weinberg}]{Hernquist:1995p717}
Hernquist L., Katz N., Weinberg D.~H., 1995, ApJ, 442, 57

\bibitem[{Hernquist \& Mihos(1995)}]{Hernquist:1995p917}
Hernquist L., Mihos J.~C., 1995, ApJ, 448, 41

\bibitem[{Hickson(1982)}]{Hickson:1982p571}
Hickson P., 1982, ApJ, 255, 382

\bibitem[{Hickson(1997)}]{Hickson:1997p693}
---, 1997, Annual Review of Astronomy and Astrophysics, 35, 357

\bibitem[{Hickson {et~al.}(1989)Hickson, Kindl, \& Auman}]{Hickson:1989p573}
Hickson P., Kindl E., Auman J.~R., 1989, ApJS, 70, 687

\bibitem[{Hickson {et~al.}(1988)Hickson, Kindl, \& Huchra}]{Hickson:1988p358}
Hickson P., Kindl E., Huchra J.~P., 1988, ApJ, 329, L65

\bibitem[{Hinshaw {et~al.}(2009)Hinshaw, Weiland, Hill, Odegard, Larson,
  Bennett, Dunkley, Gold, Greason, Jarosik, Komatsu, Nolta, Page, Spergel,
  Wollack, Halpern, Kogut, Limon, Meyer, Tucker, \& Wright}]{Hinshaw:2009p709}
Hinshaw G., Weiland J.~L., Hill R.~S., Odegard N., Larson D., Bennett C.~L.,
  Dunkley J., Gold B., Greason M.~R., Jarosik N., Komatsu E., Nolta M.~R., Page
  L., Spergel D.~N., Wollack E., Halpern M., Kogut A., Limon M., Meyer S.~S.,
  Tucker G.~S., Wright E.~L., 2009, ApJS, 180, 225

\bibitem[{{Hoekstra} {et~al.}(2004){Hoekstra}, {Yee}, \&
  {Gladders}}]{2004ApJ...606...67H}
{Hoekstra} H., {Yee} H.~K.~C., {Gladders} M.~D., 2004, ApJ, 606, 67

\bibitem[{Hogg {et~al.}(2002)Hogg, Blanton, Strateva, Bahcall, Brinkmann,
  Csabai, Doi, Fukugita, Hennessy, Ivezi{\'c}, Knapp, Lamb, Lupton, Munn,
  Nichol, Schlegel, Schneider, \& York}]{Hogg:2002p788}
Hogg D.~W., Blanton M., Strateva I., Bahcall N.~A., Brinkmann J., Csabai I.,
  Doi M., Fukugita M., Hennessy G., Ivezi{\'c} {\v Z}., Knapp G.~R., Lamb
  D.~Q., Lupton R., Munn J.~A., Nichol R., Schlegel D.~J., Schneider D.~P.,
  York D.~G., 2002, AJ, 124, 646

\bibitem[{{Holmberg}(1950)}]{1950MeLu2.128....1H}
{Holmberg} E., 1950, Meddelanden fran Lunds Astronomiska Observatorium Serie
  II, 128, 1

\bibitem[{Hoyle(1953)}]{Hoyle:1953p1023}
Hoyle F., 1953, ApJ, 118, 513

\bibitem[{Hoyle \& Vogeley(2004)}]{Hoyle:2004p902}
Hoyle F., Vogeley M.~S., 2004, ApJ, 607, 751

\bibitem[{Hu \& Dodelson(2002)}]{Hu:2002p864}
Hu W., Dodelson S., 2002, Annual Review of Astronomy and Astrophysics, 40, 171

\bibitem[{Hubble \& Humason(1931)}]{Hubble:1931p741}
Hubble E., Humason M.~L., 1931, ApJ, 74, 43

\bibitem[{Huchra \& Geller(1982)}]{Huchra:1982p1}
Huchra J.~P., Geller M.~J., 1982, ApJ, 257, 423

\bibitem[{{Humason} {et~al.}(1956){Humason}, {Mayall}, \&
  {Sandage}}]{1956AJ.....61...97H}
{Humason} M.~L., {Mayall} N.~U., {Sandage} A.~R., 1956, AJ, 61, 97

\bibitem[{Iovino \& Hickson(1997)}]{Iovino:1997p338}
Iovino A., Hickson P., 1997, MNRAS, 287, 21

\bibitem[{Irwin {et~al.}(2007)Irwin, Belokurov, Evans, Ryan-Weber, de~Jong,
  Koposov, Zucker, Hodgkin, Gilmore, Prema, Hebb, Begum, Fellhauer, Hewett,
  Kennicutt, Wilkinson, Bramich, Vidrih, Rix, Beers, Barentine, Brewington,
  Harvanek, Krzesinski, Long, Nitta, \& Snedden}]{Irwin:2007p707}
Irwin M.~J., Belokurov V., Evans N.~W., Ryan-Weber E.~V., de~Jong J. T.~A.,
  Koposov S., Zucker D.~B., Hodgkin S.~T., Gilmore G., Prema P., Hebb L., Begum
  A., Fellhauer M., Hewett P.~C., Kennicutt R.~C., Wilkinson M.~I., Bramich
  D.~M., Vidrih S., Rix H.-W., Beers T.~C., Barentine J.~C., Brewington H.,
  Harvanek M., Krzesinski J., Long D., Nitta A., Snedden S.~A., 2007, ApJ, 656,
  L13

\bibitem[{{Jaakkola}(1971)}]{1971Natur.234..534J}
{Jaakkola} T., 1971, Nature, 234, 534

\bibitem[{{Jarosik} {et~al.}(2011){Jarosik}, {Bennett}, {Dunkley}, {Gold},
  {Greason}, {Halpern}, {Hill}, {Hinshaw}, {Kogut}, {Komatsu}, {Larson},
  {Limon}, {Meyer}, {Nolta}, {Odegard}, {Page}, {Smith}, {Spergel}, {Tucker},
  {Weiland}, {Wollack}, \& {Wright}}]{Jarosik:2010p770}
{Jarosik} N., {Bennett} C.~L., {Dunkley} J., {Gold} B., {Greason} M.~R.,
  {Halpern} M., {Hill} R.~S., {Hinshaw} G., {Kogut} A., {Komatsu} E., {Larson}
  D., {Limon} M., {Meyer} S.~S., {Nolta} M.~R., {Odegard} N., {Page} L.,
  {Smith} K.~M., {Spergel} D.~N., {Tucker} G.~S., {Weiland} J.~L., {Wollack}
  E., {Wright} E.~L., 2011, ApJS, 192, 14

\bibitem[{{Jeans}(1902)}]{1902RSPTA.199....1J}
{Jeans} J.~H., 1902, Royal Society of London Philosophical Transactions Series
  A, 199, 1

\bibitem[{Jenkins {et~al.}(2001)Jenkins, Frenk, White, Colberg, Cole, Evrard,
  Couchman, \& Yoshida}]{Jenkins:2001p1011}
Jenkins A., Frenk C.~S., White S. D.~M., Colberg J.~M., Cole S., Evrard A.~E.,
  Couchman H. M.~P., Yoshida N., 2001, MNRAS, 321, 372

\bibitem[{Jerjen {et~al.}(2001)Jerjen, Rekola, Takalo, Coleman, \&
  Valtonen}]{Jerjen:2001p583}
Jerjen H., Rekola R., Takalo L.~O., Coleman M., Valtonen M., 2001, A{\&}A, 380,
  90

\bibitem[{Jones {et~al.}(1988)Jones, Martinez, Saar, \&
  Einasto}]{Jones:1988p944}
Jones B. J.~T., Martinez V.~J., Saar E., Einasto J., 1988, ApJ, 332, L1

\bibitem[{Jones {et~al.}(2000)Jones, Ponman, \& Forbes}]{Jones:2000p624}
Jones L.~R., Ponman T.~J., Forbes D.~A., 2000, MNRAS, 312, 139

\bibitem[{Jones {et~al.}(2003)Jones, Ponman, Horton, Babul, Ebeling, \&
  Burke}]{Jones:2003p594}
Jones L.~R., Ponman T.~J., Horton A., Babul A., Ebeling H., Burke D.~J., 2003,
  MNRAS, 343, 627

\bibitem[{Kang {et~al.}(2005)Kang, Jing, Mo, \& B{\"o}rner}]{Kang:2005p724}
Kang X., Jing Y.~P., Mo H.~J., B{\"o}rner G., 2005, ApJ, 631, 21

\bibitem[{{Kannappan} {et~al.}(2009){Kannappan}, {Guie}, \&
  {Baker}}]{2009AJ....138..579K}
{Kannappan} S.~J., {Guie} J.~M., {Baker} A.~J., 2009, AJ, 138, 579

\bibitem[{{Kaplan} \& {Pikelner}(1970)}]{1970inme.book.....K}
{Kaplan} S.~A., {Pikelner} S.~B., 1970, {The interstellar medium, Cambridge:
  Harvard University Press}

\bibitem[{Karachentsev(2005)}]{Karachentsev:2005p447}
Karachentsev I.~D., 2005, AJ, 129, 178

\bibitem[{Karachentsev {et~al.}(1997)Karachentsev, Drozdovsky, Kajsin, Takalo,
  Hein{\"a}m{\"a}ki, \& Valtonen}]{Karachentsev:1997p584}
Karachentsev I.~D., Drozdovsky I., Kajsin S., Takalo L.~O., Hein{\"a}m{\"a}ki
  P., Valtonen M., 1997, A{\&}A Supplement Series, 124, 559

\bibitem[{Katz(1992)}]{Katz:1992p1082}
Katz N., 1992, ApJ, 391, 502

\bibitem[{Katz \& Gunn(1991)}]{Katz:1991p1084}
Katz N., Gunn J.~E., 1991, ApJ, 377, 365

\bibitem[{Kauffmann \& Charlot(1998)}]{Kauffmann:1998p960}
Kauffmann G., Charlot S., 1998, MNRAS, 294, 705

\bibitem[{Kauffmann \& White(1993)}]{Kauffmann:1993p946}
Kauffmann G., White S. D.~M., 1993, MNRAS, 261, 921

\bibitem[{Kauffmann {et~al.}(2004)Kauffmann, White, Heckman, M{\'e}nard,
  Brinchmann, Charlot, Tremonti, \& Brinkmann}]{Kauffmann:2004p751}
Kauffmann G., White S. D.~M., Heckman T.~M., M{\'e}nard B., Brinchmann J.,
  Charlot S., Tremonti C., Brinkmann J., 2004, MNRAS, 353, 713

\bibitem[{Kaufmann {et~al.}(2006)Kaufmann, Mayer, Wadsley, Stadel, \&
  Moore}]{Kaufmann:2006p884}
Kaufmann T., Mayer L., Wadsley J., Stadel J., Moore B., 2006, MNRAS, 370, 1612

\bibitem[{Kennicutt(1989)}]{Kennicutt:1989p976}
Kennicutt R.~C., 1989, ApJ, 344, 685

\bibitem[{{Kennicutt}(1998{\natexlab{a}})}]{1998ARA&A..36..189K}
{Kennicutt} Jr. R.~C., 1998{\natexlab{a}}, Annual Review of Astronomy and
  Astrophysics, 36, 189

\bibitem[{{Kennicutt}(1998{\natexlab{b}})}]{1998ApJ...498..541K}
---, 1998{\natexlab{b}}, ApJ, 498, 541

\bibitem[{Kere{\v s} {et~al.}(2009)Kere{\v s}, Katz, Fardal, Dav{\'e}, \&
  Weinberg}]{Keres:2009p957}
Kere{\v s} D., Katz N., Fardal M., Dav{\'e} R., Weinberg D.~H., 2009, MNRAS,
  395, 160

\bibitem[{Kere{\v s} {et~al.}(2005)Kere{\v s}, Katz, Weinberg, \&
  Dav{\'e}}]{Keres:2005p836}
Kere{\v s} D., Katz N., Weinberg D.~H., Dav{\'e} R., 2005, MNRAS, 363, 2

\bibitem[{Khalatyan {et~al.}(2008)Khalatyan, Cattaneo, Schramm, Gottl{\"o}ber,
  Steinmetz, \& Wisotzki}]{Khalatyan:2008p429}
Khalatyan A., Cattaneo A., Schramm M., Gottl{\"o}ber S., Steinmetz M., Wisotzki
  L., 2008, MNRAS, 387, 13

\bibitem[{Khochfar \& Silk(2009)}]{Khochfar:2009p725}
Khochfar S., Silk J., 2009, MNRAS, 397, 506

\bibitem[{{Khosroshahi} {et~al.}(2007){Khosroshahi}, {Ponman}, \&
  {Jones}}]{Khosroshahi:2007p97}
{Khosroshahi} H.~G., {Ponman} T.~J., {Jones} L.~R., 2007, MNRAS, 377, 595

\bibitem[{Kim \& Park(2006)}]{Kim:2006p424}
Kim J., Park C., 2006, ApJ, 639, 600

\bibitem[{Kim {et~al.}(2002)Kim, Kepner, Postman, Strauss, Bahcall, Gunn,
  Lupton, Annis, Nichol, Castander, Brinkmann, Brunner, Connolly, Csabai,
  Hindsley, Ivezic, Vogeley, \& York}]{Kim:2002p498}
Kim R. S.~J., Kepner J.~V., Postman M., Strauss M.~A., Bahcall N.~A., Gunn
  J.~E., Lupton R.~H., Annis J., Nichol R.~C., Castander F.~J., Brinkmann J.,
  Brunner R.~J., Connolly A., Csabai I., Hindsley R.~B., Ivezic Z., Vogeley
  M.~S., York D.~G., 2002, AJ, 123, 20

\bibitem[{Kitzbichler \& White(2007)}]{Kitzbichler:2007p767}
Kitzbichler M.~G., White S. D.~M., 2007, MNRAS, 376, 2

\bibitem[{Knobel {et~al.}(2009)Knobel, Lilly, Iovino, Porciani, Kova{\v c},
  Cucciati, Finoguenov, Kitzbichler, Carollo, Contini, Kneib, F{\`e}vre,
  Mainieri, Renzini, Scodeggio, Zamorani, Bardelli, Bolzonella, Bongiorno,
  Caputi, Coppa, de~La~Torre, de~Ravel, Franzetti, Garilli, Kampczyk,
  Lamareille, Borgne, Brun, Maier, Mignoli, Pello, Peng, Montero, Ricciardelli,
  Silverman, Tanaka, Tasca, Tresse, Vergani, Zucca, Abbas, Bottini, Cappi,
  Cassata, Cimatti, Fumana, Guzzo, Koekemoer, Leauthaud, Maccagni, Marinoni,
  McCracken, Memeo, Meneux, Oesch, Pozzetti, \& Scaramella}]{Knobel:2009p1075}
Knobel C., Lilly S.~J., Iovino A., Porciani C., Kova{\v c} K., Cucciati O.,
  Finoguenov A., Kitzbichler M.~G., Carollo C.~M., Contini T., Kneib J.-P.,
  F{\`e}vre O.~L., Mainieri V., Renzini A., Scodeggio M., Zamorani G., Bardelli
  S., Bolzonella M., Bongiorno A., Caputi K., Coppa G., de~La~Torre S.,
  de~Ravel L., Franzetti P., Garilli B., Kampczyk P., Lamareille F., Borgne
  J.-F.~L., Brun V.~L., Maier C., Mignoli M., Pello R., Peng Y., Montero E.~P.,
  Ricciardelli E., Silverman J.~D., Tanaka M., Tasca L., Tresse L., Vergani D.,
  Zucca E., Abbas U., Bottini D., Cappi A., Cassata P., Cimatti A., Fumana M.,
  Guzzo L., Koekemoer A.~M., Leauthaud A., Maccagni D., Marinoni C., McCracken
  H.~J., Memeo P., Meneux B., Oesch P., Pozzetti L., Scaramella R., 2009, ApJ,
  697, 1842

\bibitem[{Knollmann \& Knebe(2009)}]{Knollmann:2009p1032}
Knollmann S.~R., Knebe A., 2009, ApJS, 182, 608

\bibitem[{Koester {et~al.}(2007)Koester, McKay, Annis, Wechsler, Evrard, Bleem,
  Becker, Johnston, Sheldon, Nichol, Miller, Scranton, Bahcall, Barentine,
  Brewington, Brinkmann, Harvanek, Kleinman, Krzesinski, Long, Nitta,
  Schneider, Sneddin, Voges, \& York}]{Koester:2007p588}
Koester B.~P., McKay T.~A., Annis J., Wechsler R.~H., Evrard A., Bleem L.,
  Becker M., Johnston D., Sheldon E., Nichol R.~C., Miller C.~J., Scranton R.,
  Bahcall N.~A., Barentine J., Brewington H., Brinkmann J., Harvanek M.,
  Kleinman S.~J., Krzesinski J., Long D., Nitta A., Schneider D.~P., Sneddin
  S., Voges W., York D.~G., 2007, ApJ, 660, 239

\bibitem[{Komatsu {et~al.}(2009)Komatsu, Dunkley, Nolta, Bennett, Gold,
  Hinshaw, Jarosik, Larson, Limon, Page, Spergel, Halpern, Hill, Kogut, Meyer,
  Tucker, Weiland, Wollack, \& Wright}]{Komatsu:2009p710}
Komatsu E., Dunkley J., Nolta M.~R., Bennett C.~L., Gold B., Hinshaw G.,
  Jarosik N., Larson D., Limon M., Page L., Spergel D.~N., Halpern M., Hill
  R.~S., Kogut A., Meyer S.~S., Tucker G.~S., Weiland J.~L., Wollack E., Wright
  E.~L., 2009, ApJS, 180, 330

\bibitem[{Kormendy {et~al.}(2009)Kormendy, Fisher, Cornell, \&
  Bender}]{Kormendy:2009p659}
Kormendy J., Fisher D.~B., Cornell M.~E., Bender R., 2009, ApJS, 182, 216

\bibitem[{Kormendy \& Kennicutt(2004)}]{Kormendy:2004p958}
Kormendy J., Kennicutt R.~C., 2004, Annual Review of Astronomy and
  Astrophysics, 42, 603

\bibitem[{Kravtsov {et~al.}(1997)Kravtsov, Klypin, \&
  Khokhlov}]{Kravtsov:1997p1013}
Kravtsov A.~V., Klypin A.~A., Khokhlov A.~M., 1997, ApJS, 111, 73

\bibitem[{Kroupa(2001)}]{Kroupa:2001p974}
Kroupa P., 2001, MNRAS, 322, 231

\bibitem[{{Krumholz}(2011)}]{2011arXiv1101.5172K}
{Krumholz} M.~R., 2011, arXiv:astro-ph/1101.5172

\bibitem[{Labb{\'e} {et~al.}(2010)Labb{\'e}, Gonz{\'a}lez, Bouwens,
  Illingworth, Oesch, Dokkum, Carollo, Franx, Stiavelli, Trenti, Magee, \&
  Kriek}]{Labbe:2010p776}
Labb{\'e} I., Gonz{\'a}lez V., Bouwens R.~J., Illingworth G.~D., Oesch P.~A.,
  Dokkum P. G.~V., Carollo C.~M., Franx M., Stiavelli M., Trenti M., Magee D.,
  Kriek M., 2010, ApJL, 708, L26

\bibitem[{Lacey \& Cole(1993)}]{Lacey:1993p898}
Lacey C., Cole S., 1993, MNRAS, 262, 627

\bibitem[{Lacey \& Silk(1991)}]{Lacey:1991p842}
Lacey C., Silk J., 1991, ApJ, 381, 14

\bibitem[{Lacey {et~al.}(2010)Lacey, Baugh, Frenk, Benson, Orsi, Silva,
  Granato, \& Bressan}]{Lacey:2010p1041}
Lacey C.~G., Baugh C.~M., Frenk C.~S., Benson A.~J., Orsi A., Silva L., Granato
  G.~L., Bressan A., 2010, MNRAS, 405, 2

\bibitem[{{Lahav} \& {Suto}(2004)}]{2004LRR.....7....8L}
{Lahav} O., {Suto} Y., 2004, Living Reviews in Relativity, 7, 8

\bibitem[{Landy \& Szalay(1993)}]{Landy:1993p1081}
Landy S.~D., Szalay A.~S., 1993, ApJ, 412, 64

\bibitem[{Leroy {et~al.}(2008)Leroy, Walter, Brinks, Bigiel, de~Blok, Madore,
  \& Thornley}]{Leroy:2008p975}
Leroy A.~K., Walter F., Brinks E., Bigiel F., de~Blok W. J.~G., Madore B.,
  Thornley M.~D., 2008, AJ, 136, 2782

\bibitem[{Li {et~al.}(2007)Li, Mo, van~den Bosch, \& Lin}]{Li:2007p1098}
Li Y., Mo H.~J., van~den Bosch F.~C., Lin W.~P., 2007, MNRAS, 379, 689

\bibitem[{{Liddle} \& {Lyth}(2000)}]{2000cils.book.....L}
{Liddle} A.~R., {Lyth} D.~H., 2000, Cosmological Inflation and Large-Scale
  Structure, by Andrew R.~Liddle and David H.~Lyth, pp.~414.~ISBN
  052166022X.~Cambridge, UK: Cambridge University Press, April 2000.

\bibitem[{Lin {et~al.}(2008)Lin, Patton, Koo, Casteels, Conselice, Faber, Lotz,
  Willmer, Hsieh, Chiueh, Newman, Novak, Weiner, \& Cooper}]{Lin:2008p924}
Lin L., Patton D.~R., Koo D.~C., Casteels K., Conselice C.~J., Faber S.~M.,
  Lotz J., Willmer C. N.~A., Hsieh B.~C., Chiueh T., Newman J.~A., Novak G.~S.,
  Weiner B.~J., Cooper M.~C., 2008, ApJ, 681, 232

\bibitem[{{Linde}(1982)}]{1982PhLB..108..389L}
{Linde} A.~D., 1982, Physics Letters B, 108, 389

\bibitem[{Lindner {et~al.}(1995)Lindner, Einasto, Einasto, Freudling, Fricke,
  \& Tago}]{Lindner:1995p900}
Lindner U., Einasto J., Einasto M., Freudling W., Fricke K., Tago E., 1995,
  A{\&}A, 301, 329

\bibitem[{Lineweaver(2003)}]{Lineweaver:2003p797}
Lineweaver C.~H., 2003, arXiv:astro-ph/0305179

\bibitem[{{Longair}(2008)}]{TheGreatBook}
{Longair} M.~S., 2008, Galaxy Formation, by Malcolm S.~Longair Berlin:
  Springer, 2008.~ ISBN 978-3-540-73477-2

\bibitem[{Lotz {et~al.}(2008)Lotz, Jonsson, Cox, \& Primack}]{Lotz:2008p1076}
Lotz J.~M., Jonsson P., Cox T.~J., Primack J.~R., 2008, MNRAS, 391, 1137

\bibitem[{{Lucchin} \& {Matarrese}(1985)}]{1985PhRvD..32.1316L}
{Lucchin} F., {Matarrese} S., 1985, Physical Review D, 32, 1316

\bibitem[{Lucia \& Blaizot(2007)}]{DeLucia:2007p414}
Lucia G.~D., Blaizot J., 2007, MNRAS, 375, 2

\bibitem[{Lutz {et~al.}(1998)Lutz, Spoon, Rigopoulou, Moorwood, \&
  Genzel}]{Lutz:1998p1152}
Lutz D., Spoon H. W.~W., Rigopoulou D., Moorwood A. F.~M., Genzel R., 1998,
  ApJ, 505, L103

\bibitem[{Lynden-Bell(1967)}]{LyndenBell:1967p1030}
Lynden-Bell D., 1967, MNRAS, 136, 101

\bibitem[{Macci{\`o} {et~al.}(2005)Macci{\`o}, Governato, \&
  Horellou}]{Maccio:2005p971}
Macci{\`o} A.~V., Governato F., Horellou C., 2005, MNRAS, 359, 941

\bibitem[{Madau {et~al.}(1996)Madau, Ferguson, Dickinson, Giavalisco, Steidel,
  \& Fruchter}]{Madau:1996p891}
Madau P., Ferguson H.~C., Dickinson M.~E., Giavalisco M., Steidel C.~C.,
  Fruchter A., 1996, MNRAS, 283, 1388

\bibitem[{Madau {et~al.}(1998)Madau, Pozzetti, \& Dickinson}]{Madau:1998p752}
Madau P., Pozzetti L., Dickinson M., 1998, ApJ, 498, 106

\bibitem[{Magdis {et~al.}(2011)Magdis, Elbaz, Hwang, the PEP~team, \& the
  HERMES~team}]{Magdis:2011p1146}
Magdis G.~E., Elbaz D., Hwang H.~S., the PEP~team, the HERMES~team, 2011,
  arXiv:astro-ph/1102.4236

\bibitem[{Mahdavi {et~al.}(1999)Mahdavi, Geller, B{\"o}hringer, Kurtz, \&
  Ramella}]{Mahdavi:1999p581}
Mahdavi A., Geller M.~J., B{\"o}hringer H., Kurtz M.~J., Ramella M., 1999, ApJ,
  518, 69

\bibitem[{Mamon(1986)}]{Mamon:1986p701}
Mamon G.~A., 1986, ApJ, 307, 426

\bibitem[{Mamon(1987)}]{Mamon:1987p1095}
---, 1987, ApJ, 321, 622

\bibitem[{Marcum {et~al.}(2004)Marcum, Aars, \& Fanelli}]{Marcum:2004p572}
Marcum P.~M., Aars C.~E., Fanelli M.~N., 2004, AJ, 127, 3213

\bibitem[{Martel(2005)}]{Martel:2005p1002}
Martel H., 2005, arXiv:astro-ph/0506540

\bibitem[{Masjedi {et~al.}(2006)Masjedi, Hogg, Cool, Eisenstein, Blanton,
  Zehavi, Berlind, Bell, Schneider, Warren, \& Brinkmann}]{Masjedi:2006p920}
Masjedi M., Hogg D.~W., Cool R.~J., Eisenstein D.~J., Blanton M.~R., Zehavi I.,
  Berlind A.~A., Bell E.~F., Schneider D.~P., Warren M.~S., Brinkmann J., 2006,
  ApJ, 644, 54

\bibitem[{Materne(1978)}]{Materne:1978p353}
Materne J., 1978, A{\&}A, 63, 401

\bibitem[{Mateus {et~al.}(2007)Mateus, Sodr{\'e}, Fernandes, \&
  Stasi{\'n}ska}]{Mateus:2007p735}
Mateus A., Sodr{\'e} L., Fernandes R.~C., Stasi{\'n}ska G., 2007, MNRAS, 374,
  1457

\bibitem[{Mathews \& Brighenti(2003)}]{Mathews:2003p787}
Mathews W.~G., Brighenti F., 2003, Annual Review of Astronomy and Astrophysics,
  41, 191

\bibitem[{Matteucci \& Greggio(1986)}]{Matteucci:1986p956}
Matteucci F., Greggio L., 1986, A\&A, 154, 279

\bibitem[{McKee \& Ostriker(2007)}]{McKee:2007p763}
McKee C.~F., Ostriker E.~C., 2007, Annual Review of Astronomy and Astrophysics,
  45, 565

\bibitem[{{Memola} {et~al.}(2009){Memola}, {Trinchieri}, {Wolter}, {Focardi},
  \& {Kelm}}]{Memola:2009p642}
{Memola} E., {Trinchieri} G., {Wolter} A., {Focardi} P., {Kelm} B., 2009, A\&A,
  497, 359

\bibitem[{Mendel {et~al.}(2008)Mendel, Proctor, Forbes, \&
  Brough}]{Mendel:2008p1040}
Mendel J.~T., Proctor R.~N., Forbes D.~A., Brough S., 2008, MNRAS, 389, 749

\bibitem[{Merch{\'a}n \& Zandivarez(2005)}]{Merchan:2005p495}
Merch{\'a}n M.~E., Zandivarez A., 2005, ApJ, 630, 759

\bibitem[{Mihos(1995)}]{Mihos:1995p596}
Mihos J.~C., 1995, ApJ, 438, L75

\bibitem[{{Mihos}(2004)}]{2004cgpc.symp..277M}
{Mihos} J.~C., 2004, in Clusters of Galaxies: Probes of Cosmological Structure
  and Galaxy Evolution, {J.~S.~Mulchaey, A.~Dressler, \& A.~Oemler}, ed., pp.
  277--

\bibitem[{Mihos \& Hernquist(1996)}]{Mihos:1996p1093}
Mihos J.~C., Hernquist L., 1996, ApJ, 464, 641

\bibitem[{{Milne}(1933)}]{1933ZA......6....1M}
{Milne} E.~A., 1933, Zeitschrift f\"ur Astrophysik, 6, 1

\bibitem[{{Mo} {et~al.}(2010){Mo}, {van den Bosch}, \& {White}}]{Mo_big_book}
{Mo} H., {van den Bosch} F.~C., {White} S., 2010, Galaxy Formation and
  Evolution by Houjun Mo, Frank van den Bosch and Simon White.~Cambridge
  University Press, 2010.~ISBN: 9780521857932

\bibitem[{Moller \& Warren(1998)}]{Moller:1998p1039}
Moller P., Warren S.~J., 1998, MNRAS, 299, 661

\bibitem[{Moore {et~al.}(1996)Moore, Katz, Lake, Dressler, \&
  Oemler}]{Moore:1996p716}
Moore B., Katz N., Lake G., Dressler A., Oemler A., 1996, Nature, 379, 613

\bibitem[{Moore {et~al.}(1998)Moore, Lake, \& Katz}]{Moore:1998p792}
Moore B., Lake G., Katz N., 1998, ApJ, 495, 139

\bibitem[{{Mould}(1982)}]{1982ARA&A..20...91M}
{Mould} J.~R., 1982, Annual Review of Astronomy and Astrophysics, 20, 91

\bibitem[{{Mu{\~n}oz}(2004)}]{2004IJMPA..19.3093M}
{Mu{\~n}oz} C., 2004, International Journal of Modern Physics A, 19, 3093

\bibitem[{Mukhanov \& Chibisov(1981)}]{Mukhanov:1981p1009}
Mukhanov V.~F., Chibisov G.~V., 1981, ZHETF Pis'ma v Redaktsiiu, 33, 549

\bibitem[{Mulchaey \& Zabludoff(1999)}]{Mulchaey:1999p563}
Mulchaey J.~S., Zabludoff A.~I., 1999, ApJ, 514, 133

\bibitem[{Naab {et~al.}(2007)Naab, Johansson, Ostriker, \&
  Efstathiou}]{Naab:2007p757}
Naab T., Johansson P.~H., Ostriker J.~P., Efstathiou G., 2007, ApJ, 658, 710

\bibitem[{Naab {et~al.}(2006)Naab, Khochfar, \& Burkert}]{Naab:2006p923}
Naab T., Khochfar S., Burkert A., 2006, ApJ, 636, L81

\bibitem[{Narlikar \& Padmanabhan(1991)}]{Narlikar:1991p858}
Narlikar J.~V., Padmanabhan T., 1991, Annual Review of Astronomy and
  Astrophysics, 29, 325

\bibitem[{Narlikar \& Padmanabhan(2001)}]{Narlikar:2001p870}
---, 2001, Annual Review of Astronomy and Astrophysics, 39, 211

\bibitem[{Navarro \& Benz(1991)}]{Navarro:1991p1083}
Navarro J.~F., Benz W., 1991, ApJ, 380, 320

\bibitem[{Neyrinck {et~al.}(2005)Neyrinck, Gnedin, \&
  Hamilton}]{Neyrinck:2005p1034}
Neyrinck M.~C., Gnedin N.~Y., Hamilton A. J.~S., 2005, MNRAS, 356, 1222

\bibitem[{Nolthenius \& White(1987)}]{Nolthenius:1987p357}
Nolthenius R., White S. D.~M., 1987, MNRAS, 225, 505

\bibitem[{Norberg {et~al.}(2001)Norberg, Baugh, Hawkins, Maddox, Peacock, Cole,
  Frenk, Bland-Hawthorn, Bridges, Cannon, Colless, Collins, Couch, Dalton,
  Propris, Driver, Efstathiou, Ellis, Glazebrook, Jackson, Lahav, Lewis,
  Lumsden, Madgwick, Peterson, Sutherland, \& Taylor}]{Norberg:2001p1006}
Norberg P., Baugh C.~M., Hawkins E., Maddox S., Peacock J.~A., Cole S., Frenk
  C.~S., Bland-Hawthorn J., Bridges T., Cannon R., Colless M., Collins C.,
  Couch W., Dalton G., Propris R.~D., Driver S.~P., Efstathiou G., Ellis R.~S.,
  Glazebrook K., Jackson C., Lahav O., Lewis I., Lumsden S., Madgwick D.,
  Peterson B.~A., Sutherland W., Taylor K., 2001, MNRAS, 328, 64

\bibitem[{Norberg {et~al.}(2008)Norberg, Frenk, \& Cole}]{Norberg:2008p652}
Norberg P., Frenk C.~S., Cole S., 2008, MNRAS, 383, 646

\bibitem[{Norman {et~al.}(1996)Norman, Sellwood, \& Hasan}]{Norman:1996p950}
Norman C.~A., Sellwood J.~A., Hasan H., 1996, ApJ, 462, 114

\bibitem[{Oemler(1974)}]{Oemler:1974p646}
Oemler A., 1974, ApJ, 194, 1

\bibitem[{Oemler(1988)}]{Oemler:1988p699}
---, 1988, The Minnesota lectures on clusters of galaxies and large-scale
  structure (A90-36758 15-90), 5, 19

\bibitem[{Oguri(2006)}]{Oguri:2006p863}
Oguri M., 2006, MNRAS, 367, 1241

\bibitem[{Oort(1983)}]{Oort:1983p984}
Oort J.~H., 1983, Annual Review of Astronomy and Astrophysics, 21, 373

\bibitem[{Oppenheimer \& Dav{\'e}(2006)}]{Oppenheimer:2006p952}
Oppenheimer B.~D., Dav{\'e} R., 2006, MNRAS, 373, 1265

\bibitem[{{Osterbrock}(1989)}]{1989agna.book.....O}
{Osterbrock} D.~E., 1989, Astrophysics of gaseous nebulae and active galactic
  nuclei, University of Minnesota, et al.~Mill Valley, CA, University Science
  Books, 1989, 422 p.

\bibitem[{Padmanabhan {et~al.}(2007)Padmanabhan, Schlegel, Seljak, Makarov,
  Bahcall, Blanton, Brinkmann, Eisenstein, Finkbeiner, Gunn, Hogg, Ivezi{\'c},
  Knapp, Loveday, Lupton, Nichol, Schneider, Strauss, Tegmark, \&
  York}]{Padmanabhan:2007p1036}
Padmanabhan N., Schlegel D.~J., Seljak U., Makarov A., Bahcall N.~A., Blanton
  M.~R., Brinkmann J., Eisenstein D.~J., Finkbeiner D.~P., Gunn J.~E., Hogg
  D.~W., Ivezi{\'c} {\v Z}., Knapp G.~R., Loveday J., Lupton R.~H., Nichol
  R.~C., Schneider D.~P., Strauss M.~A., Tegmark M., York D.~G., 2007, MNRAS,
  378, 852

\bibitem[{{Padmanabhan}(1993)}]{1993sfu..book.....P}
{Padmanabhan} T., 1993, Structure Formation in the Universe, by T.~Padmanabhan,
  pp.~499.~ISBN 0521424860.~Cambridge, UK: Cambridge University Press, June
  1993.

\bibitem[{{Pagel}(1997)}]{1997nceg.book.....P}
{Pagel} B.~E.~J., 1997, Nucleosynthesis and Chemical Evolution of Galaxies, by
  Bernard E.~J.~Pagel, pp.~392.~ISBN 0521550610.~Cambridge, UK: Cambridge
  University Press, October 1997.

\bibitem[{{Papadopoulos} {et~al.}(2001){Papadopoulos}, {Ivison}, {Carilli}, \&
  {Lewis}}]{2001Natur.409...58P}
{Papadopoulos} P., {Ivison} R., {Carilli} C., {Lewis} G., 2001, Nature, 409, 58

\bibitem[{Paz {et~al.}(2006)Paz, Lambas, Padilla, \&
  Merch{\'a}n}]{Paz:2006p489}
Paz D.~J., Lambas D.~G., Padilla N.~D., Merch{\'a}n M.~E., 2006, MNRAS, 366,
  1503

\bibitem[{Peacock(2002)}]{Peacock:2002p995}
Peacock J.~A., 2002, A New Era in Cosmology, 283, 19

\bibitem[{Peebles(1968)}]{Peebles:1968p1018}
Peebles P. J.~E., 1968, ApJ, 153, 1

\bibitem[{{Peebles}(1980)}]{1980lssu.book.....P}
{Peebles} P.~J.~E., 1980, The large-scale structure of the universe, Princeton
  University Press, 1980.~435 p.

\bibitem[{Peebles(1982)}]{Peebles:1982p1000}
Peebles P. J.~E., 1982, ApJ, 263, L1

\bibitem[{Peebles \& Ratra(1988)}]{Peebles:1988p1017}
Peebles P. J.~E., Ratra B., 1988, ApJ, 325, L17

\bibitem[{Pei \& Fall(1995)}]{Pei:1995p981}
Pei Y.~C., Fall S.~M., 1995, AJ, 454, 69

\bibitem[{{Penzias} \& {Wilson}(1965)}]{1965ApJ...142..419P}
{Penzias} A.~A., {Wilson} R.~W., 1965, ApJ, 142, 419

\bibitem[{Percival {et~al.}(2001)Percival, Baugh, Bland-Hawthorn, Bridges,
  Cannon, Cole, Colless, Collins, Couch, Dalton, Propris, Driver, Efstathiou,
  Ellis, Frenk, Glazebrook, Jackson, Lahav, Lewis, Lumsden, Maddox, Moody,
  Norberg, Peacock, Peterson, Sutherland, \& Taylor}]{Percival:2001p803}
Percival W.~J., Baugh C.~M., Bland-Hawthorn J., Bridges T., Cannon R., Cole S.,
  Colless M., Collins C., Couch W., Dalton G., Propris R.~D., Driver S.~P.,
  Efstathiou G., Ellis R.~S., Frenk C.~S., Glazebrook K., Jackson C., Lahav O.,
  Lewis I., Lumsden S., Maddox S., Moody S., Norberg P., Peacock J.~A.,
  Peterson B.~A., Sutherland W., Taylor K., 2001, MNRAS, 327, 1297

\bibitem[{Percival {et~al.}(2007)Percival, Nichol, Eisenstein, Frieman,
  Fukugita, Loveday, Pope, Schneider, Szalay, Tegmark, Vogeley, Weinberg,
  Zehavi, Bahcall, Brinkmann, Connolly, \& Meiksin}]{Percival:2007p810}
Percival W.~J., Nichol R.~C., Eisenstein D.~J., Frieman J.~A., Fukugita M.,
  Loveday J., Pope A.~C., Schneider D.~P., Szalay A.~S., Tegmark M., Vogeley
  M.~S., Weinberg D.~H., Zehavi I., Bahcall N.~A., Brinkmann J., Connolly
  A.~J., Meiksin A., 2007, ApJ, 657, 645

\bibitem[{Pipino \& Matteucci(2004)}]{Pipino:2004p947}
Pipino A., Matteucci F., 2004, MNRAS, 347, 968

\bibitem[{{Ponman} {et~al.}(1994){Ponman}, {Allan}, {Jones}, {Merrifield},
  {McHardy}, {Lehto}, \& {Luppino}}]{1994Natur.369..462P}
{Ponman} T.~J., {Allan} D.~J., {Jones} L.~R., {Merrifield} M., {McHardy} I.~M.,
  {Lehto} H.~J., {Luppino} G.~A., 1994, Nature, 369, 462

\bibitem[{Postman \& Geller(1984)}]{Postman:1984p719}
Postman M., Geller M.~J., 1984, ApJ, 281, 95

\bibitem[{{Prada} {et~al.}(2003){Prada}, {Vitvitska}, {Klypin}, {Holtzman},
  {Schlegel}, {Grebel}, {Rix}, {Brinkmann}, {McKay}, \&
  {Csabai}}]{2003ApJ...598..260P}
{Prada} F., {Vitvitska} M., {Klypin} A., {Holtzman} J.~A., {Schlegel} D.~J.,
  {Grebel} E.~K., {Rix} H., {Brinkmann} J., {McKay} T.~A., {Csabai} I., 2003,
  ApJ, 598, 260

\bibitem[{Press \& Schechter(1974)}]{Press:1974p694}
Press W.~H., Schechter P., 1974, ApJ, 187, 425

\bibitem[{Proctor {et~al.}(2005)Proctor, Forbes, Forestell, \&
  Gebhardt}]{Proctor:2005p558}
Proctor R.~N., Forbes D.~A., Forestell A., Gebhardt K., 2005, MNRAS, 362, 857

\bibitem[{Raha {et~al.}(1991)Raha, Sellwood, James, \& Kahn}]{Raha:1991p943}
Raha N., Sellwood J.~A., James R.~A., Kahn F.~D., 1991, Nature, 352, 411

\bibitem[{Ramella {et~al.}(1994)Ramella, Diaferio, Geller, \&
  Huchra}]{Ramella:1994p703}
Ramella M., Diaferio A., Geller M.~J., Huchra J.~P., 1994, AJ, 107, 1623

\bibitem[{{Ramella} {et~al.}(1989){Ramella}, {Geller}, \&
  {Huchra}}]{Ramella:1989p4}
{Ramella} M., {Geller} M.~J., {Huchra} J.~P., 1989, ApJ, 344, 57

\bibitem[{Ramella {et~al.}(1995)Ramella, Geller, Huchra, \&
  Thorstensen}]{Ramella:1995p562}
Ramella M., Geller M.~J., Huchra J.~P., Thorstensen J.~R., 1995, ApJ, 109, 1469

\bibitem[{Ramella {et~al.}(1997)Ramella, Pisani, \& Geller}]{Ramella:1997p718}
Ramella M., Pisani A., Geller M.~J., 1997, AJ, 113, 483

\bibitem[{Ratra \& Vogeley(2008)}]{Ratra:2008p1016}
Ratra B., Vogeley M.~S., 2008, PASP, 120, 235

\bibitem[{Razoumov \& Sommer-Larsen(2010)}]{Razoumov:2010p777}
Razoumov A.~O., Sommer-Larsen J., 2010, ApJ, 710, 1239

\bibitem[{Recchi {et~al.}(2001)Recchi, Matteucci, \&
  D'Ercole}]{Recchi:2001p959}
Recchi S., Matteucci F., D'Ercole A., 2001, MNRAS, 322, 800

\bibitem[{Reda {et~al.}(2004)Reda, Forbes, Beasley, O'Sullivan, \&
  Goudfrooij}]{Reda:2004p502}
Reda F.~M., Forbes D.~A., Beasley M.~A., O'Sullivan E.~J., Goudfrooij P., 2004,
  MNRAS, 354, 851

\bibitem[{Reda {et~al.}(2005)Reda, Forbes, \& Hau}]{Reda:2005p561}
Reda F.~M., Forbes D.~A., Hau G. K.~T., 2005, MNRAS, 360, 693

\bibitem[{Reda {et~al.}(2007)Reda, Proctor, Forbes, Hau, \&
  Larsen}]{Reda:2007p559}
Reda F.~M., Proctor R.~N., Forbes D.~A., Hau G. K.~T., Larsen S.~S., 2007,
  MNRAS, 377, 1772

\bibitem[{Reduzzi {et~al.}(1996)Reduzzi, Longhetti, \&
  Rampazzo}]{Reduzzi:1996p622}
Reduzzi L., Longhetti M., Rampazzo R., 1996, MNRAS, 282, 149

\bibitem[{Rees \& Ostriker(1977)}]{Rees:1977p1024}
Rees M.~J., Ostriker J.~P., 1977, MNRAS, 179, 541

\bibitem[{Rekola {et~al.}(2005)Rekola, Jerjen, \& Flynn}]{Rekola:2005p586}
Rekola R., Jerjen H., Flynn C., 2005, A{\&}A, 437, 823

\bibitem[{{Rekola} {et~al.}(2005){Rekola}, {Richer}, {McCall}, {Valtonen},
  {Kotilainen}, \& {Flynn}}]{Rekola:361p330R}
{Rekola} R., {Richer} M.~G., {McCall} M.~L., {Valtonen} M.~J., {Kotilainen}
  J.~K., {Flynn} C., 2005, MNRAS, 361, 330

\bibitem[{{Ricciardelli} \& {Franceschini}(2010)}]{2010arXiv1004.3289R}
{Ricciardelli} E., {Franceschini} A., 2010, A\&A, 518, A14

\bibitem[{Rieke {et~al.}(2009)Rieke, Alonso-Herrero, Weiner,
  P{\'e}rez-Gonz{\'a}lez, Blaylock, Donley, \& Marcillac}]{Rieke:2009p1088}
Rieke G.~H., Alonso-Herrero A., Weiner B.~J., P{\'e}rez-Gonz{\'a}lez P.~G.,
  Blaylock M., Donley J.~L., Marcillac D., 2009, ApJ, 692, 556

\bibitem[{Rieke \& Low(1972)}]{Rieke:1972p1101}
Rieke G.~H., Low F.~J., 1972, ApJ, 176, L95

\bibitem[{Rigopoulou {et~al.}(1999)Rigopoulou, Spoon, Genzel, Lutz, Moorwood,
  \& Tran}]{Rigopoulou:1999p1114}
Rigopoulou D., Spoon H. W.~W., Genzel R., Lutz D., Moorwood A. F.~M., Tran
  Q.~D., 1999, AJ, 118, 2625

\bibitem[{Rood(1988)}]{Rood:1988p912}
Rood H.~J., 1988, Annual Review of Astronomy and Astrophysics, 26, 245

\bibitem[{Rose(1977)}]{Rose:1977p692}
Rose J.~A., 1977, ApJ, 211, 311

\bibitem[{{Rosswog}(2009)}]{Rosswog:2009p772}
{Rosswog} S., 2009, New Astronomy Reviews, 53, 78

\bibitem[{Rowan-Robinson(2000)}]{RowanRobinson:2000p1143}
Rowan-Robinson M., 2000, MNRAS, 316, 885

\bibitem[{Rownd \& Young(1999)}]{Rownd:1999p977}
Rownd B.~K., Young J.~S., 1999, AJ, 118, 670

\bibitem[{Sales {et~al.}(2007)Sales, Navarro, Lambas, White, \&
  Croton}]{Sales:2007p380}
Sales L.~V., Navarro J.~F., Lambas D.~G., White S. D.~M., Croton D.~J., 2007,
  MNRAS, 382, 1901

\bibitem[{S{\'a}nchez {et~al.}(2009)S{\'a}nchez, Crocce, Cabr{\'e}, Baugh, \&
  Gazta{\~n}aga}]{Sanchez:2009p1035}
S{\'a}nchez A.~G., Crocce M., Cabr{\'e} A., Baugh C.~M., Gazta{\~n}aga E.,
  2009, MNRAS, 400, 1643

\bibitem[{Sanders \& Mirabel(1996)}]{Sanders:1996p1085}
Sanders D.~B., Mirabel I.~F., 1996, Annual Review of Astronomy and
  Astrophysics, 34, 749

\bibitem[{Sanders {et~al.}(1991)Sanders, Scoville, \&
  Soifer}]{Sanders:1991p1151}
Sanders D.~B., Scoville N.~Z., Soifer B.~T., 1991, ApJ, 370, 158

\bibitem[{Sanders {et~al.}(1988)Sanders, Soifer, Elias, Madore, Matthews,
  Neugebauer, \& Scoville}]{Sanders:1988p1131}
Sanders D.~B., Soifer B.~T., Elias J.~H., Madore B.~F., Matthews K., Neugebauer
  G., Scoville N.~Z., 1988, ApJ, 325, 74

\bibitem[{{Sanders} {et~al.}(1984){Sanders}, {Solomon}, \&
  {Scoville}}]{1984ApJ...276..182S}
{Sanders} D.~B., {Solomon} P.~M., {Scoville} N.~Z., 1984, ApJ, 276, 182

\bibitem[{Santos {et~al.}(2007)Santos, de~Oliveira, \&
  Sodr{\'e}}]{Santos:2007p217}
Santos W.~A., de~Oliveira C. L.~M., Sodr{\'e} L., 2007, AJ, 134, 1551

\bibitem[{Schechter(1976)}]{Schechter:1976p592}
Schechter P.~L., 1976, ApJ, 203, 297

\bibitem[{{Schmidt}(1959)}]{1959ApJ...129..243S}
{Schmidt} M., 1959, ApJ, 129, 243

\bibitem[{Scranton {et~al.}(2002)Scranton, Johnston, Dodelson, Frieman,
  Connolly, Eisenstein, Gunn, Hui, Jain, Kent, Loveday, Narayanan, Nichol,
  O'Connell, Scoccimarro, Sheth, Stebbins, Strauss, Szalay, Szapudi, Tegmark,
  Vogeley, Zehavi, Annis, Bahcall, Brinkman, Csabai, Hindsley, Ivezic, Kim,
  Knapp, Lamb, Lee, Lupton, McKay, Munn, Peoples, Pier, Richards, Rockosi,
  Schlegel, Schneider, Stoughton, Tucker, Yanny, \& York}]{Scranton:2002p809}
Scranton R., Johnston D., Dodelson S., Frieman J.~A., Connolly A., Eisenstein
  D.~J., Gunn J.~E., Hui L., Jain B., Kent S., Loveday J., Narayanan V., Nichol
  R.~C., O'Connell L., Scoccimarro R., Sheth R.~K., Stebbins A., Strauss M.~A.,
  Szalay A.~S., Szapudi I., Tegmark M., Vogeley M., Zehavi I., Annis J.,
  Bahcall N.~A., Brinkman J., Csabai I., Hindsley R., Ivezic Z., Kim R. S.~J.,
  Knapp G.~R., Lamb D.~Q., Lee B.~C., Lupton R.~H., McKay T., Munn J., Peoples
  J., Pier J., Richards G.~T., Rockosi C., Schlegel D., Schneider D.~P.,
  Stoughton C., Tucker D.~L., Yanny B., York D.~G., 2002, ApJ, 579, 48

\bibitem[{Seljak {et~al.}(2006)Seljak, Slosar, \& McDonald}]{Seljak:2006p1025}
Seljak U., Slosar A., McDonald P., 2006, Journal of Cosmology and Astroparticle
  Physics, 10, 014

\bibitem[{Seljak \& Zaldarriaga(1996)}]{Seljak:1996p990}
Seljak U., Zaldarriaga M., 1996, AJ, 469, 437

\bibitem[{Sellwood \& Merritt(1994)}]{Sellwood:1994p935}
Sellwood J.~A., Merritt D., 1994, ApJ, 425, 530

\bibitem[{{Shakhbazyan}(1973)}]{1973Afz.....9..495S}
{Shakhbazyan} R.~K., 1973, Astrofizika, 9, 495

\bibitem[{Shandarin \& Zeldovich(1989)}]{Shandarin:1989p1001}
Shandarin S.~F., Zeldovich Y.~B., 1989, Reviews of Modern Physics, 61, 185

\bibitem[{Sheth \& Tormen(1999)}]{Sheth:1999p1027}
Sheth R.~K., Tormen G., 1999, MNRAS, 308, 119

\bibitem[{{Sijacki} {et~al.}(2007){Sijacki}, {Springel}, {Di Matteo}, \&
  {Hernquist}}]{2007MNRAS.380..877S}
{Sijacki} D., {Springel} V., {Di Matteo} T., {Hernquist} L., 2007, MNRAS, 380,
  877

\bibitem[{Silk(1968)}]{Silk:1968p796}
Silk J., 1968, ApJ, 151, 459

\bibitem[{Silk(1977)}]{Silk:1977p1022}
---, 1977, ApJ, 211, 638

\bibitem[{Silk \& Rees(1998)}]{Silk:1998p996}
Silk J., Rees M.~J., 1998, A{\&}A, 331, L1

\bibitem[{Smith {et~al.}(2004)Smith, Martinez, \& Graham}]{Smith:2004p250}
Smith R.~M., Martinez V.~J., Graham M.~J., 2004, ApJ, 617, 1017

\bibitem[{Smoot {et~al.}(1992)Smoot, Bennett, Kogut, Wright, Aymon, Boggess,
  Cheng, de~Amici, Gulkis, Hauser, Hinshaw, Jackson, Janssen, Kaita, Kelsall,
  Keegstra, Lineweaver, Loewenstein, Lubin, Mather, Meyer, Moseley, Murdock,
  Rokke, Silverberg, Tenorio, Weiss, \& Wilkinson}]{Smoot:1992p860}
Smoot G.~F., Bennett C.~L., Kogut A., Wright E.~L., Aymon J., Boggess N.~W.,
  Cheng E.~S., de~Amici G., Gulkis S., Hauser M.~G., Hinshaw G., Jackson P.~D.,
  Janssen M., Kaita E., Kelsall T., Keegstra P., Lineweaver C., Loewenstein K.,
  Lubin P., Mather J., Meyer S.~S., Moseley S.~H., Murdock T., Rokke L.,
  Silverberg R.~F., Tenorio L., Weiss R., Wilkinson D.~T., 1992, ApJ, 396, L1

\bibitem[{{Solomon} \& {Sanders}(1980)}]{1980gmcg.work...41S}
{Solomon} P.~M., {Sanders} D.~B., 1980, in Giant Molecular Clouds in the
  Galaxy, {P.~M.~Solomon \& M.~G.~Edmunds}, ed., pp. 41--73

\bibitem[{Somerville {et~al.}(2008)Somerville, Hopkins, Cox, Robertson, \&
  Hernquist}]{Somerville:2008p759}
Somerville R.~S., Hopkins P.~F., Cox T.~J., Robertson B.~E., Hernquist L.,
  2008, MNRAS, 391, 481

\bibitem[{Somerville \& Primack(1999)}]{Somerville:1999p762}
Somerville R.~S., Primack J.~R., 1999, MNRAS, 310, 1087

\bibitem[{Somerville {et~al.}(2001)Somerville, Primack, \&
  Faber}]{Somerville:2001p760}
Somerville R.~S., Primack J.~R., Faber S.~M., 2001, MNRAS, 320, 504

\bibitem[{Sommer-Larsen {et~al.}(2003)Sommer-Larsen, G{\"o}tz, \&
  Portinari}]{SommerLarsen:2003p889}
Sommer-Larsen J., G{\"o}tz M., Portinari L., 2003, ApJ, 596, 47

\bibitem[{Springel \& Hernquist(2003)}]{Springel:2003p867}
Springel V., Hernquist L., 2003, MNRAS, 339, 289

\bibitem[{Springel \& Hernquist(2005)}]{Springel:2005p916}
---, 2005, ApJ, 622, L9

\bibitem[{Springel {et~al.}(2005)Springel, White, Jenkins, Frenk, Yoshida, Gao,
  Navarro, Thacker, Croton, Helly, Peacock, Cole, Thomas, Couchman, Evrard,
  Colberg, \& Pearce}]{Springel:2005p595}
Springel V., White S. D.~M., Jenkins A., Frenk C.~S., Yoshida N., Gao L.,
  Navarro J.~F., Thacker R., Croton D.~J., Helly J., Peacock J.~A., Cole S.,
  Thomas P., Couchman H., Evrard A., Colberg J.~M., Pearce F., 2005, Nature,
  435, 629

\bibitem[{Springel {et~al.}(2001)Springel, Yoshida, \&
  White}]{Springel:2001p591}
Springel V., Yoshida N., White S. D.~M., 2001, New Astronomy, 6, 79

\bibitem[{Stewart {et~al.}(2009)Stewart, Bullock, Barton, \&
  Wechsler}]{Stewart:2009p614}
Stewart K.~R., Bullock J.~S., Barton E.~J., Wechsler R.~H., 2009, ApJ, 702,
  1005

\bibitem[{Stoughton {et~al.}(2002)Stoughton, Lupton, Bernardi, Blanton, Burles,
  Castander, Connolly, Eisenstein, Frieman, Hennessy, Hindsley, Ivezi{\'c},
  Kent, Kunszt, Lee, Meiksin, Munn, Newberg, Nichol, Nicinski, Pier, Richards,
  Richmond, Schlegel, Smith, Strauss, SubbaRao, Szalay, Thakar, Tucker, Berk,
  Yanny, Adelman, Anderson, Anderson, Annis, Bahcall, Bakken, Bartelmann,
  Bastian, Bauer, Berman, B{\"o}hringer, Boroski, Bracker, Briegel, Briggs,
  Brinkmann, Brunner, Carey, Carr, Chen, Christian, Colestock, Crocker, Csabai,
  Czarapata, Dalcanton, Davidsen, Davis, Dehnen, Dodelson, Doi, Dombeck,
  Donahue, Ellman, Elms, Evans, Eyer, Fan, Federwitz, Friedman, Fukugita, Gal,
  Gillespie, Glazebrook, Gray, Grebel, Greenawalt, Greene, Gunn, de~Haas,
  Haiman, Haldeman, Hall, Hamabe, Hansen, Harris, Harris, Harvanek, Hawley,
  Hayes, Heckman, Helmi, Henden, Hogan, Hogg, Holmgren, Holtzman, Huang, Hull,
  Ichikawa, Ichikawa, Johnston, Kauffmann, Kim, Kimball, Kinney, Klaene,
  Kleinman, Klypin, Knapp, Korienek, Krolik, Kron, Krzesi{\'n}ski, Lamb, Leger,
  Limmongkol, Lindenmeyer, Long, Loomis, Loveday, MacKinnon, Mannery, Mantsch,
  Margon, McGehee, McKay, McLean, Menou, Merelli, Mo, Monet, Nakamura,
  Narayanan, Nash, Neilsen, Newman, Nitta, Odenkirchen, Okada, Okamura,
  Ostriker, Owen, Pauls, Peoples, Peterson, Petravick, Pope, Pordes, Postman,
  Prosapio, Quinn, Rechenmacher, Rivetta, Rix, Rockosi, Rosner, Ruthmansdorfer,
  Sandford, Schneider, Scranton, Sekiguchi, Sergey, Sheth, Shimasaku, Smee,
  Snedden, Stebbins, Stubbs, Szapudi, Szkody, Szokoly, Tabachnik, Tsvetanov,
  Uomoto, Vogeley, Voges, Waddell, Walterbos, i~Wang, Watanabe, Weinberg,
  White, White, Wilhite, Wolfe, Yasuda, York, Zehavi, \&
  Zheng}]{Stoughton:2002p800}
Stoughton C., Lupton R.~H., Bernardi M., Blanton M.~R., Burles S., Castander
  F.~J., Connolly A.~J., Eisenstein D.~J., Frieman J.~A., Hennessy G.~S.,
  Hindsley R.~B., Ivezi{\'c} {\v Z}., Kent S., Kunszt P.~Z., Lee B.~C., Meiksin
  A., Munn J.~A., Newberg H.~J., Nichol R.~C., Nicinski T., Pier J.~R.,
  Richards G.~T., Richmond M.~W., Schlegel D.~J., Smith J.~A., Strauss M.~A.,
  SubbaRao M., Szalay A.~S., Thakar A.~R., Tucker D.~L., Berk D. E.~V., Yanny
  B., Adelman J.~K., Anderson J.~E., Anderson S.~F., Annis J., Bahcall N.~A.,
  Bakken J.~A., Bartelmann M., Bastian S., Bauer A., Berman E., B{\"o}hringer
  H., Boroski W.~N., Bracker S., Briegel C., Briggs J.~W., Brinkmann J.,
  Brunner R., Carey L., Carr M.~A., Chen B., Christian D., Colestock P.~L.,
  Crocker J.~H., Csabai I., Czarapata P.~C., Dalcanton J., Davidsen A.~F.,
  Davis J.~E., Dehnen W., Dodelson S., Doi M., Dombeck T., Donahue M., Ellman
  N., Elms B.~R., Evans M.~L., Eyer L., Fan X., Federwitz G.~R., Friedman S.,
  Fukugita M., Gal R., Gillespie B., Glazebrook K., Gray J., Grebel E.~K.,
  Greenawalt B., Greene G., Gunn J.~E., de~Haas E., Haiman Z., Haldeman M.,
  Hall P.~B., Hamabe M., Hansen B., Harris F.~H., Harris H., Harvanek M.,
  Hawley S.~L., Hayes J. J.~E., Heckman T.~M., Helmi A., Henden A., Hogan
  C.~J., Hogg D.~W., Holmgren D.~J., Holtzman J., Huang C.-H., Hull C.,
  Ichikawa S.-I., Ichikawa T., Johnston D.~E., Kauffmann G., Kim R. S.~J.,
  Kimball T., Kinney E., Klaene M., Kleinman S.~J., Klypin A., Knapp G.~R.,
  Korienek J., Krolik J., Kron R.~G., Krzesi{\'n}ski J., Lamb D.~Q., Leger
  R.~F., Limmongkol S., Lindenmeyer C., Long D.~C., Loomis C., Loveday J.,
  MacKinnon B., Mannery E.~J., Mantsch P.~M., Margon B., McGehee P., McKay
  T.~A., McLean B., Menou K., Merelli A., Mo H.~J., Monet D.~G., Nakamura O.,
  Narayanan V.~K., Nash T., Neilsen E.~H., Newman P.~R., Nitta A., Odenkirchen
  M., Okada N., Okamura S., Ostriker J.~P., Owen R., Pauls A.~G., Peoples J.,
  Peterson R.~S., Petravick D., Pope A., Pordes R., Postman M., Prosapio A.,
  Quinn T.~R., Rechenmacher R., Rivetta C.~H., Rix H.-W., Rockosi C.~M., Rosner
  R., Ruthmansdorfer K., Sandford D., Schneider D.~P., Scranton R., Sekiguchi
  M., Sergey G., Sheth R., Shimasaku K., Smee S., Snedden S.~A., Stebbins A.,
  Stubbs C., Szapudi I., Szkody P., Szokoly G.~P., Tabachnik S., Tsvetanov Z.,
  Uomoto A., Vogeley M.~S., Voges W., Waddell P., Walterbos R., i~Wang S.,
  Watanabe M., Weinberg D.~H., White R.~L., White S. D.~M., Wilhite B., Wolfe
  D., Yasuda N., York D.~G., Zehavi I., Zheng W., 2002, AJ, 123, 485

\bibitem[{Strateva {et~al.}(2001)Strateva, Ivezi{\'c}, Knapp, Narayanan,
  Strauss, Gunn, Lupton, Schlegel, Bahcall, Brinkmann, Brunner, Budav{\'a}ri,
  Csabai, Castander, Doi, Fukugita, Gy{\H o}ry, Hamabe, Hennessy, Ichikawa,
  Kunszt, Lamb, McKay, Okamura, Racusin, Sekiguchi, Schneider, Shimasaku, \&
  York}]{Strateva:2001p746}
Strateva I., Ivezi{\'c} {\v Z}., Knapp G.~R., Narayanan V.~K., Strauss M.~A.,
  Gunn J.~E., Lupton R.~H., Schlegel D., Bahcall N.~A., Brinkmann J., Brunner
  R.~J., Budav{\'a}ri T., Csabai I., Castander F.~J., Doi M., Fukugita M.,
  Gy{\H o}ry Z., Hamabe M., Hennessy G., Ichikawa T., Kunszt P.~Z., Lamb D.~Q.,
  McKay T.~A., Okamura S., Racusin J., Sekiguchi M., Schneider D.~P., Shimasaku
  K., York D., 2001, AJ, 122, 1861

\bibitem[{Sturm {et~al.}(2010)Sturm, Verma, Graci{\'a}-Carpio,
  Hailey-Dunsheath, Contursi, Fischer, Gonz{\'a}lez-Alfonso, Poglitsch,
  Sternberg, Genzel, Lutz, Tacconi, Christopher, \& Jong}]{Sturm:2010p1055}
Sturm E., Verma A., Graci{\'a}-Carpio J., Hailey-Dunsheath S., Contursi A.,
  Fischer J., Gonz{\'a}lez-Alfonso E., Poglitsch A., Sternberg A., Genzel R.,
  Lutz D., Tacconi L., Christopher N., Jong J.~D., 2010, A{\&}A, 518, L36

\bibitem[{Sulentic(1984)}]{Sulentic:1984p341}
Sulentic J.~W., 1984, ApJ, 286, 442

\bibitem[{Tacconi {et~al.}(2008)Tacconi, Genzel, Smail, Neri, Chapman, Ivison,
  Blain, Cox, Omont, Bertoldi, Greve, Schreiber, Genel, Lutz, Swinbank,
  Shapley, Erb, Cimatti, Daddi, \& Baker}]{Tacconi:2008p914}
Tacconi L.~J., Genzel R., Smail I., Neri R., Chapman S.~C., Ivison R.~J., Blain
  A., Cox P., Omont A., Bertoldi F., Greve T., Schreiber N. M.~F., Genel S.,
  Lutz D., Swinbank A.~M., Shapley A.~E., Erb D.~K., Cimatti A., Daddi E.,
  Baker A.~J., 2008, ApJ, 680, 246

\bibitem[{Tago {et~al.}(2006)Tago, Einasto, Saar, Einasto, Suhhonenko,
  J{\~o}eveer, Vennik, Hein{\"a}m{\"a}ki, \& Tucker}]{Tago:2006p490}
Tago E., Einasto J., Saar E., Einasto M., Suhhonenko I., J{\~o}eveer M., Vennik
  J., Hein{\"a}m{\"a}ki P., Tucker D.~L., 2006, Astronomische Nachrichten, 327,
  365

\bibitem[{{Tantalo} {et~al.}(2010){Tantalo}, {Chinellato}, {Merlin}, {Piovan},
  \& {Chiosi}}]{Tantalo:2010p813}
{Tantalo} R., {Chinellato} S., {Merlin} E., {Piovan} L., {Chiosi} C., 2010,
  A\&A, 518, A43

\bibitem[{Tasca {et~al.}(2009)Tasca, Kneib, Iovino, F{\`e}vre, Kova{\v c},
  Bolzonella, Lilly, Abraham, Cassata, Cucciati, Guzzo, Tresse, Zamorani,
  Capak, Garilli, Scodeggio, Sheth, Zucca, Carollo, Contini, Mainieri, Renzini,
  Bardelli, Bongiorno, Caputi, Coppa, de~La~Torre, de~Ravel, Franzetti,
  Kampczyk, Knobel, Koekemoer, Lamareille, Borgne, Brun, Maier, Mignoli, Pello,
  Peng, Montero, Ricciardelli, Silverman, Vergani, Tanaka, Abbas, Bottini,
  Cappi, Cimatti, Ilbert, Leauthaud, Maccagni, Marinoni, McCracken, Memeo,
  Meneux, Oesch, Porciani, Pozzetti, Scaramella, \& Scarlata}]{Tasca:2009p737}
Tasca L. A.~M., Kneib J.-P., Iovino A., F{\`e}vre O.~L., Kova{\v c} K.,
  Bolzonella M., Lilly S.~J., Abraham R.~G., Cassata P., Cucciati O., Guzzo L.,
  Tresse L., Zamorani G., Capak P., Garilli B., Scodeggio M., Sheth K., Zucca
  E., Carollo C.~M., Contini T., Mainieri V., Renzini A., Bardelli S.,
  Bongiorno A., Caputi K., Coppa G., de~La~Torre S., de~Ravel L., Franzetti P.,
  Kampczyk P., Knobel C., Koekemoer A.~M., Lamareille F., Borgne J.-F.~L., Brun
  V.~L., Maier C., Mignoli M., Pello R., Peng Y., Montero E.~P., Ricciardelli
  E., Silverman J.~D., Vergani D., Tanaka M., Abbas U., Bottini D., Cappi A.,
  Cimatti A., Ilbert O., Leauthaud A., Maccagni D., Marinoni C., McCracken
  H.~J., Memeo P., Meneux B., Oesch P., Porciani C., Pozzetti L., Scaramella
  R., Scarlata C., 2009, A{\&}A, 503, 379

\bibitem[{Teerikorpi {et~al.}(2008)Teerikorpi, Chernin, Karachentsev, \&
  Valtonen}]{Teerikorpi:2008p589}
Teerikorpi P., Chernin A.~D., Karachentsev I.~D., Valtonen M., 2008, A{\&}A,
  483, 383

\bibitem[{Tegmark {et~al.}(2004{\natexlab{a}})Tegmark, Blanton, Strauss, Hoyle,
  Schlegel, Scoccimarro, Vogeley, Weinberg, Zehavi, Berlind, Budavari,
  Connolly, Eisenstein, Finkbeiner, Frieman, Gunn, Hamilton, Hui, Jain,
  Johnston, Kent, Lin, Nakajima, Nichol, Ostriker, Pope, Scranton, Seljak,
  Sheth, Stebbins, Szalay, Szapudi, Verde, Xu, Annis, Bahcall, Brinkmann,
  Burles, Castander, Csabai, Loveday, Doi, Fukugita, Gott, Hennessy, Hogg,
  Ivezi{\'c}, Knapp, Lamb, Lee, Lupton, McKay, Kunszt, Munn, O'Connell,
  Peoples, Pier, Richmond, Rockosi, Schneider, Stoughton, Tucker, Berk, Yanny,
  \& York}]{Tegmark:2004p812}
Tegmark M., Blanton M.~R., Strauss M.~A., Hoyle F., Schlegel D., Scoccimarro
  R., Vogeley M.~S., Weinberg D.~H., Zehavi I., Berlind A., Budavari T.,
  Connolly A., Eisenstein D.~J., Finkbeiner D., Frieman J.~A., Gunn J.~E.,
  Hamilton A. J.~S., Hui L., Jain B., Johnston D., Kent S., Lin H., Nakajima
  R., Nichol R.~C., Ostriker J.~P., Pope A., Scranton R., Seljak U., Sheth
  R.~K., Stebbins A., Szalay A.~S., Szapudi I., Verde L., Xu Y., Annis J.,
  Bahcall N.~A., Brinkmann J., Burles S., Castander F.~J., Csabai I., Loveday
  J., Doi M., Fukugita M., Gott J.~R., Hennessy G., Hogg D.~W., Ivezi{\'c} {\v
  Z}., Knapp G.~R., Lamb D.~Q., Lee B.~C., Lupton R.~H., McKay T.~A., Kunszt
  P., Munn J.~A., O'Connell L., Peoples J., Pier J.~R., Richmond M., Rockosi
  C., Schneider D.~P., Stoughton C., Tucker D.~L., Berk D. E.~V., Yanny B.,
  York D.~G., 2004{\natexlab{a}}, ApJ, 606, 702

\bibitem[{Tegmark {et~al.}(2004{\natexlab{b}})Tegmark, Strauss, Blanton,
  Abazajian, Dodelson, Sandvik, Wang, Weinberg, Zehavi, Bahcall, Hoyle,
  Schlegel, Scoccimarro, Vogeley, Berlind, Budavari, Connolly, Eisenstein,
  Finkbeiner, Frieman, Gunn, Hui, Jain, Johnston, Kent, Lin, Nakajima, Nichol,
  Ostriker, Pope, Scranton, Seljak, Sheth, Stebbins, Szalay, Szapudi, Xu,
  Annis, Brinkmann, Burles, Castander, Csabai, Loveday, Doi, Fukugita,
  Gillespie, Hennessy, Hogg, Ivezi{\'c}, Knapp, Lamb, Lee, Lupton, McKay,
  Kunszt, Munn, O'connell, Peoples, Pier, Richmond, Rockosi, Schneider,
  Stoughton, Tucker, Berk, Yanny, \& York}]{Tegmark:2004p885}
Tegmark M., Strauss M., Blanton M., Abazajian K., Dodelson S., Sandvik H., Wang
  X., Weinberg D., Zehavi I., Bahcall N., Hoyle F., Schlegel D., Scoccimarro
  R., Vogeley M., Berlind A., Budavari T., Connolly A., Eisenstein D.,
  Finkbeiner D., Frieman J., Gunn J., Hui L., Jain B., Johnston D., Kent S.,
  Lin H., Nakajima R., Nichol R., Ostriker J., Pope A., Scranton R., Seljak U.,
  Sheth R., Stebbins A., Szalay A., Szapudi I., Xu Y., Annis J., Brinkmann J.,
  Burles S., Castander F., Csabai I., Loveday J., Doi M., Fukugita M.,
  Gillespie B., Hennessy G., Hogg D., Ivezi{\'c} {\v Z}., Knapp G., Lamb D.,
  Lee B., Lupton R., McKay T., Kunszt P., Munn J., O'connell L., Peoples J.,
  Pier J., Richmond M., Rockosi C., Schneider D., Stoughton C., Tucker D., Berk
  D.~V., Yanny B., York D., 2004{\natexlab{b}}, Physical Review D, 69, 103501

\bibitem[{Teyssier(2002)}]{Teyssier:2002p873}
Teyssier R., 2002, A{\&}A, 385, 337

\bibitem[{Tinker \& Conroy(2009)}]{Tinker:2009p904}
Tinker J.~L., Conroy C., 2009, ApJ, 691, 633

\bibitem[{{Tinsley}(1980)}]{1980FCPh....5..287T}
{Tinsley} B.~M., 1980, Fundamentals of Cosmic Physics, 5, 287

\bibitem[{Toomre \& Toomre(1972)}]{Toomre:1972p619}
Toomre A., Toomre J., 1972, ApJ, 178, 623

\bibitem[{{Tovmassian} {et~al.}(2006){Tovmassian}, {Plionis}, \&
  {Torres-Papaqui}}]{Tovmassian:2006p286}
{Tovmassian} H., {Plionis} M., {Torres-Papaqui} J.~P., 2006, A\&A, 456, 839

\bibitem[{{Tristram} \& {Ganga}(2007)}]{2007RPPh...70..899T}
{Tristram} M., {Ganga} K., 2007, Reports on Progress in Physics, 70, 899

\bibitem[{Tucker {et~al.}(2000)Tucker, Oemler, Hashimoto, Shectman, Kirshner,
  Lin, Landy, Schechter, \& Allam}]{Tucker:2000p453}
Tucker D.~L., Oemler A., Hashimoto Y., Shectman S.~A., Kirshner R.~P., Lin H.,
  Landy S.~D., Schechter P.~L., Allam S.~S., 2000, ApJS, 130, 237

\bibitem[{{Tully}(1988)}]{1988ngc..book.....T}
{Tully} R.~B., 1988, {Nearby galaxies catalog}. Cambridge and New York,
  Cambridge University Press, 1988, 221 p.

\bibitem[{Tumlinson(2010)}]{Tumlinson:2010p868}
Tumlinson J., 2010, ApJ, 708, 1398

\bibitem[{Turner \& Gott(1976)}]{Turner:1976p356}
Turner E.~L., Gott J.~R., 1976, ApJ, 32, 409

\bibitem[{{Vale} \& {Ostriker}(2004)}]{Vale:2004p26}
{Vale} A., {Ostriker} J.~P., 2004, MNRAS, 353, 189

\bibitem[{Valtonen \& Byrd(1986)}]{Valtonen:1986p340}
Valtonen M., Byrd G.~G., 1986, ApJ, 303, 523

\bibitem[{van Dokkum(2005)}]{vanDokkum:2005p854}
van Dokkum P.~G., 2005, AJ, 130, 2647

\bibitem[{van Dokkum(2008)}]{vanDokkum:2008p913}
---, 2008, ApJ, 674, 29

\bibitem[{Vecchia \& Schaye(2008)}]{DallaVecchia:2008p396}
Vecchia C.~D., Schaye J., 2008, MNRAS, 387, 1431

\bibitem[{Vennik {et~al.}(1993)Vennik, Richter, \& Longo}]{Vennik:1993p702}
Vennik J., Richter G.~M., Longo G., 1993, Astronomische Nachrichten, 314, 393

\bibitem[{Vikhlinin {et~al.}(1999)Vikhlinin, McNamara, Hornstrup, Quintana,
  Forman, Jones, \& Way}]{Vikhlinin:1999p696}
Vikhlinin A., McNamara B.~R., Hornstrup A., Quintana H., Forman W., Jones C.,
  Way M., 1999, ApJ, 520, L1

\bibitem[{Vogeley {et~al.}(1994)Vogeley, Geller, Park, \&
  Huchra}]{Vogeley:1994p901}
Vogeley M.~S., Geller M.~J., Park C., Huchra J.~P., 1994, AJ, 108, 745

\bibitem[{von Benda-Beckmann {et~al.}(2008)von Benda-Beckmann, D'Onghia,
  Gottloeber, Hoeft, Khalatyan, Klypin, \&
  M{\"u}ller}]{vonBendaBeckmann:2008p598}
von Benda-Beckmann A.~M., D'Onghia E., Gottloeber S., Hoeft M., Khalatyan A.,
  Klypin A.~A., M{\"u}ller V., 2008, MNRAS, 386, 2345

\bibitem[{von Benda-Beckmann \& M{\"u}ller(2008)}]{vonBendaBeckmann:2008p988}
von Benda-Beckmann A.~M., M{\"u}ller V., 2008, MNRAS, 384, 1189

\bibitem[{Walker {et~al.}(2010)Walker, Johnson, Gallagher, Hibbard,
  Hornschemeier, Tzanavaris, Charlton, \& Jarrett}]{Walker:2010p1074}
Walker L.~M., Johnson K.~E., Gallagher S.~C., Hibbard J.~E., Hornschemeier
  A.~E., Tzanavaris P., Charlton J.~C., Jarrett T.~H., 2010, AJ, 140, 1254

\bibitem[{Walsh {et~al.}(2007)Walsh, Jerjen, \& Willman}]{Walsh:2007p706}
Walsh S.~M., Jerjen H., Willman B., 2007, ApJ, 662, L83

\bibitem[{Wang {et~al.}(2008)Wang, Yang, Mo, van~den Bosch, Weinmann, \&
  Chu}]{Wang:2008p1005}
Wang Y., Yang X., Mo H.~J., van~den Bosch F.~C., Weinmann S.~M., Chu Y., 2008,
  ApJ, 687, 919

\bibitem[{{Weinberg}(2008)}]{WeinbergCosmology}
{Weinberg} S., 2008, Cosmology, by Steven Weinberg.~ISBN
  978-0-19-852682-7.~Published by Oxford University Press, Oxford, UK, 2008.

\bibitem[{Weymann(1966)}]{Weymann:1966p1019}
Weymann R., 1966, ApJ, 145, 560

\bibitem[{White \& Frenk(1991)}]{White:1991p851}
White S. D.~M., Frenk C.~S., 1991, ApJ, 379, 52

\bibitem[{White {et~al.}(1987)White, Frenk, Davis, \&
  Efstathiou}]{White:1987p906}
White S. D.~M., Frenk C.~S., Davis M., Efstathiou G., 1987, ApJ, 313, 505

\bibitem[{White \& Rees(1978)}]{White:1978p755}
White S. D.~M., Rees M.~J., 1978, MNRAS, 183, 341

\bibitem[{Whitmore {et~al.}(1993)Whitmore, Gilmore, \&
  Jones}]{Whitmore:1993p720}
Whitmore B.~C., Gilmore D.~M., Jones C., 1993, ApJ, 407, 489

\bibitem[{Willman {et~al.}(2005)Willman, Dalcanton, Martinez-Delgado, West,
  Blanton, Hogg, Barentine, Brewington, Harvanek, Kleinman, Krzesinski, Long,
  Neilsen, Nitta, \& Snedden}]{Willman:2005p1072}
Willman B., Dalcanton J.~J., Martinez-Delgado D., West A.~A., Blanton M.~R.,
  Hogg D.~W., Barentine J.~C., Brewington H.~J., Harvanek M., Kleinman S.~J.,
  Krzesinski J., Long D., Neilsen E.~H., Nitta A., Snedden S.~A., 2005, ApJ,
  626, L85

\bibitem[{Xu(1995)}]{Xu:1995p1015}
Xu G., 1995, ApJS, 98, 355

\bibitem[{{Yan} {et~al.}(2010){Yan}, {Windhorst}, {Hathi}, {Cohen}, {Ryan},
  {O'Connell}, \& {McCarthy}}]{Yan:2009p782}
{Yan} H.-J., {Windhorst} R.~A., {Hathi} N.~P., {Cohen} S.~H., {Ryan} R.~E.,
  {O'Connell} R.~W., {McCarthy} P.~J., 2010, Research in Astronomy and
  Astrophysics, 10, 867

\bibitem[{Yang {et~al.}(2007)Yang, Mo, van~den Bosch, Pasquali, Li, \&
  Barden}]{Yang:2007p506}
Yang X., Mo H.~J., van~den Bosch F.~C., Pasquali A., Li C., Barden M., 2007,
  ApJ, 671, 153

\bibitem[{York {et~al.}(2000)York, Adelman, Anderson, Anderson, Annis, Bahcall,
  Bakken, Barkhouser, Bastian, Berman, Boroski, Bracker, Briegel, Briggs,
  Brinkmann, Brunner, Burles, Carey, Carr, Castander, Chen, Colestock,
  Connolly, Crocker, Csabai, Czarapata, Davis, Doi, Dombeck, Eisenstein,
  Ellman, Elms, Evans, Fan, Federwitz, Fiscelli, Friedman, Frieman, Fukugita,
  Gillespie, Gunn, Gurbani, de~Haas, Haldeman, Harris, Hayes, Heckman,
  Hennessy, Hindsley, Holm, Holmgren, hao Huang, Hull, Husby, Ichikawa,
  Ichikawa, Ivezi{\'c}, Kent, Kim, Kinney, Klaene, Kleinman, Kleinman, Knapp,
  Korienek, Kron, Kunszt, Lamb, Lee, Leger, Limmongkol, Lindenmeyer, Long,
  Loomis, Loveday, Lucinio, Lupton, MacKinnon, Mannery, Mantsch, Margon,
  McGehee, McKay, Meiksin, Merelli, Monet, Munn, Narayanan, Nash, Neilsen,
  Neswold, Newberg, Nichol, Nicinski, Nonino, Okada, Okamura, Ostriker, Owen,
  Pauls, Peoples, Peterson, Petravick, Pier, Pope, Pordes, Prosapio,
  Rechenmacher, Quinn, Richards, Richmond, Rivetta, Rockosi, Ruthmansdorfer,
  Sandford, Schlegel, Schneider, Sekiguchi, Sergey, Shimasaku, Siegmund, Smee,
  Smith, Snedden, Stone, Stoughton, Strauss, Stubbs, SubbaRao, Szalay, Szapudi,
  Szokoly, Thakar, Tremonti, Tucker, Uomoto, Berk, Vogeley, Waddell, i~Wang,
  Watanabe, Weinberg, Yanny, \& Yasuda}]{York:2000p799}
York D.~G., Adelman J., Anderson J.~E., Anderson S.~F., Annis J., Bahcall
  N.~A., Bakken J.~A., Barkhouser R., Bastian S., Berman E., Boroski W.~N.,
  Bracker S., Briegel C., Briggs J.~W., Brinkmann J., Brunner R., Burles S.,
  Carey L., Carr M.~A., Castander F.~J., Chen B., Colestock P.~L., Connolly
  A.~J., Crocker J.~H., Csabai I., Czarapata P.~C., Davis J.~E., Doi M.,
  Dombeck T., Eisenstein D., Ellman N., Elms B.~R., Evans M.~L., Fan X.,
  Federwitz G.~R., Fiscelli L., Friedman S., Frieman J.~A., Fukugita M.,
  Gillespie B., Gunn J.~E., Gurbani V.~K., de~Haas E., Haldeman M., Harris
  F.~H., Hayes J., Heckman T.~M., Hennessy G.~S., Hindsley R.~B., Holm S.,
  Holmgren D.~J., hao Huang C., Hull C., Husby D., Ichikawa S.-I., Ichikawa T.,
  Ivezi{\'c} {\v Z}., Kent S., Kim R. S.~J., Kinney E., Klaene M., Kleinman
  A.~N., Kleinman S., Knapp G.~R., Korienek J., Kron R.~G., Kunszt P.~Z., Lamb
  D.~Q., Lee B., Leger R.~F., Limmongkol S., Lindenmeyer C., Long D.~C., Loomis
  C., Loveday J., Lucinio R., Lupton R.~H., MacKinnon B., Mannery E.~J.,
  Mantsch P.~M., Margon B., McGehee P., McKay T.~A., Meiksin A., Merelli A.,
  Monet D.~G., Munn J.~A., Narayanan V.~K., Nash T., Neilsen E., Neswold R.,
  Newberg H.~J., Nichol R.~C., Nicinski T., Nonino M., Okada N., Okamura S.,
  Ostriker J.~P., Owen R., Pauls A.~G., Peoples J., Peterson R.~L., Petravick
  D., Pier J.~R., Pope A., Pordes R., Prosapio A., Rechenmacher R., Quinn
  T.~R., Richards G.~T., Richmond M.~W., Rivetta C.~H., Rockosi C.~M.,
  Ruthmansdorfer K., Sandford D., Schlegel D.~J., Schneider D.~P., Sekiguchi
  M., Sergey G., Shimasaku K., Siegmund W.~A., Smee S., Smith J.~A., Snedden
  S., Stone R., Stoughton C., Strauss M.~A., Stubbs C., SubbaRao M., Szalay
  A.~S., Szapudi I., Szokoly G.~P., Thakar A.~R., Tremonti C., Tucker D.~L.,
  Uomoto A., Berk D.~V., Vogeley M.~S., Waddell P., i~Wang S., Watanabe M.,
  Weinberg D.~H., Yanny B., Yasuda N., 2000, AJ, 120, 1579

\bibitem[{Zabludoff \& Mulchaey(1998)}]{Zabludoff:1998p578}
Zabludoff A.~I., Mulchaey J.~S., 1998, ApJ, 496, 39

\bibitem[{Zabludoff {et~al.}(1996)Zabludoff, Zaritsky, Lin, Tucker, Hashimoto,
  Shectman, Oemler, \& Kirshner}]{Zabludoff:1996p722}
Zabludoff A.~I., Zaritsky D., Lin H., Tucker D., Hashimoto Y., Shectman S.~A.,
  Oemler A., Kirshner R.~P., 1996, ApJ, 466, 104

\bibitem[{Zaritsky(1992)}]{Zaritsky:1992p516}
Zaritsky D., 1992, ApJ, 400, 74

\bibitem[{Zehavi {et~al.}(2004)Zehavi, Weinberg, Zheng, Berlind, Frieman,
  Scoccimarro, Sheth, Blanton, Tegmark, Mo, Bahcall, Brinkmann, Burles, Csabai,
  Fukugita, Gunn, Lamb, Loveday, Lupton, Meiksin, Munn, Nichol, Schlegel,
  Schneider, SubbaRao, Szalay, Uomoto, \& York}]{Zehavi:2004p805}
Zehavi I., Weinberg D.~H., Zheng Z., Berlind A.~A., Frieman J.~A., Scoccimarro
  R., Sheth R.~K., Blanton M.~R., Tegmark M., Mo H.~J., Bahcall N.~A.,
  Brinkmann J., Burles S., Csabai I., Fukugita M., Gunn J.~E., Lamb D.~Q.,
  Loveday J., Lupton R.~H., Meiksin A., Munn J.~A., Nichol R.~C., Schlegel D.,
  Schneider D.~P., SubbaRao M., Szalay A.~S., Uomoto A., York D.~G., 2004, ApJ,
  608, 16

\bibitem[{Zehavi {et~al.}(2005)Zehavi, Zheng, Weinberg, Frieman, Berlind,
  Blanton, Scoccimarro, Sheth, Strauss, Kayo, Suto, Fukugita, Nakamura,
  Bahcall, Brinkmann, Gunn, Hennessy, Ivezi{\'c}, Knapp, Loveday, Meiksin,
  Schlegel, Schneider, Szapudi, Tegmark, Vogeley, \& York}]{Zehavi:2005p1007}
Zehavi I., Zheng Z., Weinberg D.~H., Frieman J.~A., Berlind A.~A., Blanton
  M.~R., Scoccimarro R., Sheth R.~K., Strauss M.~A., Kayo I., Suto Y., Fukugita
  M., Nakamura O., Bahcall N.~A., Brinkmann J., Gunn J.~E., Hennessy G.~S.,
  Ivezi{\'c} {\v Z}., Knapp G.~R., Loveday J., Meiksin A., Schlegel D.~J.,
  Schneider D.~P., Szapudi I., Tegmark M., Vogeley M.~S., York D.~G., 2005,
  ApJ, 630, 1

\bibitem[{Zeldovich {et~al.}(1982)Zeldovich, Einasto, \&
  Shandarin}]{Zeldovich:1982p910}
Zeldovich I.~B., Einasto J., Shandarin S.~F., 1982, Nature, 300, 407

\bibitem[{Zel'Dovich(1970)}]{ZelDovich:1970p999}
Zel'Dovich Y.~B., 1970, A\&A, 5, 84

\bibitem[{{Zeldovich}(1972)}]{1972MNRAS.160P...1Z}
{Zeldovich} Y.~B., 1972, MNRAS, 160, 1P

\bibitem[{Zepf \& Whitmore(1993)}]{Zepf:1993p704}
Zepf S.~E., Whitmore B.~C., 1993, ApJ, 418, 72

\bibitem[{{Zwicky}(1933)}]{1933AcHPh...6..110Z}
{Zwicky} F., 1933, Helvetica Physica Acta, 6, 110

\end{thebibliography}
